\documentclass[aps,prd,reprint,longbibliography,nofootinbib,superscriptaddress]{revtex4-2}

\usepackage{lmodern}
\usepackage[T1]{fontenc}
\usepackage{amsmath}
\usepackage{amssymb}

\usepackage{url}
\usepackage{multirow}
\usepackage{caption}
\usepackage[labelformat=simple]{subcaption}

\usepackage[nodayofweek]{datetime}
\usepackage{enumerate}
\usepackage{xspace}
\usepackage{xcolor}
\usepackage{graphicx}
\usepackage[pdftex,hidelinks]{hyperref}
\usepackage{bookmark}

\usepackage{float}
\usepackage[capitalise]{cleveref}

\usepackage{xcolor}

\newcommand{\ctrue}{\ensuremath{\{\vec c_{true}\}}\xspace}
\newcommand{\cpred}{\ensuremath{\{\vec c_{pred}\}}\xspace}
\newcommand{\cobs}{\ensuremath{\{\vec c_{obs}\}}\xspace}
\newcommand{\jobs}{\ensuremath{j_{obs}}\xspace}

\newcommand{\jtrue}{\ensuremath{j_{true}}\xspace}
\newcommand{\zcut}{\ensuremath{z_\mathrm{cut}}\xspace}

\newcommand{\pt}{\ensuremath{{p}_\mathrm{T}}\xspace}

\newcommand{\Ntrue}{\ensuremath{N_\mathrm{true}}\xspace}

\newcommand{\softkiller}{\textsc{SoftKiller}\xspace}
\newcommand{\softdrop}{\textsc{SoftDrop}\xspace}
\newcommand{\puppi}{\textsc{Puppi}\xspace}
\newcommand{\puppiml}{\textsc{PuppiML}\xspace}
\newcommand{\antikt}{anti-$k_t$\xspace}

\newcommand*\normal[2]{\ensuremath{\mathcal{N}(\mu = {#1}, \sigma = {#2})}\xspace}

\newcommand{\ttbar}{\ensuremath{t\bar{t}}\xspace}
\newcommand{\pythia}{\textsc{Pythia}\xspace}

\newcommand{\vipr}{\textsc{Vipr}\xspace}
\newcommand{\avgmu}{\ensuremath{\langle \mu \rangle}\xspace}

\usepackage{todonotes}
\usepackage{graphicx}
\usepackage{subcaption}
\usepackage{pgffor}
\usepackage{placeins}

\makeatletter\def\frontmatter@affiliationfont{\it\footnotesize}\makeatother

\begin{document}

\title{Variational inference for pile-up removal at hadron colliders with diffusion models}
\author{Malte Algren}
\email{malte.algren@unige.ch}
    \affiliation{Département de physique nucléaire et corpusculaire, University of Geneva, Geneva 1211 Switzerland}
    \author{Christopher Pollard}
    \email{christopher.pollard@warwick.ac.uk}
    \affiliation{Department of physics, University of Warwick, Coventry CV4 7AL, United Kingdom}
    \author{John Andrew Raine}
    \email{john.raine@unige.ch}
    \affiliation{Département de physique nucléaire et corpusculaire, University of Geneva, Geneva 1211 Switzerland}
    \author{Tobias Golling}
    \affiliation{Département de physique nucléaire et corpusculaire, University of Geneva, Geneva 1211 Switzerland}
    \email{tobias.golling@unige.ch}

    \begin{abstract}
        In this paper, we present a novel method for pile-up removal of $pp$
        interactions using variational inference with diffusion models, called
        \vipr.
        Instead of using classification methods to identify which particles are
        from the primary collision, a generative model is trained to predict
        the constituents of the hard-scatter particle jets with pile-up removed.
        This results in an estimate of the full posterior over hard-scatter jet
        constituents, which has not yet been explored in the context of pile-up
        removal, yielding a clear advantage over existing methods especially in
        the presence of imperfect detector efficiency.
        We evaluate the performance of \vipr in a sample of jets from simulated
        \ttbar events overlain with pile-up contamination.
        \vipr outperforms \softdrop and has comparable performance to \puppiml
        in predicting the substructure of the hard-scatter jets over a wide
        range of pile-up scenarios.
    \end{abstract}

    \maketitle

    \section{Introduction}
    
Hadron colliders, such as the Large Hadron Collider~\cite{LHC}~(LHC) at
CERN, deliver unrivalled centre of mass energies for partonic collisions (at
least by contemporary standards) and therefore provide access to processes at previously
unexplored kinematic extremes.
The intensity of proton bunches at the LHC yields many $pp$ collisions per beam
crossing~\cite{ATLAS:2022hro,CMS:2023pad}; this is necessary to produce 
rare and energetic processes at an appreciable rate.
Most $pp$ interactions occur at low center-of-mass
energies~\cite{ATLAS:2016ygv,CMS:2018mlc} -- well below those required to
produce $W$ and $Z$ bosons, the Higgs boson, and the top
quark~\cite{ParticleDataGroup:2024cfk} -- and recording all such collisions is
not feasible with current detector readout systems and computing
infrastructure~\cite{ATLAS:2024xna,CMS:2016ngn}.

When a collision of interest (the ``hard-scatter'' interaction), e.g.\ one
involving a final-state top quark, is produced in a bunch crossing, additional
$pp$ interactions (known as ``pile-up'' interactions) generally produce
relatively low-momentum hadrons and photons, resulting in extra energy
depositions in the detector unrelated to the physics of the hard-scatter
process.
Currently, the ATLAS~\cite{ATLAS} and CMS~\cite{CMS} experiments record
collisions at a maximum instantaneous number of interactions per crossing
($\mu$) of approximately $\mu=80$, with a maximum $\mu$ averaged over several  
minutes of about $\avgmu = 60$~\cite{ATL-DAPR-PUB-2024-001}.
However, as the LHC enters the high luminosity phase, \avgmu is expected to
increase to between 130 and
200~\cite{ZurbanoFernandez:2020cco}.

Having a large number of pile-up interactions coincident with a
collision of interest yields a challenging environment for certain signatures.
For example, the reconstruction of charged particle tracks and primary
interaction vertices degrades with the density and multiplicity of pile-up
interactions~\cite{ATLAS:2016nnj,ATLAS-CONF-2012-042}.
Particles from pile-up interactions also contaminate hadronic jets (``jets'')
originating from the hard-scatter interaction, and the mitigation of this
contamination is critical for achieving acceptable resolution of jet
observables relevant for measurements of Standard Model (SM) processes as well
as searches for beyond-SM physics.\footnote{See Ref.~\cite{Salam:2010nqg} for an
overview of jet clustering algorithms used in present-day collider experiments
and their theoretical motivation.}

Several techniques have been developed to reduce the impact of pile-up particles
on jet observables.
The \softkiller and \puppi algorithms attempt to identify and remove individual
pile-up particles before jet reconstruction
begins~\cite{softkiller,puppi},
while other strategies involve removing jet constituents from pile-up
interactions only after clustering has taken
place~\cite{softdrop,Soyez:2012hv,Krohn:2013lba,ATLAS:2017pfq,CMS:2020ebo}.
Methods have also been developed to identify and remove entire jets that are
determined to originate primarily from
pile-up~\cite{PERF-2014-03,ATLAS:2017ywy}.
Some of these contemporary strategies, such as \emph{constituent
subtraction}~\cite{Berta:2014eza}, may not fully remove particle candidates but
rather alter their energies or momenta to compensate for pile-up contamination.
In recent years, machine learning-based approaches have been deployed and
studied in a variety of these
contexts~\cite{pumml,puppiml,Carrazza:2019efs,Maier:2021ymx,theonemaltefound,Kim:2023koz}.

Hadronic jets come in many shapes and sizes, and several approaches to jet
reconstruction are currently in use by LHC experiments.
The ATLAS and CMS collaborations primarily utilize the \antikt jet clustering
algorithm with radius parameter $R = 0.4$ to collect hadrons from the
fragmentation of high-momentum gluons and quarks with low masses relative to the
$pp$ collision energy ($u, d, c, s, b$)~\cite{ParticleDataGroup:2024cfk}.
Jet algorithms with larger radius parameters have proven successful for the
reconstruction and identification of hadronic decays of particles whose momenta
transverse to the proton beam (\pt) are on the order of the decaying
particle's mass ($m$): $m / \pt \sim 1$.  
This approach allows decay products to be treated coherently as part of a single
object (sometimes referred to as a ``boosted'' reconstruction strategy) rather
than using several smaller-radius jets as proxies for a subset of partonic
decay products (a ``resolved'' reconstruction strategy).
The boosted reconstruction strategy has been shown to achieve strong rejection
of background processes at a given signal efficiency, taking advantage 
of the discriminating power of the internal structure (``substructure'') of the
large-radius jet.
This approach has been studied in depth for high-momentum vector bosons,
Higgs bosons, and top quarks in both phenomenological and experimental
settings~\cite{Seymour:1993mx,Butterworth:2008iy,Kaplan:2008ie,ATLAS:2012muc,CMS:2024ddc}.

The effective area of \antikt jets in the detector, and therefore the amount of
pile-up contamination, grows with the jet algorithm radius
parameter~\cite{Salam:2010nqg}; the presence of pile-up can substantially alter 
measurements of large-radius jets in particular.
Observables that depend strongly on the angular separation of jet
constituents relative to the jet axis, such as the jet
mass~\cite{Larkoski:2013eya}, are sensitive to pile-up since pile-up particles
tend to be more evenly spread throughout the jet area than those from a
resonance decay.
Dedicated algorithms have been developed to mitigate this pile-up contribution,
and many of these techniques also work to remove contributions from the
underlying event.
Section 4.2 of Ref~\cite{ATLAS:2020gwe} in particular provides a useful resource
for a comparison across several of them.

In this work we introduce a new method, called variational inference for pile-up
removal (\vipr), which exploits diffusion
models~\cite{Song2020,yang2023diffusion,EDM} in order to infer the true
constituents of a jet originating from the hard-scatter interaction based on an
observed jet that has been contaminated by pile-up.
Diffusion models are well suited to generate unordered set of
constituents which comprise jets, and have already been demonstrated to provide
state-of-the-art performance in high energy
physics~\cite{CaloScore,CaloClouds,Acosta:2023zik,pcjedi,FPCD,Butter:2023fov,
Shmakov:2023kjj,Mikuni:2023tok,pcdroid,epicjedi,Heimel:2023ngj,Buhmann:2023acn,
drapes,nuflows,v2flows,Shmakov:2023kjj,Heimel:2023ngj}.
Rather than produce a single estimate of jet observables, we approximate the
full posterior distribution over jet constituents, from which a wide array of
observables may be built; this posterior includes variations over jet 
constituent multiplicities.
\vipr generates samples of pile-up-free hard-scatter jets from this posterior,
complete with individual constituent information, that are consistent with an
\emph{observed jet} containing a combination of hard-scatter and pile-up
contributions.
A population of these samples faithfully represents the relative probabilities
of pure hard-scatter jets that may have produced the observed jet.

To illustrate the method and to quantify its performance for a concrete process
of interest at the LHC, we overlay particles from pile-up collisions onto
large-radius jets in \ttbar events from $pp$ collisions at 14~TeV.
Searches and measurements of high-$Q^2$ \ttbar production at the LHC often
involve a high-purity sample of jets initiated by top-quark decays, with a signal
fraction often well above 80\%~\cite{CMS:2018rkg,ATLAS:2018rvc}.
As such we focus here on algorithm's performance in a population made up
exclusively of top-quark jets, but we note that \vipr can be used for another
topology by simply changing the training sample appropriately.

We use \vipr to approximate the posterior over particle-level constituents of
large-radius jets from high-\pt top-quark decays, given an observation composed
of jets contaminated by the pile-up overlay.
We benchmark our method against the standard \softdrop grooming
algorithm~\cite{softdrop} and a version of the \puppiml algorithm~\cite{puppiml}
using the transformer architecture~\cite{attn,preLN} instead of a graph neural
network~\cite{li2017gatedgraphsequenceneural}.
We show that \vipr yields an unbiased estimate of the hard-scatter jet mass,
\pt, and substructure observables over a wide range of pile-up conditions for
$50 \leq \mu \leq 300$, whereas \softdrop and \puppiml do not.
We also find that the resolution of the \vipr posterior tends to be comparable
to that of \puppiml, but both are considerably narrower than that of \softdrop
across all evaluated $\mu$ values.
In the presence of imperfect detector reconstruction efficiency, we find that
an unavoidable bias in these quantities is introduced by \puppiml, while this
bias is automatically corrected for by \vipr.
Additionally, we find the coverage of \vipr in these quantities
to be a conservative estimate of ground truth jet observables.
         
    \section{Simulated dataset}
    In this work we study the impact of pile-up and performance of pile-up removal
techniques on large radius jets and their substructure using large radius jets
initiated by boosted top quarks.
Samples of simulated \ttbar events in which both top quarks decay hadronically
are generated using MadGraph5 aMC@NLO~\cite{Alwall:2014hca}~(v3.1.0).
The decays of the top quarks and $W$ bosons are performed using
MadSpin~\cite{MadSpin}, with the partonic top quarks required to have
\mbox{\pt > 450~GeV}.
The parton shower and hadronisation is subsequently performed with
\pythia~\cite{Sjostrand:2014zea}~(v8.243) using the \mbox{NNPDF2.3LO} PDF
set~\cite{Ball:2012cx} with $\alpha_S(m_Z) = 0.130$ using the
LHAPDF~\cite{Buckley:2014ana} framework\footnote{The dataset can be found in
Refs.~\cite{Zoch:2024eyp, zoch2024rodemjetdatasets}.}.

To avoid costly resimulation and reclustering of the jets for different pile-up
conditions, interactions with the detector are not performed. This is in line
with many previous
studies~\cite{softkiller,pumml,theonemaltefound}, 
whereas others~\cite{puppi, puppiml} use simple detector simulations that are fast to run. 
Instead we consider all visible final state particles after the parton shower
arising from the hard scatter event and pile-up collisions.

The particles in the \ttbar events are clustered into jets using the anti-$k_t$
algorithm~\cite{Cacciari:2008gp} with $R=1.0$ using the
\texttt{FastJet}~\cite{Cacciari:2011ma,deFavereau:2013fsa} implementation, and each of the two
resulting jets (``top jets'') are considered independently.
A minimum \pt requirement of 500~MeV is applied to the jet constituents before
clustering.
These top jets without pile-up contamination are considered the ground truth for
all studies.

To simulate various scenarios for the number of interactions in a crossing of
LHC bunches, $\mu$, inclusive inelastic proton-proton collisions (IICs)
produced with \pythia are overlain on the simulated \ttbar events.
A collection of 500,000 IICs was produced, and to simulate $\mu$ interactions
in a bunch crossing, $\mu - 1$ IICs are drawn from this collection.
Each IIC is rotated by a random uniform azimuthal angle $\phi$ and randomly
mirrored across the $xy$ plane (i.e. the $z$-coordinate is transformed by
$z \to -z$) in order to improve the statistical power of the IIC sample. 

The resulting final-state particles which fall within the radius of the top jet
are appended to the jet constituents; only particles with $E > 500$~MeV are
included.
This process is performed independently for each jet and value of $\mu$, and
jets after this pile-up overlay are called ``observed jets''.

In total there are about 1~million simulated top jets and 500,000 IICs. The top
jets are split into train, test and validation samples, making up 70\%, 19.5\%
and 10.5\% of the total sample, respectively.

    \section{Method}
    \vipr follows the EDM diffusion scheme described in Ref.~\cite{EDM}.
We aim to derive a generative model to approximate
\begin{equation}
    p(\ctrue | S , \cobs , \mu)
\end{equation}
where the notation $\{ \cdot \}$ indicates a homogeneous unordered set,
\ctrue are the desired inferred observables for hard-scatter constituents of the
jet, $S$ represents observed summary quantities of the jet in question, \cobs
are the observed quantities for each jet constituent, and $\mu$ is the number of
pile-up interactions in a bunch-crossing.
This approach differs from deterministic machine learning-based
approaches~\cite{pumml,puppiml,Carrazza:2019efs,Maier:2021ymx,theonemaltefound,Kim:2023koz}
that formulate the problem as a classification task for removing individual
constituents directly from the observed jet.
In this study we focus on the case where \ctrue and \cobs are simply constituent
three-momenta and leave other particle quantities such as charge, species, etc.\ 
for future work.

As both \cobs and \ctrue are unordered sets, the method should be permutation
equivariant.
To achieve this, \vipr employs the transformer architecture~\cite{attn,preLN},
using both the transformer encoders and decoders to extract information from the
\cobs and infer \ctrue.  
For learning $p(\ctrue|\Ntrue, S, \cobs, \mu)$, where \Ntrue represents the
number of ground-truth jet constituents, a conditional diffusion
scheme~\cite{yang2023diffusion,Song2020,EDM,pcjedi} is used.
To obtain the full density $p(\ctrue | S, \cobs, \mu)$, a conditional
normalizing flow is trained to approximate $p(\Ntrue | S, \cobs, \mu)$.

Diffusion models are trained by adding known gaussian noise with a strength
$\sigma$ to corrupt the input, and then the network $F_\theta$ is trained to 
remove the added gaussian noise given $\sigma$ and the noisy data.
We follow the training and sampling scheme described in Ref.~\cite{EDM}.
The number of iterative denoising steps is chosen beforehand, and the number of
steps defines the linearly spaced values of $\sigma$.
Using the same observed properties, multiple samples can be generated from noise
to obtain the posterior distribution. 

During training, the maximum number of \ctrue and \cobs are set to 175 and 400
respectively.
Summary variables for the entire observed jet (\jobs) are used as conditions $S$ to the
network; these are the observed jet pseudorapidity $\eta$, azimuthal angle
$\phi$, momentum transverse to the proton beams $p_T$, and the mass $m$ of the
observed jet.

\begin{figure}[htbp]
    \includegraphics[width=0.49\textwidth]{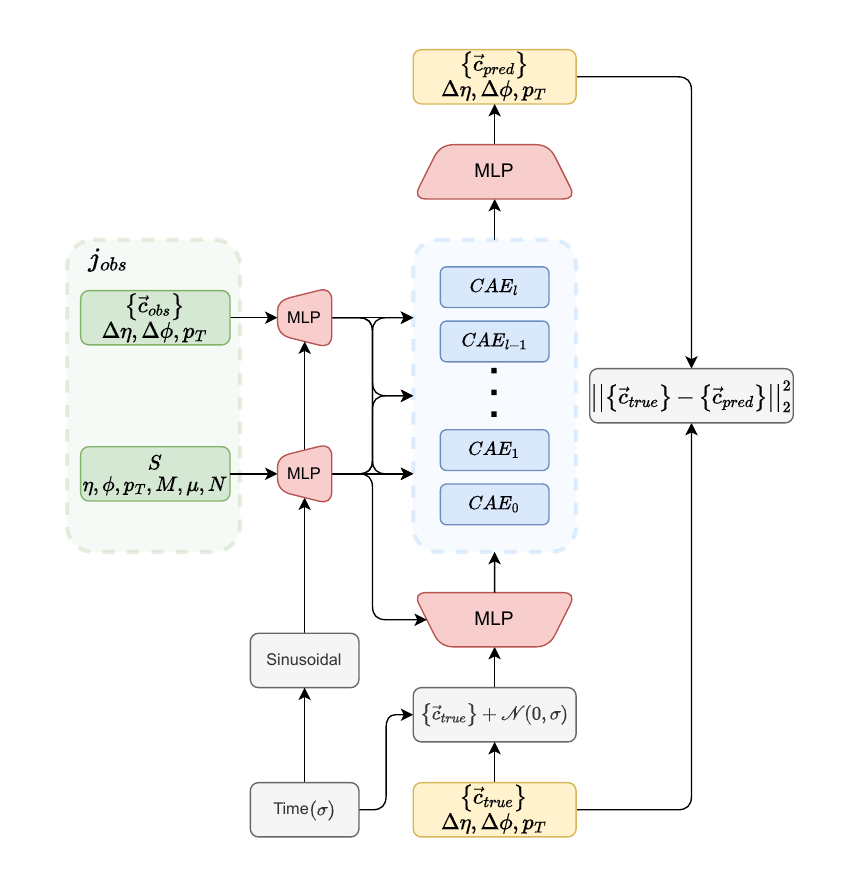}
    \caption{
    The training scheme used in \vipr with \cobs in the green box,
    \ctrue and \cpred in yellow boxes.
    Cross attention encoders are shown in blue with standard MLPs in red. Both with trainable parameters.
    The gray boxes indicate transformations that are non-trainable.
    $\left(\cobs+\mathcal{N}(0, \sigma)\right)$ follows from Ref.~\cite{EDM} and is only relevant during the training process.
    }
    \label{fig:arch_diff}
\end{figure}

\subsubsection*{Architecture}

The full diffusion architecture for \vipr is depicted in \cref{fig:arch_diff},
where multiple cross attention encoder (CAE) blocks are stacked together to
improve the fidelity of the transformation and to facilitate conditional message
passing from \cobs to \ctrue.
The Mean Squared Error (MSE) is used as the loss function, which minimises the
difference between \ctrue and \cpred.
All features are projected into a higher-dimensional space using multi-layer
perceptrons (MLPs).
The embedded features are then passed to the CAE blocks.

\begin{figure}[htbp]
    \centering
    \includegraphics[width=0.45\textwidth]{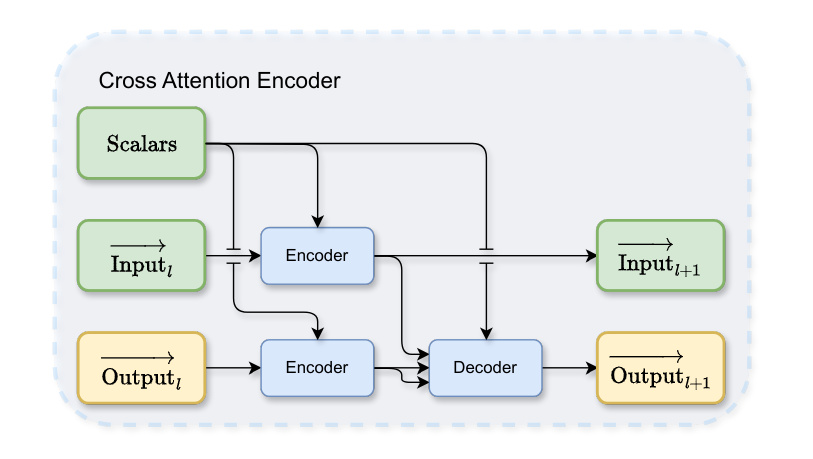}
    \caption{
    The cross attention encoder (CAE) used iteratively within \vipr
    (\cref{fig:arch_diff}).
    It consists of two transformer encoders and a single transformer decoder
    following Ref~\cite{preLN}.
    Scalar variables are concatenated into the MLPs within each transformer
    layer.
    The color of the boxes indicates the type of information and follows the
    coloring from  \cref{fig:arch_diff}} 
    \label{fig:CAE}
\end{figure}

A CAE block is shown in \cref{fig:CAE} and comprises two transformer encoders
used to derive corrections within each point-cloud and a transformer decoder
used for message passing between the two point-clouds. 
Both encoder and decoders follow the architecture described in Ref~\cite{preLN}.
Additional scalar information about the jet properties are concatenated
internally in the MLP within the transformer layers. 

The \puppiml algorithm in benchmarks that follow consists of only
transformer encoders~\cite{attn,preLN} trained using the Binary Cross
Entropy loss~\cite{068cb646-8936-3df0-aa16-e97573eda53c} to classify the
constituents of the jet as either originating from pile-up or hard-scatter
interactions.
This differs from the original \puppiml algorithm~\cite{puppiml} which utilized
Graph Neural Networks (GNN) to classify the constituents\cite{li2017gatedgraphsequenceneural}.

    \section{Results} \label{sec:results}
    
\begin{figure*}[ht]
    \centering
    \subfloat[Observed jet]{{\includegraphics[width=0.24\textwidth]{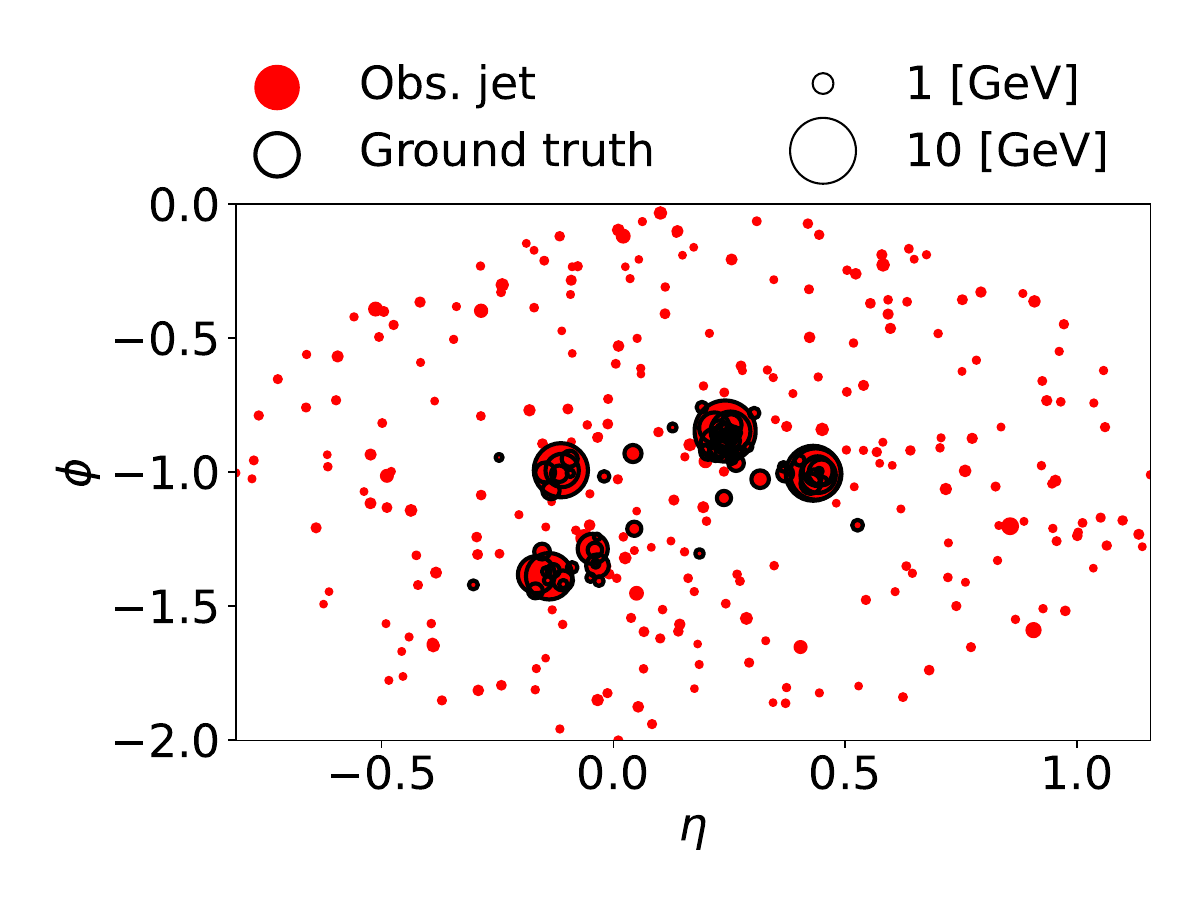}}}
    \subfloat[\softdrop jet]{{\includegraphics[width=0.24\textwidth]{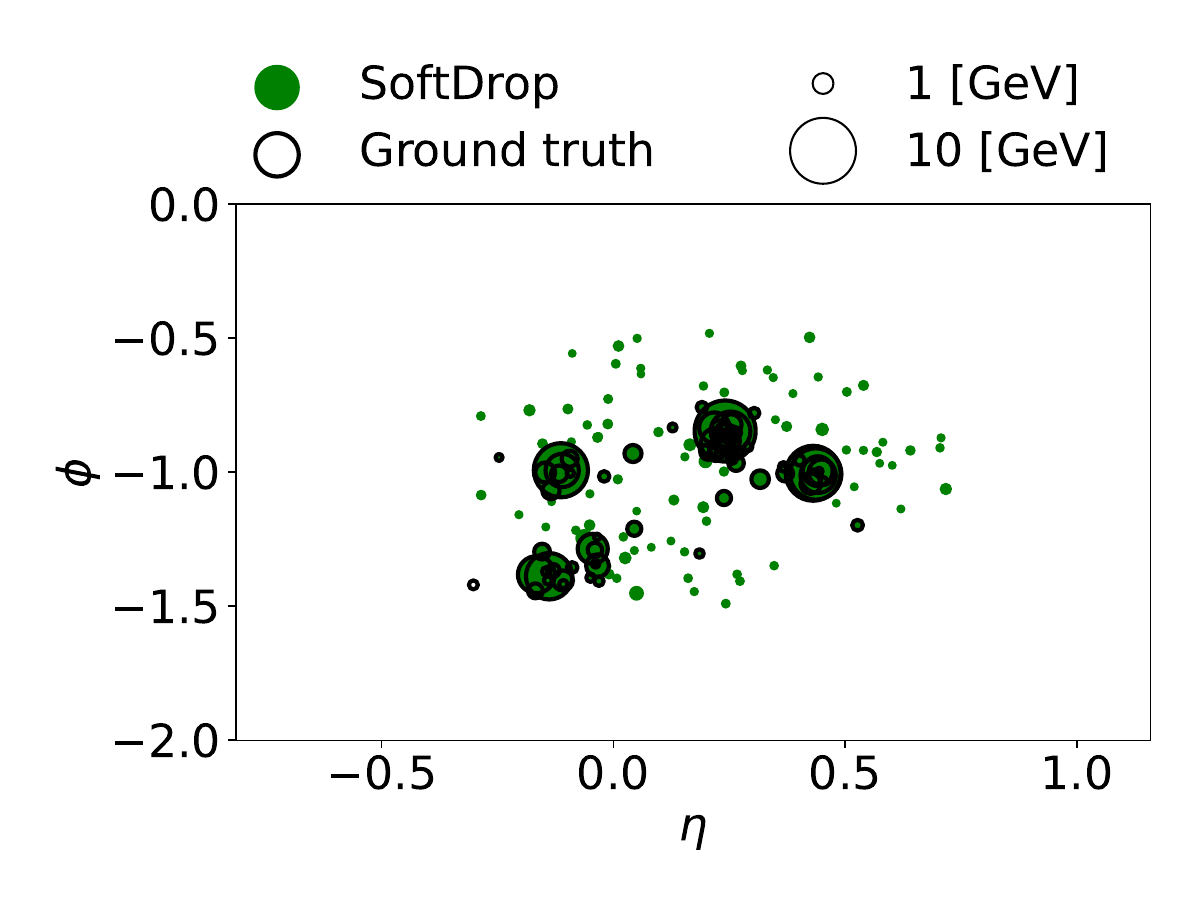}}}
    \subfloat[\puppiml jet]{{\includegraphics[width=0.24\textwidth]{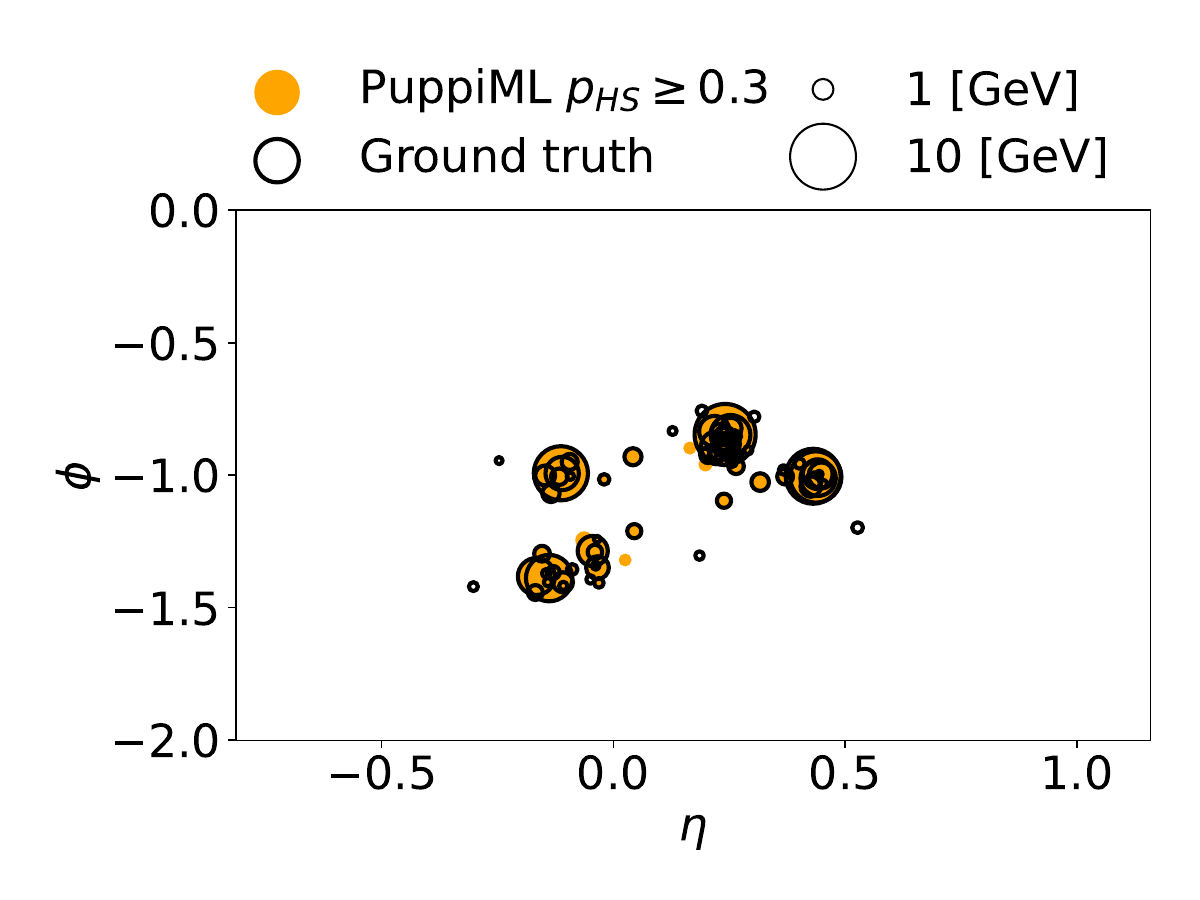}}}
    \subfloat[\vipr jet]{{\includegraphics[width=0.24\textwidth]{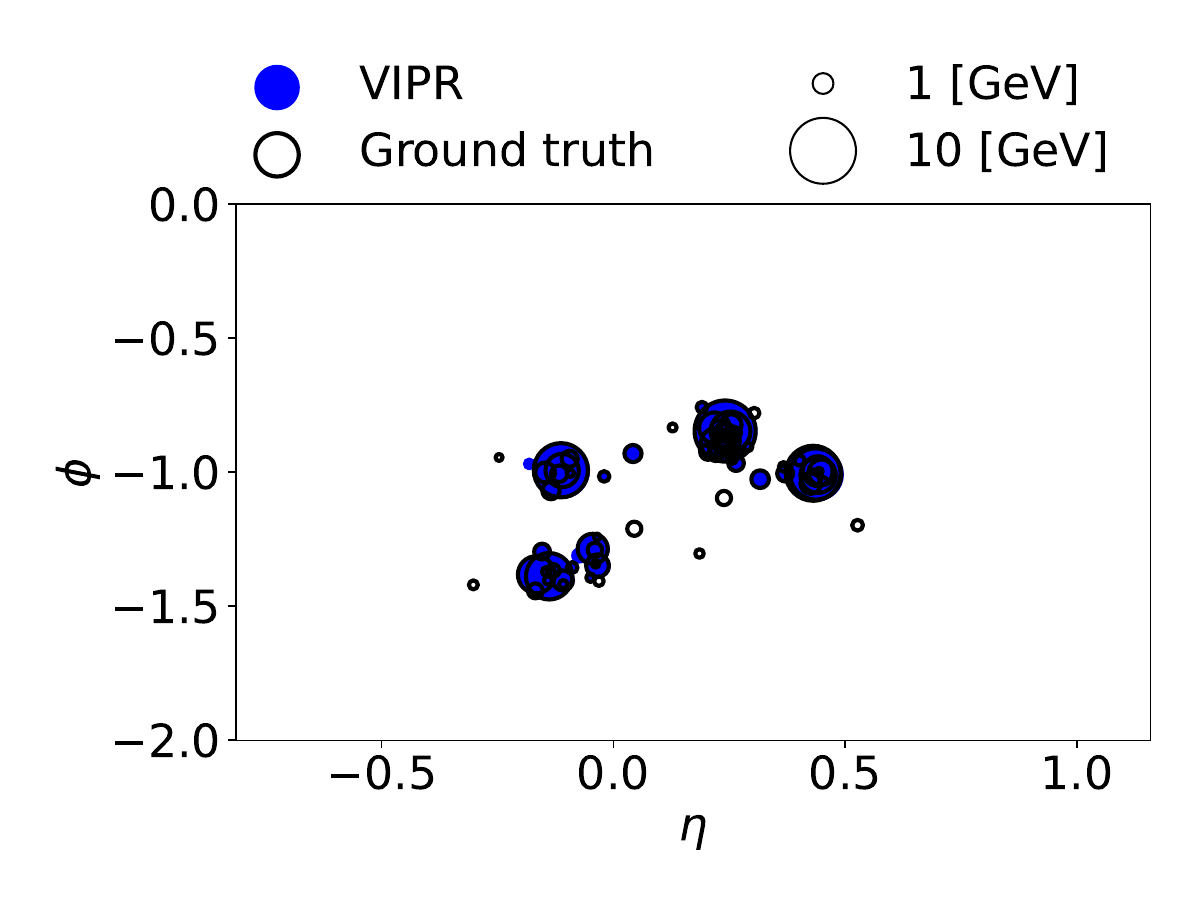}}}
    \caption{
    Constituent locations in the $\eta \times \phi$ plane for a jet at
    $\mu = 200$: the (a) observed \cobs, (b) \softdrop, (c) \puppiml, and (d)
    \vipr predictions (filled circles) are compared to the ground truth
    (unfilled circles).
    Constituent \pt{}s are indicated by circle areas.
    The \vipr jet shown is a single sample taken from the predicted
    $p(\ctrue | S, \cobs, \mu)$.
    } 
    \label{fig:denoised_scatter}
\end{figure*}

To assess the performance of \vipr , we inspect its ability to obtain the
correct jet \pt, invariant mass, and several jet substructure observables that
are known to be useful for top-jet identification at a hadron collider: the
ratio of the jet 2-subjettiness to the 1-subjettiness ($\tau_{21}$) as well as
the 3- to 2-subjettiness ratio ($\tau_{32}$)~\cite{jet_sub}; the ratio of the
three-point energy correlation function to the third power of the 2-point energy
correlation function, as suggested in Ref.~\cite{dij_jet_sub} ($D_2$); and the
square-root of the scale of the first and second $k_t$ splittings ($d_{12}$ and
$d_{23}$)~\cite{dij_jet_sub_ratio}.
Only \pt, invariant mass, $D_2$ and $\tau_{32}$ are shown in the main text, 
whereas the rest can be found in \ref{sec:app_results} in the appendix.
The \antikt jet clustering algorithm with radius parameter of $R = 1.0$ is used
to build jets of stable particles that fall within $|\eta| < 2.5$ and have jet
$p_{\mathrm{T}} \geq 250$ GeV.
\begin{figure*}[htpb]
    \centering
    \subfloat[]{{\includegraphics[width=0.33\textwidth]{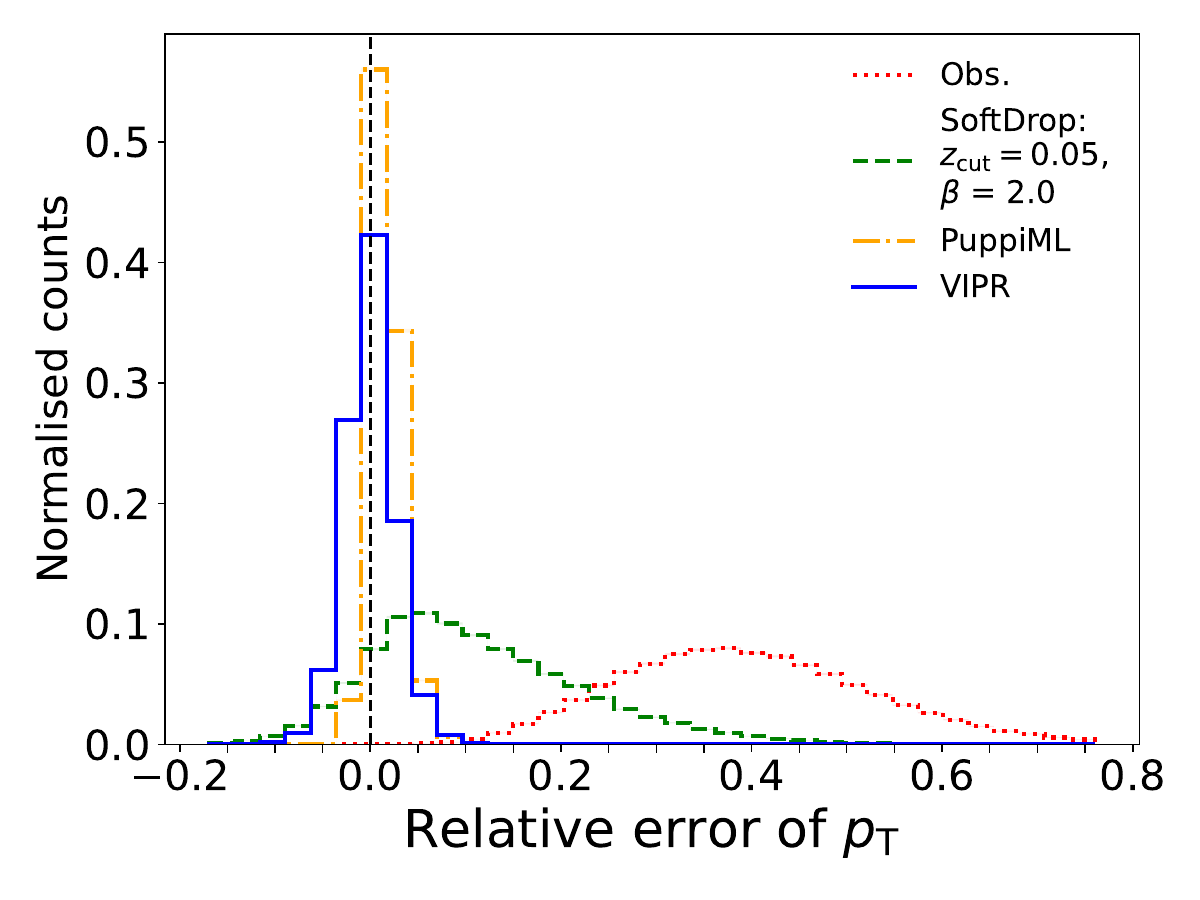}}}
    \subfloat[]{{\includegraphics[width=0.33\textwidth]{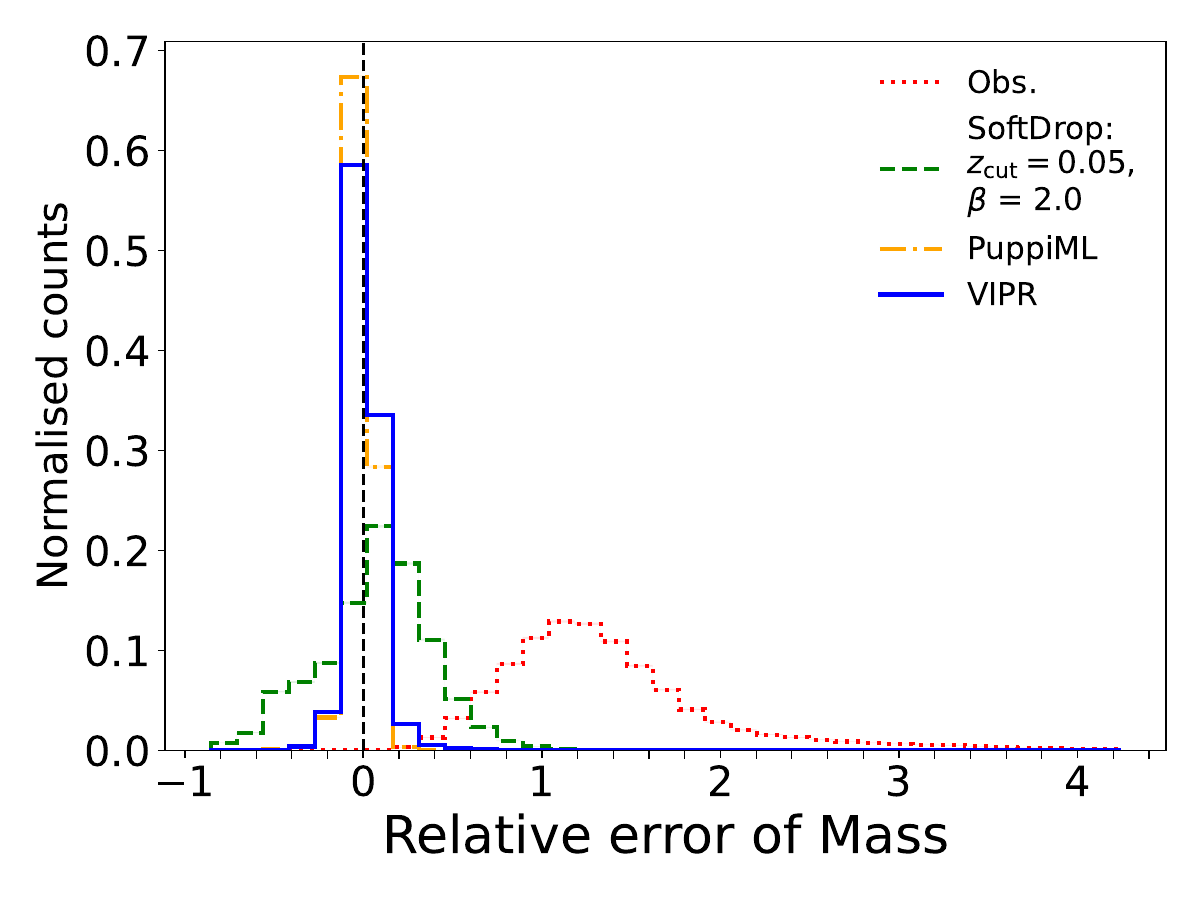}}}
    \\
    \subfloat[]{{\includegraphics[width=0.33\textwidth]{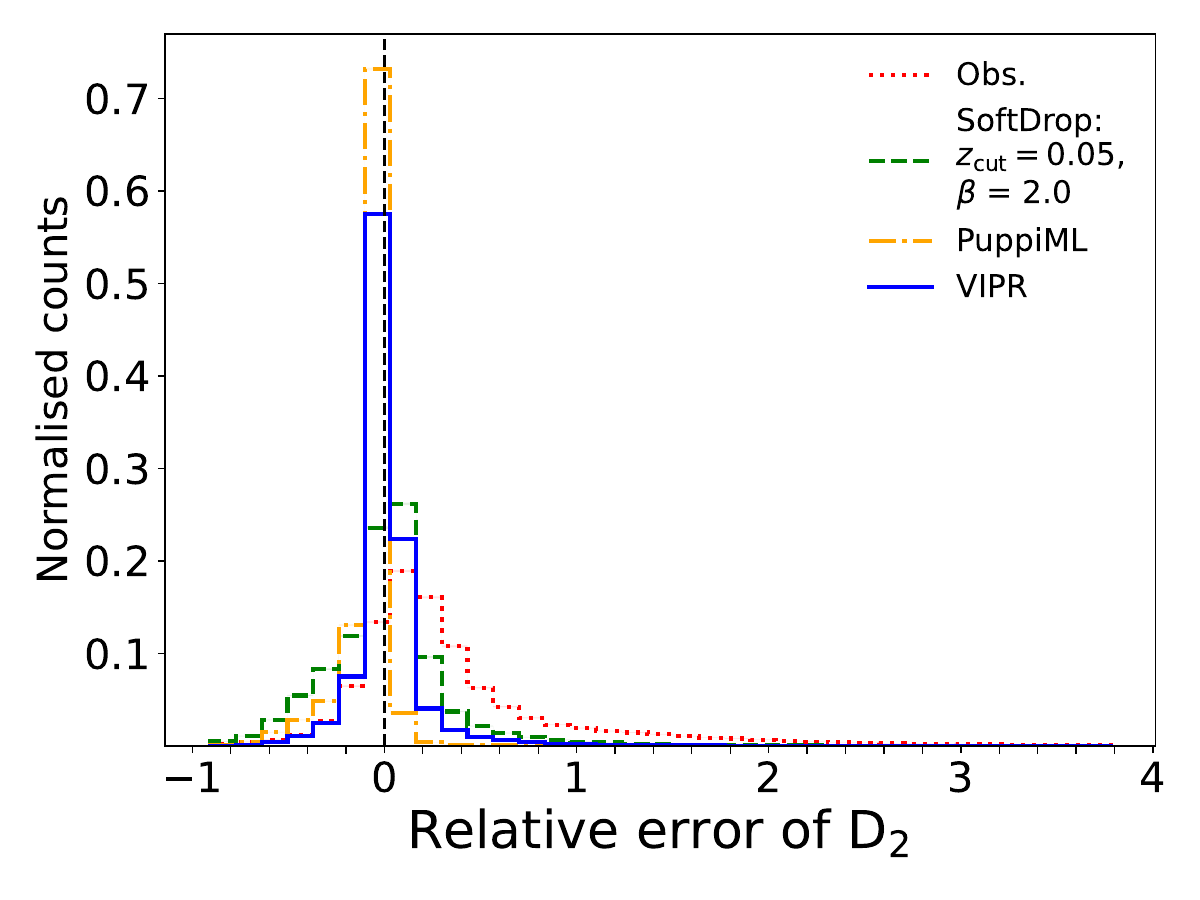}}}
    \subfloat[]{{\includegraphics[width=0.33\textwidth]{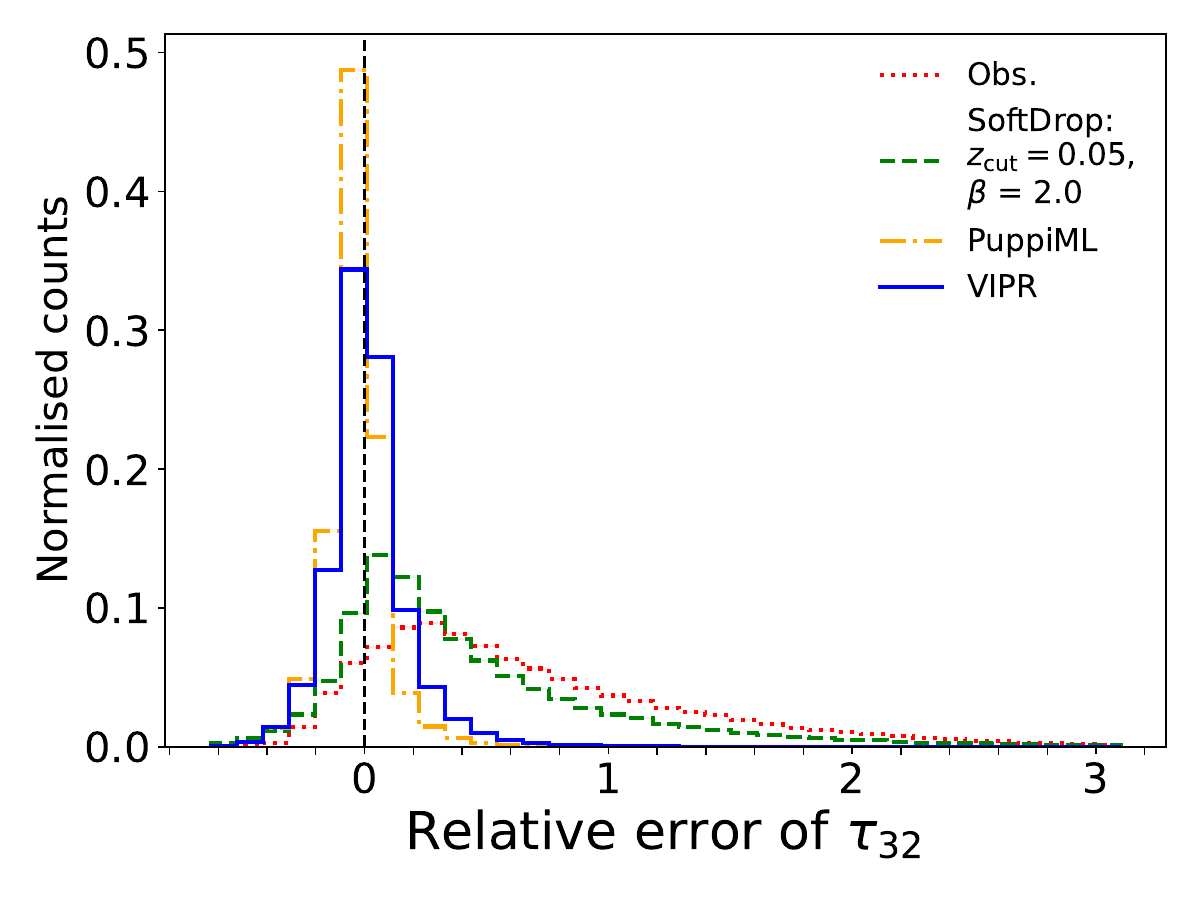}}}
    \caption{
    Comparisons of the RE distributions between observed (red), \softdrop
    (green) and \vipr (blue) jets for jet \pt, mass, $D_2$ and $\tau_{32}$.
    The observed jet is generated using a pile-up distribution of 
    \normal{200}{50}.
    }
    \label{fig:marginal_resolution_of_single}
\end{figure*}

In order to assess its viability in an experimental setting, we compare \vipr to
two established algorithms for pile-up mitigation: \softdrop and \puppiml.
As \vipr is designed to remove pile-up from a single jet, we do not make
comparisons to event-level approaches, although it should be noted that these
could be combined with \vipr by applying them before jet clustering. 
\begin{figure*}[htb]
    \centering
    \subfloat[]{{\includegraphics[width=0.33\textwidth]{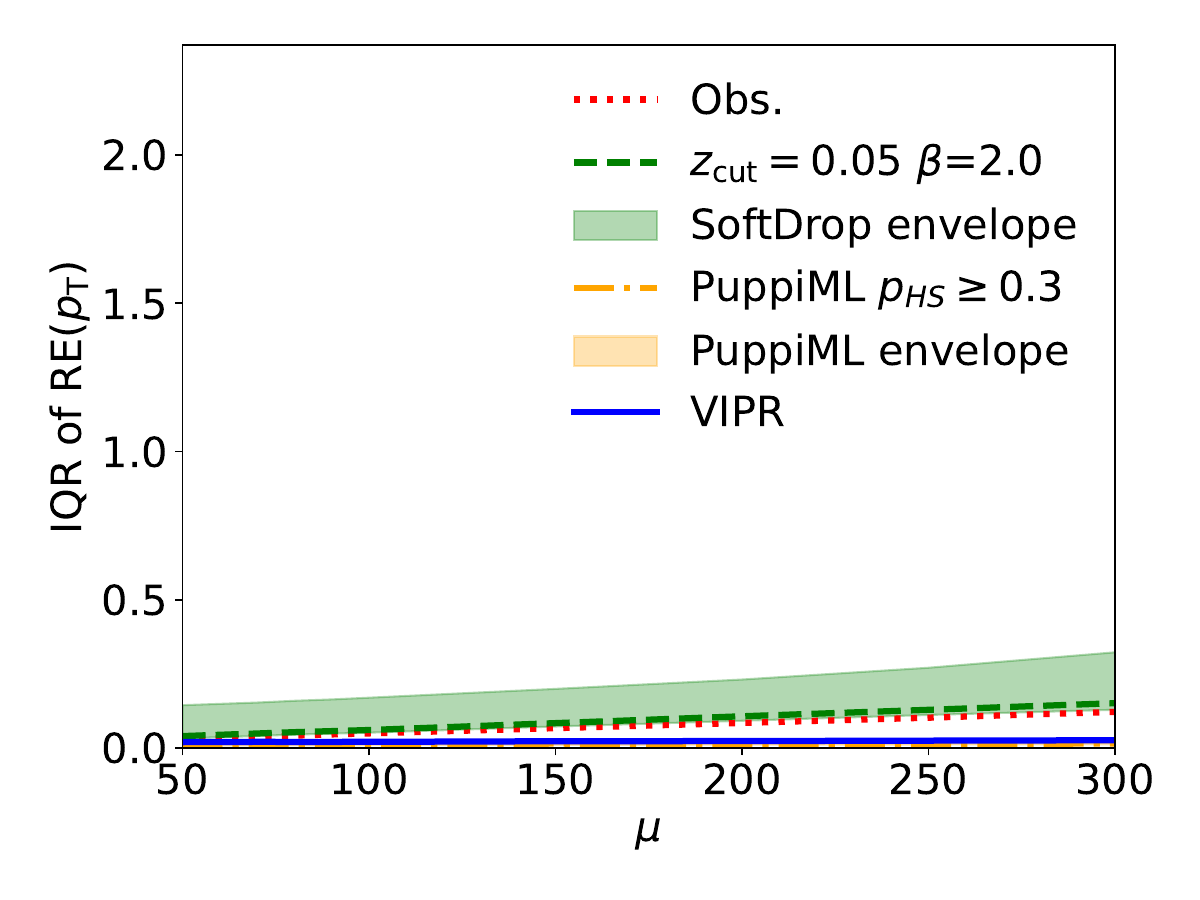}}}
    \subfloat[]{{\includegraphics[width=0.33\textwidth]{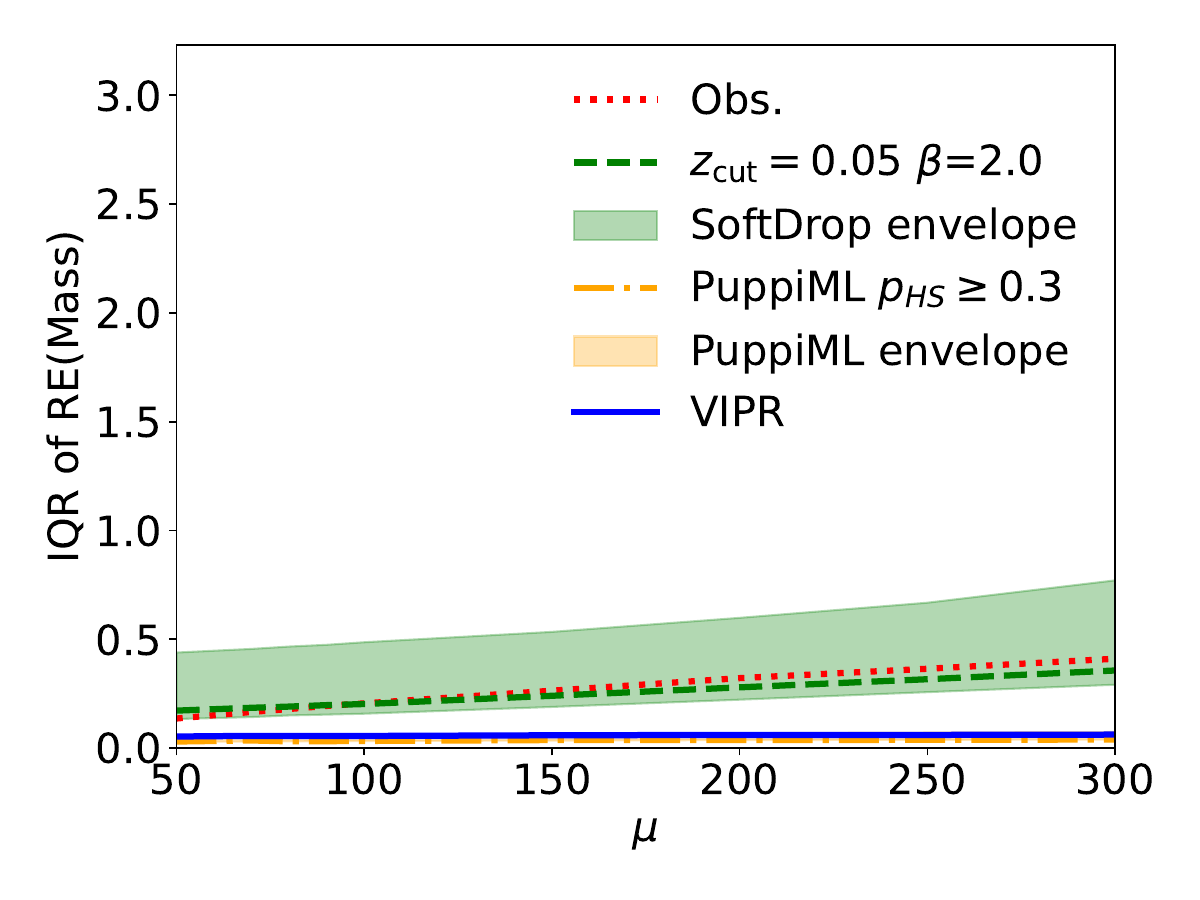}}}
    \\
    \subfloat[]{{\includegraphics[width=0.33\textwidth]{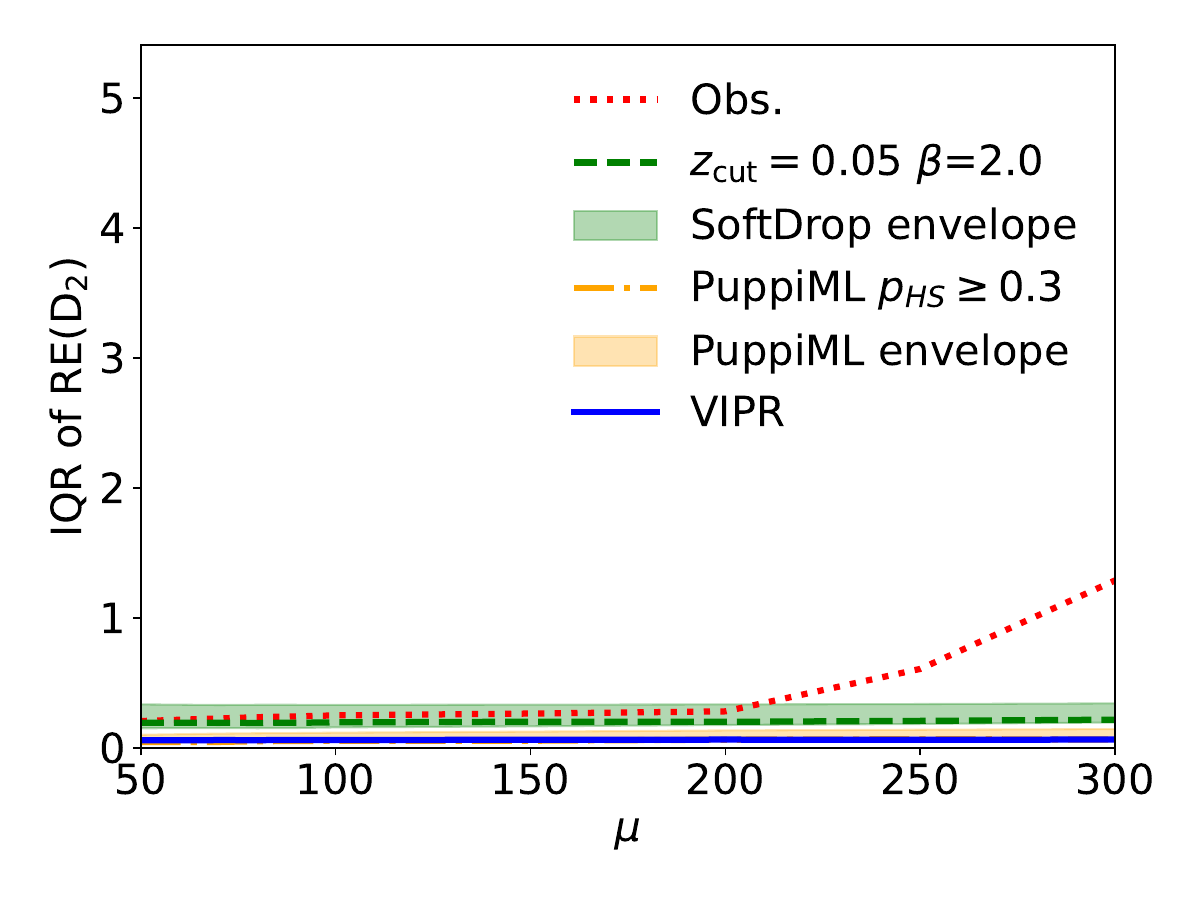}}}
    \subfloat[]{{\includegraphics[width=0.33\textwidth]{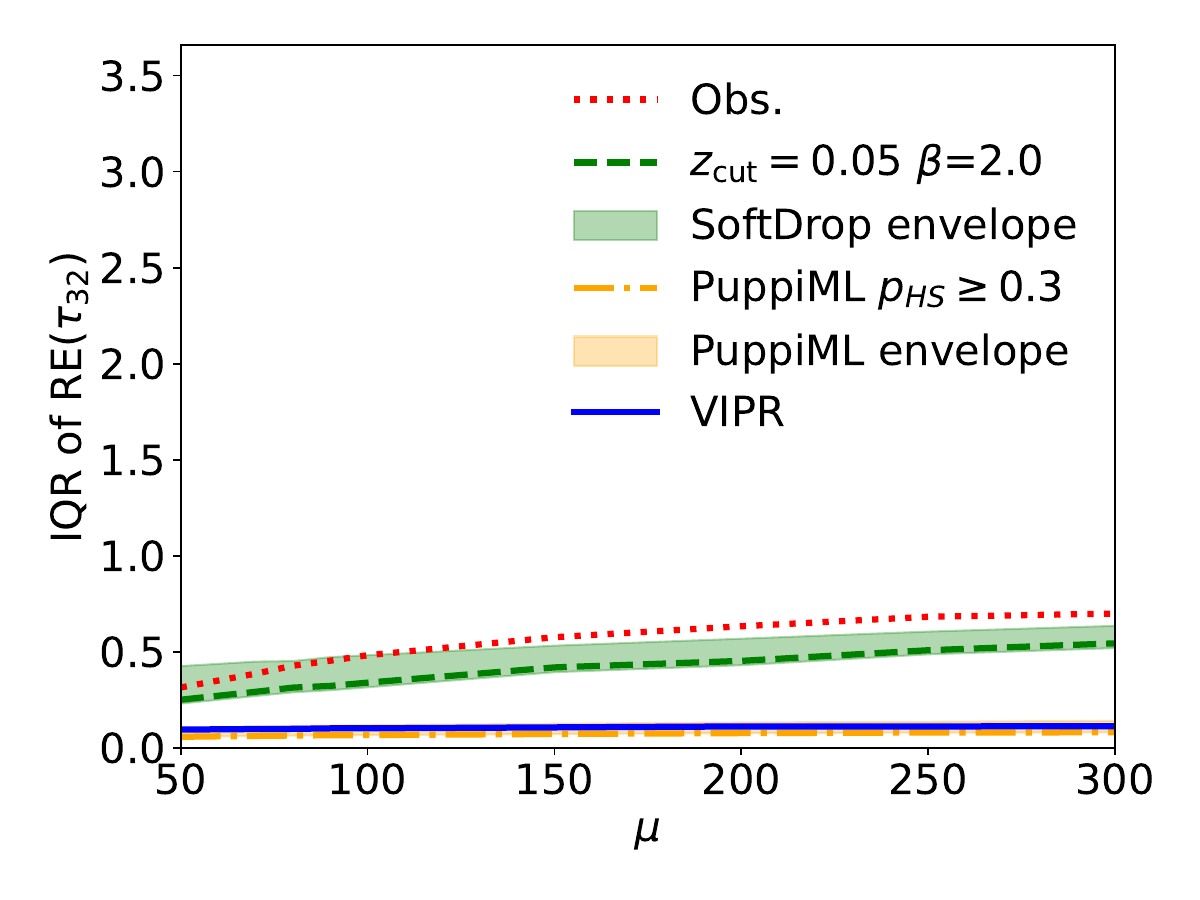}}}
    \caption{
    Comparisons of relative error IQRs as a function of $\mu$
    for jet \pt, mass, $D_2$, and $\tau_{32}$.
    A constant IQR as a function of $\mu$ indicates robustness against
    increasing pile-up.
    Envelopes from scans over \softdrop parameters and \puppiml cuts are also
    shown.
    }
    \label{fig:response_at_different_iqr}
\end{figure*}

The \softdrop algorithm has two hyperparameters, \zcut and $\beta$, controlling
the sensitivity to soft and wide-angle radiation.
Soft radiation is removed by increasing \zcut whereas decreasing $\beta$ removes
wide-angle radiation.
We sweep $z_\mathrm{cut}=[0.05, 0.1, 0.15]$ and $\beta=[0, 0.5, 1,
2]$, motivated by choices in Ref.~\cite{ATLAS:2017zda,softdrop, ATLAS:2019mgf}
and find $z_\mathrm{cut}=0.05$ and $\beta=2$ to minimize the bias
and achieve the narrowest distribution across most quantities tested.
The \softdrop implementation from Ref.~\cite{energyflow} is used throughout.

The \puppiml algorithm learns to predict the probability $p_{HS}$ for each
constituent to originate from the hard-scatter interaction; as such, unlike
\vipr, does not reconstruct the posterior over jet constituents.
The $p_{HS}$ threshold controls the strictness of the pile-up removal;
we evaluate different thresholds: $[0.2, 0.25, 0.3, 0.35, 0.4, 0.5, 0.75, 0.8,
0.825, 0.85]$ and find that the $p_{HS} \geq 0.3$ threshold generally performs
well across relevant substructure quantities over a wide range of $\mu$.
\begin{figure*}[htpb]
    \centering
    \subfloat[]{{\includegraphics[width=0.33\textwidth]{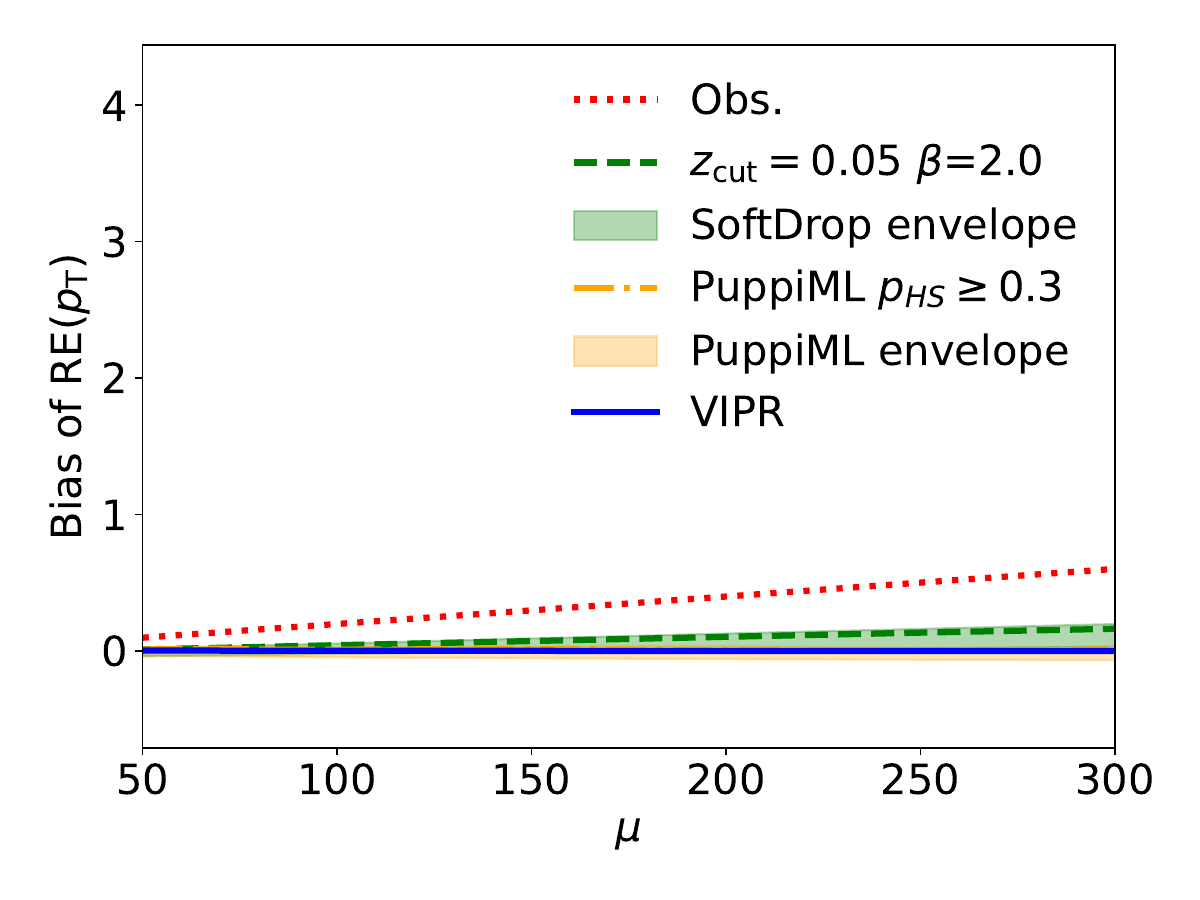}}}
    \subfloat[]{{\includegraphics[width=0.33\textwidth]{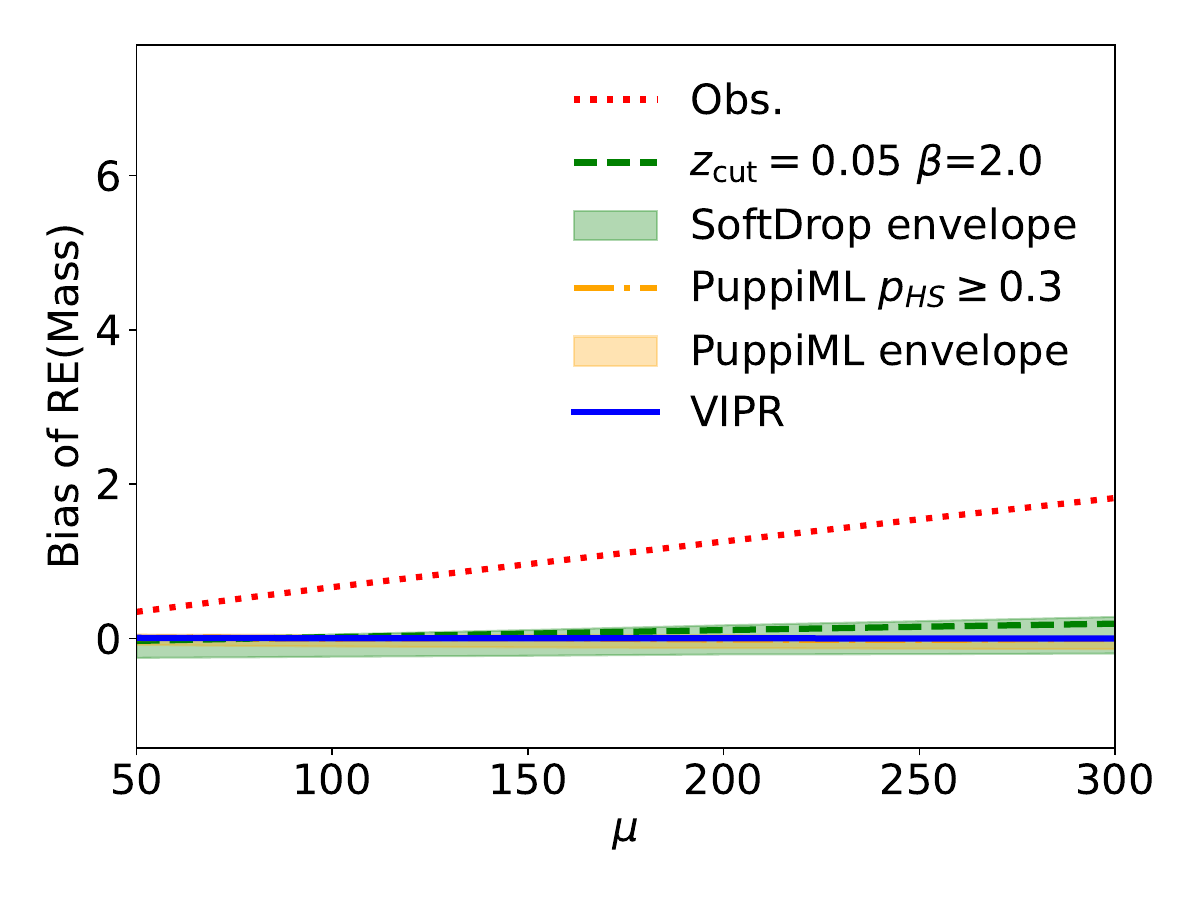}}}
    \\
    \subfloat[]{{\includegraphics[width=0.33\textwidth]{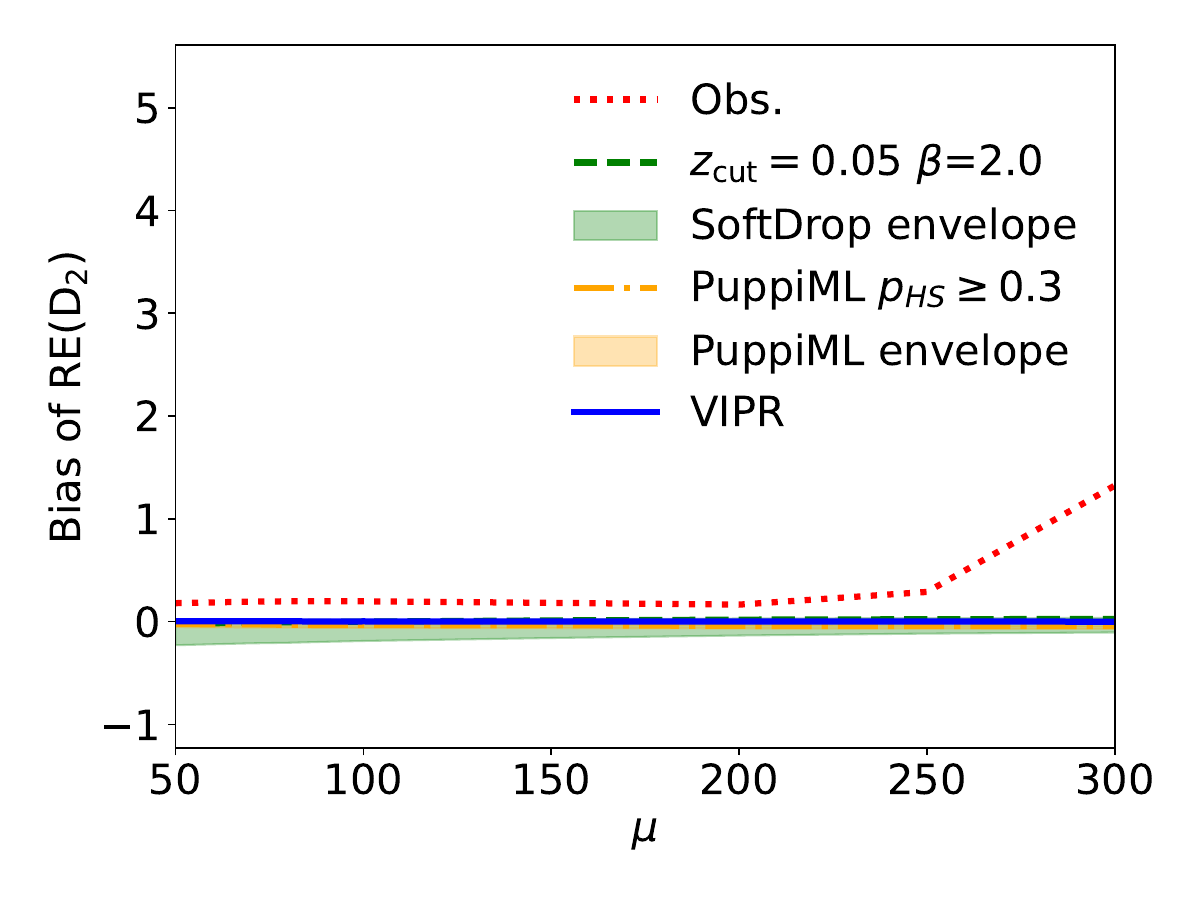}}}
    \subfloat[]{{\includegraphics[width=0.33\textwidth]{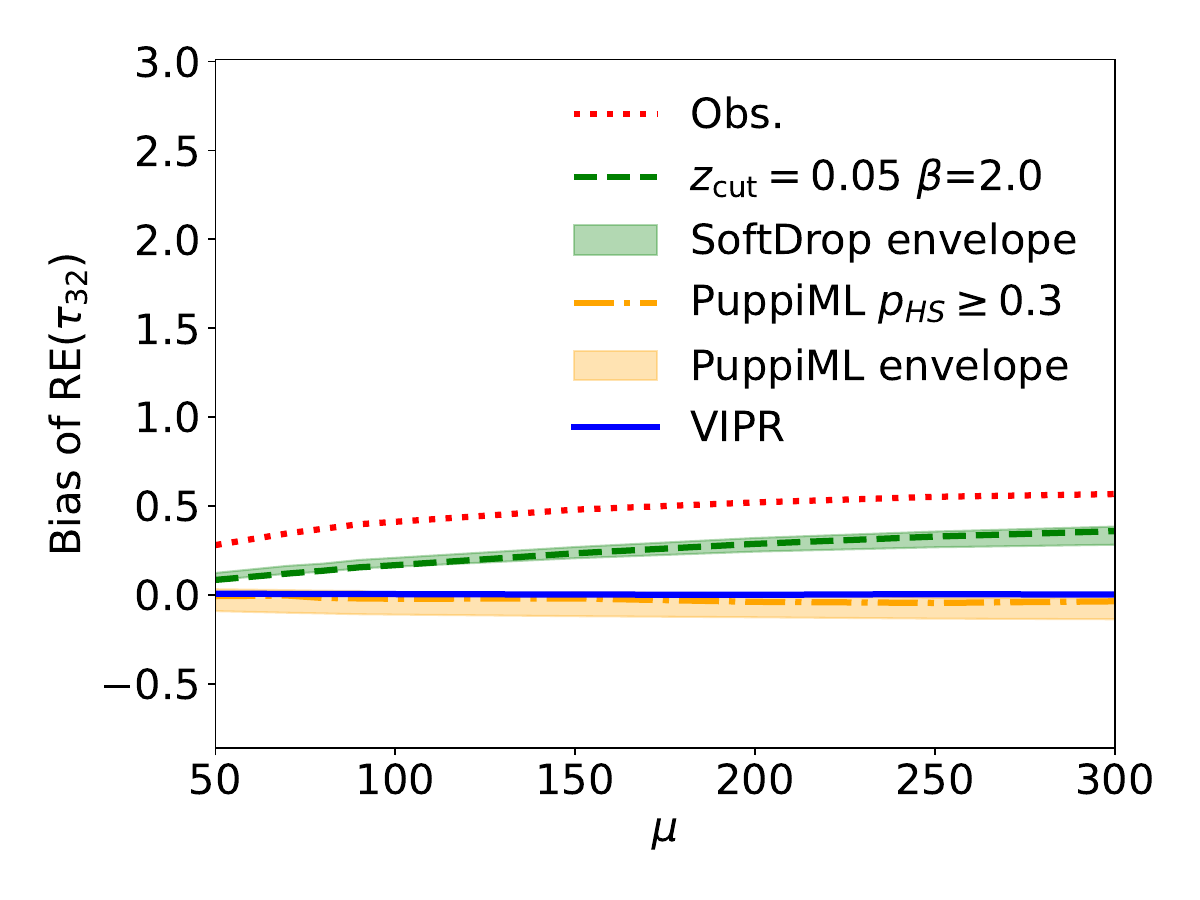}}}
    \caption{
    Comparisons of RE bias as a function of $\mu$
    for jet \pt, mass, $D_2$, and $\tau_{32}$.
    Zero bias as a function of $\mu$ indicates robustness against increasing
    pile-up.
    Envelopes from scans over \softdrop parameters and \puppiml cuts are also
    shown.
    }
    \label{fig:response_at_different_mu}
\end{figure*}

\cref{fig:denoised_scatter} shows an example observed jet as well as the output
of the \softdrop, \puppiml, and \vipr pile-up mitigation algorithms compared to
the hard-scatter jet constituents.
For this jet, only one sample from the \vipr posterior is shown, but more
samples for this observed jet can be found in 
\ref{fig:denoised_scatter_posterior} in the appendix.
In this single example, we already observe that the ML-methods, \puppiml and
\vipr, approximate the ground-truth much more accurately than the original
observation and \softdrop.

\subsection{Performance integrated over $\mu$}

To assess the predictive power of the different pile-up mitigation
algorithms, we construct the distribution of the relative error (RE) between the
prediction and ground-truth, defined as $\mathrm{RE} = \frac{\hat{x}-x}{x}$,
where $\hat{x}$ is the predicted value of some relevant quantity, and $x$ is the
ground-truth.
The RE distributions are built by drawing $100,000$ individual \jobs instances,
where the pile-up distribution follows a normal distribution over $\mu$ with
mean and standard deviation of 200 and 50, respectively: \normal{200}{50}.
For \vipr, a single sample is drawn from the posterior for each \jobs instance.
In general, a high-performance algorithm should result in a small bias (i.e.~the
median of the RE is close to zero) and good resolution (i.e.~the RE width is
small).

The RE distribution of the jet \pt, invariant mass, $D_2$ and $\tau_{32}$
compared between pile-up mitigation strategies can be seen in
\cref{fig:marginal_resolution_of_single}.
Across all observables, \vipr has a substantially better resolution than
the original observation and \softdrop.
The RE distribution of \vipr is also centered at zero, whereas \softdrop and the
original observation tend to be relatively biased.
\puppiml exhibits lower resolution than \vipr across the quantities but, unlike
\vipr, cannot estimate the full posterior over the jet constituents.

\begin{figure*}[htbp]
    \centering
    \subfloat[]{{\includegraphics[width=0.45\textwidth]{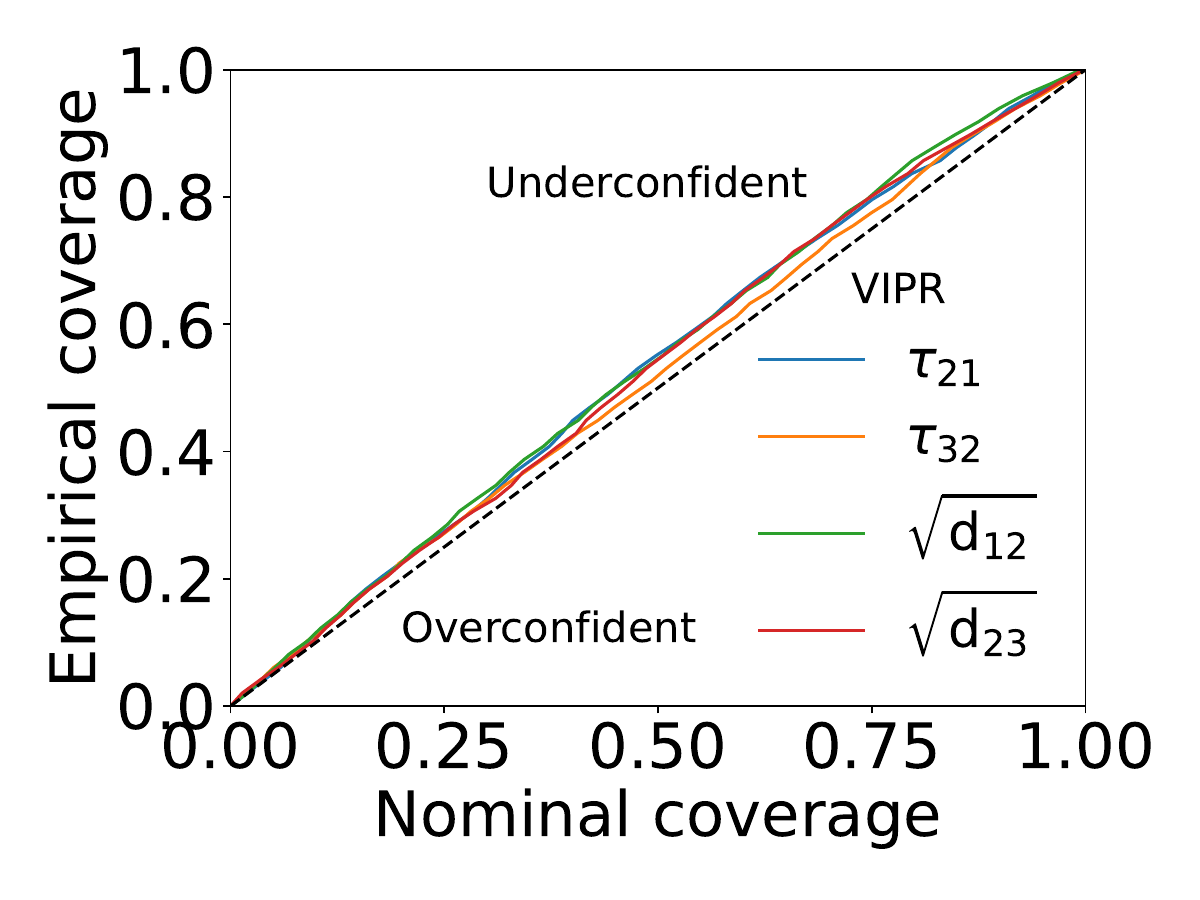}}}
    \subfloat[]{{\includegraphics[width=0.45\textwidth]{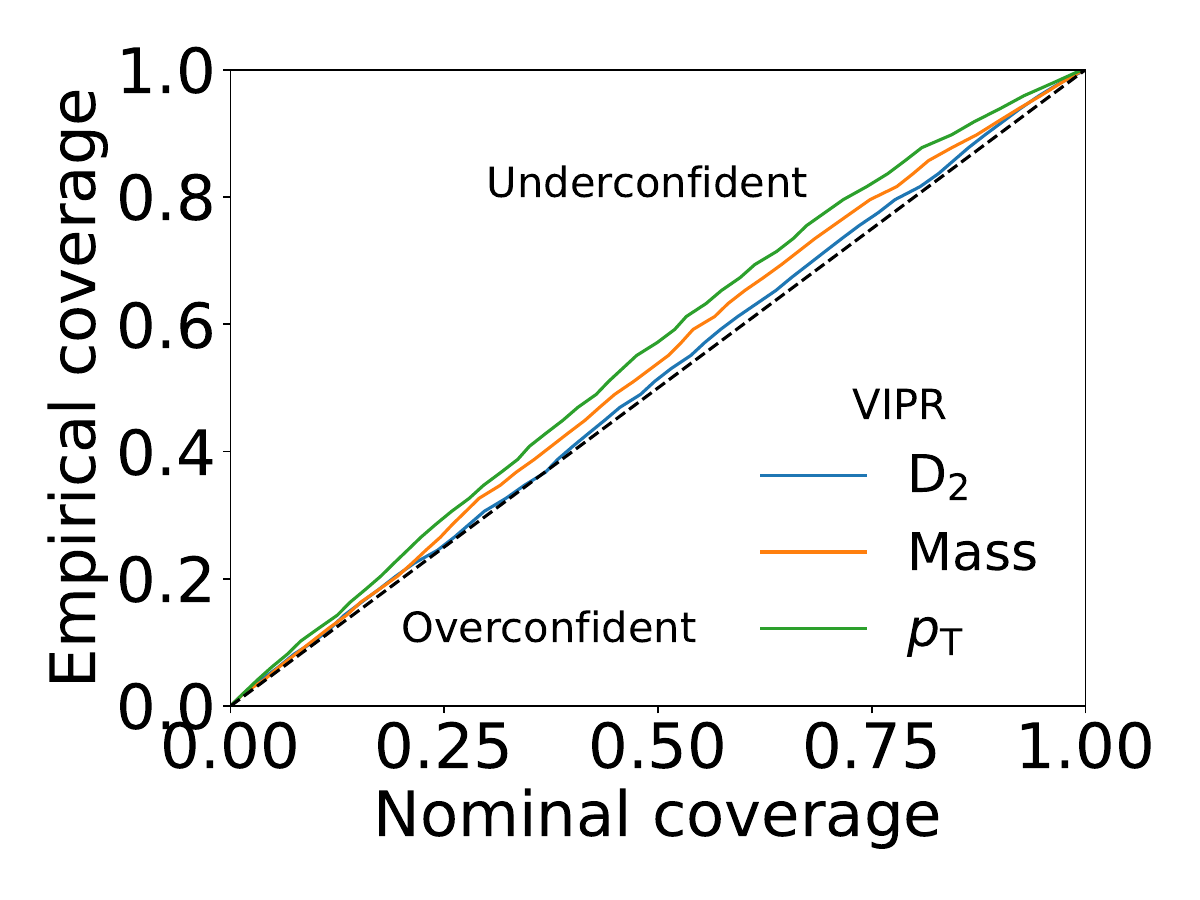}}}
    \caption{
    Comparison between the ideal coverage in the dashed black line and the \vipr 
    coverage for jet \pt, mass and substructure variables. 
    The coverage is calculated by integrating from the median and out 
    over the posterior truth quantiles.
    Correct coverage is indicated by the black dashed line.
    Coverage curves above the black line indicate underconfidence of the model,
    while those below it indicate overconfidence.
    }
    \label{fig:coverage_plots}
\end{figure*}

\subsection{Performance vs $\mu$}

We also evaluate the performance as a function of $\mu$ by calculating the bias
(median) and interquartile range ($\text{IQR}=\frac{Q_{75\%}-Q_{25\%}}{1.349}$)
of the RE distributions, where $Q_{75\%}$ and $Q_{25\%}$ are the $25\%$ and
$75\%$ quantiles, respectively. 
We show the envelope of the \softdrop and \puppiml options, with the
best-performing \softdrop (\puppiml) choice of $\zcut = 0.05$ and $\beta = 2$
($p_\text{HS} \geq 0.3$) drawn separately.
The IQR of the RE as a function of $\mu$ for the tested algorithms can be seen
in \cref{fig:response_at_different_iqr}.
\vipr and \puppiml both appear robust to pile-up: their IQRs remain relatively
constant across the tested $\mu$ range.
\puppiml exhibits a small but consistently non-zero bias for several quantities,
while \vipr is directly well-calibrated.
The bias as a function of $\mu$ can be seen in \cref{fig:response_at_different_mu}.
Across all observables, \vipr is centered at zero and remains consistent as a
function of $\mu$.
On the other hand, \softdrop increases with increasing $\mu$ and has larger
biases than \vipr across various $\mu$ values.
\puppiml results in a small but noticeable slope in the bias versus $\mu$.
\subsection{Coverage}
\vipr distinguishes itself from other pile-up removal methods due to its
variational inference nature.
Consequently, for each \jobs, \vipr can generate a posterior to establish 
empirical coverage of the ground truth and verify whether \vipr is
underconfident, overconfident, or correctly calibrated.

To obtain \vipr's posterior of a single \jobs, we sample from \vipr 512 times
conditioned on the same \jobs to generate its posterior.
We repeat this procedure for 2,000 different \jobs to generate a total of 2,000
posteriors.
To assess the coverage, we calculate the quantile of the \jtrue observables
for each of the generated posteriors.

These truth quantiles across the generated posteriors can be found in \ref{fig:posterior_distributions_app}.
\cref{fig:coverage_plots} shows the integral of the posterior truth quantiles,
which indicate an unbiased estimate if they are linear.
Across the observables, we see a slight under-confidence in \vipr, meaning the
coverage properties will be slightly conservative; this is usually preferable to
an overconfident procedure.

\subsection{Performance vs $\mu$ with a simulated detector efficiency of $\epsilon_{\mathrm{det}} = 90\%$}

While the previous analysis evaluated performance under idealized conditions, 
realistic particle reconstruction in collider experiments exhibits inherent
inefficiencies, e.g. the ATLAS detector achieves an approximately 90\% track
reconstruction efficiency~\cite{reconstruction_performance_2024}, 
resulting in a small fraction of particle trajectories remaining undetected.
This reconstruction inefficiency constitutes an additional challenge for pile-up
mitigation techniques and provides an opportunity to further differentiate
methods.

\begin{figure}[htpb]
    \centering
    \subfloat[]{{\includegraphics[width=0.30\textwidth]{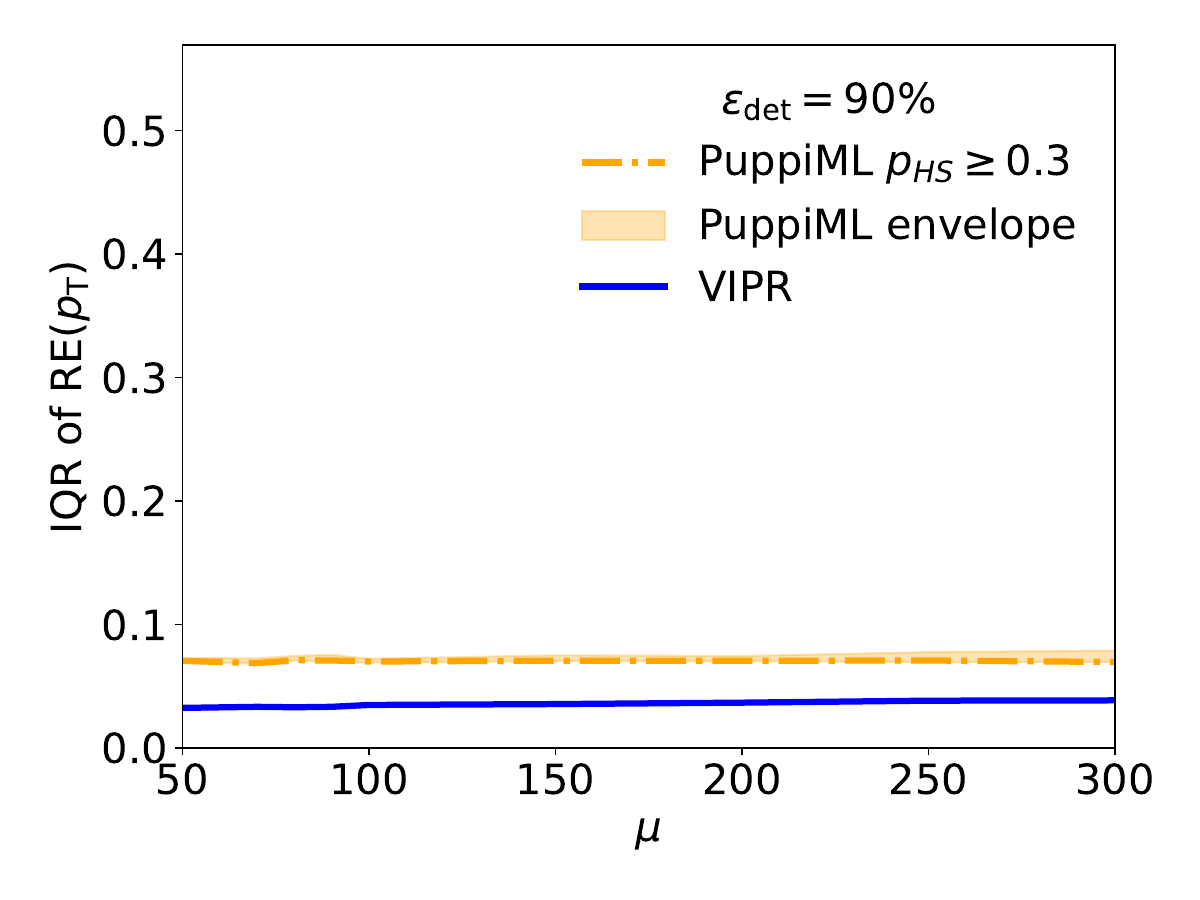}}}
    \\
    \subfloat[]{{\includegraphics[width=0.30\textwidth]{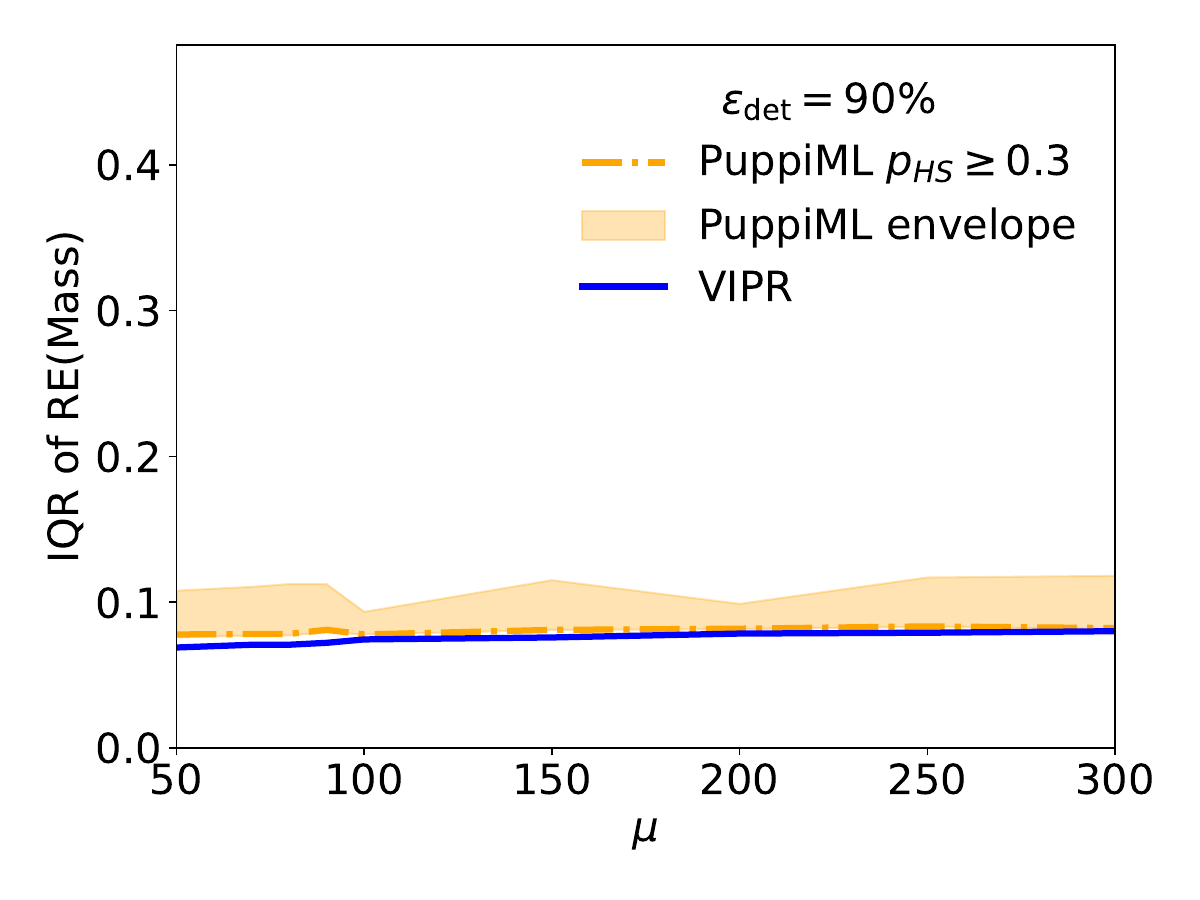}}}
    \\
    \subfloat[]{{\includegraphics[width=0.30\textwidth]{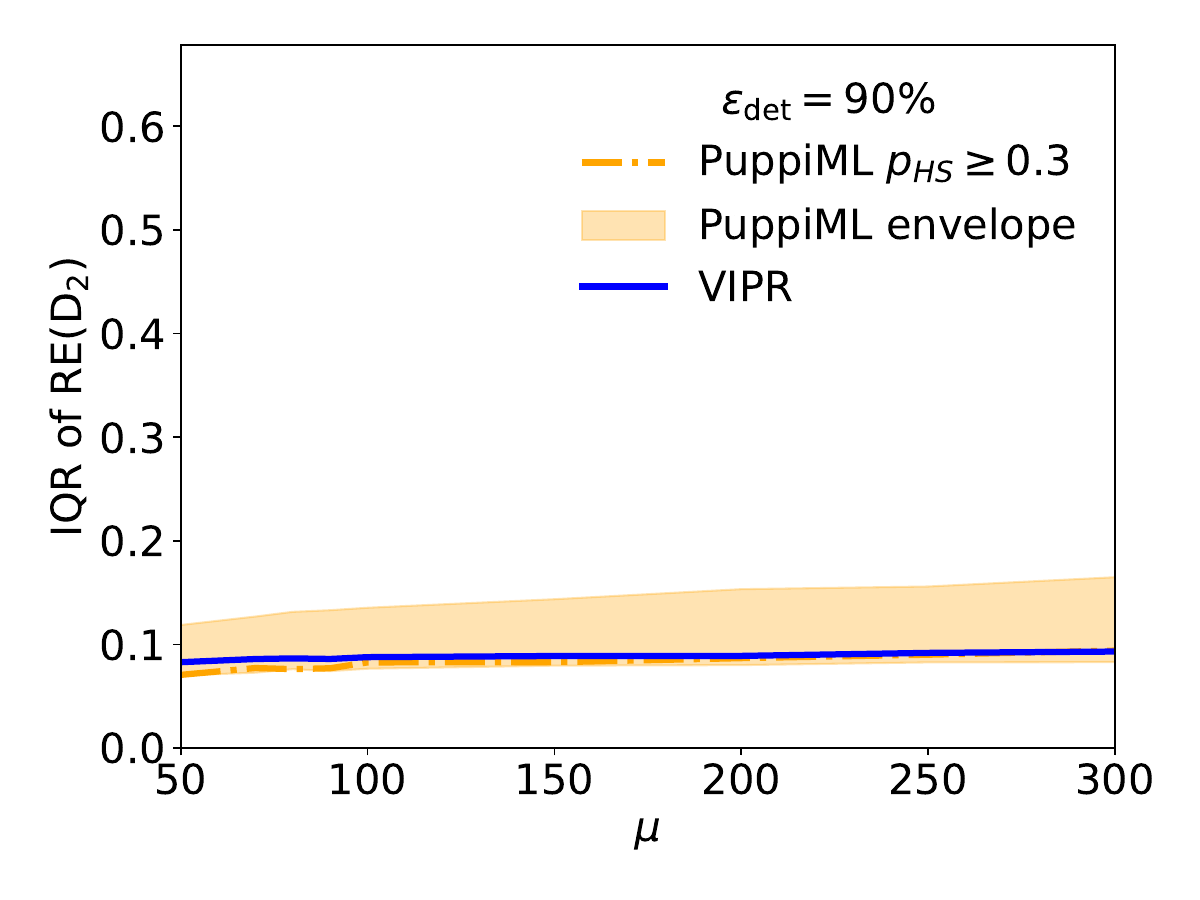}}}
    \\
    \subfloat[]{{\includegraphics[width=0.30\textwidth]{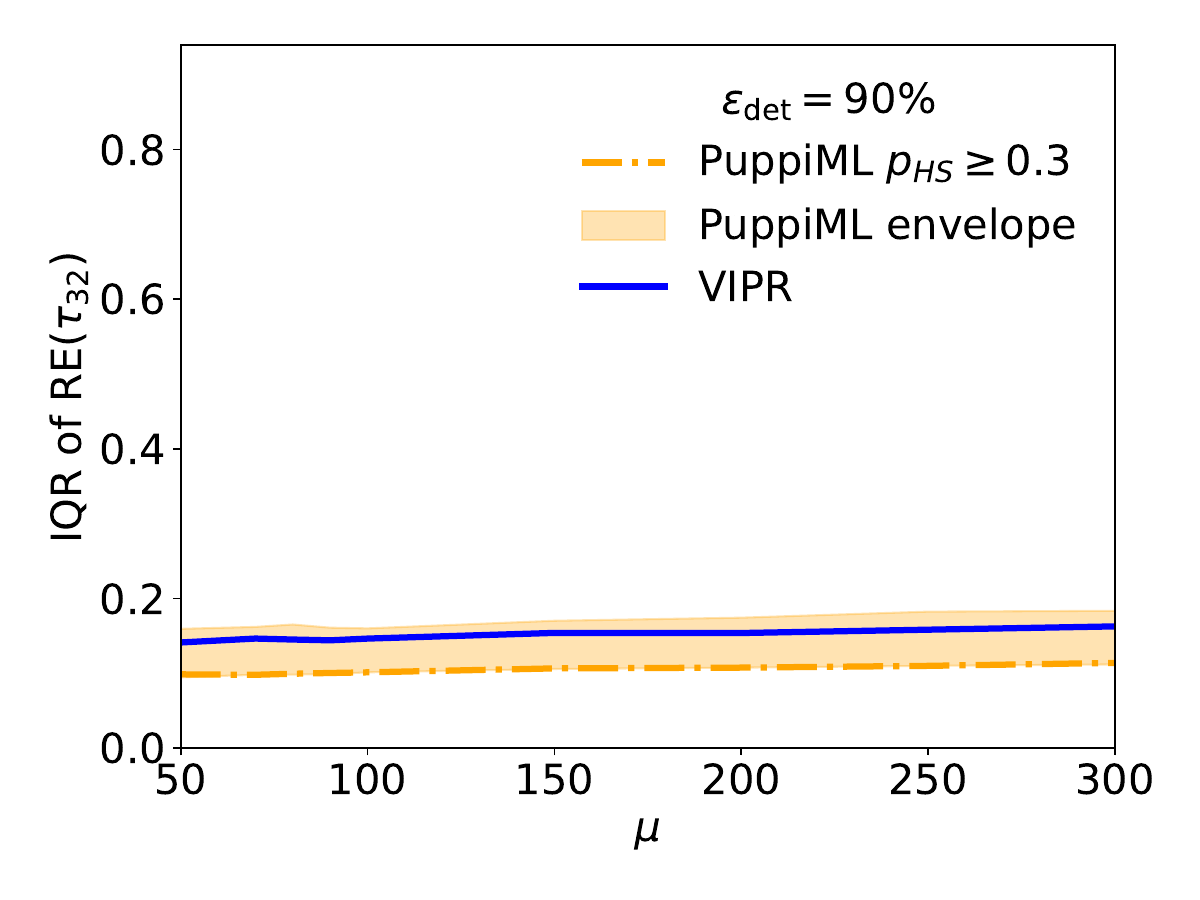}}}
    \caption{
    Comparisons of relative error IQRs as a function of $\mu$
    for jet \pt, mass, $D_2$, and $\tau_{32}$, with a constituent
    reconstruction efficiency of $\epsilon_{\mathrm{det}} = 90\%$.
    A constant IQR as a function of $\mu$ indicates robustness against
    increasing pile-up.
    The envelope of a scan of \puppiml cuts is also shown, with the best
    resulting parameters in dashed orange.
    }
    \label{fig:response_at_different_iqr_ineff}
\end{figure}

\begin{figure}[htpb]
    \centering
    \subfloat[]{{\includegraphics[width=0.30\textwidth]{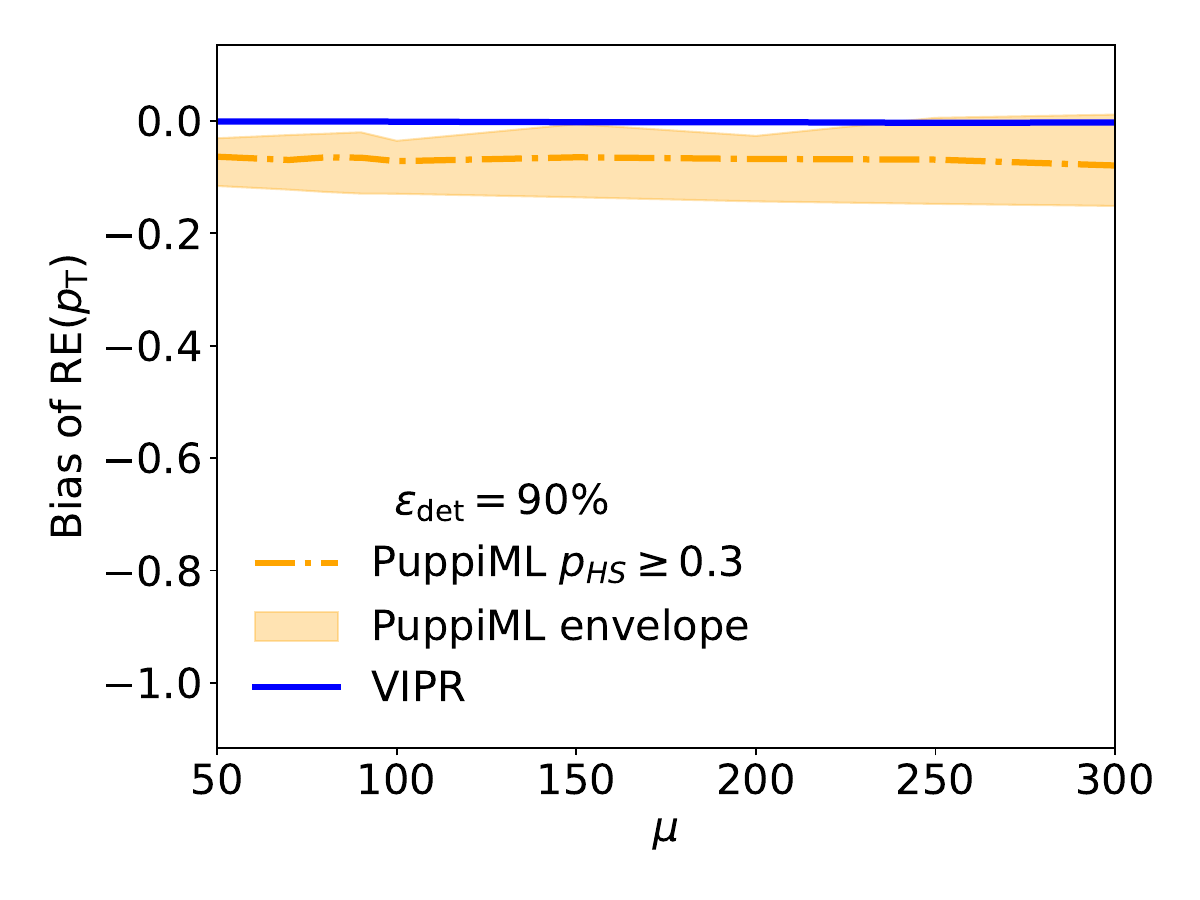}}}
    \\
    \subfloat[]{{\includegraphics[width=0.30\textwidth]{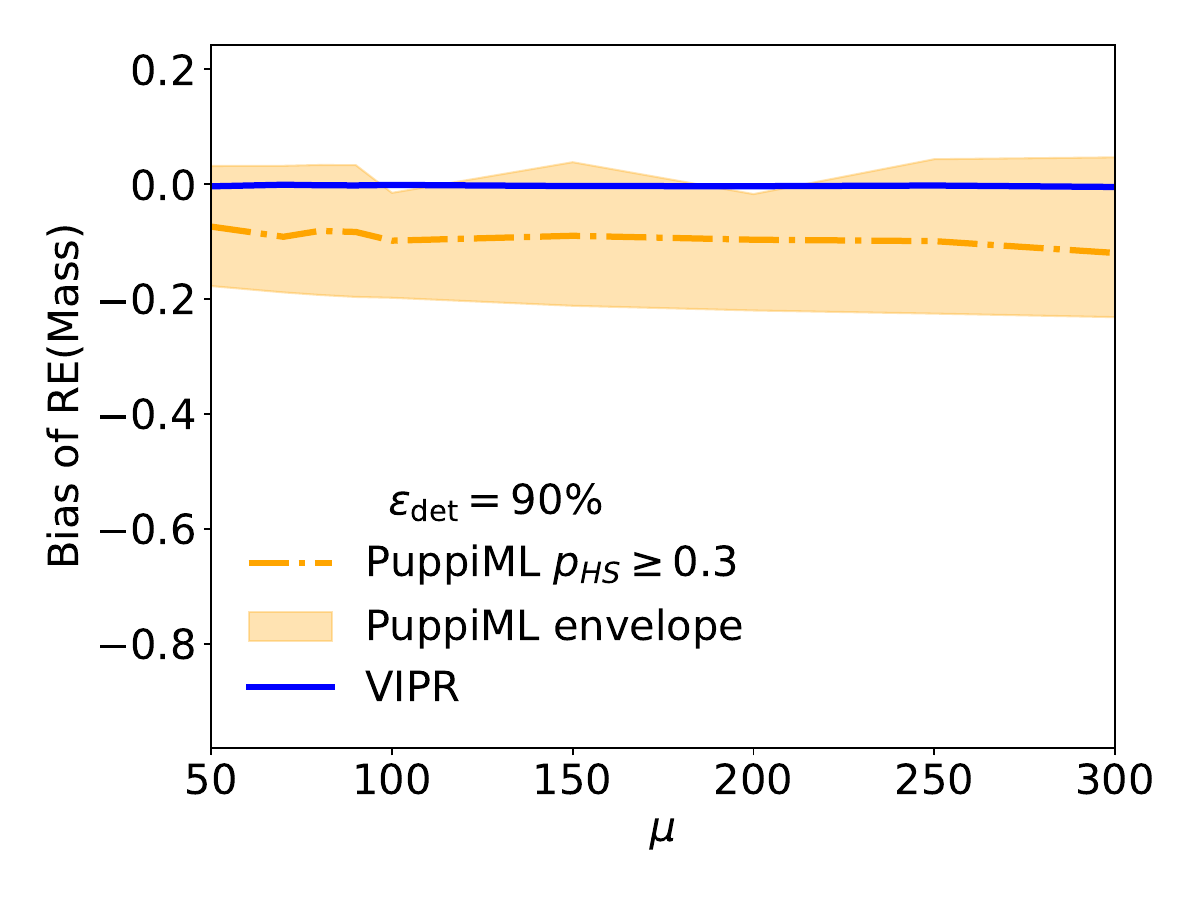}}}
    \\
    \subfloat[]{{\includegraphics[width=0.30\textwidth]{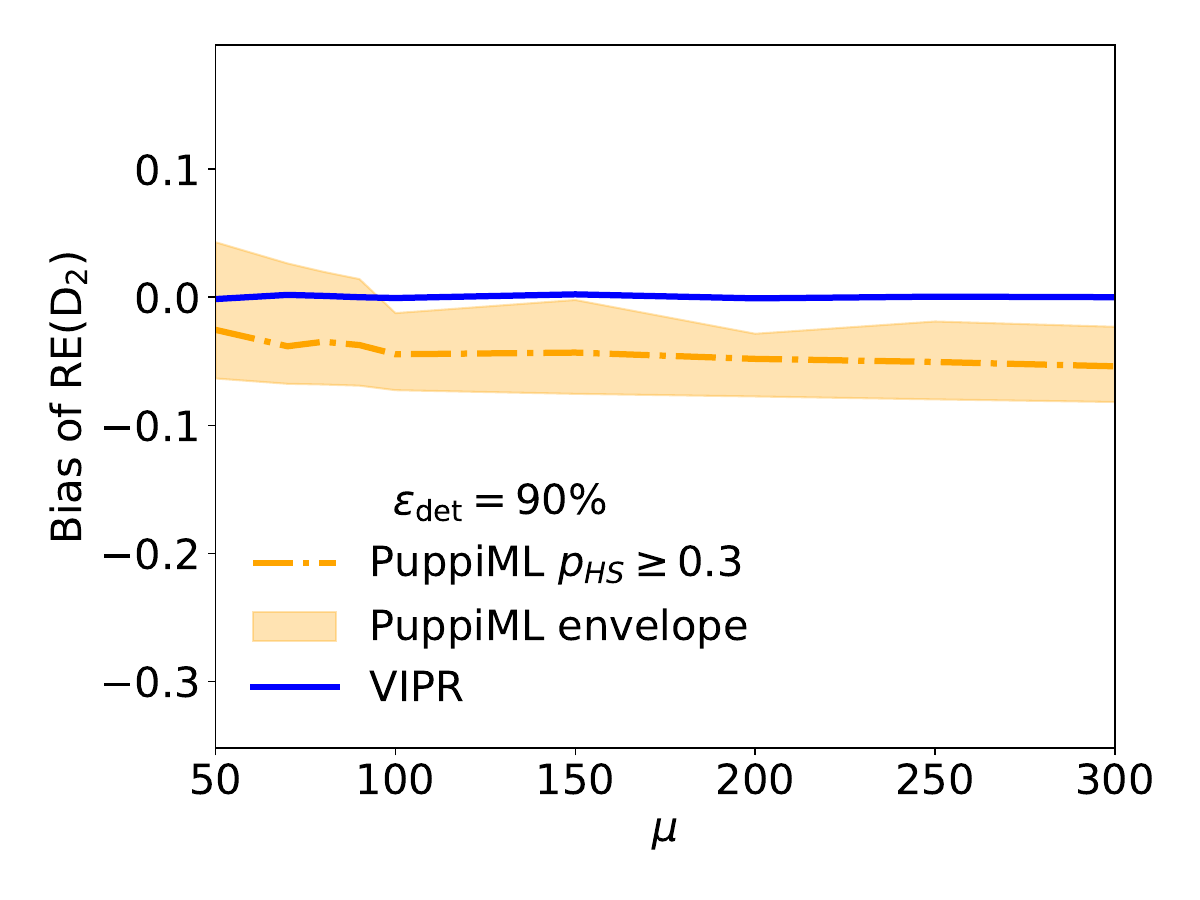}}}
    \\
    \subfloat[]{{\includegraphics[width=0.30\textwidth]{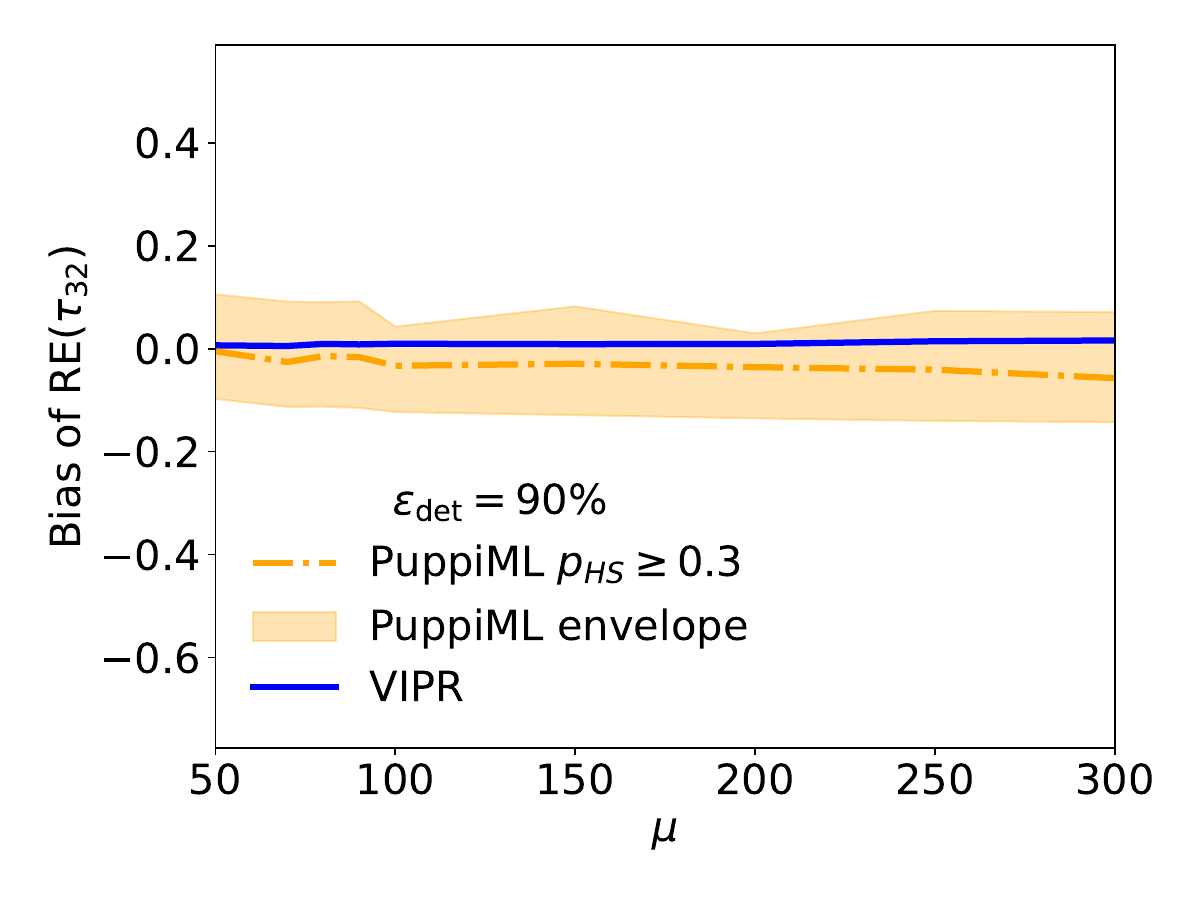}}}
    \caption{
    Comparisons of relative error bias as a function of $\mu$
    for jet \pt, mass, $D_2$, and $\tau_{32}$, with a constituent
    reconstruction efficiency of $\epsilon_{\mathrm{det}} = 90\%$.
    Zero bias as a function of $\mu$ indicates robustness against
    increasing pile-up.
    The envelope of a scan of \puppiml cuts is also shown, with the best
    resulting parameters in dashed orange.
    }
    \label{fig:response_at_different_mu_ineff}
\end{figure}

The generative nature of \vipr offers significant advantages in this context, as
\vipr models the underlying distribution of complete jets rather than
classifying reconstructed constituents.
This generative approach enables \vipr to infer missing information and
reconstruct complete jet signatures even when presented with partial
observations.
On the other hand, \puppiml uses classification to remove likely pile-up
contamination, rendering it unable to account for true constituents that were
not reconstructed.

To assess performance under these more realistic conditions, we reduce the
detector efficiency to $\epsilon_{\mathrm{det}} = 90\%$ by stochastically
removing 10\% of constituents from each observed jet, applying this inefficiency
uniformly across the \pt, $\eta$, and $\mu$ parameter space.
Results are shown in
\cref{fig:response_at_different_iqr_ineff,fig:response_at_different_mu_ineff},
which indicates that both \vipr and \puppiml have comparable performances for some
features, but \vipr is better-performing across benchmarks.
In particular, \vipr outperforms \puppiml considerably in jet \pt
accuracy and precision.
 
    \FloatBarrier

    \section{Conclusions}
    In this work we have introduced \vipr, a variational approach for pile-up
removal by solving the inverse problem using diffusion models. 
\vipr is trained to generate a \jtrue from a pile-up contaminated \jobs at the
constituent level.
The performance of \vipr has been evaluated on boosted top-quark jets under
pile-up levels consistent with the high luminosity phase at the LHC, during
which it is anticipated for bunch-crossings to reach \avgmu at the level of
200~\cite{ZurbanoFernandez:2020cco,ATL-PHYS-PUB-2021-023}.

\vipr significantly outperforms the \softdrop technique in pile-up removal and
exhibits similar performance to \puppiml in an idealized reconstruction
scenario.
When introducing a more realistic scenario with reconstruction inefficiencies,
\vipr outperforms \puppiml.

Rather than producing a single estimate of the true jet constituents given
observations, \vipr approximates the full posterior distribution of \jtrue given
\jobs, which has powerful potential use-cases.
Our results show that \vipr accurately reconstructs jet substructure, with only
a slight tendency toward under-confidence. 

The code required to reproduce these results is publicly available at
\url{https://github.com/rodem-hep/VIPR}.
 
    \section*{Acknowledgements}
    The MA, TG, and JAR would like to acknowledge funding through the SNSF Sinergia
grant CRSII5\_193716 ``Robust Deep Density Models for High-Energy Particle
Physics and Solar Flare Analysis (RODEM)'' and the SNSF project grant
200020\_212127 ``At the two upgrade frontiers: machine learning and the ITk
Pixel detector''. 
CP acknowledges support through STFC consolidated grant ST/W000571/1.
     
\bibliography{main_onedoc.bib}

\begin{thebibliography}{81}%
\makeatletter
\providecommand \@ifxundefined [1]{%
 \@ifx{#1\undefined}
}%
\providecommand \@ifnum [1]{%
 \ifnum #1\expandafter \@firstoftwo
 \else \expandafter \@secondoftwo
 \fi
}%
\providecommand \@ifx [1]{%
 \ifx #1\expandafter \@firstoftwo
 \else \expandafter \@secondoftwo
 \fi
}%
\providecommand \natexlab [1]{#1}%
\providecommand \enquote  [1]{``#1''}%
\providecommand \bibnamefont  [1]{#1}%
\providecommand \bibfnamefont [1]{#1}%
\providecommand \citenamefont [1]{#1}%
\providecommand \href@noop [0]{\@secondoftwo}%
\providecommand \href [0]{\begingroup \@sanitize@url \@href}%
\providecommand \@href[1]{\@@startlink{#1}\@@href}%
\providecommand \@@href[1]{\endgroup#1\@@endlink}%
\providecommand \@sanitize@url [0]{\catcode `\\12\catcode `\$12\catcode
  `\&12\catcode `\#12\catcode `\^12\catcode `\_12\catcode `\%12\relax}%
\providecommand \@@startlink[1]{}%
\providecommand \@@endlink[0]{}%
\providecommand \url  [0]{\begingroup\@sanitize@url \@url }%
\providecommand \@url [1]{\endgroup\@href {#1}{\urlprefix }}%
\providecommand \urlprefix  [0]{URL }%
\providecommand \Eprint [0]{\href }%
\providecommand \doibase [0]{https://doi.org/}%
\providecommand \selectlanguage [0]{\@gobble}%
\providecommand \bibinfo  [0]{\@secondoftwo}%
\providecommand \bibfield  [0]{\@secondoftwo}%
\providecommand \translation [1]{[#1]}%
\providecommand \BibitemOpen [0]{}%
\providecommand \bibitemStop [0]{}%
\providecommand \bibitemNoStop [0]{.\EOS\space}%
\providecommand \EOS [0]{\spacefactor3000\relax}%
\providecommand \BibitemShut  [1]{\csname bibitem#1\endcsname}%
\let\auto@bib@innerbib\@empty
\bibitem [{\citenamefont {Evans}\ and\ \citenamefont {Bryant}(2008)}]{LHC}%
  \BibitemOpen
  \bibfield  {author} {\bibinfo {author} {\bibfnamefont {L.}~\bibnamefont
  {Evans}}\ and\ \bibinfo {author} {\bibfnamefont {P.}~\bibnamefont {Bryant}},\
  }\bibfield  {title} {\bibinfo {title} {{LHC Machine}},\ }\href
  {https://doi.org/10.1088/1748-0221/3/08/S08001} {\bibfield  {journal}
  {\bibinfo  {journal} {JINST}\ }\textbf {\bibinfo {volume} {3}},\ \bibinfo
  {pages} {S08001}}\BibitemShut {NoStop}%
\bibitem [{\citenamefont {Aad}\ \emph {et~al.}(2023)\citenamefont {Aad} \emph
  {et~al.}}]{ATLAS:2022hro}%
  \BibitemOpen
  \bibfield  {author} {\bibinfo {author} {\bibfnamefont {G.}~\bibnamefont
  {Aad}} \emph {et~al.} (\bibinfo {collaboration} {ATLAS}),\ }\bibfield
  {title} {\bibinfo {title} {{Luminosity determination in $pp$ collisions at
  $\sqrt{s}=13$ TeV using the ATLAS detector at the LHC}},\ }\href
  {https://doi.org/10.1140/epjc/s10052-023-11747-w} {\bibfield  {journal}
  {\bibinfo  {journal} {Eur. Phys. J. C}\ }\textbf {\bibinfo {volume} {83}},\
  \bibinfo {pages} {982} (\bibinfo {year} {2023})},\ \Eprint
  {https://arxiv.org/abs/2212.09379} {arXiv:2212.09379 [hep-ex]} \BibitemShut
  {NoStop}%
\bibitem [{\citenamefont {Hayrapetyan}\ \emph {et~al.}(2024)\citenamefont
  {Hayrapetyan} \emph {et~al.}}]{CMS:2023pad}%
  \BibitemOpen
  \bibfield  {author} {\bibinfo {author} {\bibfnamefont {A.}~\bibnamefont
  {Hayrapetyan}} \emph {et~al.} (\bibinfo {collaboration} {CMS}),\ }\bibfield
  {title} {\bibinfo {title} {{Luminosity determination using Z boson production
  at the CMS experiment}},\ }\href
  {https://doi.org/10.1140/epjc/s10052-023-12268-2} {\bibfield  {journal}
  {\bibinfo  {journal} {Eur. Phys. J. C}\ }\textbf {\bibinfo {volume} {84}},\
  \bibinfo {pages} {26} (\bibinfo {year} {2024})},\ \Eprint
  {https://arxiv.org/abs/2309.01008} {arXiv:2309.01008 [hep-ex]} \BibitemShut
  {NoStop}%
\bibitem [{\citenamefont {Aaboud}\ \emph {et~al.}(2016)\citenamefont {Aaboud}
  \emph {et~al.}}]{ATLAS:2016ygv}%
  \BibitemOpen
  \bibfield  {author} {\bibinfo {author} {\bibfnamefont {M.}~\bibnamefont
  {Aaboud}} \emph {et~al.} (\bibinfo {collaboration} {ATLAS}),\ }\bibfield
  {title} {\bibinfo {title} {{Measurement of the Inelastic Proton-Proton Cross
  Section at $\sqrt{s} = 13$ TeV with the ATLAS Detector at the LHC}},\ }\href
  {https://doi.org/10.1103/PhysRevLett.117.182002} {\bibfield  {journal}
  {\bibinfo  {journal} {Phys. Rev. Lett.}\ }\textbf {\bibinfo {volume} {117}},\
  \bibinfo {pages} {182002} (\bibinfo {year} {2016})},\ \Eprint
  {https://arxiv.org/abs/1606.02625} {arXiv:1606.02625 [hep-ex]} \BibitemShut
  {NoStop}%
\bibitem [{\citenamefont {Sirunyan}\ \emph {et~al.}(2018)\citenamefont
  {Sirunyan} \emph {et~al.}}]{CMS:2018mlc}%
  \BibitemOpen
  \bibfield  {author} {\bibinfo {author} {\bibfnamefont {A.~M.}\ \bibnamefont
  {Sirunyan}} \emph {et~al.} (\bibinfo {collaboration} {CMS}),\ }\bibfield
  {title} {\bibinfo {title} {{Measurement of the inelastic proton-proton cross
  section at $ \sqrt{s}=13 $ TeV}},\ }\href
  {https://doi.org/10.1007/JHEP07(2018)161} {\bibfield  {journal} {\bibinfo
  {journal} {JHEP}\ }\textbf {\bibinfo {volume} {07}},\ \bibinfo {pages}
  {161}},\ \Eprint {https://arxiv.org/abs/1802.02613} {arXiv:1802.02613
  [hep-ex]} \BibitemShut {NoStop}%
\bibitem [{\citenamefont {Navas}\ \emph {et~al.}(2024)\citenamefont {Navas}
  \emph {et~al.}}]{ParticleDataGroup:2024cfk}%
  \BibitemOpen
  \bibfield  {author} {\bibinfo {author} {\bibfnamefont {S.}~\bibnamefont
  {Navas}} \emph {et~al.} (\bibinfo {collaboration} {Particle Data Group}),\
  }\bibfield  {title} {\bibinfo {title} {{Review of particle physics}},\ }\href
  {https://doi.org/10.1103/PhysRevD.110.030001} {\bibfield  {journal} {\bibinfo
   {journal} {Phys. Rev. D}\ }\textbf {\bibinfo {volume} {110}},\ \bibinfo
  {pages} {030001} (\bibinfo {year} {2024})}\BibitemShut {NoStop}%
\bibitem [{\citenamefont {Aad}\ \emph {et~al.}(2024)\citenamefont {Aad} \emph
  {et~al.}}]{ATLAS:2024xna}%
  \BibitemOpen
  \bibfield  {author} {\bibinfo {author} {\bibfnamefont {G.}~\bibnamefont
  {Aad}} \emph {et~al.} (\bibinfo {collaboration} {ATLAS}),\ }\bibfield
  {title} {\bibinfo {title} {{The ATLAS trigger system for LHC Run 3 and
  trigger performance in 2022}},\ }\href
  {https://doi.org/10.1088/1748-0221/19/06/P06029} {\bibfield  {journal}
  {\bibinfo  {journal} {JINST}\ }\textbf {\bibinfo {volume} {19}}\bibfield
  {number} {\bibinfo  {number} { (06)},\ \bibinfo {pages} {P06029}},\ }\Eprint
  {https://arxiv.org/abs/2401.06630} {arXiv:2401.06630 [hep-ex]} \BibitemShut
  {NoStop}%
\bibitem [{\citenamefont {Khachatryan}\ \emph {et~al.}(2017)\citenamefont
  {Khachatryan} \emph {et~al.}}]{CMS:2016ngn}%
  \BibitemOpen
  \bibfield  {author} {\bibinfo {author} {\bibfnamefont {V.}~\bibnamefont
  {Khachatryan}} \emph {et~al.} (\bibinfo {collaboration} {CMS}),\ }\bibfield
  {title} {\bibinfo {title} {{The CMS trigger system}},\ }\href
  {https://doi.org/10.1088/1748-0221/12/01/P01020} {\bibfield  {journal}
  {\bibinfo  {journal} {JINST}\ }\textbf {\bibinfo {volume} {12}}\bibfield
  {number} {\bibinfo  {number} { (01)},\ \bibinfo {pages} {P01020}},\ }\Eprint
  {https://arxiv.org/abs/1609.02366} {arXiv:1609.02366 [physics.ins-det]}
  \BibitemShut {NoStop}%
\bibitem [{\citenamefont {{The ATLAS Collaboration}}(2008)}]{ATLAS}%
  \BibitemOpen
  \bibfield  {author} {\bibinfo {author} {\bibnamefont {{The ATLAS
  Collaboration}}},\ }\bibfield  {title} {\bibinfo {title} {{The ATLAS
  Experiment at the CERN Large Hadron Collider}},\ }\href
  {https://doi.org/{10.1088/1748-0221/3/08/S08003}} {\bibfield  {journal}
  {\bibinfo  {journal} {{Journal of Instrumentation}}\ }\textbf {\bibinfo
  {volume} {{3}}},\ \bibinfo {pages} {{S08003}} (\bibinfo {year}
  {{2008}})}\BibitemShut {NoStop}%
\bibitem [{\citenamefont {collaboration}(2008)}]{CMS}%
  \BibitemOpen
  \bibfield  {author} {\bibinfo {author} {\bibfnamefont {T.~C.}\ \bibnamefont
  {collaboration}} (\bibinfo {collaboration} {CMS}),\ }\bibfield  {title}
  {\bibinfo {title} {{The CMS experiment at the CERN LHC. The Compact Muon
  Solenoid experiment}},\ }\href
  {https://doi.org/10.1088/1748-0221/3/08/S08004} {\bibfield  {journal}
  {\bibinfo  {journal} {JINST}\ }\textbf {\bibinfo {volume} {3}},\ \bibinfo
  {pages} {S08004}},\ \bibinfo {note} {also published by CERN Geneva in
  2010}\BibitemShut {NoStop}%
\bibitem [{ATL(2024)}]{ATL-DAPR-PUB-2024-001}%
  \BibitemOpen
  \href {https://cds.cern.ch/record/2900949} {\emph {\bibinfo {title}
  {{Preliminary analysis of the luminosity calibration for the ATLAS 13.6 TeV
  data recorded in 2023}}}},\ \bibinfo {type} {Tech. Rep.}\ (\bibinfo
  {institution} {CERN},\ \bibinfo {address} {Geneva},\ \bibinfo {year}
  {2024})\BibitemShut {NoStop}%
\bibitem [{\citenamefont {Zurbano~Fernandez}\ \emph {et~al.}(2020)\citenamefont
  {Zurbano~Fernandez} \emph {et~al.}}]{ZurbanoFernandez:2020cco}%
  \BibitemOpen
  \bibfield  {author} {\bibinfo {author} {\bibfnamefont {I.}~\bibnamefont
  {Zurbano~Fernandez}} \emph {et~al.},\ }\href
  {https://doi.org/10.23731/CYRM-2020-0010} {\emph {\bibinfo {title}
  {{High-Luminosity Large Hadron Collider (HL-LHC): Technical design
  report}}}},\ edited by\ \bibinfo {editor} {\bibfnamefont {I.}~\bibnamefont
  {B\'ejar~Alonso}}, \bibinfo {editor} {\bibfnamefont {O.}~\bibnamefont
  {Br\"uning}}, \bibinfo {editor} {\bibfnamefont {P.}~\bibnamefont {Fessia}},
  \bibinfo {editor} {\bibfnamefont {L.}~\bibnamefont {Rossi}}, \bibinfo
  {editor} {\bibfnamefont {L.}~\bibnamefont {Tavian}},\ and\ \bibinfo {editor}
  {\bibfnamefont {M.}~\bibnamefont {Zerlauth}},\ Vol.\ \bibinfo {volume}
  {10/2020}\ (\bibinfo {year} {2020})\BibitemShut {NoStop}%
\bibitem [{\citenamefont {Aaboud}\ \emph {et~al.}(2017)\citenamefont {Aaboud}
  \emph {et~al.}}]{ATLAS:2016nnj}%
  \BibitemOpen
  \bibfield  {author} {\bibinfo {author} {\bibfnamefont {M.}~\bibnamefont
  {Aaboud}} \emph {et~al.} (\bibinfo {collaboration} {ATLAS}),\ }\bibfield
  {title} {\bibinfo {title} {{Reconstruction of primary vertices at the ATLAS
  experiment in Run 1 proton\textendash{}proton collisions at the LHC}},\
  }\href {https://doi.org/10.1140/epjc/s10052-017-4887-5} {\bibfield  {journal}
  {\bibinfo  {journal} {Eur. Phys. J. C}\ }\textbf {\bibinfo {volume} {77}},\
  \bibinfo {pages} {332} (\bibinfo {year} {2017})},\ \Eprint
  {https://arxiv.org/abs/1611.10235} {arXiv:1611.10235 [physics.ins-det]}
  \BibitemShut {NoStop}%
\bibitem [{ATL(2012)}]{ATLAS-CONF-2012-042}%
  \BibitemOpen
  \href {https://cds.cern.ch/record/1435196} {\emph {\bibinfo {title}
  {{Performance of the ATLAS Inner Detector Track and Vertex Reconstruction in
  the High Pile-Up LHC Environment}}}},\ \bibinfo {type} {Tech. Rep.}\
  (\bibinfo  {institution} {CERN},\ \bibinfo {address} {Geneva},\ \bibinfo
  {year} {2012})\BibitemShut {NoStop}%
\bibitem [{\citenamefont {Salam}(2010)}]{Salam:2010nqg}%
  \BibitemOpen
  \bibfield  {author} {\bibinfo {author} {\bibfnamefont {G.~P.}\ \bibnamefont
  {Salam}},\ }\bibfield  {title} {\bibinfo {title} {{Towards Jetography}},\
  }\href {https://doi.org/10.1140/epjc/s10052-010-1314-6} {\bibfield  {journal}
  {\bibinfo  {journal} {Eur. Phys. J. C}\ }\textbf {\bibinfo {volume} {67}},\
  \bibinfo {pages} {637} (\bibinfo {year} {2010})},\ \Eprint
  {https://arxiv.org/abs/0906.1833} {arXiv:0906.1833 [hep-ph]} \BibitemShut
  {NoStop}%
\bibitem [{\citenamefont {Cacciari}\ \emph {et~al.}(2015)\citenamefont
  {Cacciari}, \citenamefont {Salam},\ and\ \citenamefont {Soyez}}]{softkiller}%
  \BibitemOpen
  \bibfield  {author} {\bibinfo {author} {\bibfnamefont {M.}~\bibnamefont
  {Cacciari}}, \bibinfo {author} {\bibfnamefont {G.~P.}\ \bibnamefont
  {Salam}},\ and\ \bibinfo {author} {\bibfnamefont {G.}~\bibnamefont {Soyez}},\
  }\bibfield  {title} {\bibinfo {title} {{SoftKiller, a particle-level pileup
  removal method}},\ }\href
  {{http://dx.doi.org/10.1140/epjc/s10052-015-3267-2}} {\bibfield  {journal}
  {\bibinfo  {journal} {{The European Physical Journal C}}\ }\textbf {\bibinfo
  {volume} {{75}}} (\bibinfo {year} {{2015}})}\BibitemShut {NoStop}%
\bibitem [{\citenamefont {Bertolini}\ \emph {et~al.}(2014)\citenamefont
  {Bertolini}, \citenamefont {Harris}, \citenamefont {Low},\ and\ \citenamefont
  {Tran}}]{puppi}%
  \BibitemOpen
  \bibfield  {author} {\bibinfo {author} {\bibfnamefont {D.}~\bibnamefont
  {Bertolini}}, \bibinfo {author} {\bibfnamefont {P.}~\bibnamefont {Harris}},
  \bibinfo {author} {\bibfnamefont {M.}~\bibnamefont {Low}},\ and\ \bibinfo
  {author} {\bibfnamefont {N.}~\bibnamefont {Tran}},\ }\bibfield  {title}
  {\bibinfo {title} {{Pileup Per Particle Identification}},\ }\href
  {https://doi.org/10.1007/JHEP10(2014)059} {\bibfield  {journal} {\bibinfo
  {journal} {JHEP}\ }\textbf {\bibinfo {volume} {10}},\ \bibinfo {pages}
  {059}},\ \Eprint {https://arxiv.org/abs/1407.6013} {arXiv:1407.6013 [hep-ph]}
  \BibitemShut {NoStop}%
\bibitem [{\citenamefont {Larkoski}\ \emph
  {et~al.}(2014{\natexlab{a}})\citenamefont {Larkoski}, \citenamefont
  {Marzani}, \citenamefont {Soyez},\ and\ \citenamefont {Thaler}}]{softdrop}%
  \BibitemOpen
  \bibfield  {author} {\bibinfo {author} {\bibfnamefont {A.~J.}\ \bibnamefont
  {Larkoski}}, \bibinfo {author} {\bibfnamefont {S.}~\bibnamefont {Marzani}},
  \bibinfo {author} {\bibfnamefont {G.}~\bibnamefont {Soyez}},\ and\ \bibinfo
  {author} {\bibfnamefont {J.}~\bibnamefont {Thaler}},\ }\bibfield  {title}
  {\bibinfo {title} {{Soft drop}},\ }\href
  {{http://dx.doi.org/10.1007/JHEP05(2014)146}} {\bibfield  {journal} {\bibinfo
   {journal} {{Journal of High Energy Physics}}\ }\textbf {\bibinfo {volume}
  {{2014}}} (\bibinfo {year} {{2014}}{\natexlab{a}})}\BibitemShut {NoStop}%
\bibitem [{\citenamefont {Soyez}\ \emph {et~al.}(2013)\citenamefont {Soyez},
  \citenamefont {Salam}, \citenamefont {Kim}, \citenamefont {Dutta},\ and\
  \citenamefont {Cacciari}}]{Soyez:2012hv}%
  \BibitemOpen
  \bibfield  {author} {\bibinfo {author} {\bibfnamefont {G.}~\bibnamefont
  {Soyez}}, \bibinfo {author} {\bibfnamefont {G.~P.}\ \bibnamefont {Salam}},
  \bibinfo {author} {\bibfnamefont {J.}~\bibnamefont {Kim}}, \bibinfo {author}
  {\bibfnamefont {S.}~\bibnamefont {Dutta}},\ and\ \bibinfo {author}
  {\bibfnamefont {M.}~\bibnamefont {Cacciari}},\ }\bibfield  {title} {\bibinfo
  {title} {{Pileup subtraction for jet shapes}},\ }\href
  {https://doi.org/10.1103/PhysRevLett.110.162001} {\bibfield  {journal}
  {\bibinfo  {journal} {Phys. Rev. Lett.}\ }\textbf {\bibinfo {volume} {110}},\
  \bibinfo {pages} {162001} (\bibinfo {year} {2013})},\ \Eprint
  {https://arxiv.org/abs/1211.2811} {arXiv:1211.2811 [hep-ph]} \BibitemShut
  {NoStop}%
\bibitem [{\citenamefont {Krohn}\ \emph {et~al.}(2014)\citenamefont {Krohn},
  \citenamefont {Schwartz}, \citenamefont {Low},\ and\ \citenamefont
  {Wang}}]{Krohn:2013lba}%
  \BibitemOpen
  \bibfield  {author} {\bibinfo {author} {\bibfnamefont {D.}~\bibnamefont
  {Krohn}}, \bibinfo {author} {\bibfnamefont {M.~D.}\ \bibnamefont {Schwartz}},
  \bibinfo {author} {\bibfnamefont {M.}~\bibnamefont {Low}},\ and\ \bibinfo
  {author} {\bibfnamefont {L.-T.}\ \bibnamefont {Wang}},\ }\bibfield  {title}
  {\bibinfo {title} {{Jet Cleansing: Pileup Removal at High Luminosity}},\
  }\href {https://doi.org/10.1103/PhysRevD.90.065020} {\bibfield  {journal}
  {\bibinfo  {journal} {Phys. Rev. D}\ }\textbf {\bibinfo {volume} {90}},\
  \bibinfo {pages} {065020} (\bibinfo {year} {2014})},\ \Eprint
  {https://arxiv.org/abs/1309.4777} {arXiv:1309.4777 [hep-ph]} \BibitemShut
  {NoStop}%
\bibitem [{\citenamefont {{ATLAS
  Collaboration}}(2017{\natexlab{a}})}]{ATLAS:2017pfq}%
  \BibitemOpen
  \bibfield  {author} {\bibinfo {author} {\bibnamefont {{ATLAS
  Collaboration}}},\ }\bibfield  {title} {\bibinfo {title} {{Constituent-level
  pile-up mitigation techniques in ATLAS}}} (\bibinfo {year}
  {2017}{\natexlab{a}})\BibitemShut {NoStop}%
\bibitem [{\citenamefont {Sirunyan}\ \emph {et~al.}(2020)\citenamefont
  {Sirunyan} \emph {et~al.}}]{CMS:2020ebo}%
  \BibitemOpen
  \bibfield  {author} {\bibinfo {author} {\bibfnamefont {A.~M.}\ \bibnamefont
  {Sirunyan}} \emph {et~al.} (\bibinfo {collaboration} {CMS}),\ }\bibfield
  {title} {\bibinfo {title} {{Pileup mitigation at CMS in 13 TeV data}},\
  }\href {https://doi.org/10.1088/1748-0221/15/09/P09018} {\bibfield  {journal}
  {\bibinfo  {journal} {JINST}\ }\textbf {\bibinfo {volume} {15}}\bibfield
  {number} {\bibinfo  {number} { (09)},\ \bibinfo {pages} {P09018}},\ }\Eprint
  {https://arxiv.org/abs/2003.00503} {arXiv:2003.00503 [hep-ex]} \BibitemShut
  {NoStop}%
\bibitem [{\citenamefont {{ATLAS Collaboration}}(2016)}]{PERF-2014-03}%
  \BibitemOpen
  \bibfield  {author} {\bibinfo {author} {\bibnamefont {{ATLAS
  Collaboration}}},\ }\bibfield  {title} {\bibinfo {title} {{Performance of
  pile-up mitigation techniques for jets in \(pp\) collisions at \(\sqrt{s} =
  8\,\text{TeV}\) using the ATLAS detector}},\ }\href
  {https://doi.org/10.1140/epjc/s10052-016-4395-z} {\bibfield  {journal}
  {\bibinfo  {journal} {Eur. Phys. J. C}\ }\textbf {\bibinfo {volume} {76}},\
  \bibinfo {pages} {581} (\bibinfo {year} {2016})},\ \Eprint
  {https://arxiv.org/abs/1510.03823} {arXiv:1510.03823 [hep-ex]} \BibitemShut
  {NoStop}%
\bibitem [{\citenamefont {{ATLAS
  Collaboration}}(2017{\natexlab{b}})}]{ATLAS:2017ywy}%
  \BibitemOpen
  \bibfield  {author} {\bibinfo {author} {\bibnamefont {{ATLAS
  Collaboration}}},\ }\bibfield  {title} {\bibinfo {title} {{Identification and
  rejection of pile-up jets at high pseudorapidity with the ATLAS detector}},\
  }\href {https://doi.org/10.1140/epjc/s10052-017-5081-5} {\bibfield  {journal}
  {\bibinfo  {journal} {Eur. Phys. J. C}\ }\textbf {\bibinfo {volume} {77}},\
  \bibinfo {pages} {580} (\bibinfo {year} {2017}{\natexlab{b}})},\ \bibinfo
  {note} {[Erratum: Eur.Phys.J.C 77, 712 (2017)]},\ \Eprint
  {https://arxiv.org/abs/1705.02211} {arXiv:1705.02211 [hep-ex]} \BibitemShut
  {NoStop}%
\bibitem [{\citenamefont {Berta}\ \emph {et~al.}(2014)\citenamefont {Berta},
  \citenamefont {Spousta}, \citenamefont {Miller},\ and\ \citenamefont
  {Leitner}}]{Berta:2014eza}%
  \BibitemOpen
  \bibfield  {author} {\bibinfo {author} {\bibfnamefont {P.}~\bibnamefont
  {Berta}}, \bibinfo {author} {\bibfnamefont {M.}~\bibnamefont {Spousta}},
  \bibinfo {author} {\bibfnamefont {D.~W.}\ \bibnamefont {Miller}},\ and\
  \bibinfo {author} {\bibfnamefont {R.}~\bibnamefont {Leitner}},\ }\bibfield
  {title} {\bibinfo {title} {{Particle-level pileup subtraction for jets and
  jet shapes}},\ }\href {https://doi.org/10.1007/JHEP06(2014)092} {\bibfield
  {journal} {\bibinfo  {journal} {JHEP}\ }\textbf {\bibinfo {volume} {06}},\
  \bibinfo {pages} {092}},\ \Eprint {https://arxiv.org/abs/1403.3108}
  {arXiv:1403.3108 [hep-ex]} \BibitemShut {NoStop}%
\bibitem [{\citenamefont {Komiske}\ \emph {et~al.}(2017)\citenamefont
  {Komiske}, \citenamefont {Metodiev}, \citenamefont {Nachman},\ and\
  \citenamefont {Schwartz}}]{pumml}%
  \BibitemOpen
  \bibfield  {author} {\bibinfo {author} {\bibfnamefont {P.~T.}\ \bibnamefont
  {Komiske}}, \bibinfo {author} {\bibfnamefont {E.~M.}\ \bibnamefont
  {Metodiev}}, \bibinfo {author} {\bibfnamefont {B.}~\bibnamefont {Nachman}},\
  and\ \bibinfo {author} {\bibfnamefont {M.~D.}\ \bibnamefont {Schwartz}},\
  }\bibfield  {title} {\bibinfo {title} {{Pileup Mitigation with Machine
  Learning (PUMML)}},\ }\href {https://doi.org/10.1007/JHEP12(2017)051}
  {\bibfield  {journal} {\bibinfo  {journal} {JHEP}\ }\textbf {\bibinfo
  {volume} {12}},\ \bibinfo {pages} {051}},\ \Eprint
  {https://arxiv.org/abs/1707.08600} {arXiv:1707.08600 [hep-ph]} \BibitemShut
  {NoStop}%
\bibitem [{\citenamefont {Arjona~Mart\'\i{}nez}\ \emph
  {et~al.}(2019)\citenamefont {Arjona~Mart\'\i{}nez}, \citenamefont {Cerri},
  \citenamefont {Pierini}, \citenamefont {Spiropulu},\ and\ \citenamefont
  {Vlimant}}]{puppiml}%
  \BibitemOpen
  \bibfield  {author} {\bibinfo {author} {\bibfnamefont {J.}~\bibnamefont
  {Arjona~Mart\'\i{}nez}}, \bibinfo {author} {\bibfnamefont {O.}~\bibnamefont
  {Cerri}}, \bibinfo {author} {\bibfnamefont {M.}~\bibnamefont {Pierini}},
  \bibinfo {author} {\bibfnamefont {M.}~\bibnamefont {Spiropulu}},\ and\
  \bibinfo {author} {\bibfnamefont {J.-R.}\ \bibnamefont {Vlimant}},\
  }\bibfield  {title} {\bibinfo {title} {{Pileup mitigation at the Large Hadron
  Collider with graph neural networks}},\ }\href
  {https://doi.org/10.1140/epjp/i2019-12710-3} {\bibfield  {journal} {\bibinfo
  {journal} {Eur. Phys. J. Plus}\ }\textbf {\bibinfo {volume} {134}},\ \bibinfo
  {pages} {333} (\bibinfo {year} {2019})},\ \Eprint
  {https://arxiv.org/abs/1810.07988} {arXiv:1810.07988 [hep-ph]} \BibitemShut
  {NoStop}%
\bibitem [{\citenamefont {Carrazza}\ and\ \citenamefont
  {Dreyer}(2019)}]{Carrazza:2019efs}%
  \BibitemOpen
  \bibfield  {author} {\bibinfo {author} {\bibfnamefont {S.}~\bibnamefont
  {Carrazza}}\ and\ \bibinfo {author} {\bibfnamefont {F.~A.}\ \bibnamefont
  {Dreyer}},\ }\bibfield  {title} {\bibinfo {title} {{Jet grooming through
  reinforcement learning}},\ }\href
  {https://doi.org/10.1103/PhysRevD.100.014014} {\bibfield  {journal} {\bibinfo
   {journal} {Phys. Rev. D}\ }\textbf {\bibinfo {volume} {100}},\ \bibinfo
  {pages} {014014} (\bibinfo {year} {2019})},\ \Eprint
  {https://arxiv.org/abs/1903.09644} {arXiv:1903.09644 [hep-ph]} \BibitemShut
  {NoStop}%
\bibitem [{\citenamefont {Maier}\ \emph {et~al.}(2022)\citenamefont {Maier},
  \citenamefont {Narayanan}, \citenamefont {de~Castro}, \citenamefont
  {Goncharov}, \citenamefont {Paus},\ and\ \citenamefont
  {Schott}}]{Maier:2021ymx}%
  \BibitemOpen
  \bibfield  {author} {\bibinfo {author} {\bibfnamefont {B.}~\bibnamefont
  {Maier}}, \bibinfo {author} {\bibfnamefont {S.~M.}\ \bibnamefont
  {Narayanan}}, \bibinfo {author} {\bibfnamefont {G.}~\bibnamefont
  {de~Castro}}, \bibinfo {author} {\bibfnamefont {M.}~\bibnamefont
  {Goncharov}}, \bibinfo {author} {\bibfnamefont {C.}~\bibnamefont {Paus}},\
  and\ \bibinfo {author} {\bibfnamefont {M.}~\bibnamefont {Schott}},\
  }\bibfield  {title} {\bibinfo {title} {{Pile-up mitigation using
  attention}},\ }\href {https://doi.org/10.1088/2632-2153/ac7198} {\bibfield
  {journal} {\bibinfo  {journal} {Mach. Learn. Sci. Tech.}\ }\textbf {\bibinfo
  {volume} {3}},\ \bibinfo {pages} {025012} (\bibinfo {year} {2022})},\ \Eprint
  {https://arxiv.org/abs/2107.02779} {arXiv:2107.02779 [physics.ins-det]}
  \BibitemShut {NoStop}%
\bibitem [{\citenamefont {Li}\ \emph {et~al.}(2023)\citenamefont {Li},
  \citenamefont {Liu}, \citenamefont {Feng}, \citenamefont {Paspalaki},
  \citenamefont {Tran}, \citenamefont {Liu},\ and\ \citenamefont
  {Li}}]{theonemaltefound}%
  \BibitemOpen
  \bibfield  {author} {\bibinfo {author} {\bibfnamefont {T.}~\bibnamefont
  {Li}}, \bibinfo {author} {\bibfnamefont {S.}~\bibnamefont {Liu}}, \bibinfo
  {author} {\bibfnamefont {Y.}~\bibnamefont {Feng}}, \bibinfo {author}
  {\bibfnamefont {G.}~\bibnamefont {Paspalaki}}, \bibinfo {author}
  {\bibfnamefont {N.~V.}\ \bibnamefont {Tran}}, \bibinfo {author}
  {\bibfnamefont {M.}~\bibnamefont {Liu}},\ and\ \bibinfo {author}
  {\bibfnamefont {P.}~\bibnamefont {Li}},\ }\bibfield  {title} {\bibinfo
  {title} {{Semi-supervised graph neural networks for pileup noise removal}},\
  }\href {https://doi.org/10.1140/epjc/s10052-022-11083-5} {\bibfield
  {journal} {\bibinfo  {journal} {Eur. Phys. J. C}\ }\textbf {\bibinfo {volume}
  {83}},\ \bibinfo {pages} {99} (\bibinfo {year} {2023})},\ \Eprint
  {https://arxiv.org/abs/2203.15823} {arXiv:2203.15823 [hep-ex]} \BibitemShut
  {NoStop}%
\bibitem [{\citenamefont {Kim}\ \emph {et~al.}(2023)\citenamefont {Kim},
  \citenamefont {Ahn}, \citenamefont {Chae}, \citenamefont {Hooker},\ and\
  \citenamefont {Rogachev}}]{Kim:2023koz}%
  \BibitemOpen
  \bibfield  {author} {\bibinfo {author} {\bibfnamefont {C.~H.}\ \bibnamefont
  {Kim}}, \bibinfo {author} {\bibfnamefont {S.}~\bibnamefont {Ahn}}, \bibinfo
  {author} {\bibfnamefont {K.~Y.}\ \bibnamefont {Chae}}, \bibinfo {author}
  {\bibfnamefont {J.}~\bibnamefont {Hooker}},\ and\ \bibinfo {author}
  {\bibfnamefont {G.~V.}\ \bibnamefont {Rogachev}},\ }\bibfield  {title}
  {\bibinfo {title} {{Restoring original signals from pile-up using deep
  learning}},\ }\href {https://doi.org/10.1016/j.nima.2023.168492} {\bibfield
  {journal} {\bibinfo  {journal} {Nucl. Instrum. Meth. A}\ }\textbf {\bibinfo
  {volume} {1055}},\ \bibinfo {pages} {168492} (\bibinfo {year} {2023})},\
  \Eprint {https://arxiv.org/abs/2304.14496} {arXiv:2304.14496
  [physics.ins-det]} \BibitemShut {NoStop}%
\bibitem [{\citenamefont {Seymour}(1994)}]{Seymour:1993mx}%
  \BibitemOpen
  \bibfield  {author} {\bibinfo {author} {\bibfnamefont {M.~H.}\ \bibnamefont
  {Seymour}},\ }\bibfield  {title} {\bibinfo {title} {{Searches for new
  particles using cone and cluster jet algorithms: A Comparative study}},\
  }\href {https://doi.org/10.1007/BF01559532} {\bibfield  {journal} {\bibinfo
  {journal} {Z. Phys. C}\ }\textbf {\bibinfo {volume} {62}},\ \bibinfo {pages}
  {127} (\bibinfo {year} {1994})}\BibitemShut {NoStop}%
\bibitem [{\citenamefont {Butterworth}\ \emph {et~al.}(2008)\citenamefont
  {Butterworth}, \citenamefont {Davison}, \citenamefont {Rubin},\ and\
  \citenamefont {Salam}}]{Butterworth:2008iy}%
  \BibitemOpen
  \bibfield  {author} {\bibinfo {author} {\bibfnamefont {J.~M.}\ \bibnamefont
  {Butterworth}}, \bibinfo {author} {\bibfnamefont {A.~R.}\ \bibnamefont
  {Davison}}, \bibinfo {author} {\bibfnamefont {M.}~\bibnamefont {Rubin}},\
  and\ \bibinfo {author} {\bibfnamefont {G.~P.}\ \bibnamefont {Salam}},\
  }\bibfield  {title} {\bibinfo {title} {{Jet substructure as a new Higgs
  search channel at the LHC}},\ }\href
  {https://doi.org/10.1103/PhysRevLett.100.242001} {\bibfield  {journal}
  {\bibinfo  {journal} {Phys. Rev. Lett.}\ }\textbf {\bibinfo {volume} {100}},\
  \bibinfo {pages} {242001} (\bibinfo {year} {2008})},\ \Eprint
  {https://arxiv.org/abs/0802.2470} {arXiv:0802.2470 [hep-ph]} \BibitemShut
  {NoStop}%
\bibitem [{\citenamefont {Kaplan}\ \emph {et~al.}(2008)\citenamefont {Kaplan},
  \citenamefont {Rehermann}, \citenamefont {Schwartz},\ and\ \citenamefont
  {Tweedie}}]{Kaplan:2008ie}%
  \BibitemOpen
  \bibfield  {author} {\bibinfo {author} {\bibfnamefont {D.~E.}\ \bibnamefont
  {Kaplan}}, \bibinfo {author} {\bibfnamefont {K.}~\bibnamefont {Rehermann}},
  \bibinfo {author} {\bibfnamefont {M.~D.}\ \bibnamefont {Schwartz}},\ and\
  \bibinfo {author} {\bibfnamefont {B.}~\bibnamefont {Tweedie}},\ }\bibfield
  {title} {\bibinfo {title} {{Top Tagging: A Method for Identifying Boosted
  Hadronically Decaying Top Quarks}},\ }\href
  {https://doi.org/10.1103/PhysRevLett.101.142001} {\bibfield  {journal}
  {\bibinfo  {journal} {Phys. Rev. Lett.}\ }\textbf {\bibinfo {volume} {101}},\
  \bibinfo {pages} {142001} (\bibinfo {year} {2008})},\ \Eprint
  {https://arxiv.org/abs/0806.0848} {arXiv:0806.0848 [hep-ph]} \BibitemShut
  {NoStop}%
\bibitem [{\citenamefont {Aad}\ \emph {et~al.}(2012)\citenamefont {Aad} \emph
  {et~al.}}]{ATLAS:2012muc}%
  \BibitemOpen
  \bibfield  {author} {\bibinfo {author} {\bibfnamefont {G.}~\bibnamefont
  {Aad}} \emph {et~al.} (\bibinfo {collaboration} {ATLAS}),\ }\bibfield
  {title} {\bibinfo {title} {{A search for $t\bar{t}$ resonances in lepton+jets
  events with highly boosted top quarks collected in $pp$ collisions at
  $\sqrt{s} = 7$ TeV with the ATLAS detector}},\ }\href
  {https://doi.org/10.1007/JHEP09(2012)041} {\bibfield  {journal} {\bibinfo
  {journal} {JHEP}\ }\textbf {\bibinfo {volume} {09}},\ \bibinfo {pages}
  {041}},\ \Eprint {https://arxiv.org/abs/1207.2409} {arXiv:1207.2409 [hep-ex]}
  \BibitemShut {NoStop}%
\bibitem [{\citenamefont {Collaboration}(2024)}]{CMS:2024ddc}%
  \BibitemOpen
  \bibfield  {author} {\bibinfo {author} {\bibfnamefont {C.}~\bibnamefont
  {Collaboration}},\ }\href {https://arxiv.org/abs/2407.08012} {\bibinfo
  {title} {Measurement of boosted higgs bosons produced via vector boson fusion
  or gluon fusion in the h $\to$ $\mathrm{b\bar{b}}$ decay mode using lhc
  proton-proton collision data at $\sqrt{s}$ = 13 tev}} (\bibinfo {year}
  {2024}),\ \Eprint {https://arxiv.org/abs/2407.08012} {arXiv:2407.08012
  [hep-ex]} \BibitemShut {NoStop}%
\bibitem [{\citenamefont {Larkoski}\ \emph
  {et~al.}(2013{\natexlab{a}})\citenamefont {Larkoski}, \citenamefont {Salam},\
  and\ \citenamefont {Thaler}}]{Larkoski:2013eya}%
  \BibitemOpen
  \bibfield  {author} {\bibinfo {author} {\bibfnamefont {A.~J.}\ \bibnamefont
  {Larkoski}}, \bibinfo {author} {\bibfnamefont {G.~P.}\ \bibnamefont
  {Salam}},\ and\ \bibinfo {author} {\bibfnamefont {J.}~\bibnamefont
  {Thaler}},\ }\bibfield  {title} {\bibinfo {title} {{Energy Correlation
  Functions for Jet Substructure}},\ }\href
  {https://doi.org/10.1007/JHEP06(2013)108} {\bibfield  {journal} {\bibinfo
  {journal} {JHEP}\ }\textbf {\bibinfo {volume} {06}},\ \bibinfo {pages}
  {108}},\ \Eprint {https://arxiv.org/abs/1305.0007} {arXiv:1305.0007 [hep-ph]}
  \BibitemShut {NoStop}%
\bibitem [{\citenamefont {Aad}\ \emph {et~al.}(2021)\citenamefont {Aad} \emph
  {et~al.}}]{ATLAS:2020gwe}%
  \BibitemOpen
  \bibfield  {author} {\bibinfo {author} {\bibfnamefont {G.}~\bibnamefont
  {Aad}} \emph {et~al.} (\bibinfo {collaboration} {ATLAS}),\ }\bibfield
  {title} {\bibinfo {title} {{Optimisation of large-radius jet reconstruction
  for the ATLAS detector in 13 TeV proton\textendash{}proton collisions}},\
  }\href {https://doi.org/10.1140/epjc/s10052-021-09054-3} {\bibfield
  {journal} {\bibinfo  {journal} {Eur. Phys. J. C}\ }\textbf {\bibinfo {volume}
  {81}},\ \bibinfo {pages} {334} (\bibinfo {year} {2021})},\ \Eprint
  {https://arxiv.org/abs/2009.04986} {arXiv:2009.04986 [hep-ex]} \BibitemShut
  {NoStop}%
\bibitem [{\citenamefont {Song}\ \emph {et~al.}(2021)\citenamefont {Song} \emph
  {et~al.}}]{Song2020}%
  \BibitemOpen
  \bibfield  {author} {\bibinfo {author} {\bibfnamefont {Y.}~\bibnamefont
  {Song}} \emph {et~al.},\ }\bibfield  {title} {\bibinfo {title} {Score-based
  generative modeling through stochastic differential equations},\ }in\
  \href@noop {} {\emph {\bibinfo {booktitle} {Proceedings of the International
  Conference on Learning Representations}}}\ (\bibinfo {year} {2021})\ \Eprint
  {https://arxiv.org/abs/2011.13456} {2011.13456} \BibitemShut {NoStop}%
\bibitem [{\citenamefont {Yang}\ \emph {et~al.}(2023)\citenamefont {Yang},
  \citenamefont {Zhang}, \citenamefont {Song}, \citenamefont {Hong},
  \citenamefont {Xu}, \citenamefont {Zhao}, \citenamefont {Zhang},
  \citenamefont {Cui},\ and\ \citenamefont {Yang}}]{yang2023diffusion}%
  \BibitemOpen
  \bibfield  {author} {\bibinfo {author} {\bibfnamefont {L.}~\bibnamefont
  {Yang}}, \bibinfo {author} {\bibfnamefont {Z.}~\bibnamefont {Zhang}},
  \bibinfo {author} {\bibfnamefont {Y.}~\bibnamefont {Song}}, \bibinfo {author}
  {\bibfnamefont {S.}~\bibnamefont {Hong}}, \bibinfo {author} {\bibfnamefont
  {R.}~\bibnamefont {Xu}}, \bibinfo {author} {\bibfnamefont {Y.}~\bibnamefont
  {Zhao}}, \bibinfo {author} {\bibfnamefont {W.}~\bibnamefont {Zhang}},
  \bibinfo {author} {\bibfnamefont {B.}~\bibnamefont {Cui}},\ and\ \bibinfo
  {author} {\bibfnamefont {M.-H.}\ \bibnamefont {Yang}},\ }\bibfield  {title}
  {\bibinfo {title} {Diffusion models: A comprehensive survey of methods and
  applications},\ }\Eprint {https://arxiv.org/abs/2209.00796} {arXiv:2209.00796
  [cs.LG]}  (\bibinfo {year} {2023})\BibitemShut {NoStop}%
\bibitem [{\citenamefont {Karras}\ \emph {et~al.}(2022)\citenamefont {Karras}
  \emph {et~al.}}]{EDM}%
  \BibitemOpen
  \bibfield  {author} {\bibinfo {author} {\bibfnamefont {T.}~\bibnamefont
  {Karras}} \emph {et~al.},\ }\bibfield  {title} {\bibinfo {title} {Elucidating
  the design space of diffusion-based generative models},\ }in\ \href@noop {}
  {\emph {\bibinfo {booktitle} {Proceedings of Advances in Neural Information
  Processing Systems}}}\ (\bibinfo {year} {2022})\ \Eprint
  {https://arxiv.org/abs/2206.00364} {2206.00364} \BibitemShut {NoStop}%
\bibitem [{\citenamefont {Mikuni}\ and\ \citenamefont
  {Nachman}(2022)}]{CaloScore}%
  \BibitemOpen
  \bibfield  {author} {\bibinfo {author} {\bibfnamefont {V.}~\bibnamefont
  {Mikuni}}\ and\ \bibinfo {author} {\bibfnamefont {B.}~\bibnamefont
  {Nachman}},\ }\bibfield  {title} {\bibinfo {title} {Score-based generative
  models for calorimeter shower simulation},\ }\href
  {https://doi.org/10.1103/PhysRevD.106.092009} {\bibfield  {journal} {\bibinfo
   {journal} {Phys. Rev. D}\ }\textbf {\bibinfo {volume} {106}},\ \bibinfo
  {pages} {092009} (\bibinfo {year} {2022})}\BibitemShut {NoStop}%
\bibitem [{\citenamefont {Buhmann}\ \emph
  {et~al.}(2023{\natexlab{a}})\citenamefont {Buhmann}, \citenamefont
  {Diefenbacher}, \citenamefont {Eren}, \citenamefont {Gaede}, \citenamefont
  {Kasieczka}, \citenamefont {Korol}, \citenamefont {Korcari}, \citenamefont
  {Kr\"uger},\ and\ \citenamefont {McKeown}}]{CaloClouds}%
  \BibitemOpen
  \bibfield  {author} {\bibinfo {author} {\bibfnamefont {E.}~\bibnamefont
  {Buhmann}}, \bibinfo {author} {\bibfnamefont {S.}~\bibnamefont
  {Diefenbacher}}, \bibinfo {author} {\bibfnamefont {E.}~\bibnamefont {Eren}},
  \bibinfo {author} {\bibfnamefont {F.}~\bibnamefont {Gaede}}, \bibinfo
  {author} {\bibfnamefont {G.}~\bibnamefont {Kasieczka}}, \bibinfo {author}
  {\bibfnamefont {A.}~\bibnamefont {Korol}}, \bibinfo {author} {\bibfnamefont
  {W.}~\bibnamefont {Korcari}}, \bibinfo {author} {\bibfnamefont
  {K.}~\bibnamefont {Kr\"uger}},\ and\ \bibinfo {author} {\bibfnamefont
  {P.}~\bibnamefont {McKeown}},\ }\bibfield  {title} {\bibinfo {title}
  {{CaloClouds: Fast Geometry-Independent Highly-Granular Calorimeter
  Simulation}},\ }\Eprint {https://arxiv.org/abs/2305.04847} {arXiv:2305.04847
  [physics.ins-det]}  (\bibinfo {year} {2023}{\natexlab{a}})\BibitemShut
  {NoStop}%
\bibitem [{\citenamefont {Acosta}\ \emph {et~al.}(2023)\citenamefont {Acosta},
  \citenamefont {Mikuni}, \citenamefont {Nachman}, \citenamefont {Arratia},
  \citenamefont {Barish}, \citenamefont {Karki}, \citenamefont {Milton},
  \citenamefont {Karande},\ and\ \citenamefont {Angerami}}]{Acosta:2023zik}%
  \BibitemOpen
  \bibfield  {author} {\bibinfo {author} {\bibfnamefont {F.~T.}\ \bibnamefont
  {Acosta}}, \bibinfo {author} {\bibfnamefont {V.}~\bibnamefont {Mikuni}},
  \bibinfo {author} {\bibfnamefont {B.}~\bibnamefont {Nachman}}, \bibinfo
  {author} {\bibfnamefont {M.}~\bibnamefont {Arratia}}, \bibinfo {author}
  {\bibfnamefont {K.}~\bibnamefont {Barish}}, \bibinfo {author} {\bibfnamefont
  {B.}~\bibnamefont {Karki}}, \bibinfo {author} {\bibfnamefont
  {R.}~\bibnamefont {Milton}}, \bibinfo {author} {\bibfnamefont
  {P.}~\bibnamefont {Karande}},\ and\ \bibinfo {author} {\bibfnamefont
  {A.}~\bibnamefont {Angerami}},\ }\bibfield  {title} {\bibinfo {title}
  {{Comparison of Point Cloud and Image-based Models for Calorimeter Fast
  Simulation}},\ }\Eprint {https://arxiv.org/abs/2307.04780} {arXiv:2307.04780
  [cs.LG]}  (\bibinfo {year} {2023})\BibitemShut {NoStop}%
\bibitem [{\citenamefont {Leigh}\ \emph
  {et~al.}(2023{\natexlab{a}})\citenamefont {Leigh}, \citenamefont {Sengupta},
  \citenamefont {Quétant}, \citenamefont {Raine}, \citenamefont {Zoch},\ and\
  \citenamefont {Golling}}]{pcjedi}%
  \BibitemOpen
  \bibfield  {author} {\bibinfo {author} {\bibfnamefont {M.}~\bibnamefont
  {Leigh}}, \bibinfo {author} {\bibfnamefont {D.}~\bibnamefont {Sengupta}},
  \bibinfo {author} {\bibfnamefont {G.}~\bibnamefont {Quétant}}, \bibinfo
  {author} {\bibfnamefont {J.~A.}\ \bibnamefont {Raine}}, \bibinfo {author}
  {\bibfnamefont {K.}~\bibnamefont {Zoch}},\ and\ \bibinfo {author}
  {\bibfnamefont {T.}~\bibnamefont {Golling}},\ }\bibfield  {title} {\bibinfo
  {title} {{PC-JeDi: Diffusion for Particle Cloud Generation in High Energy
  Physics}},\ }\href {https://doi.org/10.21468/SciPostPhys.16.1.018} {\bibfield
   {journal} {\bibinfo  {journal} {SciPost Phys.}\ }\textbf {\bibinfo {volume}
  {16}},\ \bibinfo {pages} {018} (\bibinfo {year} {2023}{\natexlab{a}})},\
  \Eprint {https://arxiv.org/abs/2303.05376} {arXiv:2303.05376 [hep-ph]}
  \BibitemShut {NoStop}%
\bibitem [{\citenamefont {Mikuni}\ \emph {et~al.}(2023)\citenamefont {Mikuni},
  \citenamefont {Nachman},\ and\ \citenamefont {Pettee}}]{FPCD}%
  \BibitemOpen
  \bibfield  {author} {\bibinfo {author} {\bibfnamefont {V.}~\bibnamefont
  {Mikuni}}, \bibinfo {author} {\bibfnamefont {B.}~\bibnamefont {Nachman}},\
  and\ \bibinfo {author} {\bibfnamefont {M.}~\bibnamefont {Pettee}},\
  }\bibfield  {title} {\bibinfo {title} {{Fast Point Cloud Generation with
  Diffusion Models in High Energy Physics}},\ }\Eprint
  {https://arxiv.org/abs/2304.01266} {arXiv:2304.01266 [hep-ph]}  (\bibinfo
  {year} {2023})\BibitemShut {NoStop}%
\bibitem [{\citenamefont {Butter}\ \emph {et~al.}(2023)\citenamefont {Butter},
  \citenamefont {Huetsch}, \citenamefont {Schweitzer}, \citenamefont {Plehn},
  \citenamefont {Sorrenson},\ and\ \citenamefont {Spinner}}]{Butter:2023fov}%
  \BibitemOpen
  \bibfield  {author} {\bibinfo {author} {\bibfnamefont {A.}~\bibnamefont
  {Butter}}, \bibinfo {author} {\bibfnamefont {N.}~\bibnamefont {Huetsch}},
  \bibinfo {author} {\bibfnamefont {S.~P.}\ \bibnamefont {Schweitzer}},
  \bibinfo {author} {\bibfnamefont {T.}~\bibnamefont {Plehn}}, \bibinfo
  {author} {\bibfnamefont {P.}~\bibnamefont {Sorrenson}},\ and\ \bibinfo
  {author} {\bibfnamefont {J.}~\bibnamefont {Spinner}},\ }\bibfield  {title}
  {\bibinfo {title} {{Jet Diffusion versus JetGPT -- Modern Networks for the
  LHC}},\ }\Eprint {https://arxiv.org/abs/2305.10475} {arXiv:2305.10475
  [hep-ph]}  (\bibinfo {year} {2023})\BibitemShut {NoStop}%
\bibitem [{\citenamefont {Shmakov}\ \emph {et~al.}(2023)\citenamefont
  {Shmakov}, \citenamefont {Greif}, \citenamefont {Fenton}, \citenamefont
  {Ghosh}, \citenamefont {Baldi},\ and\ \citenamefont
  {Whiteson}}]{Shmakov:2023kjj}%
  \BibitemOpen
  \bibfield  {author} {\bibinfo {author} {\bibfnamefont {A.}~\bibnamefont
  {Shmakov}}, \bibinfo {author} {\bibfnamefont {K.}~\bibnamefont {Greif}},
  \bibinfo {author} {\bibfnamefont {M.}~\bibnamefont {Fenton}}, \bibinfo
  {author} {\bibfnamefont {A.}~\bibnamefont {Ghosh}}, \bibinfo {author}
  {\bibfnamefont {P.}~\bibnamefont {Baldi}},\ and\ \bibinfo {author}
  {\bibfnamefont {D.}~\bibnamefont {Whiteson}},\ }\bibfield  {title} {\bibinfo
  {title} {{End-To-End Latent Variational Diffusion Models for Inverse Problems
  in High Energy Physics}},\ }\Eprint {https://arxiv.org/abs/2305.10399}
  {arXiv:2305.10399 [hep-ex]}  (\bibinfo {year} {2023})\BibitemShut {NoStop}%
\bibitem [{\citenamefont {Mikuni}\ and\ \citenamefont
  {Nachman}(2023)}]{Mikuni:2023tok}%
  \BibitemOpen
  \bibfield  {author} {\bibinfo {author} {\bibfnamefont {V.}~\bibnamefont
  {Mikuni}}\ and\ \bibinfo {author} {\bibfnamefont {B.}~\bibnamefont
  {Nachman}},\ }\bibfield  {title} {\bibinfo {title} {{High-dimensional and
  Permutation Invariant Anomaly Detection}},\ }\Eprint
  {https://arxiv.org/abs/2306.03933} {arXiv:2306.03933 [hep-ph]}  (\bibinfo
  {year} {2023})\BibitemShut {NoStop}%
\bibitem [{\citenamefont {Leigh}\ \emph
  {et~al.}(2023{\natexlab{b}})\citenamefont {Leigh}, \citenamefont {Sengupta},
  \citenamefont {Raine}, \citenamefont {Quétant},\ and\ \citenamefont
  {Golling}}]{pcdroid}%
  \BibitemOpen
  \bibfield  {author} {\bibinfo {author} {\bibfnamefont {M.}~\bibnamefont
  {Leigh}}, \bibinfo {author} {\bibfnamefont {D.}~\bibnamefont {Sengupta}},
  \bibinfo {author} {\bibfnamefont {J.~A.}\ \bibnamefont {Raine}}, \bibinfo
  {author} {\bibfnamefont {G.}~\bibnamefont {Quétant}},\ and\ \bibinfo
  {author} {\bibfnamefont {T.}~\bibnamefont {Golling}},\ }\bibfield  {title}
  {\bibinfo {title} {{Faster diffusion model with improved quality for particle
  cloud generation}},\ }\bibfield  {journal} {\bibinfo  {journal} {Phys.Rev.D}\
  }\textbf {\bibinfo {volume} {109}},\ \href
  {https://doi.org/10.1103/PhysRevD.109.012010} {10.1103/PhysRevD.109.012010}
  (\bibinfo {year} {2023}{\natexlab{b}}),\ \Eprint
  {https://arxiv.org/abs/2307.06836} {arXiv:2307.06836 [hep-ex]} \BibitemShut
  {NoStop}%
\bibitem [{\citenamefont {Buhmann}\ \emph
  {et~al.}(2023{\natexlab{b}})\citenamefont {Buhmann}, \citenamefont {Ewen},
  \citenamefont {Faroughy}, \citenamefont {Golling}, \citenamefont {Kasieczka},
  \citenamefont {Leigh}, \citenamefont {Quétant}, \citenamefont {Raine},
  \citenamefont {Sengupta},\ and\ \citenamefont {Shih}}]{epicjedi}%
  \BibitemOpen
  \bibfield  {author} {\bibinfo {author} {\bibfnamefont {E.}~\bibnamefont
  {Buhmann}}, \bibinfo {author} {\bibfnamefont {C.}~\bibnamefont {Ewen}},
  \bibinfo {author} {\bibfnamefont {D.~A.}\ \bibnamefont {Faroughy}}, \bibinfo
  {author} {\bibfnamefont {T.}~\bibnamefont {Golling}}, \bibinfo {author}
  {\bibfnamefont {G.}~\bibnamefont {Kasieczka}}, \bibinfo {author}
  {\bibfnamefont {M.}~\bibnamefont {Leigh}}, \bibinfo {author} {\bibfnamefont
  {G.}~\bibnamefont {Quétant}}, \bibinfo {author} {\bibfnamefont {J.~A.}\
  \bibnamefont {Raine}}, \bibinfo {author} {\bibfnamefont {D.}~\bibnamefont
  {Sengupta}},\ and\ \bibinfo {author} {\bibfnamefont {D.}~\bibnamefont
  {Shih}},\ }\bibfield  {title} {\bibinfo {title} {{EPiC-ly Fast Particle Cloud
  Generation with Flow-Matching and Diffusion}},\ }\Eprint
  {https://arxiv.org/abs/2310.00049} {arXiv:2310.00049 [hep-ph]}  (\bibinfo
  {year} {2023}{\natexlab{b}})\BibitemShut {NoStop}%
\bibitem [{\citenamefont {Heimel}\ \emph {et~al.}(2023)\citenamefont {Heimel},
  \citenamefont {Huetsch}, \citenamefont {Maltoni}, \citenamefont {Mattelaer},
  \citenamefont {Plehn},\ and\ \citenamefont {Winterhalder}}]{Heimel:2023ngj}%
  \BibitemOpen
  \bibfield  {author} {\bibinfo {author} {\bibfnamefont {T.}~\bibnamefont
  {Heimel}}, \bibinfo {author} {\bibfnamefont {N.}~\bibnamefont {Huetsch}},
  \bibinfo {author} {\bibfnamefont {F.}~\bibnamefont {Maltoni}}, \bibinfo
  {author} {\bibfnamefont {O.}~\bibnamefont {Mattelaer}}, \bibinfo {author}
  {\bibfnamefont {T.}~\bibnamefont {Plehn}},\ and\ \bibinfo {author}
  {\bibfnamefont {R.}~\bibnamefont {Winterhalder}},\ }\bibfield  {title}
  {\bibinfo {title} {{The MadNIS Reloaded}},\ }\Eprint
  {https://arxiv.org/abs/2311.01548} {arXiv:2311.01548 [hep-ph]}  (\bibinfo
  {year} {2023})\BibitemShut {NoStop}%
\bibitem [{\citenamefont {Buhmann}\ \emph
  {et~al.}(2023{\natexlab{c}})\citenamefont {Buhmann}, \citenamefont {Ewen},
  \citenamefont {Kasieczka}, \citenamefont {Mikuni}, \citenamefont {Nachman},\
  and\ \citenamefont {Shih}}]{Buhmann:2023acn}%
  \BibitemOpen
  \bibfield  {author} {\bibinfo {author} {\bibfnamefont {E.}~\bibnamefont
  {Buhmann}}, \bibinfo {author} {\bibfnamefont {C.}~\bibnamefont {Ewen}},
  \bibinfo {author} {\bibfnamefont {G.}~\bibnamefont {Kasieczka}}, \bibinfo
  {author} {\bibfnamefont {V.}~\bibnamefont {Mikuni}}, \bibinfo {author}
  {\bibfnamefont {B.}~\bibnamefont {Nachman}},\ and\ \bibinfo {author}
  {\bibfnamefont {D.}~\bibnamefont {Shih}},\ }\bibfield  {title} {\bibinfo
  {title} {{Full Phase Space Resonant Anomaly Detection}},\ }\Eprint
  {https://arxiv.org/abs/2310.06897} {arXiv:2310.06897 [hep-ph]}  (\bibinfo
  {year} {2023}{\natexlab{c}})\BibitemShut {NoStop}%
\bibitem [{\citenamefont {Sengupta}\ \emph {et~al.}(2023)\citenamefont
  {Sengupta}, \citenamefont {Leigh}, \citenamefont {Raine}, \citenamefont
  {Klein},\ and\ \citenamefont {Golling}}]{drapes}%
  \BibitemOpen
  \bibfield  {author} {\bibinfo {author} {\bibfnamefont {D.}~\bibnamefont
  {Sengupta}}, \bibinfo {author} {\bibfnamefont {M.}~\bibnamefont {Leigh}},
  \bibinfo {author} {\bibfnamefont {J.~A.}\ \bibnamefont {Raine}}, \bibinfo
  {author} {\bibfnamefont {S.}~\bibnamefont {Klein}},\ and\ \bibinfo {author}
  {\bibfnamefont {T.}~\bibnamefont {Golling}},\ }\bibfield  {title} {\bibinfo
  {title} {{Improving new physics searches with diffusion models for event
  observables and jet constituents}},\ }\Eprint
  {https://arxiv.org/abs/2312.10130} {arXiv:2312.10130 [physics.data-an]}
  (\bibinfo {year} {2023})\BibitemShut {NoStop}%
\bibitem [{\citenamefont {Leigh}\ \emph
  {et~al.}(2023{\natexlab{c}})\citenamefont {Leigh}, \citenamefont {Raine},
  \citenamefont {Zoch},\ and\ \citenamefont {Golling}}]{nuflows}%
  \BibitemOpen
  \bibfield  {author} {\bibinfo {author} {\bibfnamefont {M.}~\bibnamefont
  {Leigh}}, \bibinfo {author} {\bibfnamefont {J.~A.}\ \bibnamefont {Raine}},
  \bibinfo {author} {\bibfnamefont {K.}~\bibnamefont {Zoch}},\ and\ \bibinfo
  {author} {\bibfnamefont {T.}~\bibnamefont {Golling}},\ }\bibfield  {title}
  {\bibinfo {title} {{$\nu$-flows: Conditional neutrino regression}},\ }\href
  {https://doi.org/10.21468/SciPostPhys.14.6.159} {\bibfield  {journal}
  {\bibinfo  {journal} {SciPost Phys.}\ }\textbf {\bibinfo {volume} {14}},\
  \bibinfo {pages} {159} (\bibinfo {year} {2023}{\natexlab{c}})},\ \Eprint
  {https://arxiv.org/abs/2207.00664} {arXiv:2207.00664 [hep-ph]} \BibitemShut
  {NoStop}%
\bibitem [{\citenamefont {Raine}\ \emph {et~al.}(2023)\citenamefont {Raine},
  \citenamefont {Leigh}, \citenamefont {Zoch},\ and\ \citenamefont
  {Golling}}]{v2flows}%
  \BibitemOpen
  \bibfield  {author} {\bibinfo {author} {\bibfnamefont {J.~A.}\ \bibnamefont
  {Raine}}, \bibinfo {author} {\bibfnamefont {M.}~\bibnamefont {Leigh}},
  \bibinfo {author} {\bibfnamefont {K.}~\bibnamefont {Zoch}},\ and\ \bibinfo
  {author} {\bibfnamefont {T.}~\bibnamefont {Golling}},\ }\bibfield  {title}
  {\bibinfo {title} {{Fast and improved neutrino reconstruction in
  multineutrino final states with conditional normalizing flows}},\ }\bibfield
  {journal} {\bibinfo  {journal} {Phys.Rev.D}\ }\textbf {\bibinfo {volume}
  {109}},\ \href {https://doi.org/10.1103/PhysRevD.109.012005}
  {10.1103/PhysRevD.109.012005} (\bibinfo {year} {2023}),\ \Eprint
  {https://arxiv.org/abs/2307.02405} {arXiv:2307.02405 [hep-ph]} \BibitemShut
  {NoStop}%
\bibitem [{\citenamefont {Sirunyan}\ \emph {et~al.}(2019)\citenamefont
  {Sirunyan} \emph {et~al.}}]{CMS:2018rkg}%
  \BibitemOpen
  \bibfield  {author} {\bibinfo {author} {\bibfnamefont {A.~M.}\ \bibnamefont
  {Sirunyan}} \emph {et~al.} (\bibinfo {collaboration} {CMS}),\ }\bibfield
  {title} {\bibinfo {title} {{Search for resonant $
  \mathrm{t}\overline{\mathrm{t}} $ production in proton-proton collisions at $
  \sqrt{s}=13 $ TeV}},\ }\href {https://doi.org/10.1007/JHEP04(2019)031}
  {\bibfield  {journal} {\bibinfo  {journal} {JHEP}\ }\textbf {\bibinfo
  {volume} {04}},\ \bibinfo {pages} {031}},\ \Eprint
  {https://arxiv.org/abs/1810.05905} {arXiv:1810.05905 [hep-ex]} \BibitemShut
  {NoStop}%
\bibitem [{\citenamefont {Aaboud}\ \emph
  {et~al.}(2018{\natexlab{a}})\citenamefont {Aaboud} \emph
  {et~al.}}]{ATLAS:2018rvc}%
  \BibitemOpen
  \bibfield  {author} {\bibinfo {author} {\bibfnamefont {M.}~\bibnamefont
  {Aaboud}} \emph {et~al.} (\bibinfo {collaboration} {ATLAS}),\ }\bibfield
  {title} {\bibinfo {title} {{Search for heavy particles decaying into
  top-quark pairs using lepton-plus-jets events in proton\textendash{}proton
  collisions at $\sqrt{s} = 13$ $\text {TeV}$ with the ATLAS detector}},\
  }\href {https://doi.org/10.1140/epjc/s10052-018-5995-6} {\bibfield  {journal}
  {\bibinfo  {journal} {Eur. Phys. J. C}\ }\textbf {\bibinfo {volume} {78}},\
  \bibinfo {pages} {565} (\bibinfo {year} {2018}{\natexlab{a}})},\ \Eprint
  {https://arxiv.org/abs/1804.10823} {arXiv:1804.10823 [hep-ex]} \BibitemShut
  {NoStop}%
\bibitem [{\citenamefont {Vaswani}\ \emph {et~al.}(2023)\citenamefont
  {Vaswani}, \citenamefont {Shazeer}, \citenamefont {Parmar}, \citenamefont
  {Uszkoreit}, \citenamefont {Jones}, \citenamefont {Gomez}, \citenamefont
  {Kaiser},\ and\ \citenamefont {Polosukhin}}]{attn}%
  \BibitemOpen
  \bibfield  {author} {\bibinfo {author} {\bibfnamefont {A.}~\bibnamefont
  {Vaswani}}, \bibinfo {author} {\bibfnamefont {N.}~\bibnamefont {Shazeer}},
  \bibinfo {author} {\bibfnamefont {N.}~\bibnamefont {Parmar}}, \bibinfo
  {author} {\bibfnamefont {J.}~\bibnamefont {Uszkoreit}}, \bibinfo {author}
  {\bibfnamefont {L.}~\bibnamefont {Jones}}, \bibinfo {author} {\bibfnamefont
  {A.~N.}\ \bibnamefont {Gomez}}, \bibinfo {author} {\bibfnamefont
  {L.}~\bibnamefont {Kaiser}},\ and\ \bibinfo {author} {\bibfnamefont
  {I.}~\bibnamefont {Polosukhin}},\ }\href {https://arxiv.org/abs/1706.03762}
  {\bibinfo {title} {{Attention Is All You Need}}} (\bibinfo {year} {{2023}}),\
  \Eprint {https://arxiv.org/abs/{1706.03762}} {{arXiv}:{1706.03762} [{cs.CL}]}
  \BibitemShut {NoStop}%
\bibitem [{\citenamefont {{Ruibin Xiong and Yunchang Yang and Di He and Kai
  Zheng and Shuxin Zheng and Chen Xing and Huishuai Zhang and Yanyan Lan and
  Liwei Wang and Tie-Yan Liu}}(2020)}]{preLN}%
  \BibitemOpen
  \bibfield  {author} {\bibinfo {author} {\bibnamefont {{Ruibin Xiong and
  Yunchang Yang and Di He and Kai Zheng and Shuxin Zheng and Chen Xing and
  Huishuai Zhang and Yanyan Lan and Liwei Wang and Tie-Yan Liu}}},\ }\href@noop
  {} {\bibinfo {title} {{On Layer Normalization in the Transformer
  Architecture}}} (\bibinfo {year} {{2020}}),\ \Eprint
  {https://arxiv.org/abs/{2002.04745}} {{arXiv}:{2002.04745} [{cs.LG}]}
  \BibitemShut {NoStop}%
\bibitem [{\citenamefont {Li}\ \emph {et~al.}(2017)\citenamefont {Li},
  \citenamefont {Tarlow}, \citenamefont {Brockschmidt},\ and\ \citenamefont
  {Zemel}}]{li2017gatedgraphsequenceneural}%
  \BibitemOpen
  \bibfield  {author} {\bibinfo {author} {\bibfnamefont {Y.}~\bibnamefont
  {Li}}, \bibinfo {author} {\bibfnamefont {D.}~\bibnamefont {Tarlow}}, \bibinfo
  {author} {\bibfnamefont {M.}~\bibnamefont {Brockschmidt}},\ and\ \bibinfo
  {author} {\bibfnamefont {R.}~\bibnamefont {Zemel}},\ }\href
  {https://arxiv.org/abs/1511.05493} {\bibinfo {title} {Gated graph sequence
  neural networks}} (\bibinfo {year} {2017}),\ \Eprint
  {https://arxiv.org/abs/1511.05493} {arXiv:1511.05493 [cs.LG]} \BibitemShut
  {NoStop}%
\bibitem [{\citenamefont {Alwall}\ \emph {et~al.}(2014)\citenamefont {Alwall},
  \citenamefont {Frederix}, \citenamefont {Frixione}, \citenamefont {Hirschi},
  \citenamefont {Maltoni}, \citenamefont {Mattelaer}, \citenamefont {Shao},
  \citenamefont {Stelzer}, \citenamefont {Torrielli},\ and\ \citenamefont
  {Zaro}}]{Alwall:2014hca}%
  \BibitemOpen
  \bibfield  {author} {\bibinfo {author} {\bibfnamefont {J.}~\bibnamefont
  {Alwall}}, \bibinfo {author} {\bibfnamefont {R.}~\bibnamefont {Frederix}},
  \bibinfo {author} {\bibfnamefont {S.}~\bibnamefont {Frixione}}, \bibinfo
  {author} {\bibfnamefont {V.}~\bibnamefont {Hirschi}}, \bibinfo {author}
  {\bibfnamefont {F.}~\bibnamefont {Maltoni}}, \bibinfo {author} {\bibfnamefont
  {O.}~\bibnamefont {Mattelaer}}, \bibinfo {author} {\bibfnamefont {H.~S.}\
  \bibnamefont {Shao}}, \bibinfo {author} {\bibfnamefont {T.}~\bibnamefont
  {Stelzer}}, \bibinfo {author} {\bibfnamefont {P.}~\bibnamefont {Torrielli}},\
  and\ \bibinfo {author} {\bibfnamefont {M.}~\bibnamefont {Zaro}},\ }\bibfield
  {title} {\bibinfo {title} {{The automated computation of tree-level and
  next-to-leading order differential cross sections, and their matching to
  parton shower simulations}},\ }\href
  {https://doi.org/10.1007/JHEP07(2014)079} {\bibfield  {journal} {\bibinfo
  {journal} {JHEP}\ }\textbf {\bibinfo {volume} {07}},\ \bibinfo {pages}
  {079}},\ \Eprint {https://arxiv.org/abs/{1405.0301}} {{1405.0301}}
  \BibitemShut {NoStop}%
\bibitem [{\citenamefont {{Artoisenet, Pierre and others}}(2013)}]{MadSpin}%
  \BibitemOpen
  \bibfield  {author} {\bibinfo {author} {\bibnamefont {{Artoisenet, Pierre and
  others}}},\ }\bibfield  {title} {\bibinfo {title} {{{Automatic spin-entangled
  decays of heavy resonances in Monte Carlo simulations}}},\ }\href
  {https://doi.org/{10.1007/jhep03(2013)015}} {\bibfield  {journal} {\bibinfo
  {journal} {{JHEP}}\ }\textbf {\bibinfo {volume} {{03}}},\ \bibinfo {pages}
  {{15}} (\bibinfo {year} {{2013}})}\BibitemShut {NoStop}%
\bibitem [{\citenamefont {Sjöstrand}\ \emph {et~al.}(2015)\citenamefont
  {Sjöstrand}, \citenamefont {Ask}, \citenamefont {Christiansen},
  \citenamefont {Corke}, \citenamefont {Desai}, \citenamefont {Ilten},
  \citenamefont {Mrenna}, \citenamefont {Prestel}, \citenamefont {Rasmussen},\
  and\ \citenamefont {Skands}}]{Sjostrand:2014zea}%
  \BibitemOpen
  \bibfield  {author} {\bibinfo {author} {\bibfnamefont {T.}~\bibnamefont
  {Sjöstrand}}, \bibinfo {author} {\bibfnamefont {S.}~\bibnamefont {Ask}},
  \bibinfo {author} {\bibfnamefont {J.~R.}\ \bibnamefont {Christiansen}},
  \bibinfo {author} {\bibfnamefont {R.}~\bibnamefont {Corke}}, \bibinfo
  {author} {\bibfnamefont {N.}~\bibnamefont {Desai}}, \bibinfo {author}
  {\bibfnamefont {P.}~\bibnamefont {Ilten}}, \bibinfo {author} {\bibfnamefont
  {S.}~\bibnamefont {Mrenna}}, \bibinfo {author} {\bibfnamefont
  {S.}~\bibnamefont {Prestel}}, \bibinfo {author} {\bibfnamefont {C.~O.}\
  \bibnamefont {Rasmussen}},\ and\ \bibinfo {author} {\bibfnamefont {P.~Z.}\
  \bibnamefont {Skands}},\ }\bibfield  {title} {\bibinfo {title} {{An
  introduction to PYTHIA 8.2}},\ }\href
  {https://doi.org/10.1016/j.cpc.2015.01.024} {\bibfield  {journal} {\bibinfo
  {journal} {Comput.Phys.Commun.}\ }\textbf {\bibinfo {volume} {191}},\
  \bibinfo {pages} {159} (\bibinfo {year} {{2015}})},\ \Eprint
  {https://arxiv.org/abs/{1410.3012}} {{1410.3012}} \BibitemShut {NoStop}%
\bibitem [{\citenamefont {Ball}\ \emph {et~al.}(2013)\citenamefont {Ball},
  \citenamefont {Bertone}, \citenamefont {Carrazza}, \citenamefont {Deans},
  \citenamefont {Del~Debbio}, \citenamefont {Forte}, \citenamefont {Guffanti},
  \citenamefont {Hartland}, \citenamefont {Latorre}, \citenamefont {Rojo},\
  and\ \citenamefont {Ubiali}}]{Ball:2012cx}%
  \BibitemOpen
  \bibfield  {author} {\bibinfo {author} {\bibfnamefont {R.~D.}\ \bibnamefont
  {Ball}}, \bibinfo {author} {\bibfnamefont {V.}~\bibnamefont {Bertone}},
  \bibinfo {author} {\bibfnamefont {S.}~\bibnamefont {Carrazza}}, \bibinfo
  {author} {\bibfnamefont {C.~S.}\ \bibnamefont {Deans}}, \bibinfo {author}
  {\bibfnamefont {L.}~\bibnamefont {Del~Debbio}}, \bibinfo {author}
  {\bibfnamefont {S.}~\bibnamefont {Forte}}, \bibinfo {author} {\bibfnamefont
  {A.}~\bibnamefont {Guffanti}}, \bibinfo {author} {\bibfnamefont {N.~P.}\
  \bibnamefont {Hartland}}, \bibinfo {author} {\bibfnamefont {J.~I.}\
  \bibnamefont {Latorre}}, \bibinfo {author} {\bibfnamefont {J.}~\bibnamefont
  {Rojo}},\ and\ \bibinfo {author} {\bibfnamefont {M.}~\bibnamefont {Ubiali}},\
  }\bibfield  {title} {\bibinfo {title} {{Parton distributions with LHC
  data}},\ }\href {https://doi.org/10.1016/j.nuclphysb.2012.10.003} {\bibfield
  {journal} {\bibinfo  {journal} {Nucl.Phys.B}\ }\textbf {\bibinfo {volume}
  {867}},\ \bibinfo {pages} {244} (\bibinfo {year} {{2013}})},\ \Eprint
  {https://arxiv.org/abs/{1207.1303}} {{1207.1303}} \BibitemShut {NoStop}%
\bibitem [{\citenamefont {Buckley}\ \emph {et~al.}(2015)\citenamefont
  {Buckley}, \citenamefont {Ferrando}, \citenamefont {Lloyd}, \citenamefont
  {Nordström}, \citenamefont {Page}, \citenamefont {Rüfenacht}, \citenamefont
  {Schönherr},\ and\ \citenamefont {Watt}}]{Buckley:2014ana}%
  \BibitemOpen
  \bibfield  {author} {\bibinfo {author} {\bibfnamefont {A.}~\bibnamefont
  {Buckley}}, \bibinfo {author} {\bibfnamefont {J.}~\bibnamefont {Ferrando}},
  \bibinfo {author} {\bibfnamefont {S.}~\bibnamefont {Lloyd}}, \bibinfo
  {author} {\bibfnamefont {K.}~\bibnamefont {Nordström}}, \bibinfo {author}
  {\bibfnamefont {B.}~\bibnamefont {Page}}, \bibinfo {author} {\bibfnamefont
  {M.}~\bibnamefont {Rüfenacht}}, \bibinfo {author} {\bibfnamefont
  {M.}~\bibnamefont {Schönherr}},\ and\ \bibinfo {author} {\bibfnamefont
  {G.}~\bibnamefont {Watt}},\ }\bibfield  {title} {\bibinfo {title} {{LHAPDF6:
  parton density access in the LHC precision era}},\ }\href
  {https://doi.org/10.1140/epjc/s10052-015-3318-8} {\bibfield  {journal}
  {\bibinfo  {journal} {Eur.Phys.J.C}\ }\textbf {\bibinfo {volume} {75}},\
  \bibinfo {pages} {132} (\bibinfo {year} {{2015}})},\ \Eprint
  {https://arxiv.org/abs/{1412.7420}} {{1412.7420}} \BibitemShut {NoStop}%
\bibitem [{\citenamefont {Zoch}\ \emph
  {et~al.}(2024{\natexlab{a}})\citenamefont {Zoch}, \citenamefont {Raine},
  \citenamefont {Sengupta},\ and\ \citenamefont {Golling}}]{Zoch:2024eyp}%
  \BibitemOpen
  \bibfield  {author} {\bibinfo {author} {\bibfnamefont {K.}~\bibnamefont
  {Zoch}}, \bibinfo {author} {\bibfnamefont {J.~A.}\ \bibnamefont {Raine}},
  \bibinfo {author} {\bibfnamefont {D.}~\bibnamefont {Sengupta}},\ and\
  \bibinfo {author} {\bibfnamefont {T.}~\bibnamefont {Golling}},\ }\href@noop
  {} {\bibinfo {title} {{RODEM Jet Datasets}}} (\bibinfo {year}
  {2024}{\natexlab{a}}),\ \bibinfo {note} {available on Zenodo:
  \href{https://doi.org/10.5281/zenodo.12793616}{10.5281/zenodo.12793616}.},\
  \Eprint {https://arxiv.org/abs/2408.11616} {arXiv:2408.11616 [hep-ph]}
  \BibitemShut {NoStop}%
\bibitem [{\citenamefont {Zoch}\ \emph
  {et~al.}(2024{\natexlab{b}})\citenamefont {Zoch}, \citenamefont {Raine},
  \citenamefont {Sengupta},\ and\ \citenamefont
  {Golling}}]{zoch2024rodemjetdatasets}%
  \BibitemOpen
  \bibfield  {author} {\bibinfo {author} {\bibfnamefont {K.}~\bibnamefont
  {Zoch}}, \bibinfo {author} {\bibfnamefont {J.~A.}\ \bibnamefont {Raine}},
  \bibinfo {author} {\bibfnamefont {D.}~\bibnamefont {Sengupta}},\ and\
  \bibinfo {author} {\bibfnamefont {T.}~\bibnamefont {Golling}},\ }\href
  {https://arxiv.org/abs/2408.11616} {\bibinfo {title} {Rodem jet datasets}}
  (\bibinfo {year} {2024}{\natexlab{b}}),\ \Eprint
  {https://arxiv.org/abs/2408.11616} {arXiv:2408.11616 [hep-ph]} \BibitemShut
  {NoStop}%
\bibitem [{\citenamefont {Cacciari}\ \emph {et~al.}(2008)\citenamefont
  {Cacciari}, \citenamefont {Salam},\ and\ \citenamefont
  {Soyez}}]{Cacciari:2008gp}%
  \BibitemOpen
  \bibfield  {author} {\bibinfo {author} {\bibfnamefont {M.}~\bibnamefont
  {Cacciari}}, \bibinfo {author} {\bibfnamefont {G.~P.}\ \bibnamefont
  {Salam}},\ and\ \bibinfo {author} {\bibfnamefont {G.}~\bibnamefont {Soyez}},\
  }\bibfield  {title} {\bibinfo {title} {{The anti-$k_t$ jet clustering
  algorithm}},\ }\href {https://doi.org/10.1088/1126-6708/2008/04/063}
  {\bibfield  {journal} {\bibinfo  {journal} {JHEP}\ }\textbf {\bibinfo
  {volume} {04}},\ \bibinfo {pages} {063}},\ \Eprint
  {https://arxiv.org/abs/{0802.1189}} {{0802.1189}} \BibitemShut {NoStop}%
\bibitem [{\citenamefont {Cacciari}\ \emph {et~al.}(2012)\citenamefont
  {Cacciari}, \citenamefont {Salam},\ and\ \citenamefont
  {Soyez}}]{Cacciari:2011ma}%
  \BibitemOpen
  \bibfield  {author} {\bibinfo {author} {\bibfnamefont {M.}~\bibnamefont
  {Cacciari}}, \bibinfo {author} {\bibfnamefont {G.~P.}\ \bibnamefont
  {Salam}},\ and\ \bibinfo {author} {\bibfnamefont {G.}~\bibnamefont {Soyez}},\
  }\bibfield  {title} {\bibinfo {title} {{FastJet User Manual}},\ }\href
  {https://doi.org/10.1140/epjc/s10052-012-1896-2} {\bibfield  {journal}
  {\bibinfo  {journal} {Eur.Phys.J.C}\ }\textbf {\bibinfo {volume} {72}},\
  \bibinfo {pages} {1896} (\bibinfo {year} {{2012}})},\ \Eprint
  {https://arxiv.org/abs/{1111.6097}} {{1111.6097}} \BibitemShut {NoStop}%
\bibitem [{\citenamefont {{DELPHES 3
  Collaboration}}(2014)}]{deFavereau:2013fsa}%
  \BibitemOpen
  \bibfield  {author} {\bibinfo {author} {\bibnamefont {{DELPHES 3
  Collaboration}}},\ }\bibfield  {title} {\bibinfo {title} {{DELPHES 3, A
  modular framework for fast simulation of a generic collider experiment}},\
  }\href {https://doi.org/10.1007/JHEP02(2014)057} {\bibfield  {journal}
  {\bibinfo  {journal} {JHEP}\ }\textbf {\bibinfo {volume} {02}},\ \bibinfo
  {pages} {057}},\ \Eprint {https://arxiv.org/abs/{1307.6346}} {{1307.6346}}
  \BibitemShut {NoStop}%
\bibitem [{\citenamefont {Good}(1952)}]{068cb646-8936-3df0-aa16-e97573eda53c}%
  \BibitemOpen
  \bibfield  {author} {\bibinfo {author} {\bibfnamefont {I.~J.}\ \bibnamefont
  {Good}},\ }\bibfield  {title} {\bibinfo {title} {Rational decisions},\ }\href
  {http://www.jstor.org/stable/2984087} {\bibfield  {journal} {\bibinfo
  {journal} {Journal of the Royal Statistical Society. Series B
  (Methodological)}\ }\textbf {\bibinfo {volume} {14}},\ \bibinfo {pages} {107}
  (\bibinfo {year} {1952})}\BibitemShut {NoStop}%
\bibitem [{\citenamefont {{Thaler, Jesse and Van Tilburg,
  Ken}}(2011)}]{jet_sub}%
  \BibitemOpen
  \bibfield  {author} {\bibinfo {author} {\bibnamefont {{Thaler, Jesse and Van
  Tilburg, Ken}}},\ }\bibfield  {title} {\bibinfo {title} {{Identifying boosted
  objects with N-subjettiness}},\ }\href
  {{http://dx.doi.org/10.1007/JHEP03(2011)015}} {\bibfield  {journal} {\bibinfo
   {journal} {{Journal of High Energy Physics}}\ }\textbf {\bibinfo {volume}
  {{2011}}} (\bibinfo {year} {{2011}})}\BibitemShut {NoStop}%
\bibitem [{\citenamefont {Larkoski}\ \emph
  {et~al.}(2013{\natexlab{b}})\citenamefont {Larkoski}, \citenamefont {Salam},\
  and\ \citenamefont {Thaler}}]{dij_jet_sub}%
  \BibitemOpen
  \bibfield  {author} {\bibinfo {author} {\bibfnamefont {A.~J.}\ \bibnamefont
  {Larkoski}}, \bibinfo {author} {\bibfnamefont {G.~P.}\ \bibnamefont
  {Salam}},\ and\ \bibinfo {author} {\bibfnamefont {J.}~\bibnamefont
  {Thaler}},\ }\bibfield  {title} {\bibinfo {title} {Energy correlation
  functions for jet substructure},\ }\bibfield  {journal} {\bibinfo  {journal}
  {Journal of High Energy Physics}\ }\textbf {\bibinfo {volume} {2013}},\ \href
  {https://doi.org/10.1007/jhep06(2013)108} {10.1007/jhep06(2013)108} (\bibinfo
  {year} {2013}{\natexlab{b}})\BibitemShut {NoStop}%
\bibitem [{\citenamefont {Larkoski}\ \emph
  {et~al.}(2014{\natexlab{b}})\citenamefont {Larkoski}, \citenamefont {Moult},\
  and\ \citenamefont {Neill}}]{dij_jet_sub_ratio}%
  \BibitemOpen
  \bibfield  {author} {\bibinfo {author} {\bibfnamefont {A.~J.}\ \bibnamefont
  {Larkoski}}, \bibinfo {author} {\bibfnamefont {I.}~\bibnamefont {Moult}},\
  and\ \bibinfo {author} {\bibfnamefont {D.}~\bibnamefont {Neill}},\ }\bibfield
   {title} {\bibinfo {title} {Power counting to better jet observables},\
  }\bibfield  {journal} {\bibinfo  {journal} {Journal of High Energy Physics}\
  }\textbf {\bibinfo {volume} {2014}},\ \href
  {https://doi.org/10.1007/jhep12(2014)009} {10.1007/jhep12(2014)009} (\bibinfo
  {year} {2014}{\natexlab{b}})\BibitemShut {NoStop}%
\bibitem [{\citenamefont {Aaboud}\ \emph
  {et~al.}(2018{\natexlab{b}})\citenamefont {Aaboud} \emph
  {et~al.}}]{ATLAS:2017zda}%
  \BibitemOpen
  \bibfield  {author} {\bibinfo {author} {\bibfnamefont {M.}~\bibnamefont
  {Aaboud}} \emph {et~al.} (\bibinfo {collaboration} {ATLAS}),\ }\bibfield
  {title} {\bibinfo {title} {{Measurement of the Soft-Drop Jet Mass in pp
  Collisions at $\sqrt{s} = 13$ TeV with the ATLAS Detector}},\ }\href
  {https://doi.org/10.1103/PhysRevLett.121.092001} {\bibfield  {journal}
  {\bibinfo  {journal} {Phys. Rev. Lett.}\ }\textbf {\bibinfo {volume} {121}},\
  \bibinfo {pages} {092001} (\bibinfo {year} {2018}{\natexlab{b}})},\ \Eprint
  {https://arxiv.org/abs/1711.08341} {arXiv:1711.08341 [hep-ex]} \BibitemShut
  {NoStop}%
\bibitem [{\citenamefont {{ATLAS Collaboration}}(2020)}]{ATLAS:2019mgf}%
  \BibitemOpen
  \bibfield  {author} {\bibinfo {author} {\bibnamefont {{ATLAS Collaboration}}}
  (\bibinfo {collaboration} {ATLAS}),\ }\bibfield  {title} {\bibinfo {title}
  {{Measurement of soft-drop jet observables in $pp$ collisions with the ATLAS
  detector at $\sqrt {s}$ =13 TeV}},\ }\href
  {https://doi.org/10.1103/PhysRevD.101.052007} {\bibfield  {journal} {\bibinfo
   {journal} {Phys. Rev. D}\ }\textbf {\bibinfo {volume} {101}},\ \bibinfo
  {pages} {052007} (\bibinfo {year} {2020})},\ \Eprint
  {https://arxiv.org/abs/1912.09837} {arXiv:1912.09837 [hep-ex]} \BibitemShut
  {NoStop}%
\bibitem [{\citenamefont {Rikab~Gambhir}\ and\ \citenamefont
  {Thaler}(2024)}]{energyflow}%
  \BibitemOpen
  \bibfield  {author} {\bibinfo {author} {\bibfnamefont {E.~M.}\ \bibnamefont
  {Rikab~Gambhir}, \bibfnamefont {Patrick~Komiske}}\ and\ \bibinfo {author}
  {\bibfnamefont {J.}~\bibnamefont {Thaler}},\ }\href
  {https://energyflow.network/} {\bibinfo {title} {\texttt{EnergyFlow}}}
  (\bibinfo {year} {2024})\BibitemShut {NoStop}%
\bibitem [{rec(2024)}]{reconstruction_performance_2024}%
  \BibitemOpen
  \bibfield  {title} {\bibinfo {title} {Software performance of the atlas track
  reconstruction for lhc run 3},\ }\bibfield  {journal} {\bibinfo  {journal}
  {Computing and Software for Big Science}\ }\textbf {\bibinfo {volume} {8}},\
  \href {https://doi.org/10.1007/s41781-023-00111-y}
  {10.1007/s41781-023-00111-y} (\bibinfo {year} {2024})\BibitemShut {NoStop}%
\bibitem [{\citenamefont {{ATLAS
  Collaboration}}(2021)}]{ATL-PHYS-PUB-2021-023}%
  \BibitemOpen
  \bibfield  {author} {\bibinfo {author} {\bibnamefont {{ATLAS
  Collaboration}}},\ }\bibfield  {title} {\bibinfo {title} {{"{Expected
  performance of the ATLAS detector under different High-Luminosity LHC
  conditions}"}}} (\bibinfo {year} {2021})\BibitemShut {NoStop}%
\bibitem [{\citenamefont {Durkan}\ \emph {et~al.}(2019)\citenamefont {Durkan},
  \citenamefont {Bekasov}, \citenamefont {Murray},\ and\ \citenamefont
  {Papamakarios}}]{spline_flows}%
  \BibitemOpen
  \bibfield  {author} {\bibinfo {author} {\bibfnamefont {C.}~\bibnamefont
  {Durkan}}, \bibinfo {author} {\bibfnamefont {A.}~\bibnamefont {Bekasov}},
  \bibinfo {author} {\bibfnamefont {I.}~\bibnamefont {Murray}},\ and\ \bibinfo
  {author} {\bibfnamefont {G.}~\bibnamefont {Papamakarios}},\ }\href@noop {}
  {\bibinfo {title} {Neural spline flows}} (\bibinfo {year} {2019}),\ \Eprint
  {https://arxiv.org/abs/1906.04032} {arXiv:1906.04032 [stat.ML]} \BibitemShut
  {NoStop}%
\end{thebibliography}%

\cleardoublepage
\FloatBarrier

\section{Appendix}
\subsection{Data distributions}
\begin{figure}[htpb]
    \centering
    \begin{subfigure}{0.35\textwidth}
        \includegraphics[width=\textwidth]{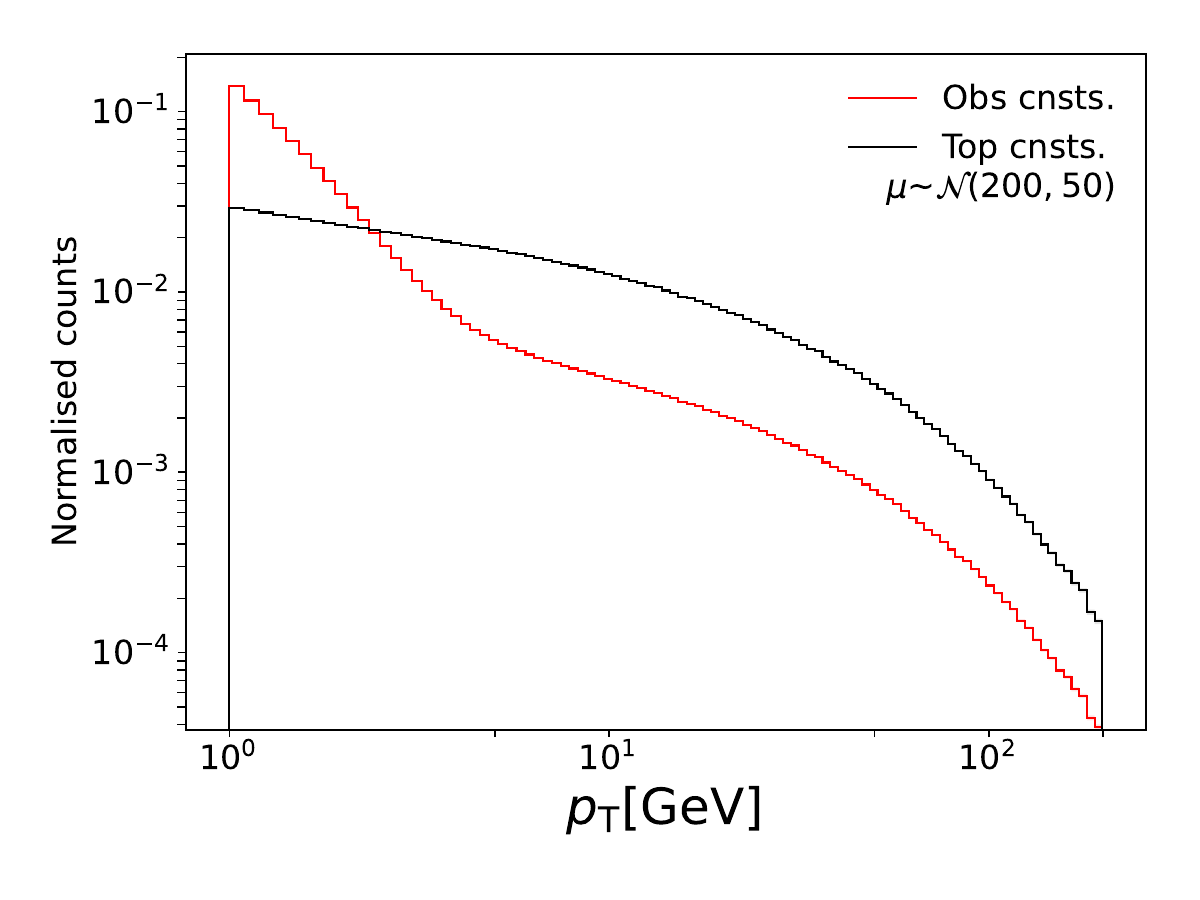}
    \end{subfigure}
    \begin{subfigure}{0.35\textwidth}
        \includegraphics[width=\textwidth]{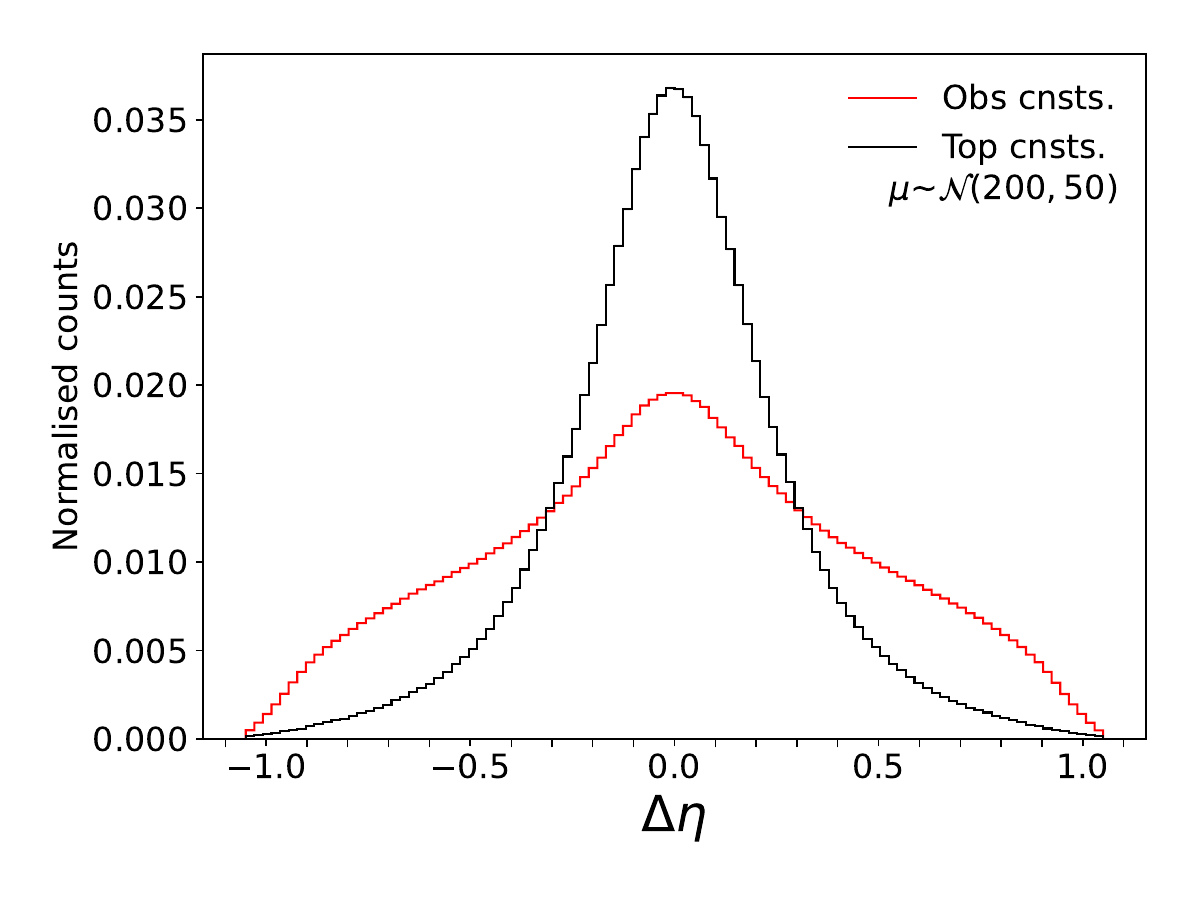}
    \end{subfigure}
    \begin{subfigure}{0.35\textwidth}
        \includegraphics[width=\textwidth]{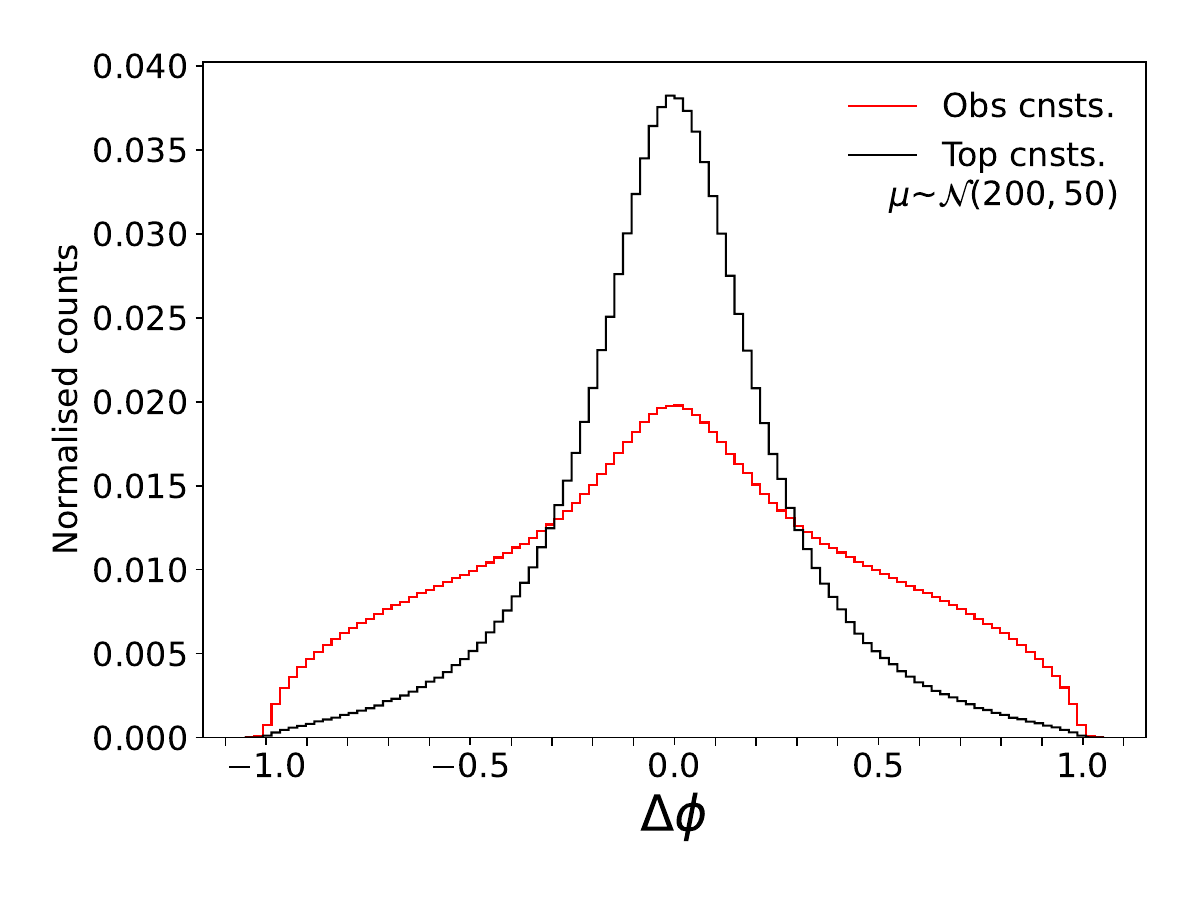}
    \end{subfigure}
    \caption{Marginal distributions of $p_\mathrm{T}$, $\Delta\eta$ and
    $\Delta\phi$ on the constituent level across events for the observed jets with
    pile-up distribution of \normal{200}{50} and the top jets.} 
    \label{fig:rel_cnsts_vars}
\end{figure}

\begin{figure*}[htpb]
    \centering
    \begin{subfigure}{0.32\textwidth}
        \includegraphics[width=\textwidth]{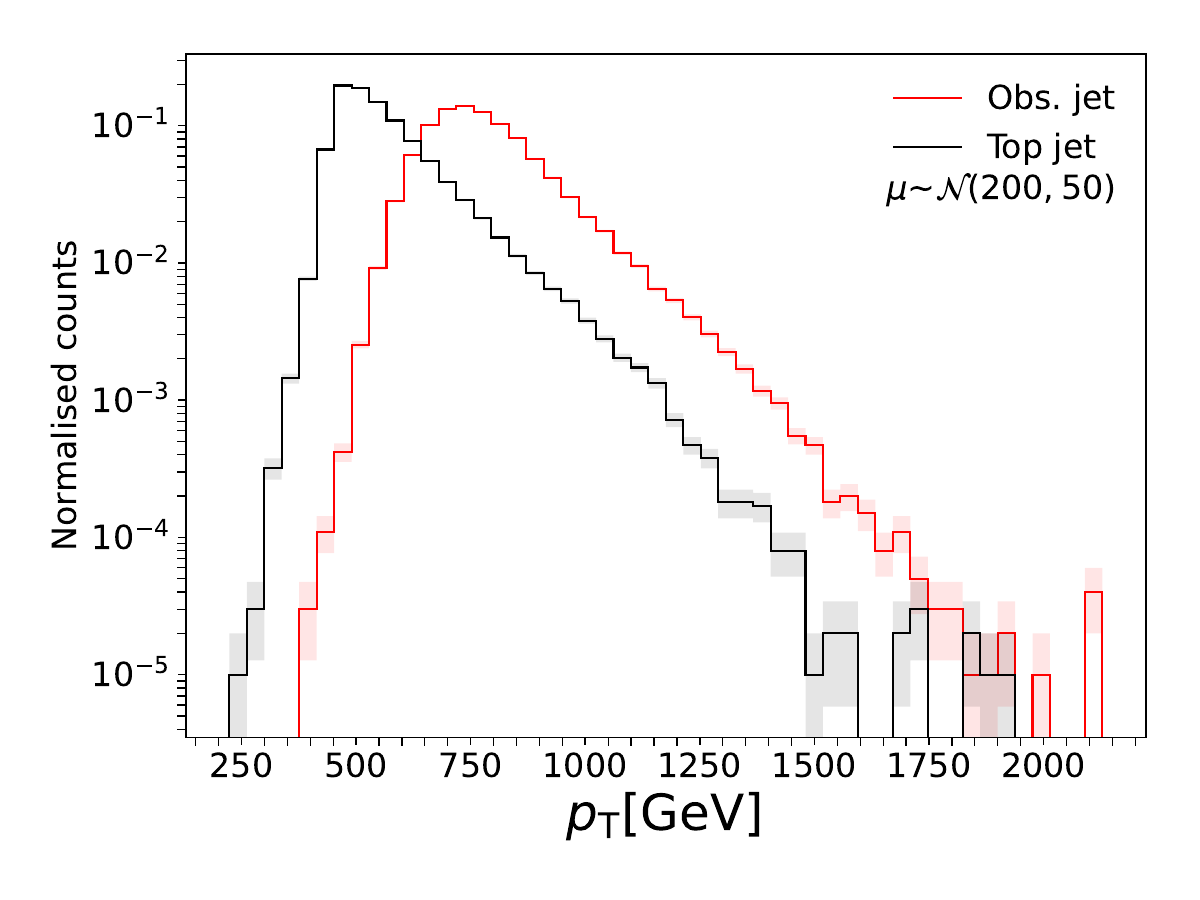}
    \end{subfigure}
    \begin{subfigure}{0.32\textwidth}
        \includegraphics[width=\textwidth]{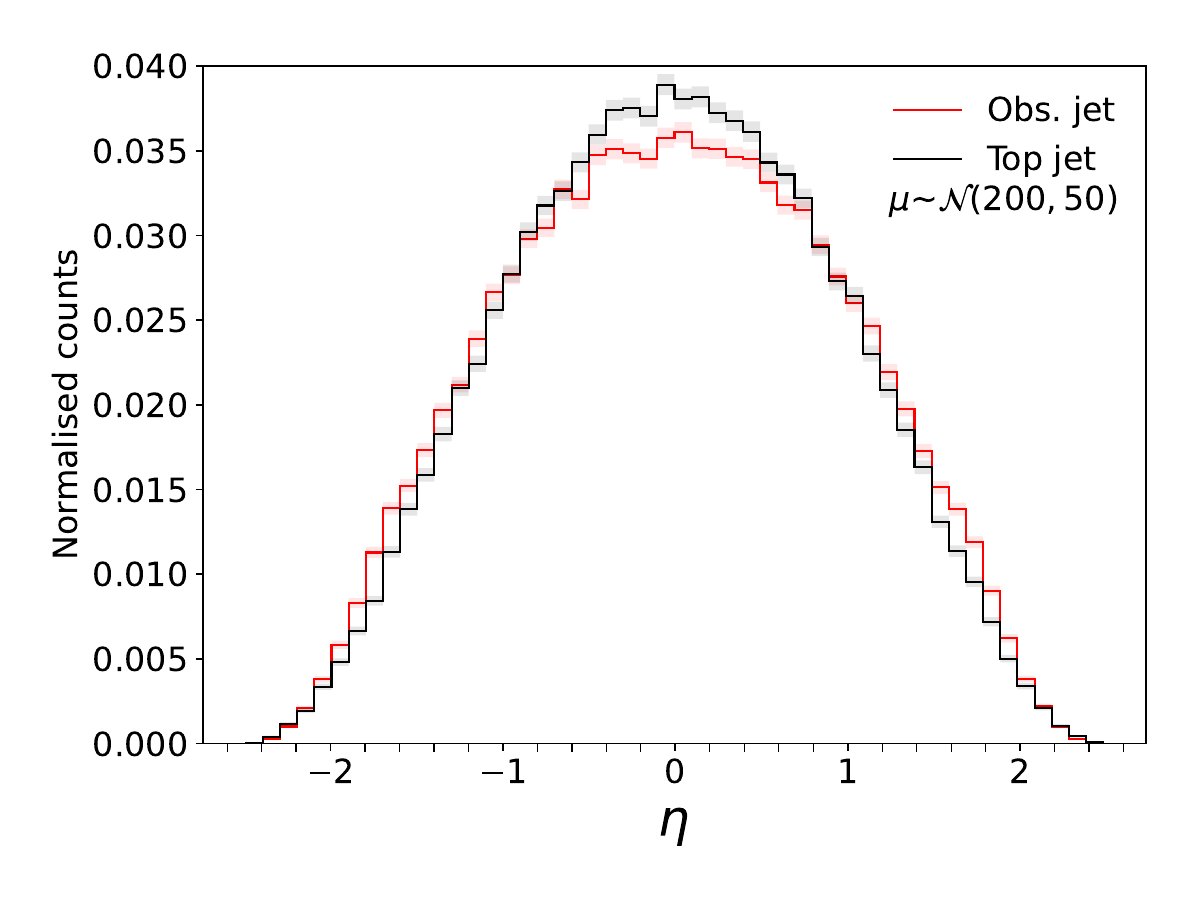}
    \end{subfigure}
    \begin{subfigure}{0.32\textwidth}
        \includegraphics[width=\textwidth]{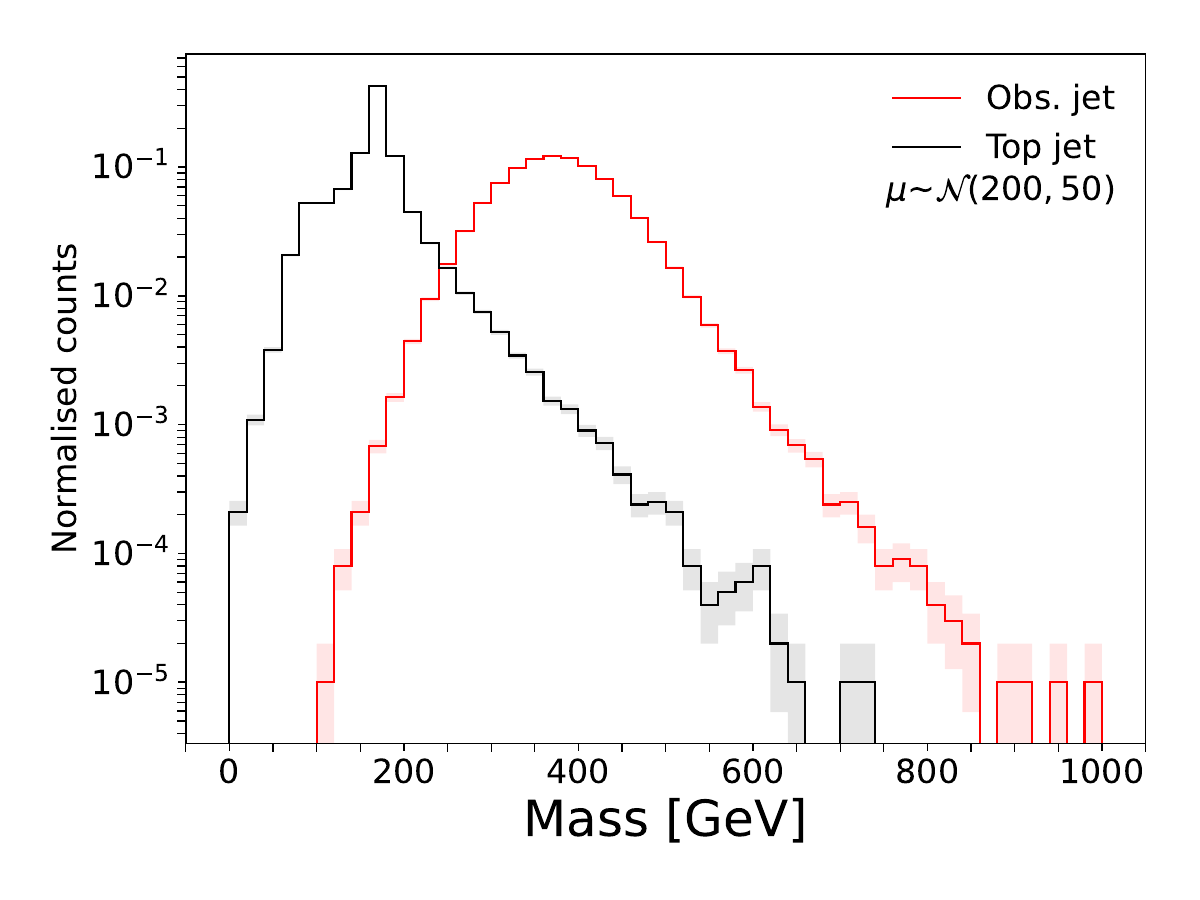}
    \end{subfigure}
    \caption{    
    Marginal distributions of $p_\mathrm{T}$, $\eta$ and mass on the jet level across events for the observed jets with pile-up distribution of \normal{200}{50} and the top jets.
    }
    \label{fig:rel_jet_vars}
\end{figure*}

\cref{fig:rel_cnsts_vars} shows the kinematic distributions of the constituents
in the observed jets with pile-up distribution of \normal{200}{50} and the
top jets.
The transverse momentum, pseudo-rapidity and invariant mass are shown in \cref{fig:rel_jet_vars}. The
$\eta$ and $\phi$ distribution of the \jtrue and \jobs are shown in
\cref{fig:rel_jet_vars}. 
Here it can be seen that the observed jets contain substantially more low momentum constituents falling in a wider distribution within the jet.
The increased constituent multiplicity also leads to a substantial increase in the observed jet \pt and invariant mass.

\subsubsection{Single event generation}
\begin{figure*}[htpb]
    \centering
    \subfloat[]{{\includegraphics[width=0.31\textwidth]{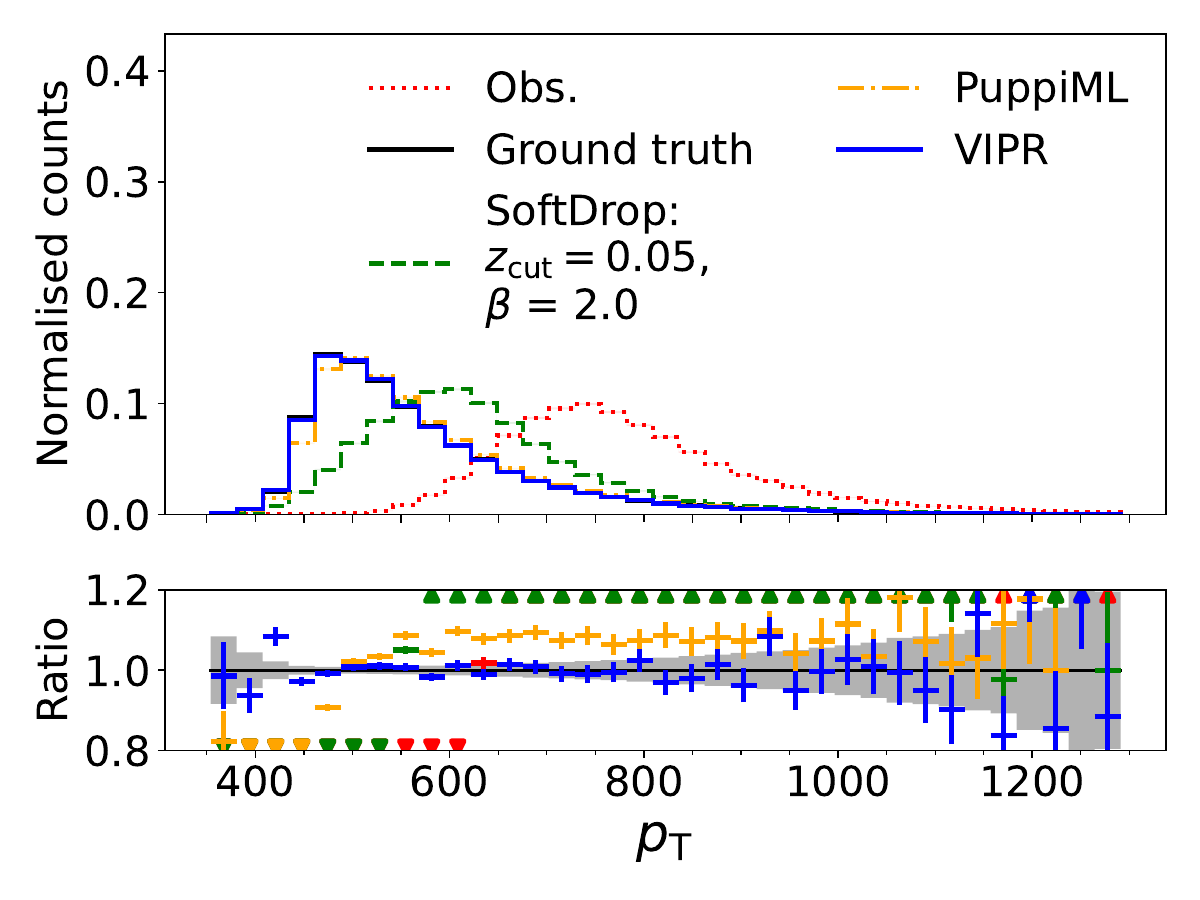}}}
    \subfloat[]{{\includegraphics[width=0.31\textwidth]{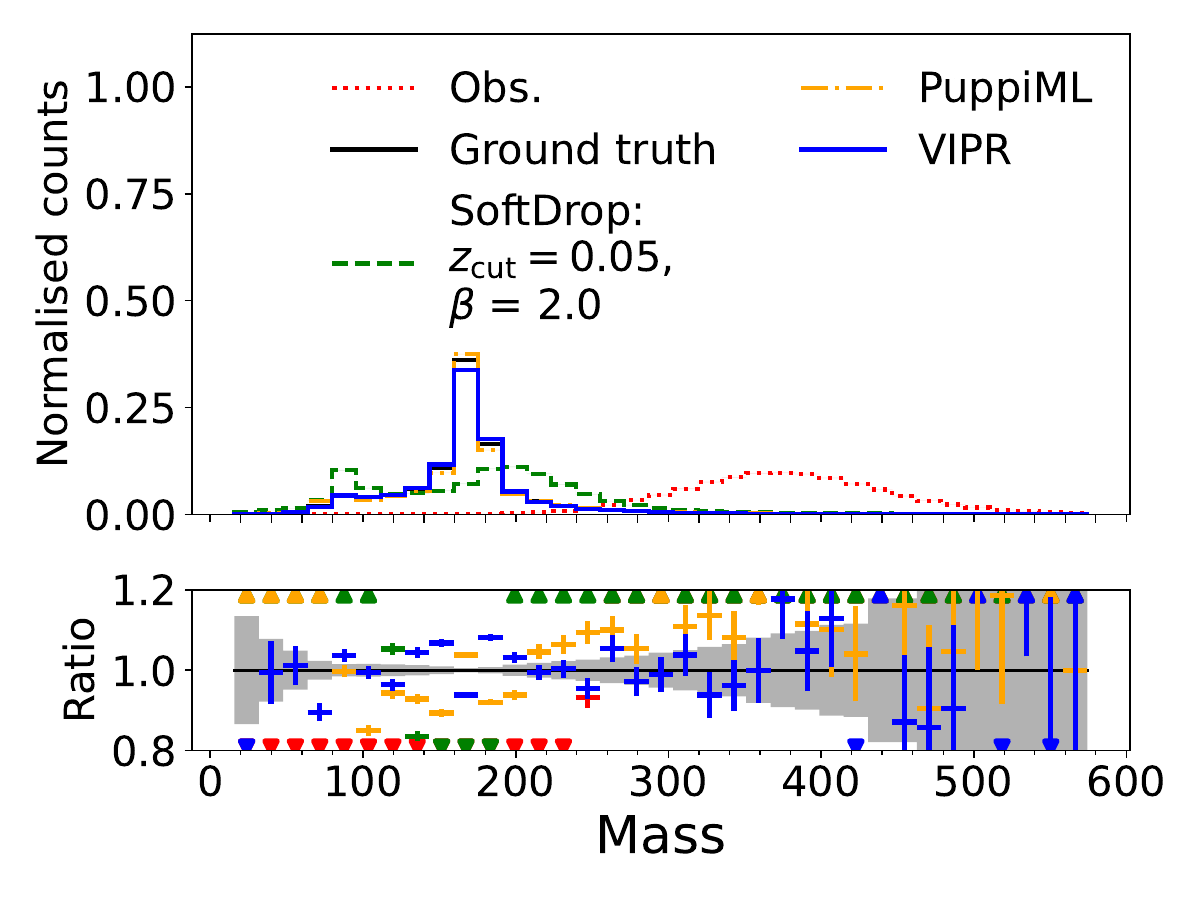}}}
    \subfloat[]{{\includegraphics[width=0.31\textwidth]{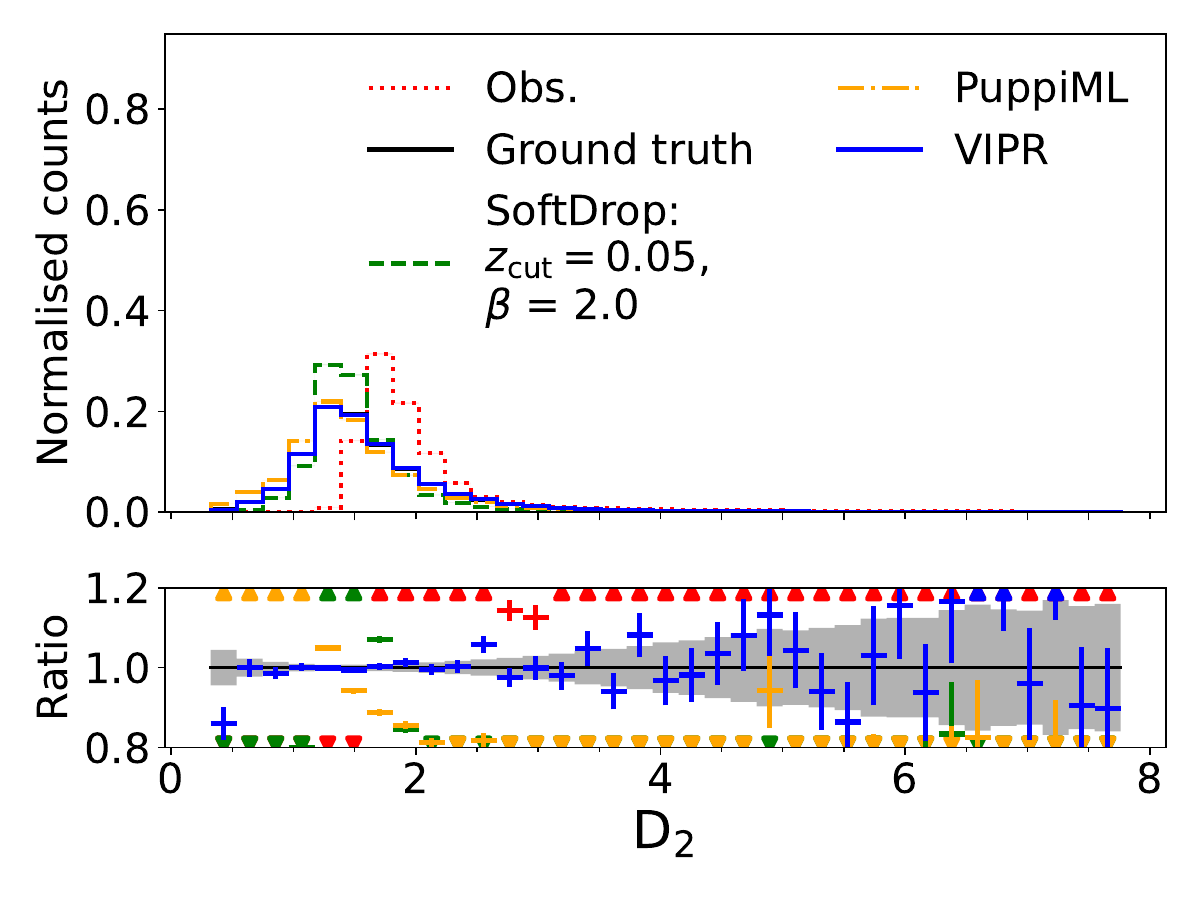}}}
    \\
    \subfloat[]{{\includegraphics[width=0.31\textwidth]{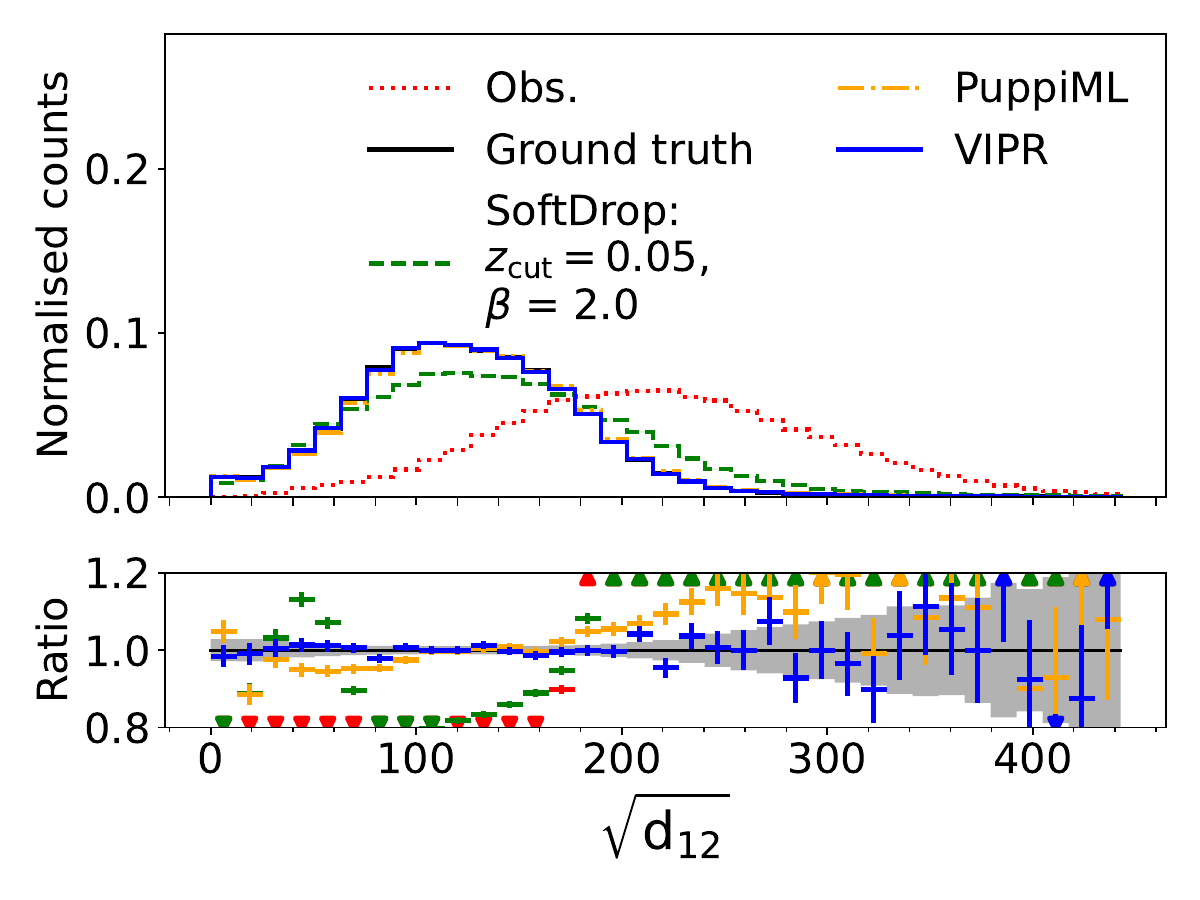}}}
    \subfloat[]{{\includegraphics[width=0.31\textwidth]{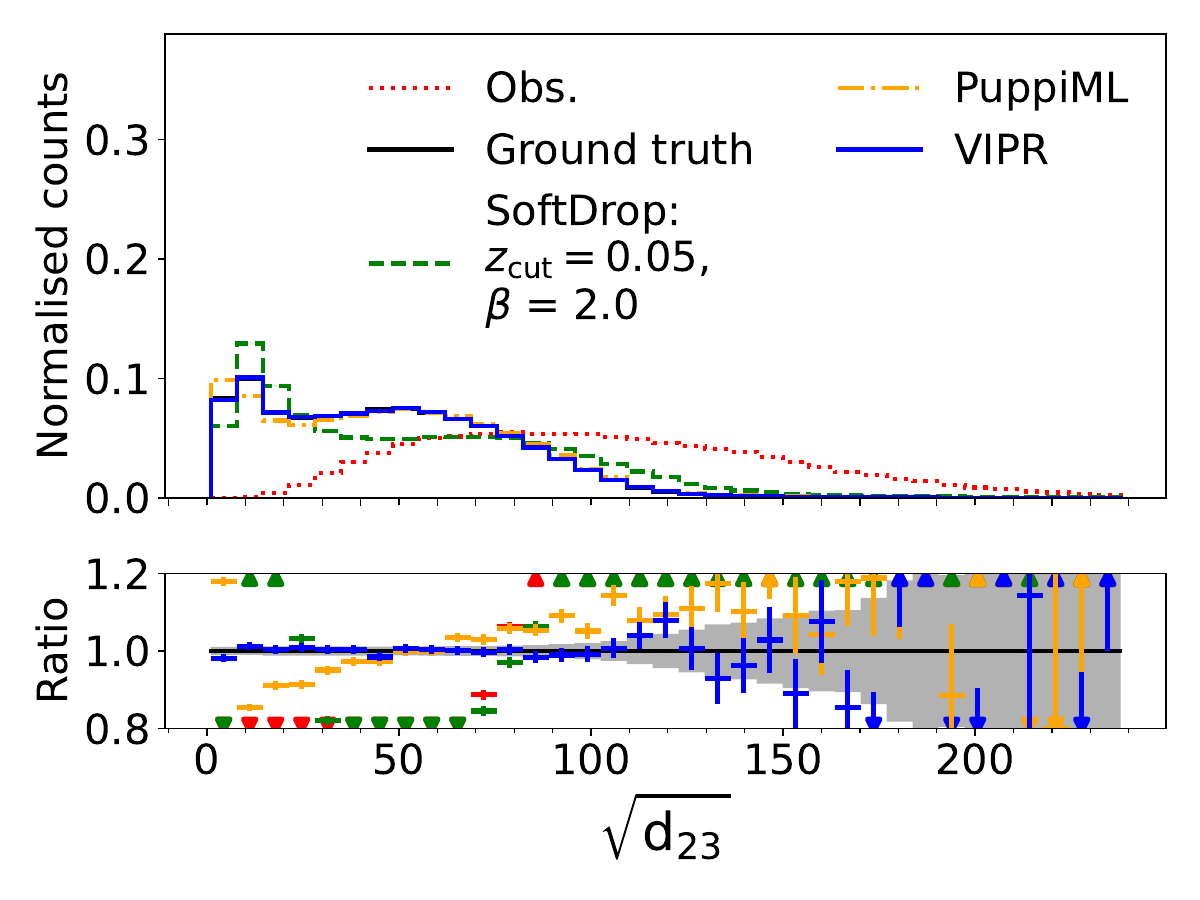}}}
    \subfloat[]{{\includegraphics[width=0.31\textwidth]{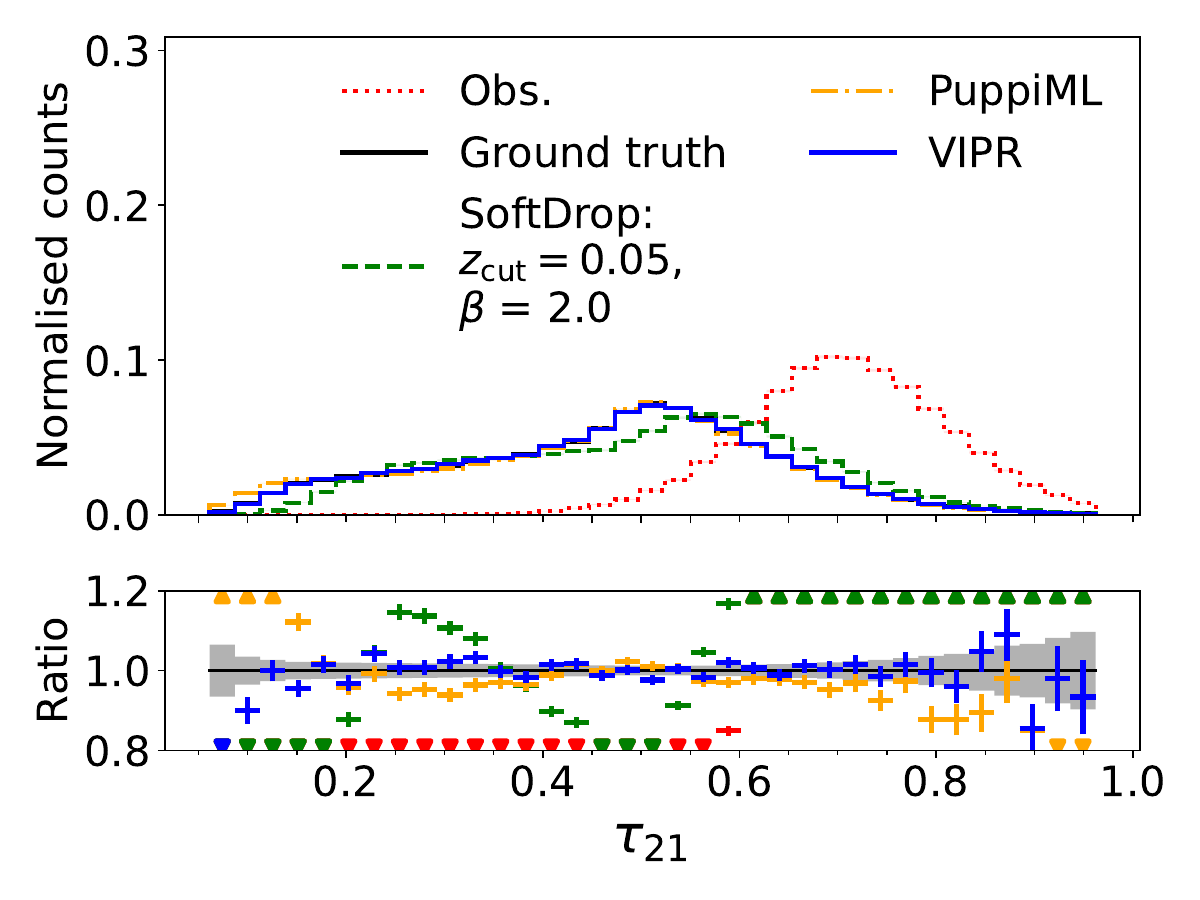}}}
    \\
    \subfloat[]{{\includegraphics[width=0.31\textwidth]{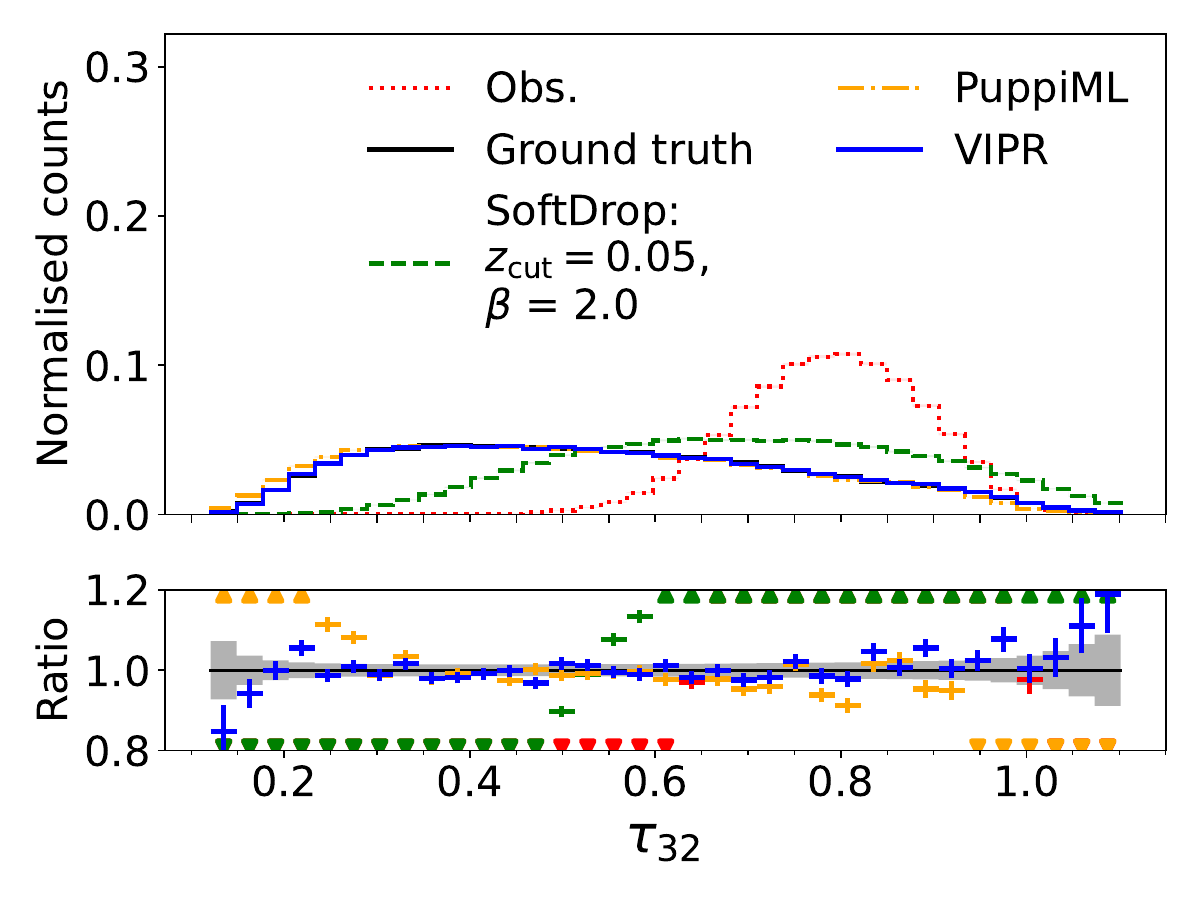}}}
    \caption{
    Marginal distributions of the jet' \pt, mass and substructures
    observables of the ground truth, \softdrop, observed, and \vipr jets.
    The \jobs have been generated using pile-up distribution at \normal{200}{50}.
    }
    \label{fig:marginal_of_single}
\end{figure*}

\cref{fig:marginal_of_single} show the marginal distribution of the substructure, \pt and invariant mass of the observed, \softdrop, \vipr and \jtrue. It is these distributions that is used to calculate the RE distribution in \cref{fig:marginal_resolution_of_single}.

\FloatBarrier

\begin{figure}[htpb]
    \centering
    \subfloat[]{{\includegraphics[width=0.4\textwidth]{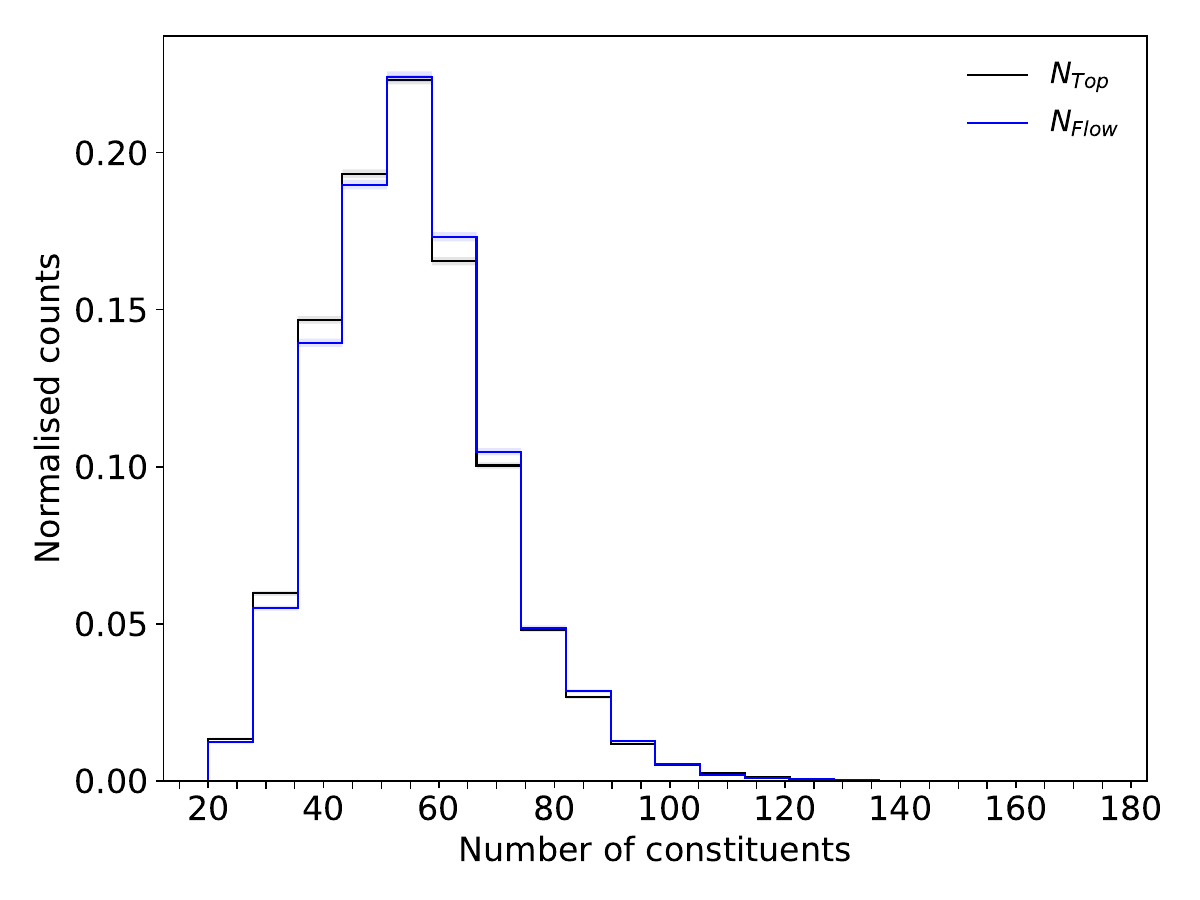}}}
    \\
    \subfloat[]{{\includegraphics[width=0.4\textwidth]{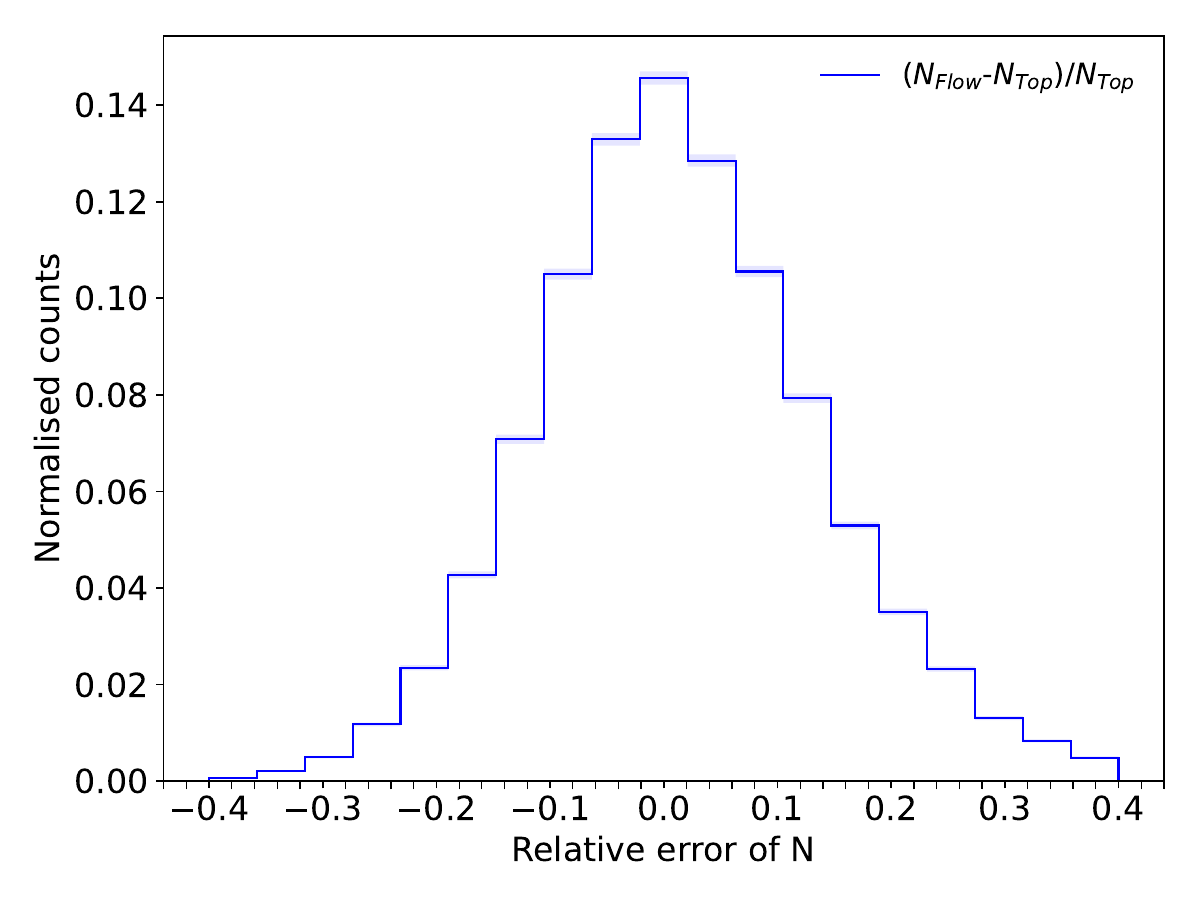}}}
    \\
    \subfloat[]{{\includegraphics[width=0.4\textwidth]{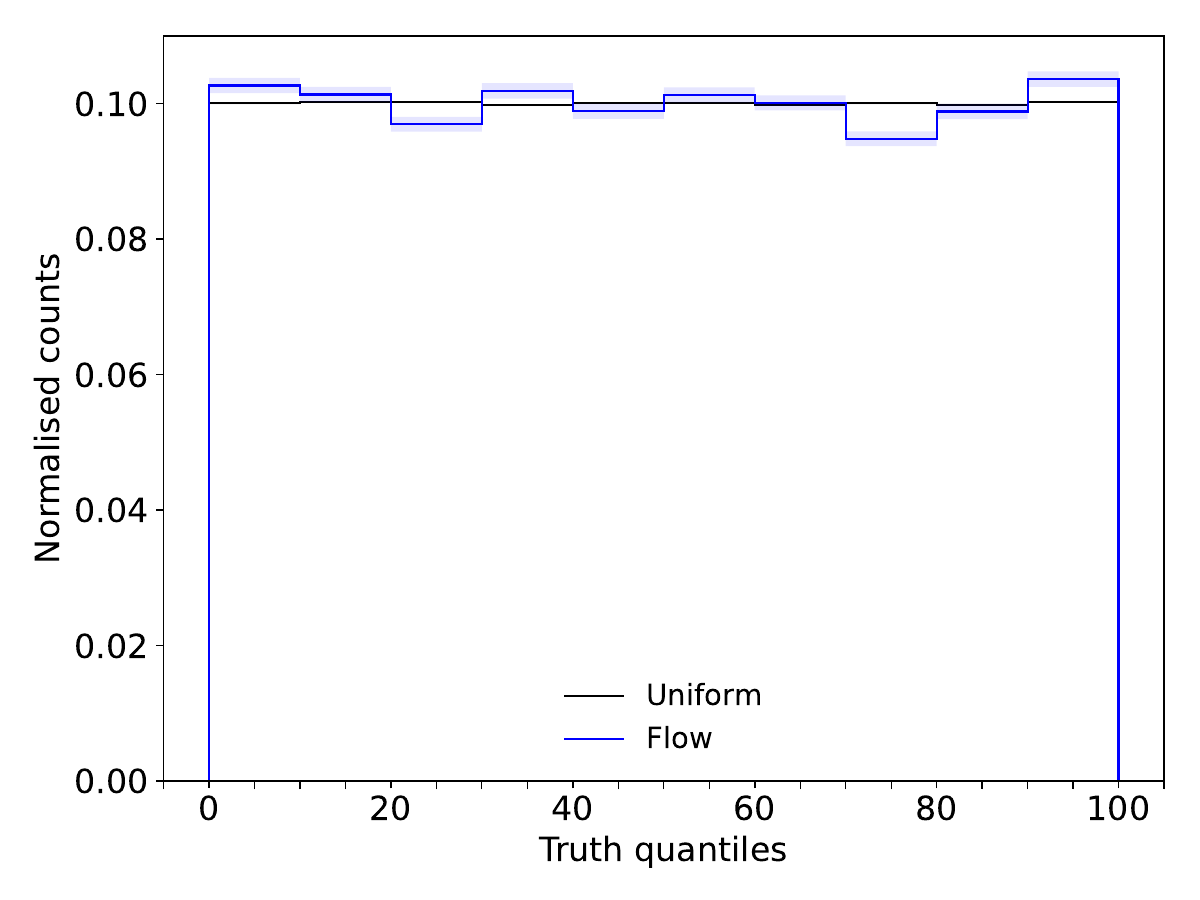}}}
    \caption{
        (a) Comparisons between the flow generated $N$ and the truth $N$.
        (b) RE of the flow generated $N$.
        (c) Posterior truth quantiles of the flow.
        In all three cases, \normal{200}{50} is used as pile-up distribution.
        }
        \label{fig:flow_performance}
\end{figure}
\subsection{Estimate of $p(N;j)$}
To estimate the number constituents of the Top jet, we train a normalising flow \cite{spline_flows} to estimate $p(\Ntrue | S, \cobs, \mu)$ from the observed jet. 
To learn the distribution of $N$, 
the normalising flow is conditioned on the observed jet, 
the number of pile-up interactions $\mu$ and summary quantities $S$. 
To dequantize the distribution of $N$, we add $\mathcal{N}(0,0.5)$ to 
the discrete $N$ and when sampling we round to the nearest integer. 
The performance of the normalising flow can be seen in \cref{fig:flow_performance}.

\subsection{Generated posteriors from \vipr}
\label{sec:posterior_samples}
\begin{figure*}
    \centering
\subfloat[]{
        \includegraphics[width=0.31\textwidth]{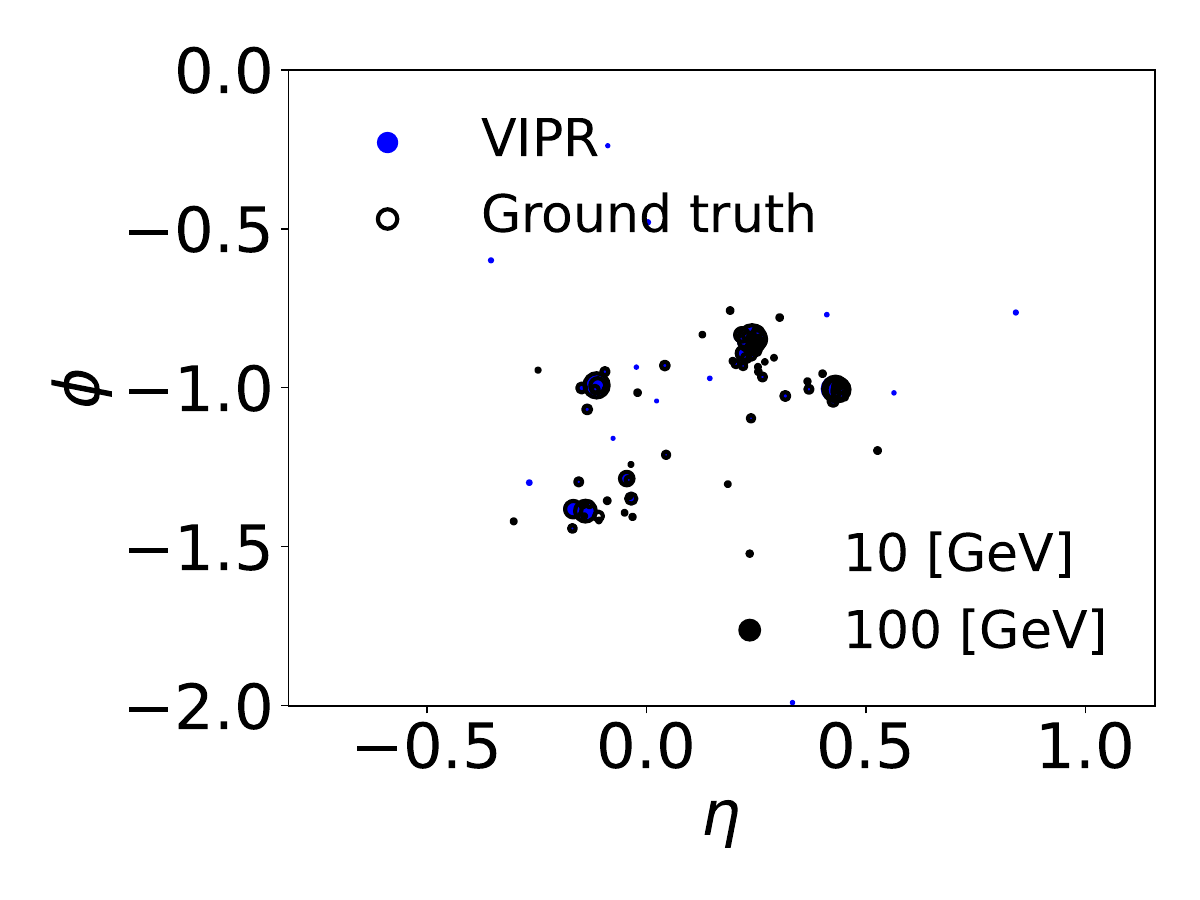}
    }
    \subfloat[]{
        \includegraphics[width=0.31\textwidth]{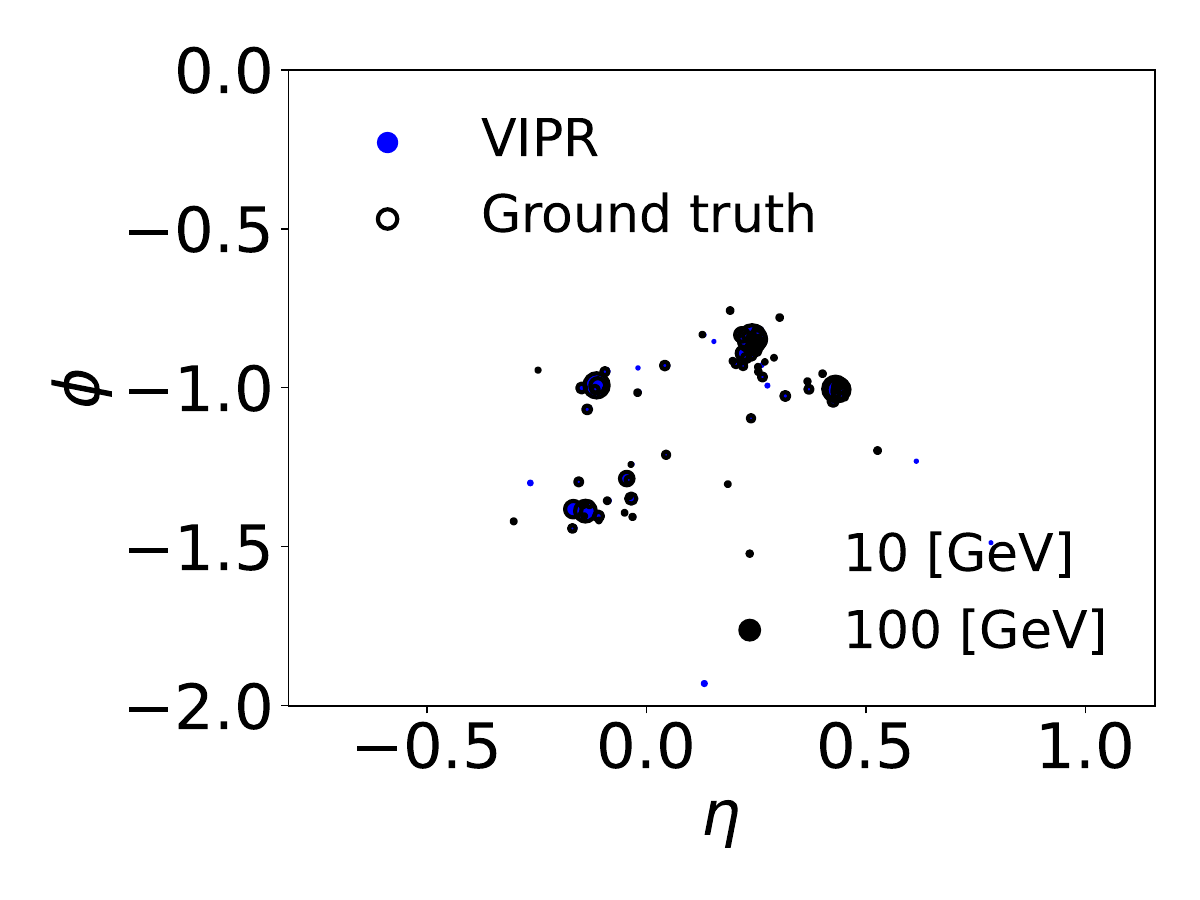}
    }
    \subfloat[]{
        \includegraphics[width=0.31\textwidth]{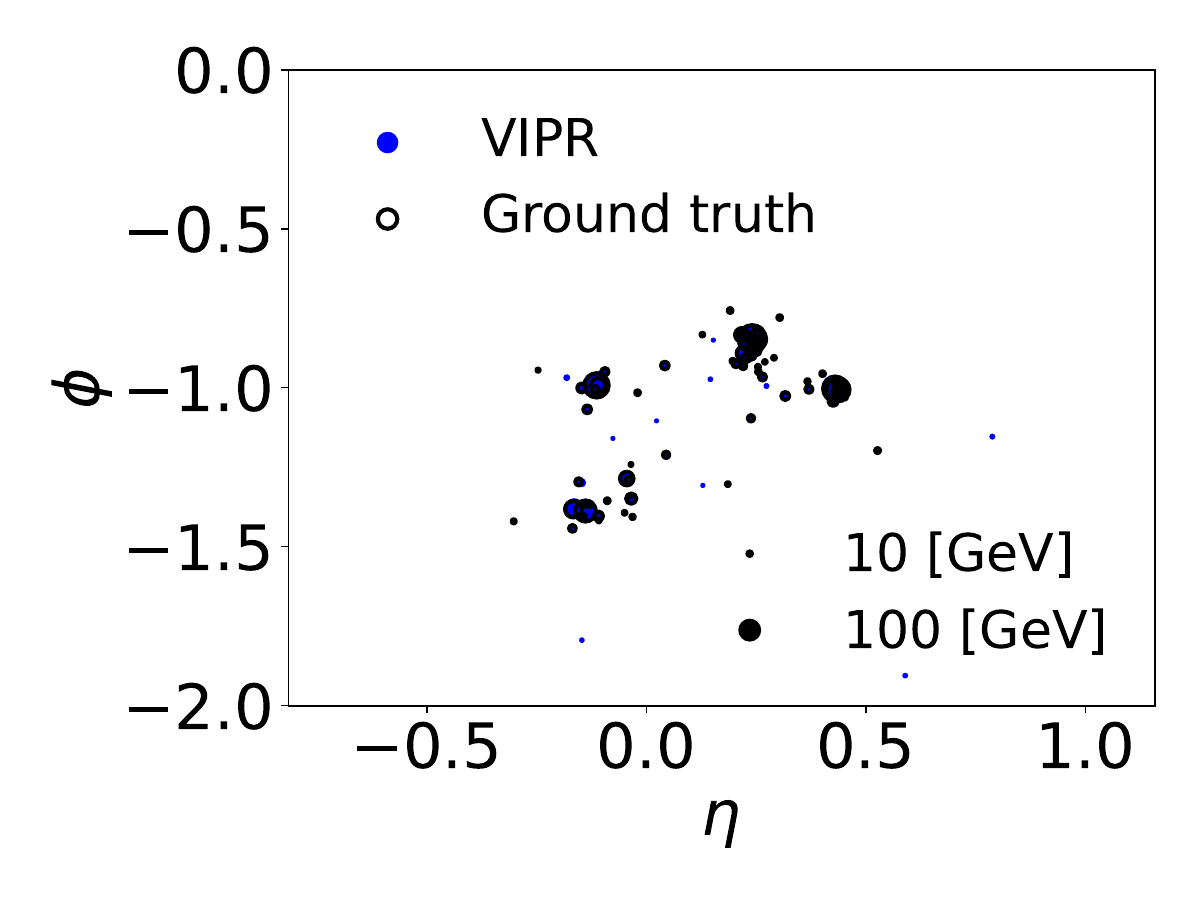}
    }
    \\
    \subfloat[]{
        \includegraphics[width=0.31\textwidth]{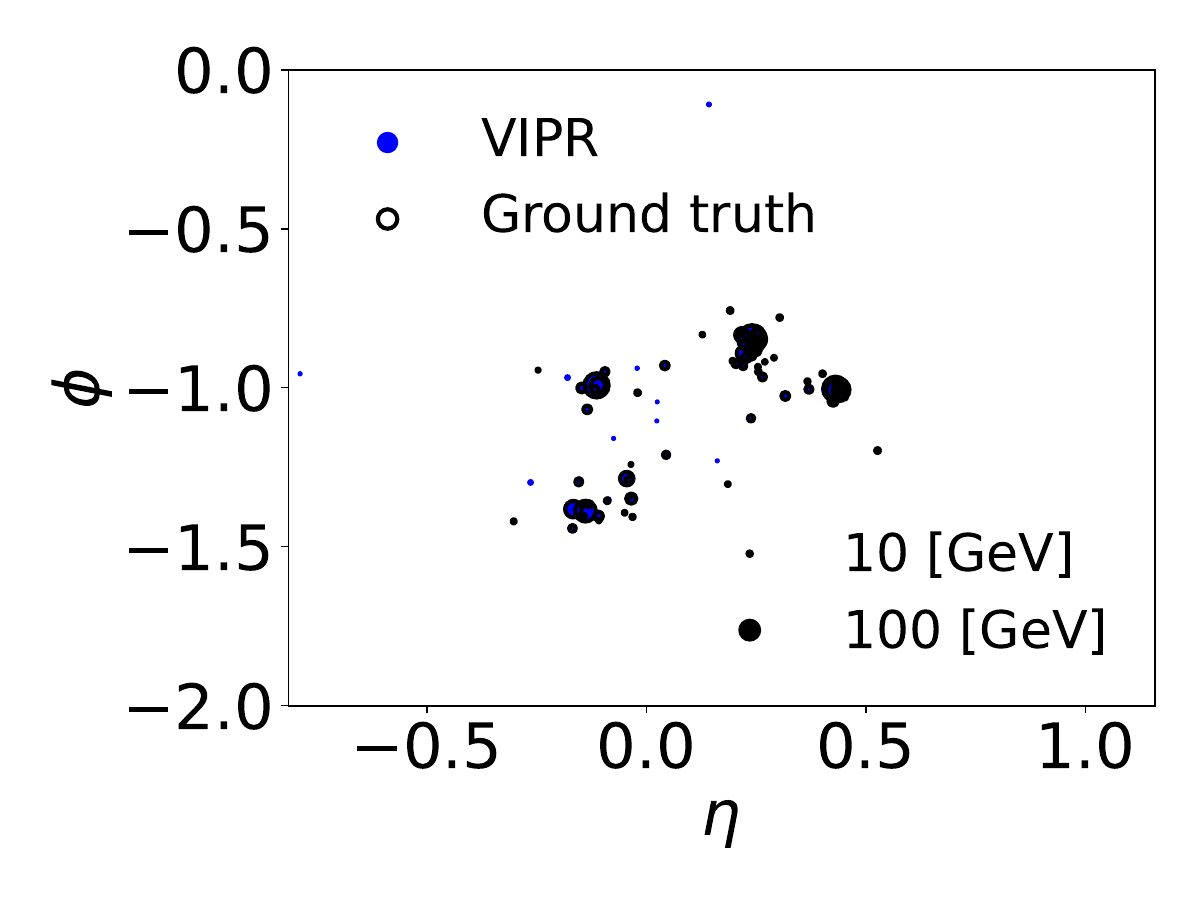}
        }
    \subfloat[]{
        \includegraphics[width=0.31\textwidth]{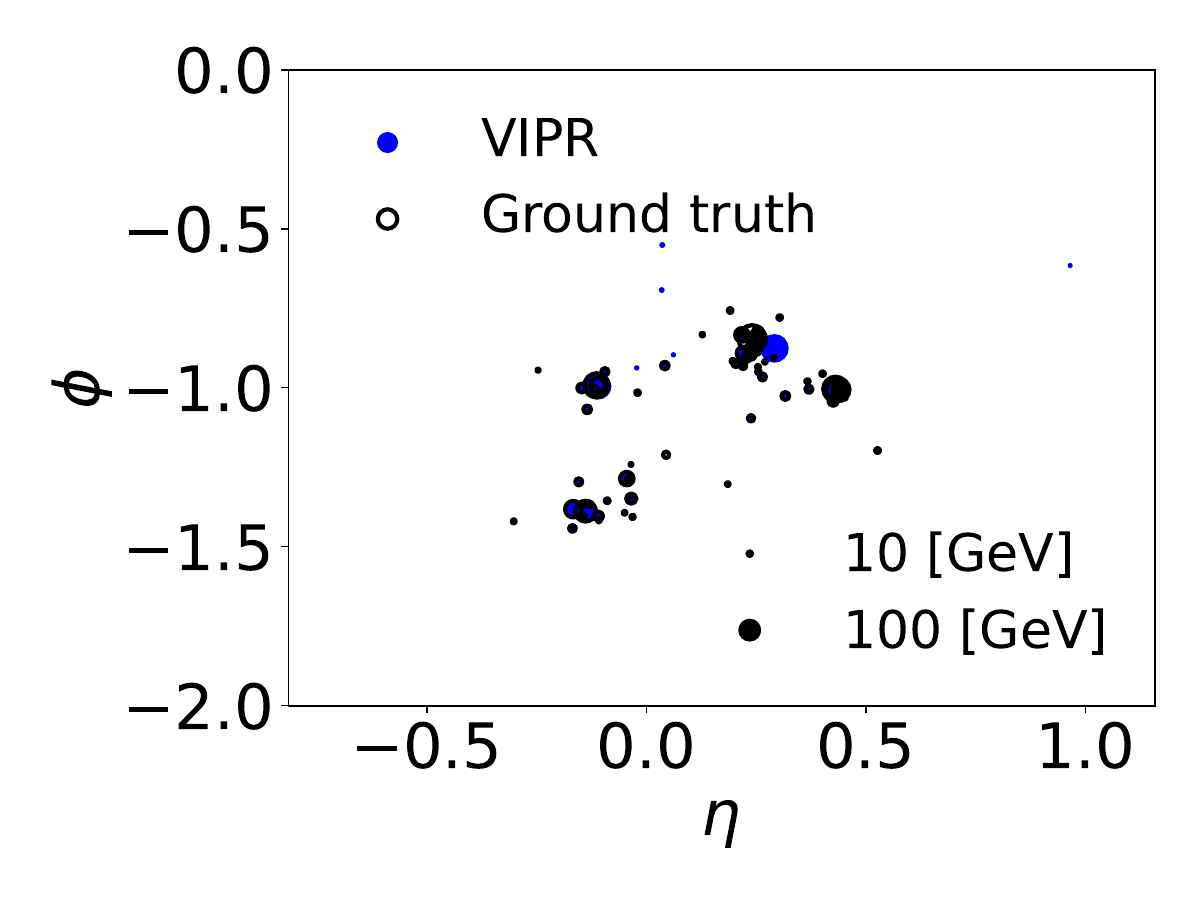}
    }
    \subfloat[]{
        \includegraphics[width=0.31\textwidth]{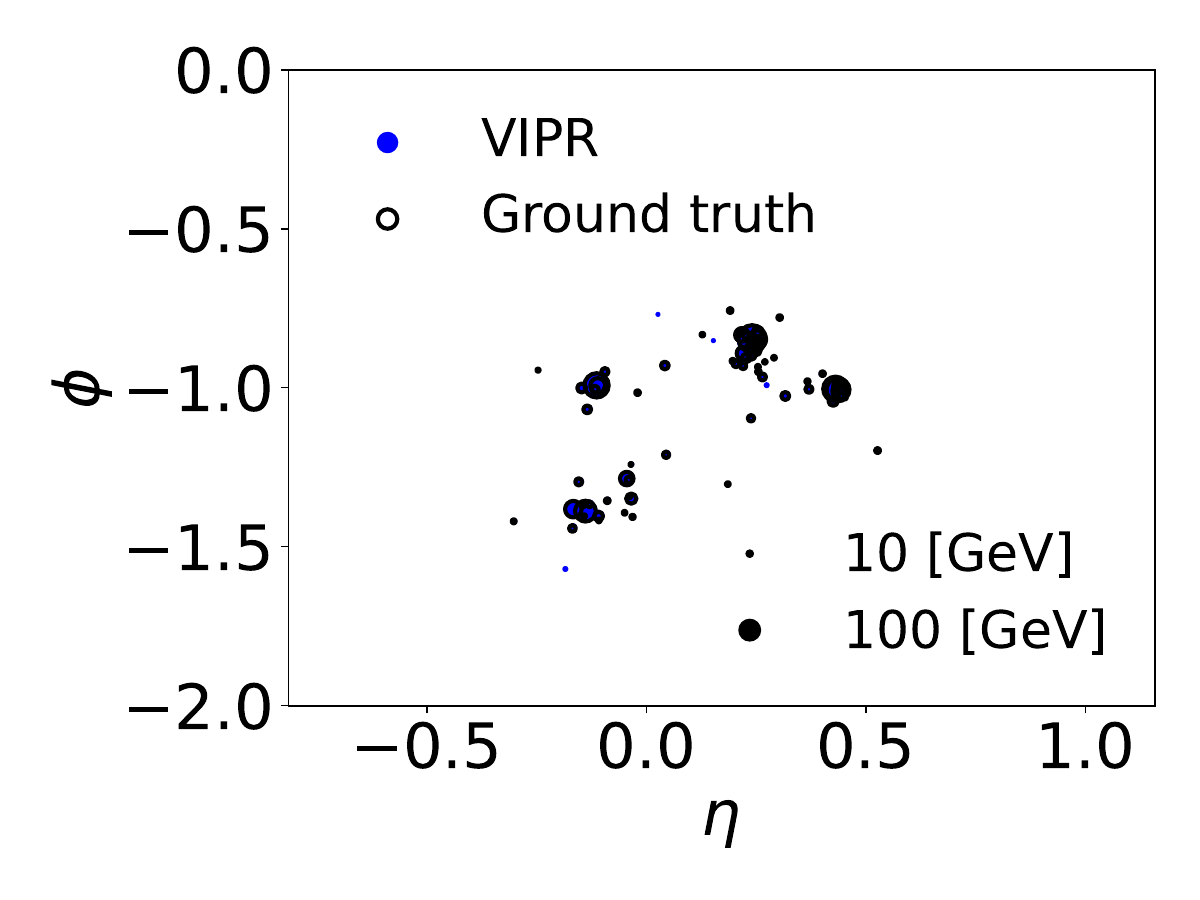}
        }
    \\
    \subfloat[]{
        \includegraphics[width=0.31\textwidth]{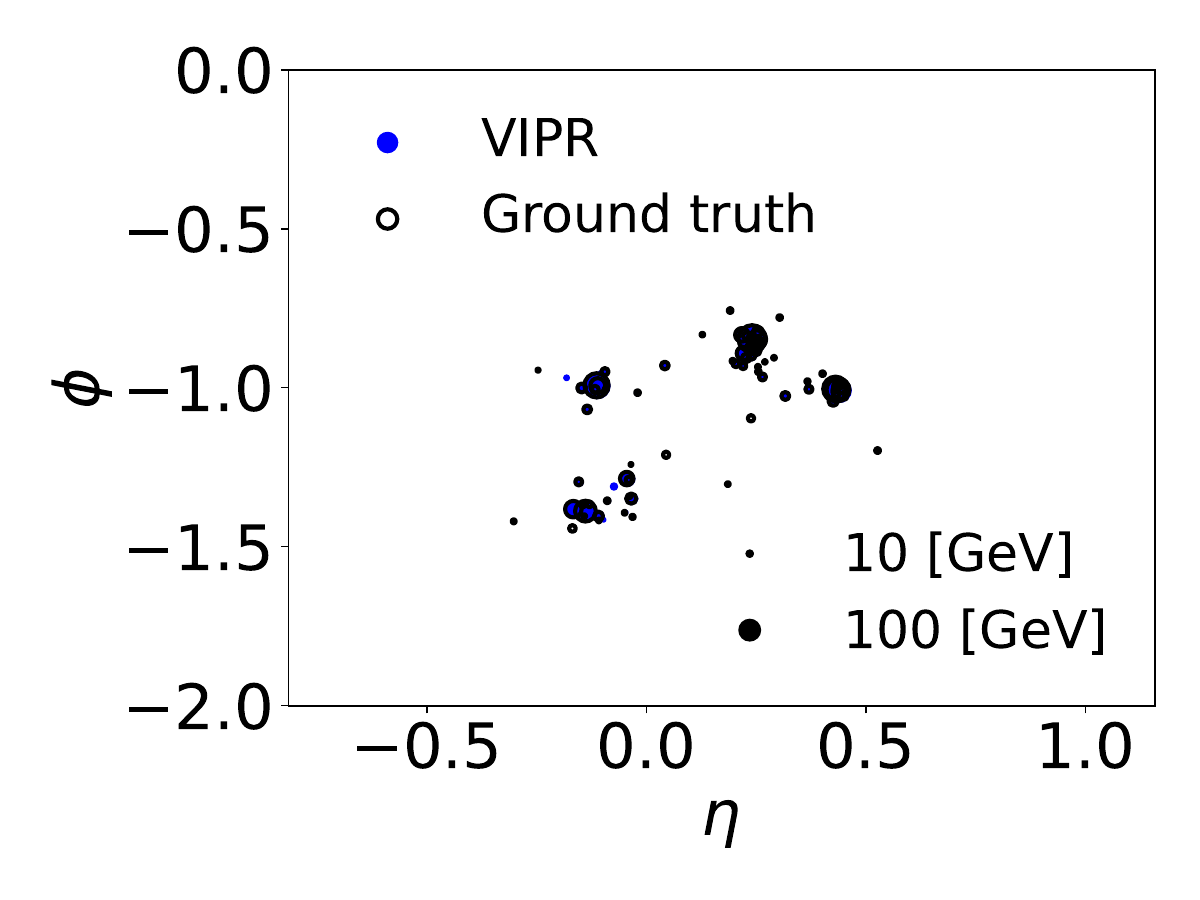}
    }

    \caption{Scatter plots shows multiple examples of generated \vipr jets on the  constituent level in $\eta$ and $\phi$ plane, with \pt indicated by the area of the circle. All the generated jets are using the same observed jet.
    The black circle represents the ground truth constituents.
    } 
    \label{fig:denoised_scatter_posterior}

\end{figure*}

\begin{figure*}[htpb]
    \centering
    \subfloat[]{{\includegraphics[width=0.33\textwidth]{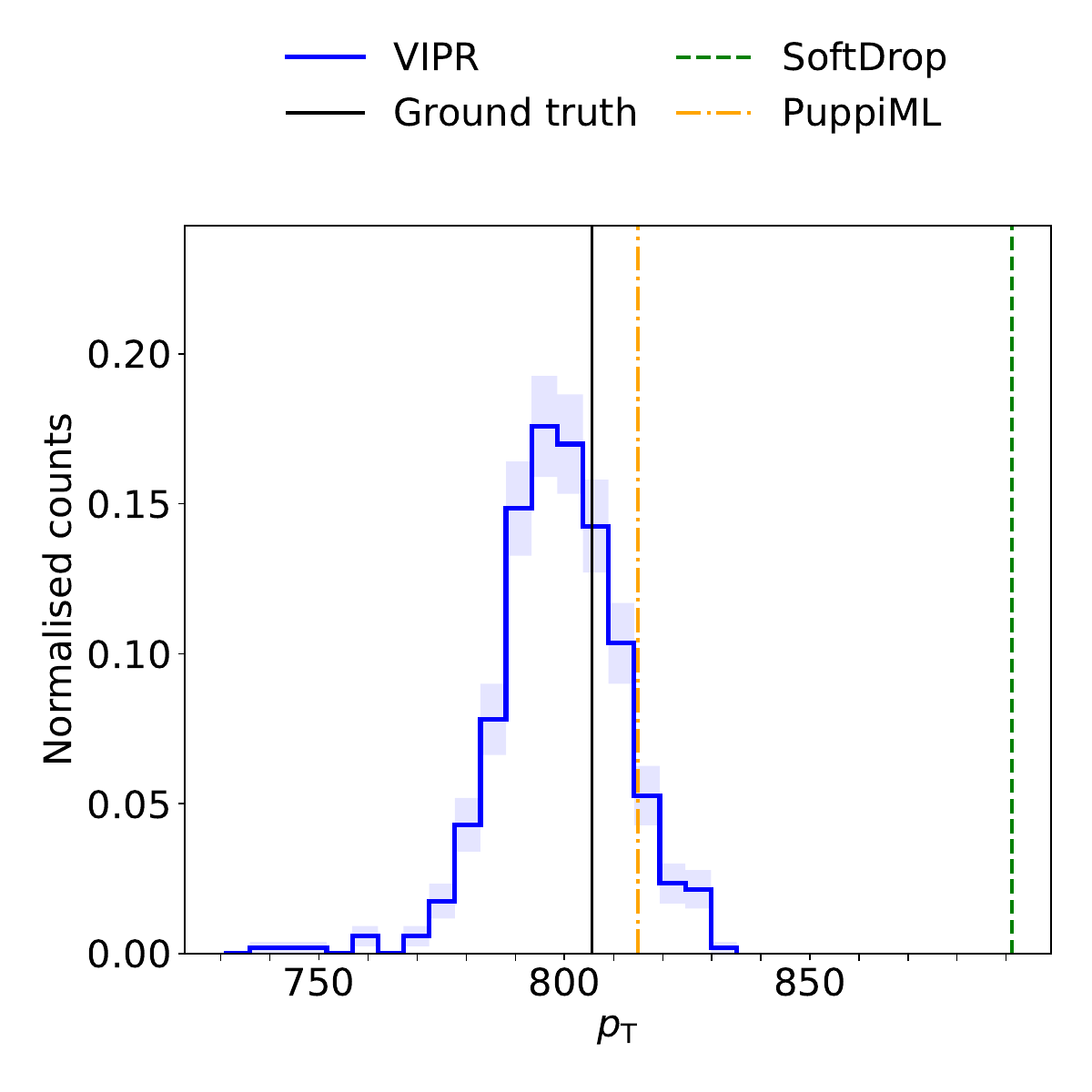}}}
    \subfloat[]{{\includegraphics[width=0.33\textwidth]{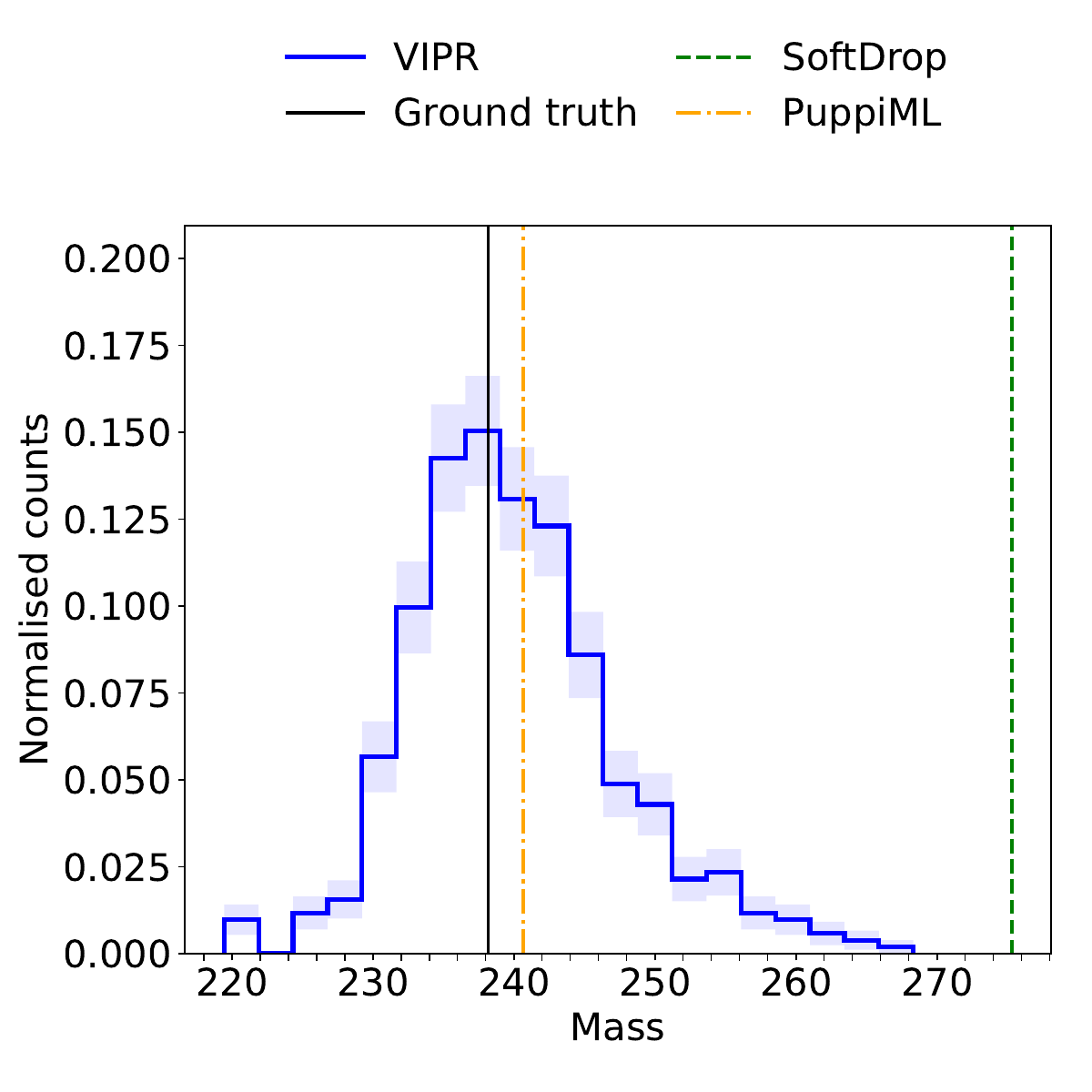}}}
    \subfloat[]{{\includegraphics[width=0.33\textwidth]{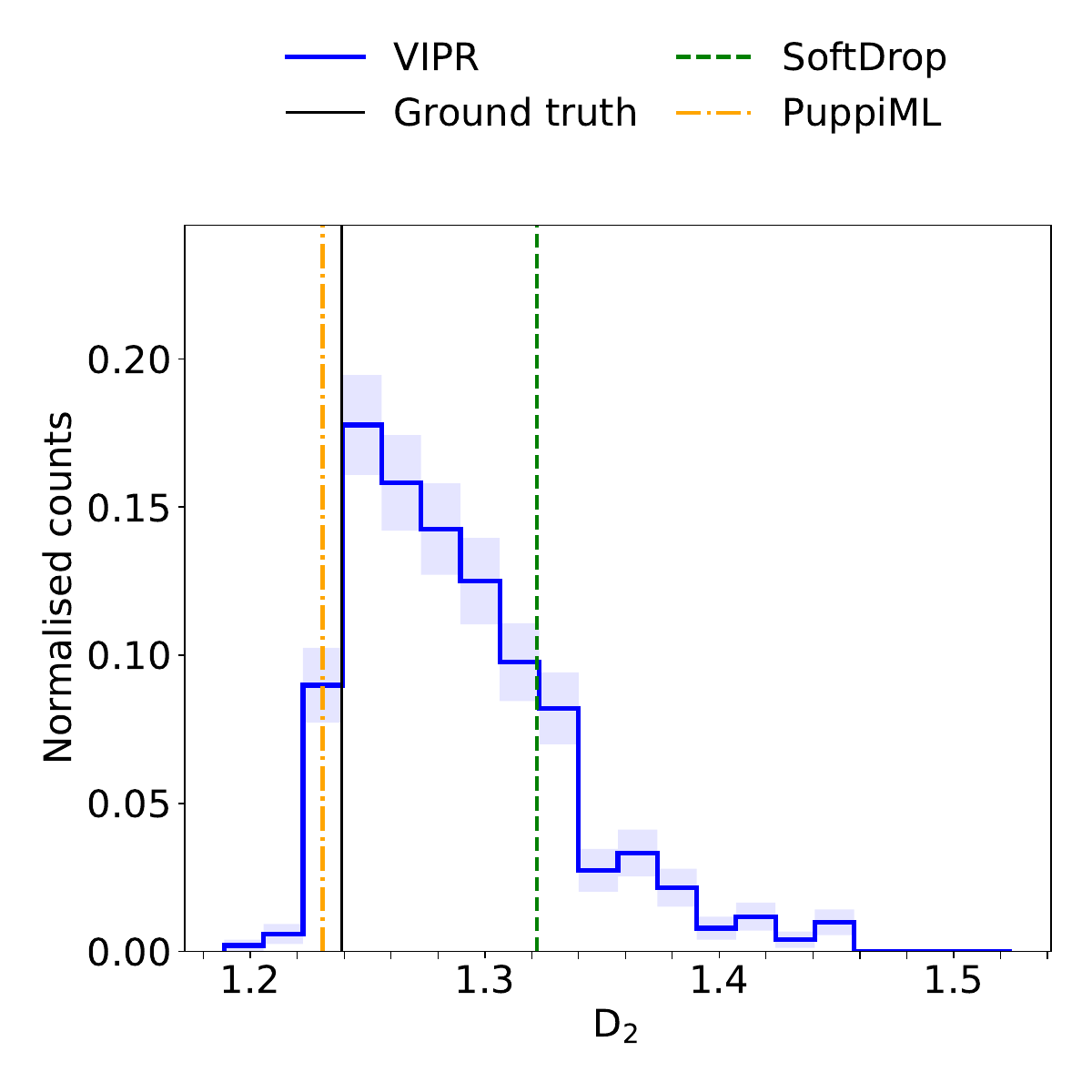}}}
    \\
    \subfloat[]{{\includegraphics[width=0.33\textwidth]{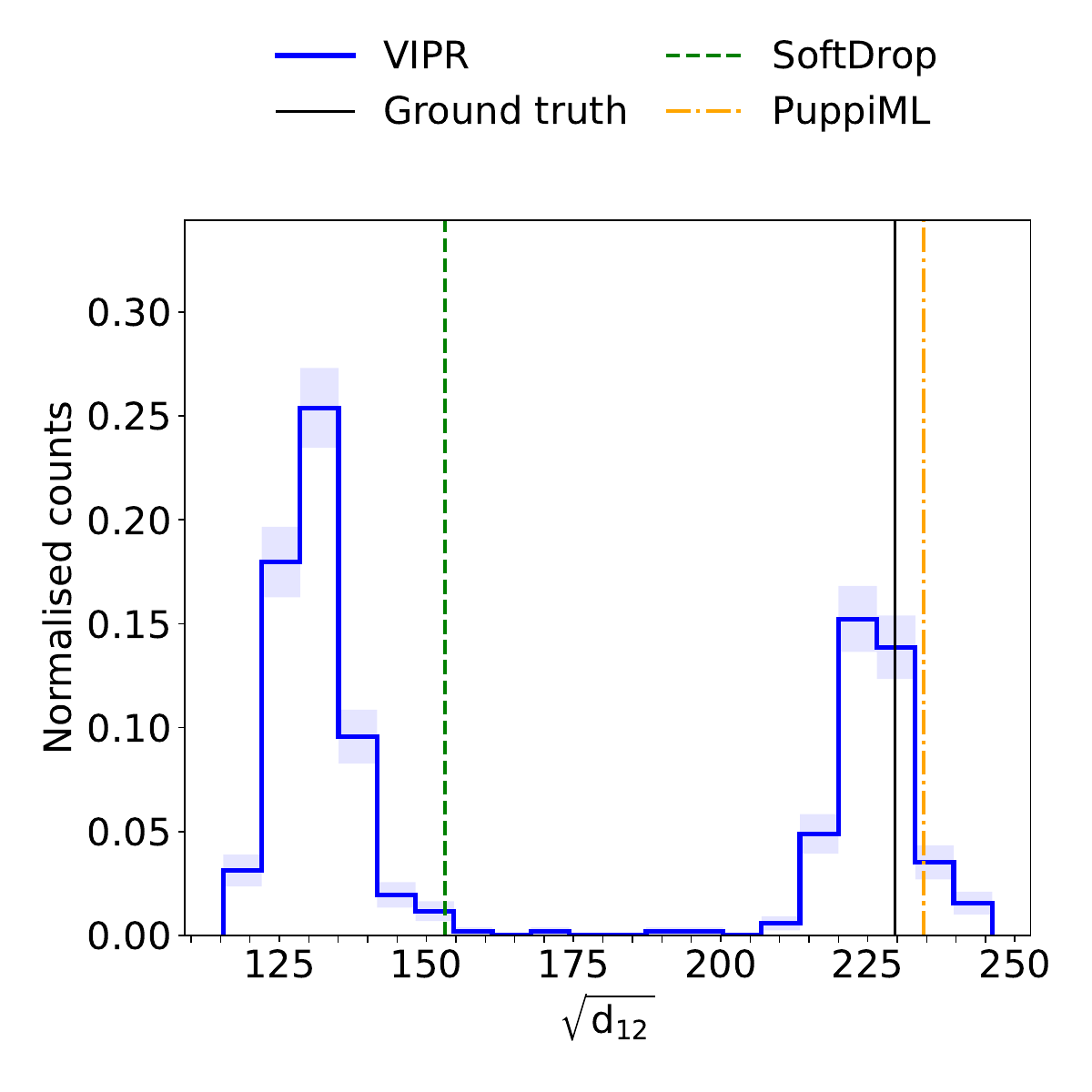}}}
    \subfloat[]{{\includegraphics[width=0.33\textwidth]{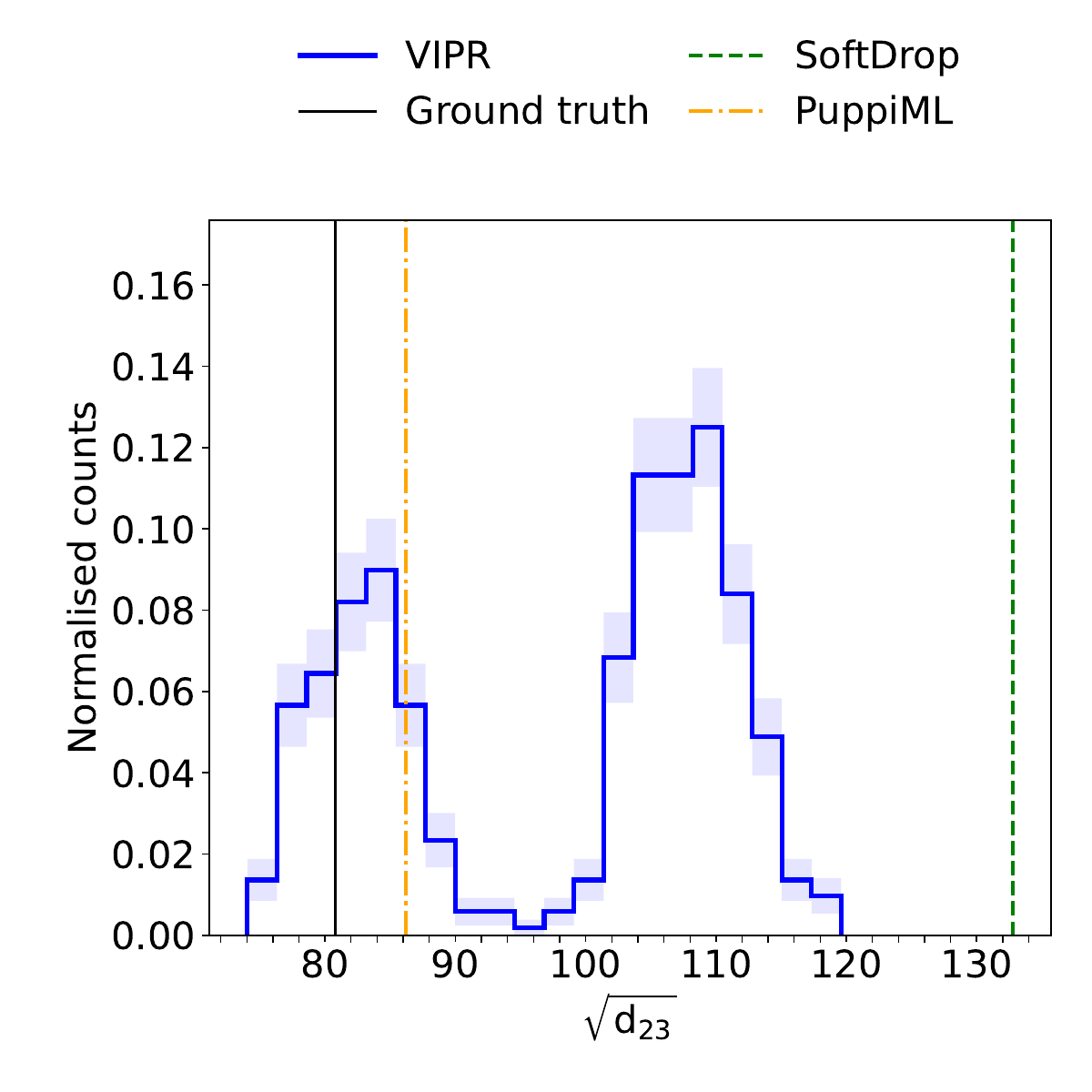}}}
    \subfloat[]{{\includegraphics[width=0.33\textwidth]{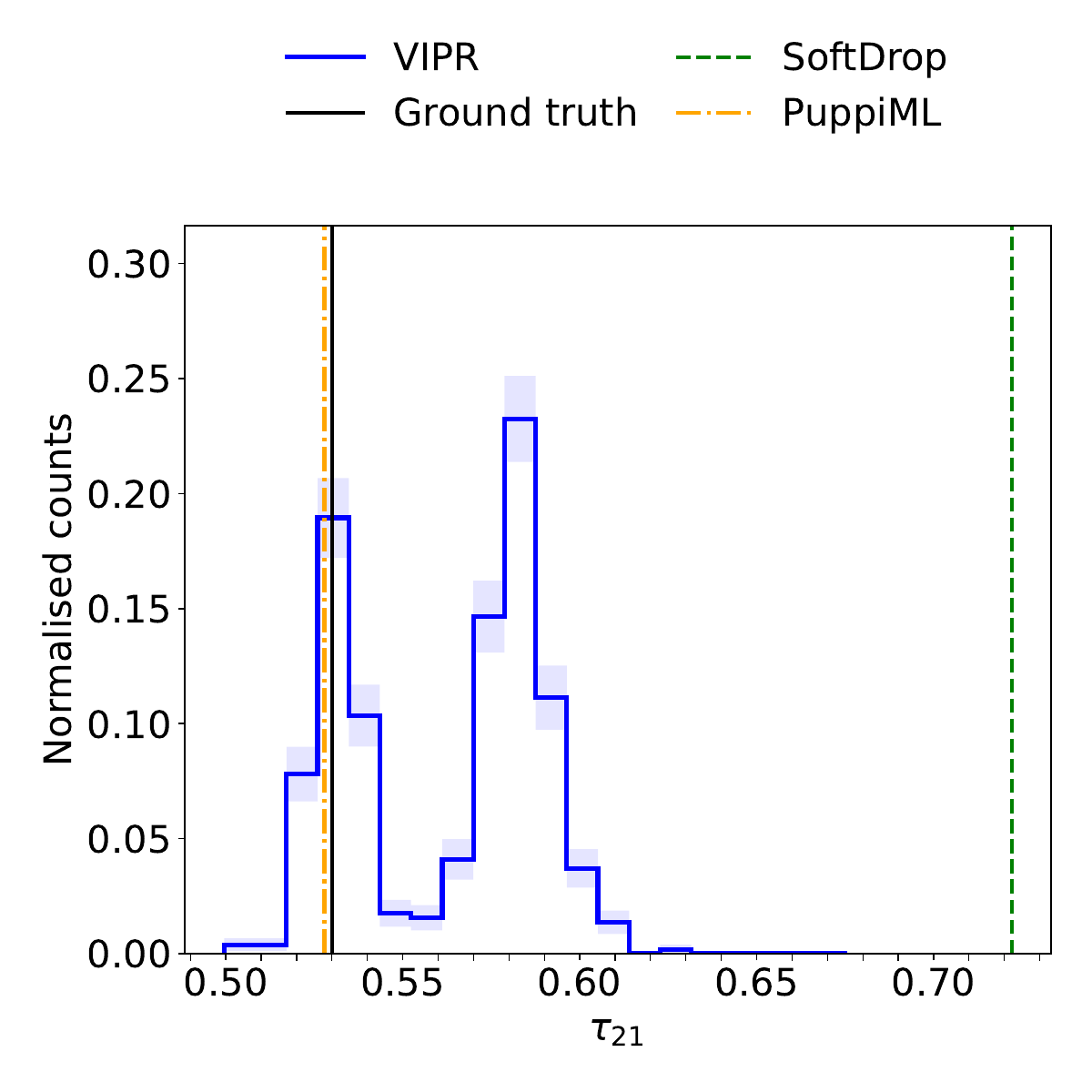}}}
    \\
    \subfloat[]{{\includegraphics[width=0.33\textwidth]{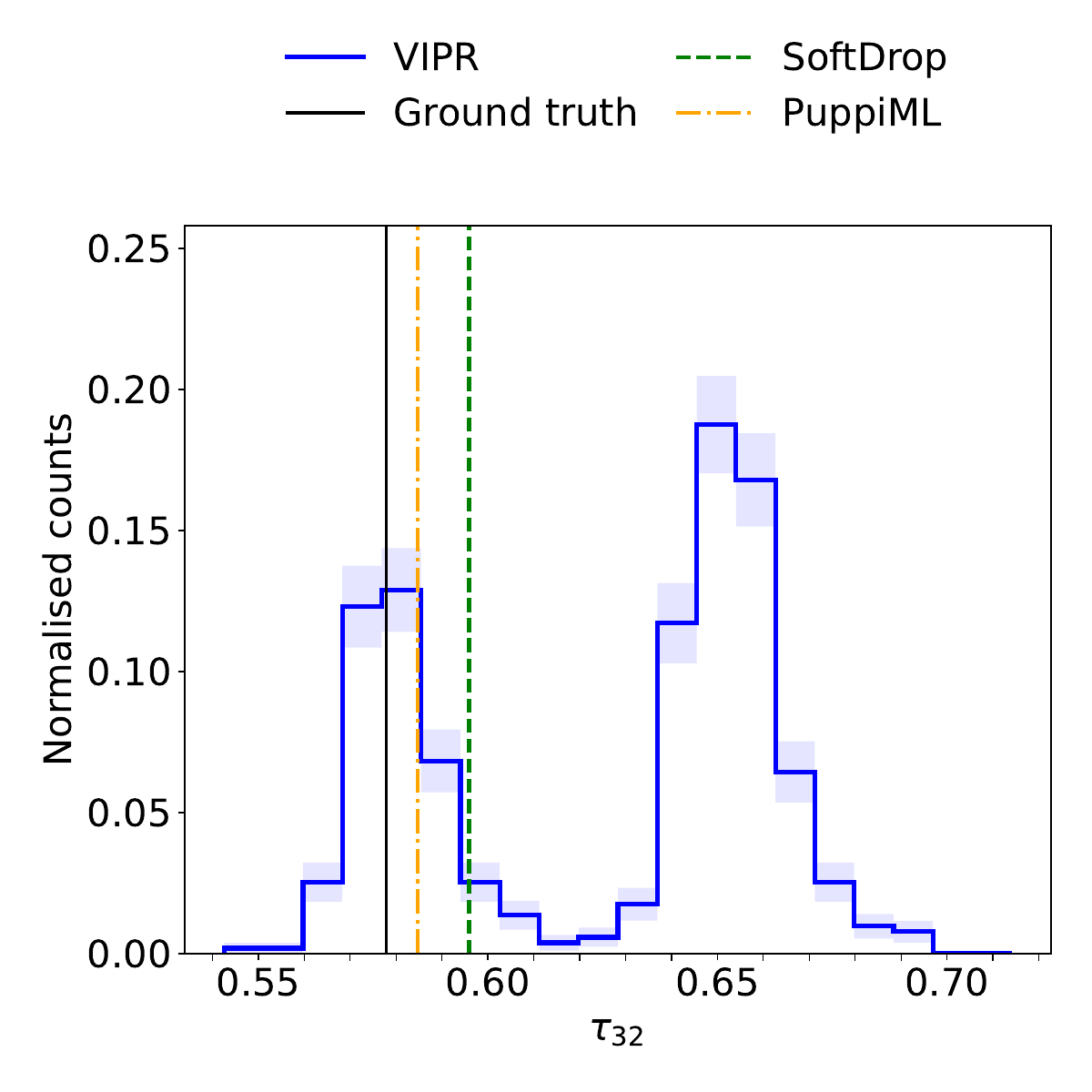}}}
    \caption{
        Posterior distributions of the \pt, mass, and substructure variables generated by \vipr.
        The black line represents the ground truth, the green line represents the \softdrop jet and orange shows the \puppiml jet.
    }
    \label{fig:posterior_distributions_app}
\end{figure*}

\begin{figure*}[htpb]
    \centering
    \subfloat[]{{\includegraphics[width=0.33\textwidth]{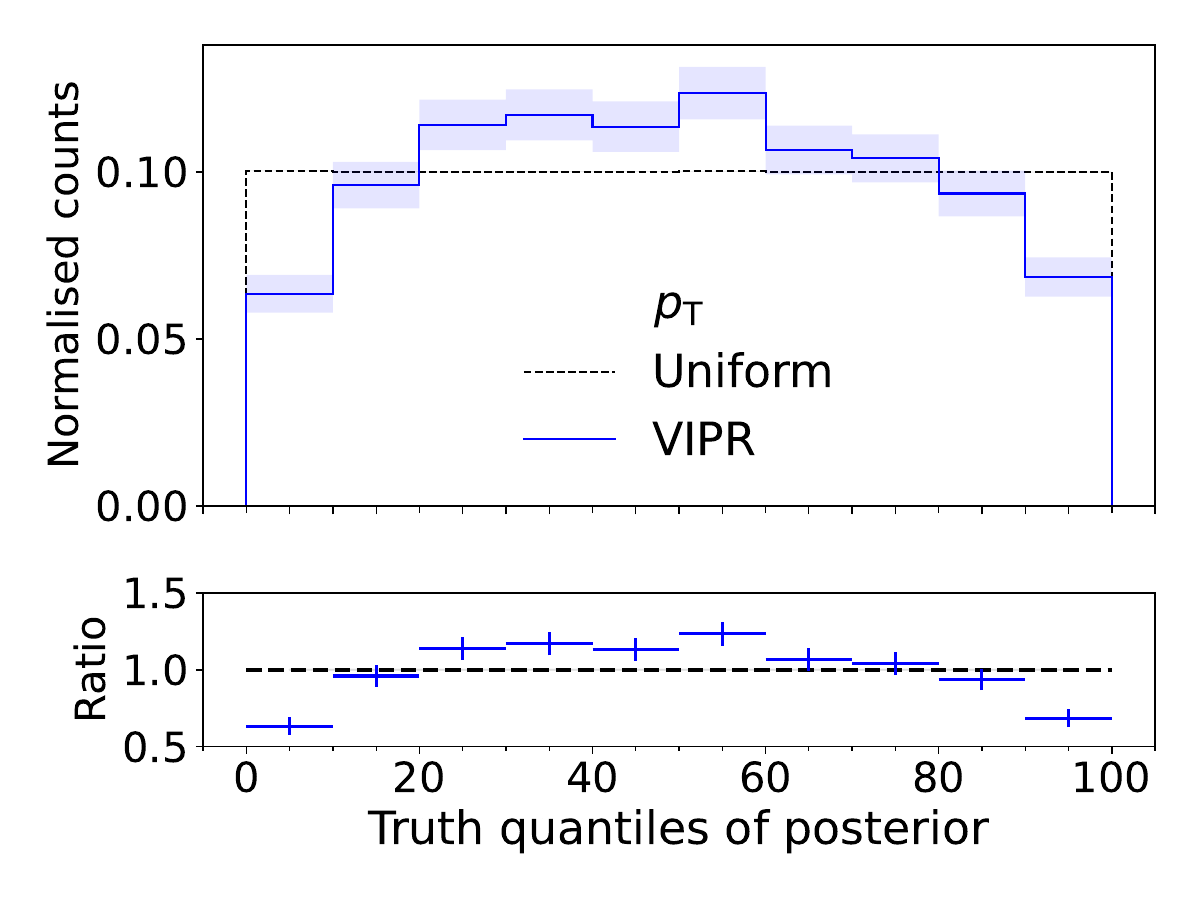}}}
    \subfloat[]{{\includegraphics[width=0.33\textwidth]{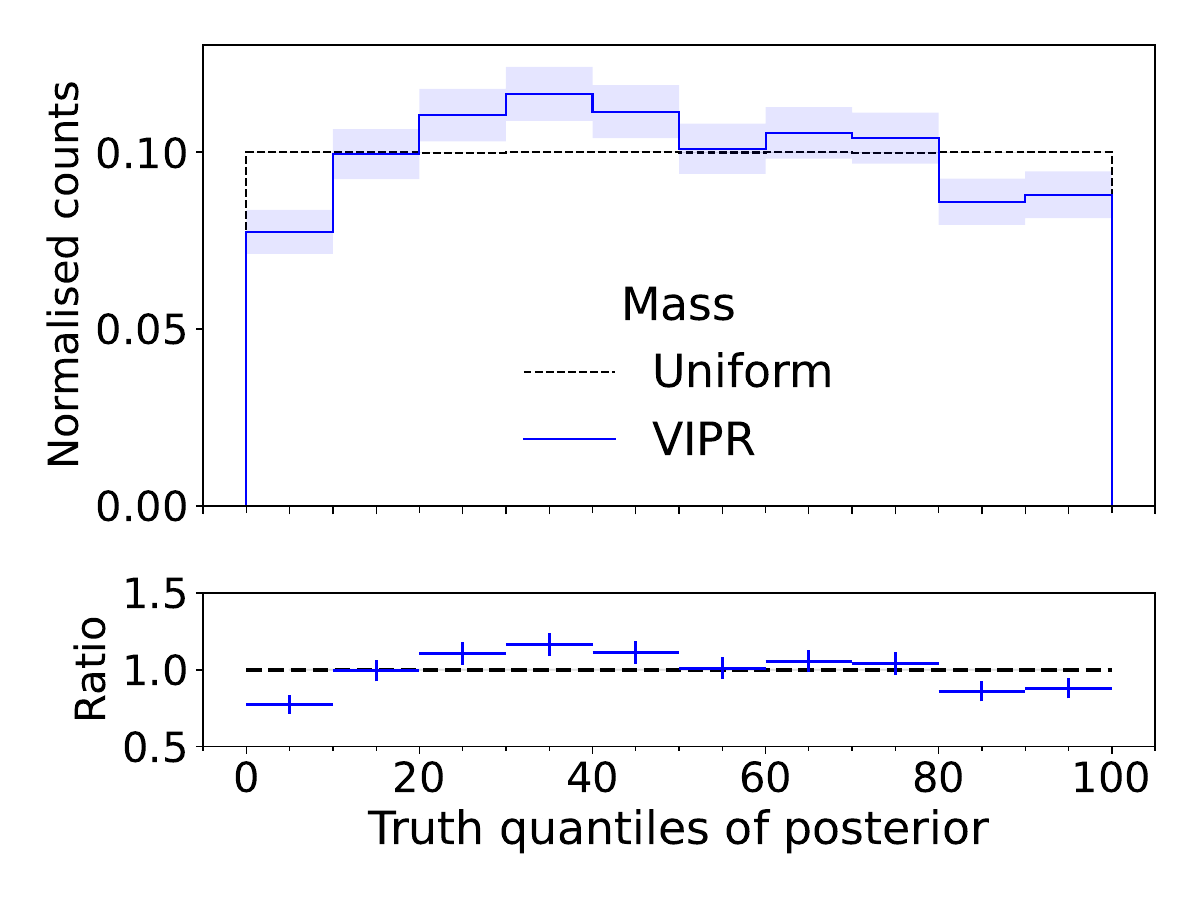}}}
    \subfloat[]{{\includegraphics[width=0.33\textwidth]{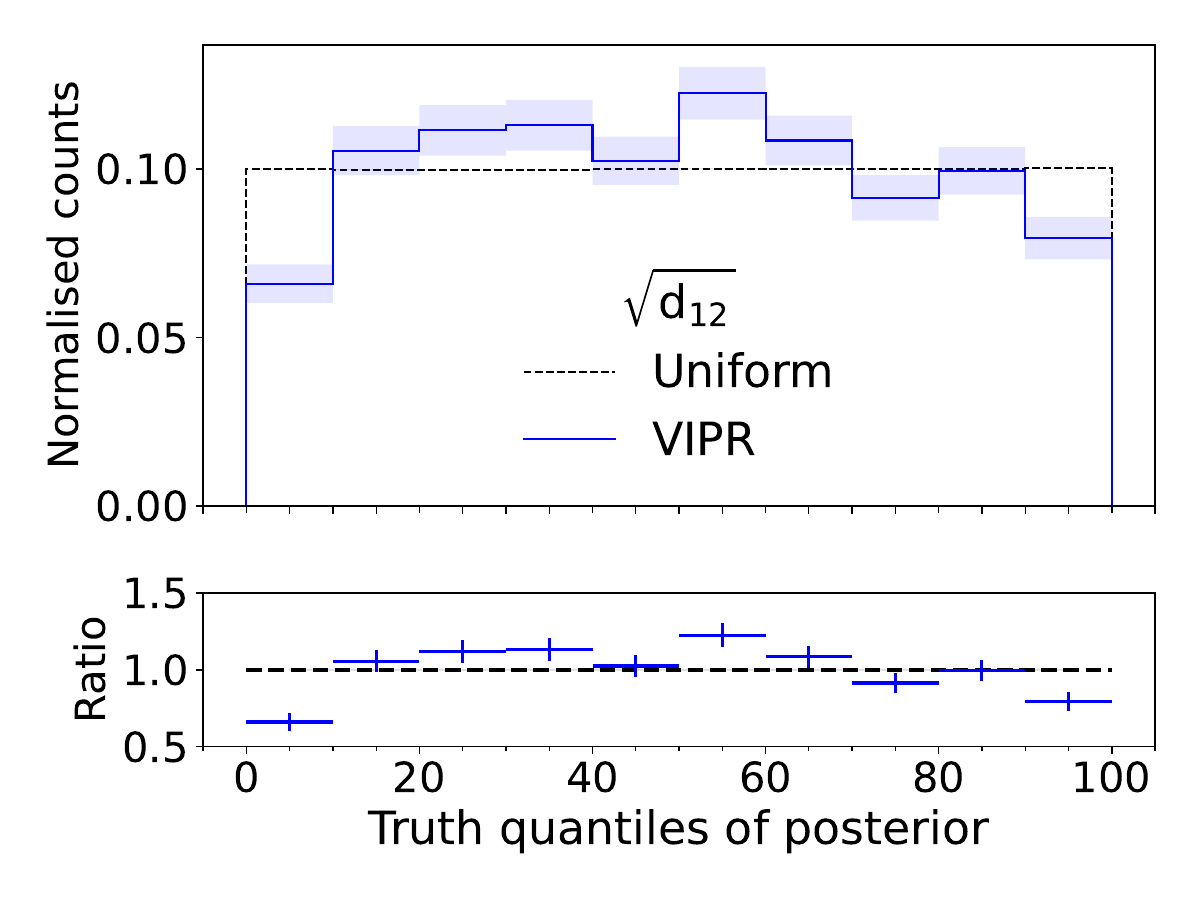}}}
    \\
    \subfloat[]{{\includegraphics[width=0.33\textwidth]{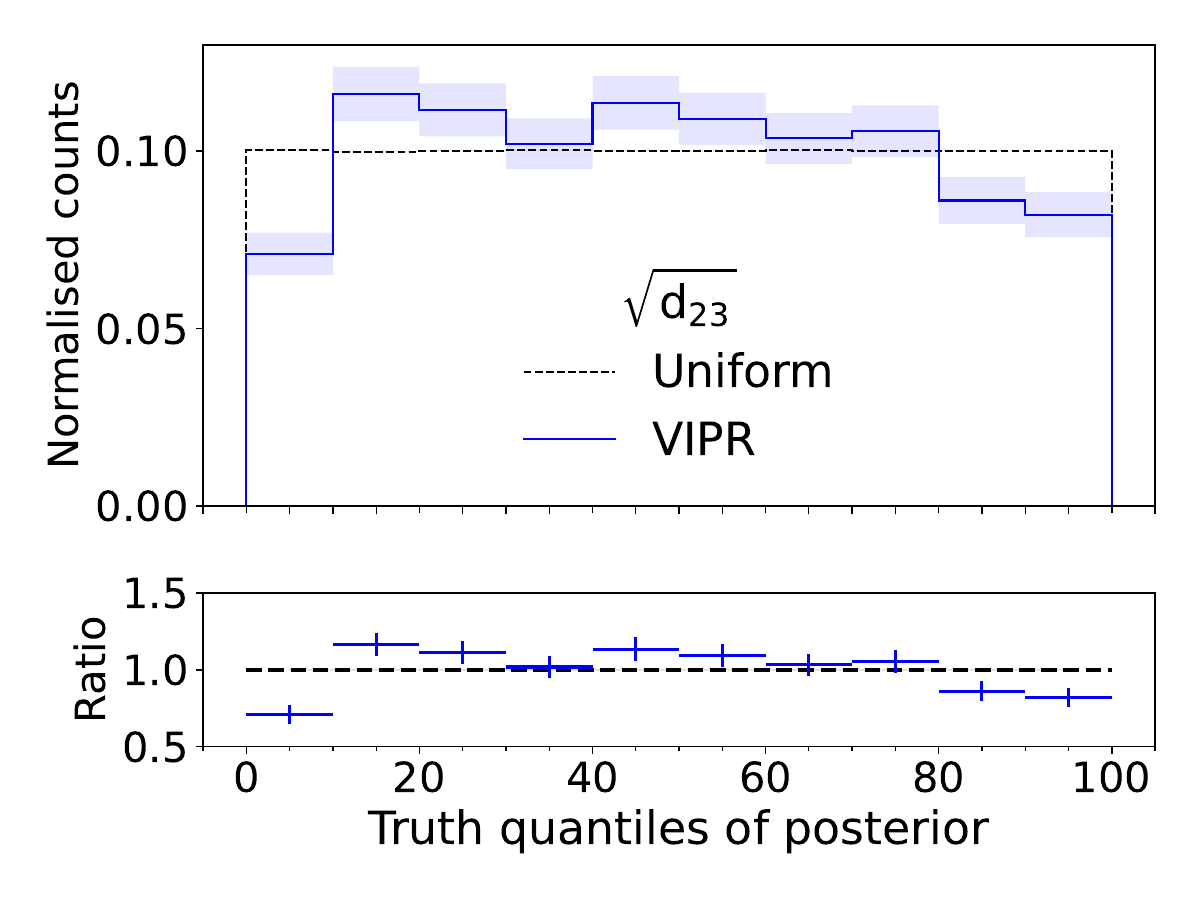}}}
    \subfloat[]{{\includegraphics[width=0.33\textwidth]{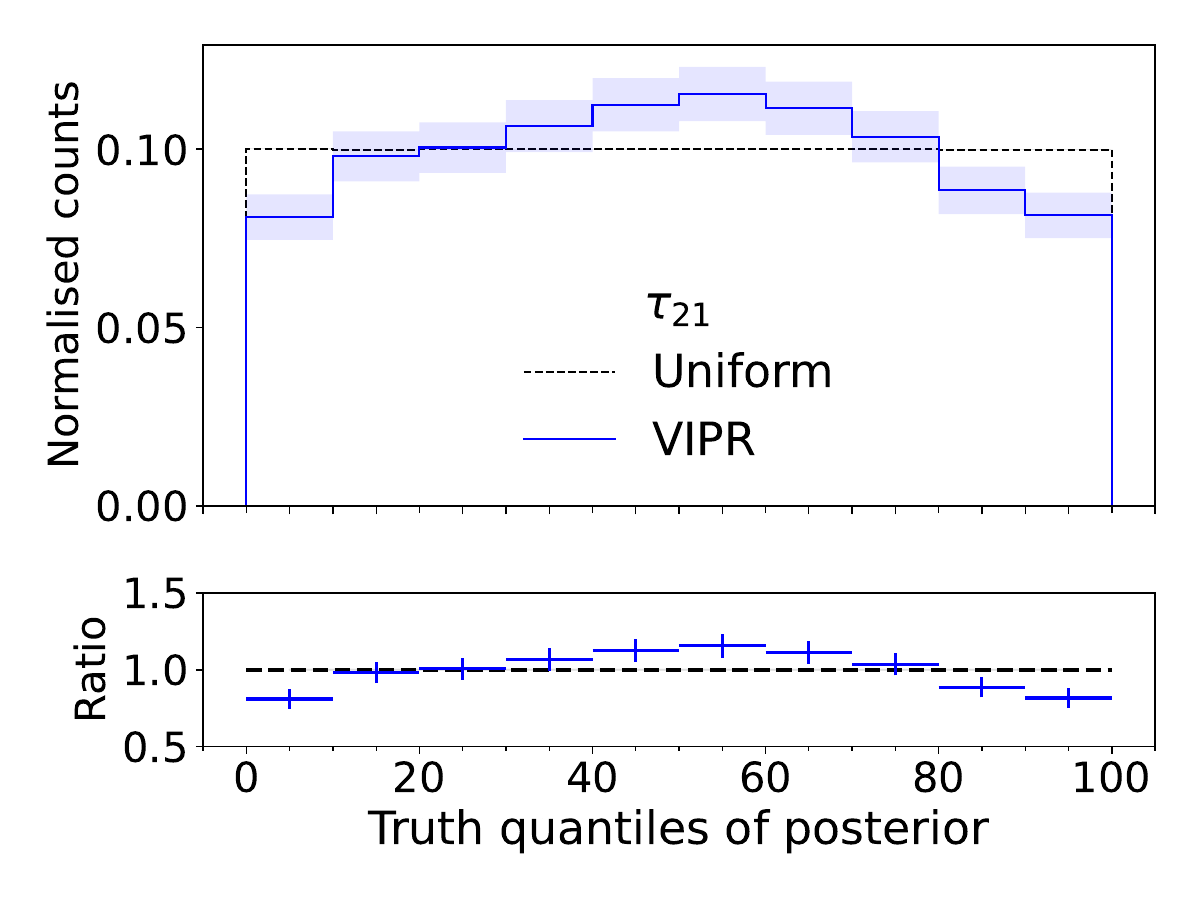}}}
    \subfloat[]{{\includegraphics[width=0.33\textwidth]{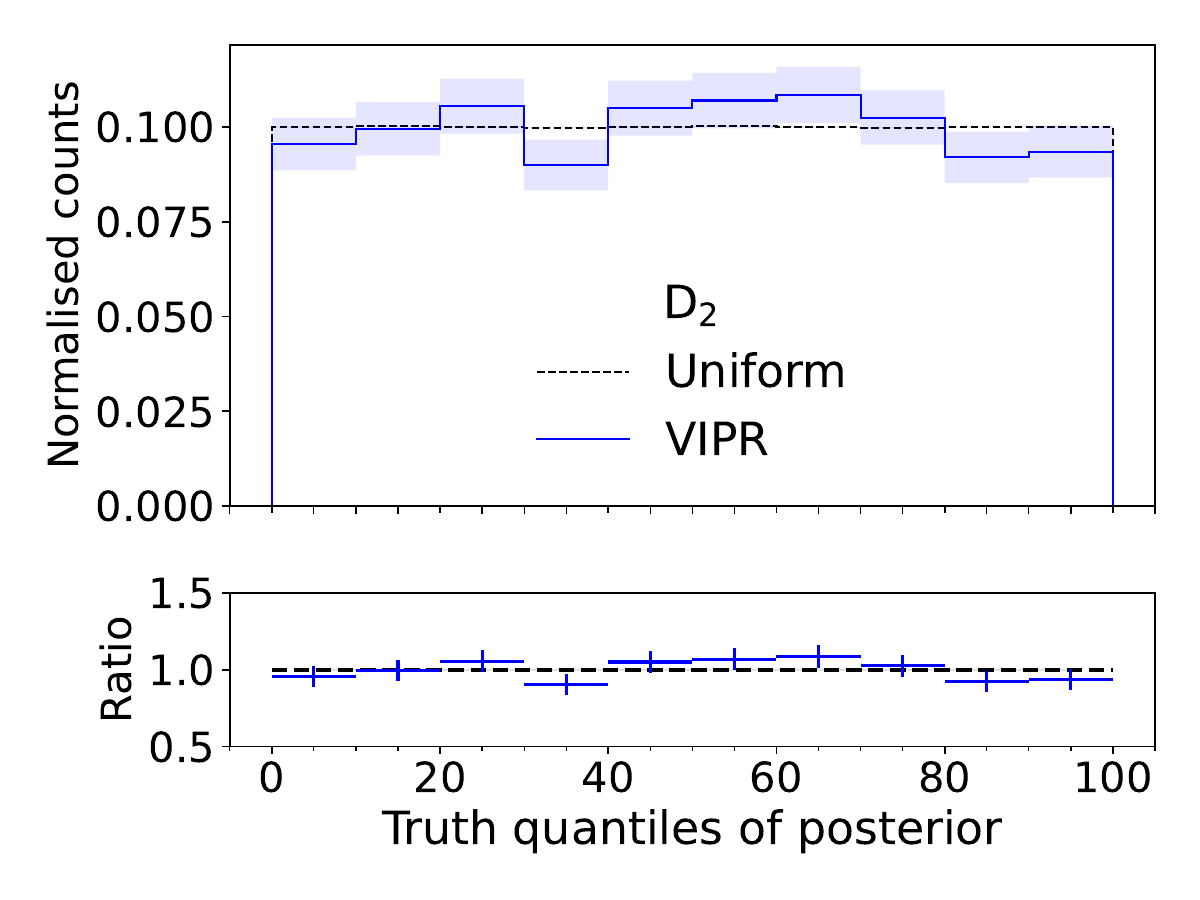}}}
    \\
    \subfloat[]{{\includegraphics[width=0.33\textwidth]{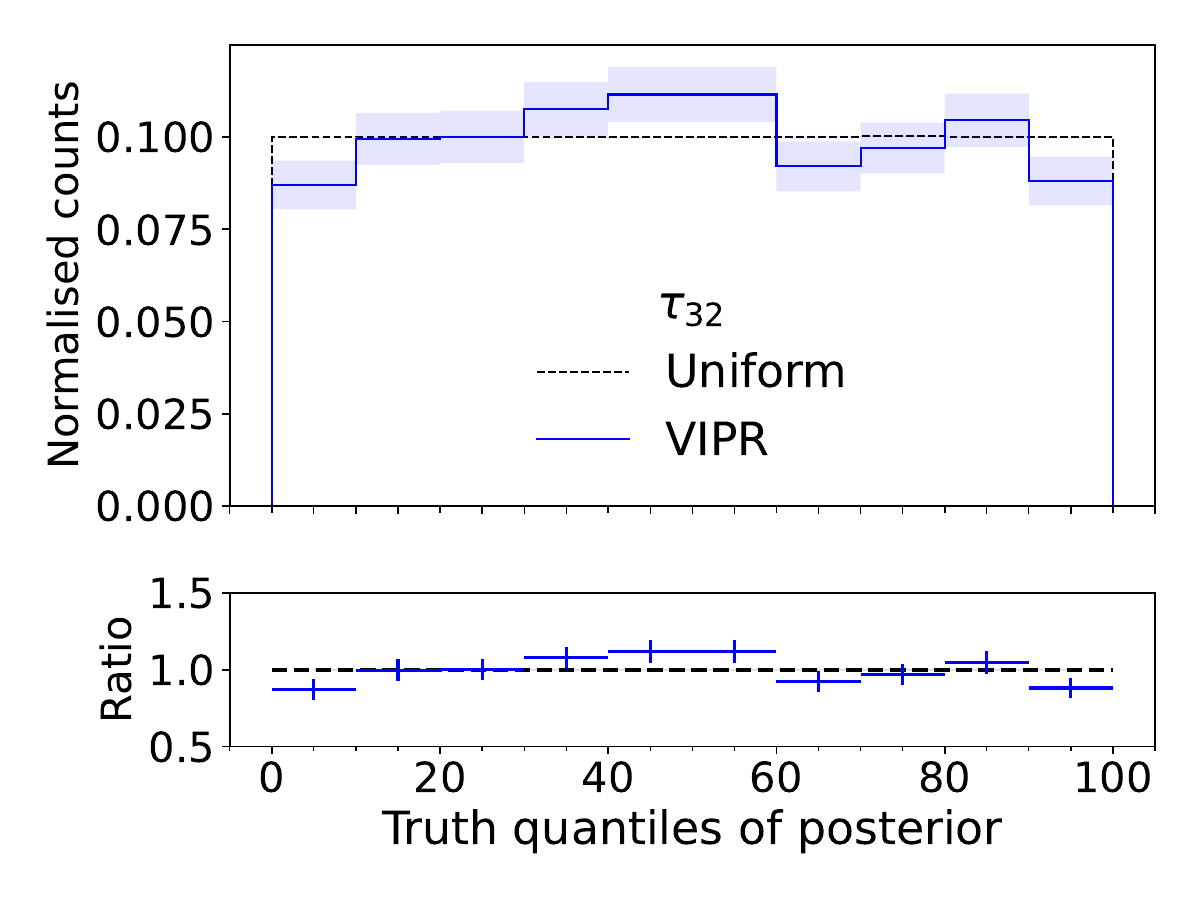}}}
    \caption{Comparisons between the ideal uniform posterior truth quantile and the posterior truth quantile of \vipr in \pt, mass, and substructure variables.
    The posteriors have been generated by sampling a single \jobs 512 times from the base distribution.
    This procedure has been repeated 2,000 times for different \jobs to generate the posterior truth quantiles. The single \jobs is generated using a pile-up distribution of \normal{200}{0}.
    Ideally, the quantiles should be distributed uniformly on the generated posteriors to indicate that the model is correctly calibrated.
    }
    \label{fig:posterior_truth_quantiles_app}
\end{figure*}

\cref{fig:denoised_scatter_posterior} shows multiple examples of possible generated \vipr jets. Each of these are used to construct the posterior distribution of the jet variables.
\cref{fig:posterior_distributions_app} shows the posterior distributions of the \pt, mass and substructure variables generated by \vipr. 
The vertical lines indicate the predictions from other methods.
An interesting feature of the \vipr posterior is that it is not symmetric 
and double-peaked in some variables, which indicates that the model is not well calibrated.
The sampled truth quantiles used for the coverage plot (\cref{fig:coverage_plots}) are shown in \cref{fig:posterior_truth_quantiles_app}.

\FloatBarrier
\subsection{Results in other variables} \label{sec:app_results}
\begin{figure}[htpb]
    \centering
    \subfloat[]{{\includegraphics[width=0.4\textwidth]{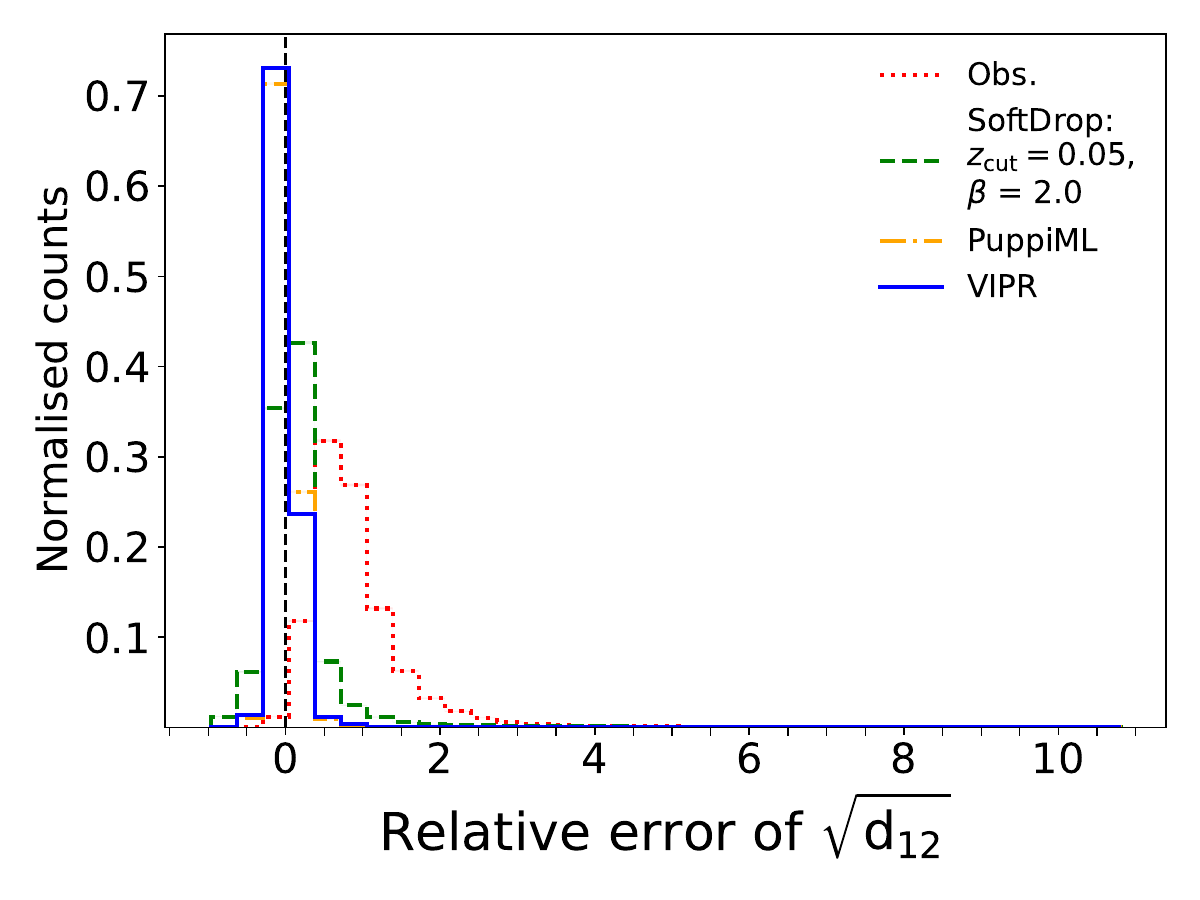}}}
    \\
    \subfloat[]{{\includegraphics[width=0.4\textwidth]{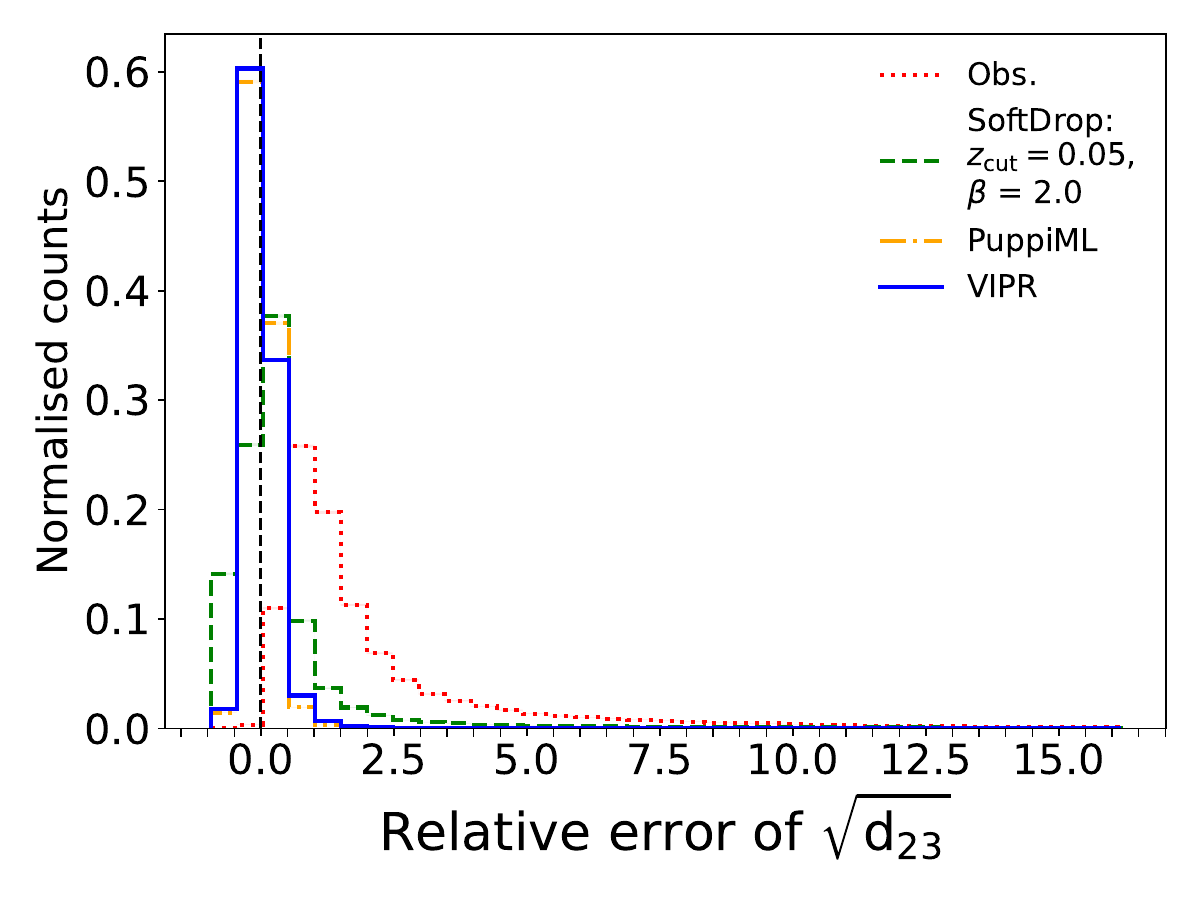}}}
    \\
    \subfloat[]{{\includegraphics[width=0.4\textwidth]{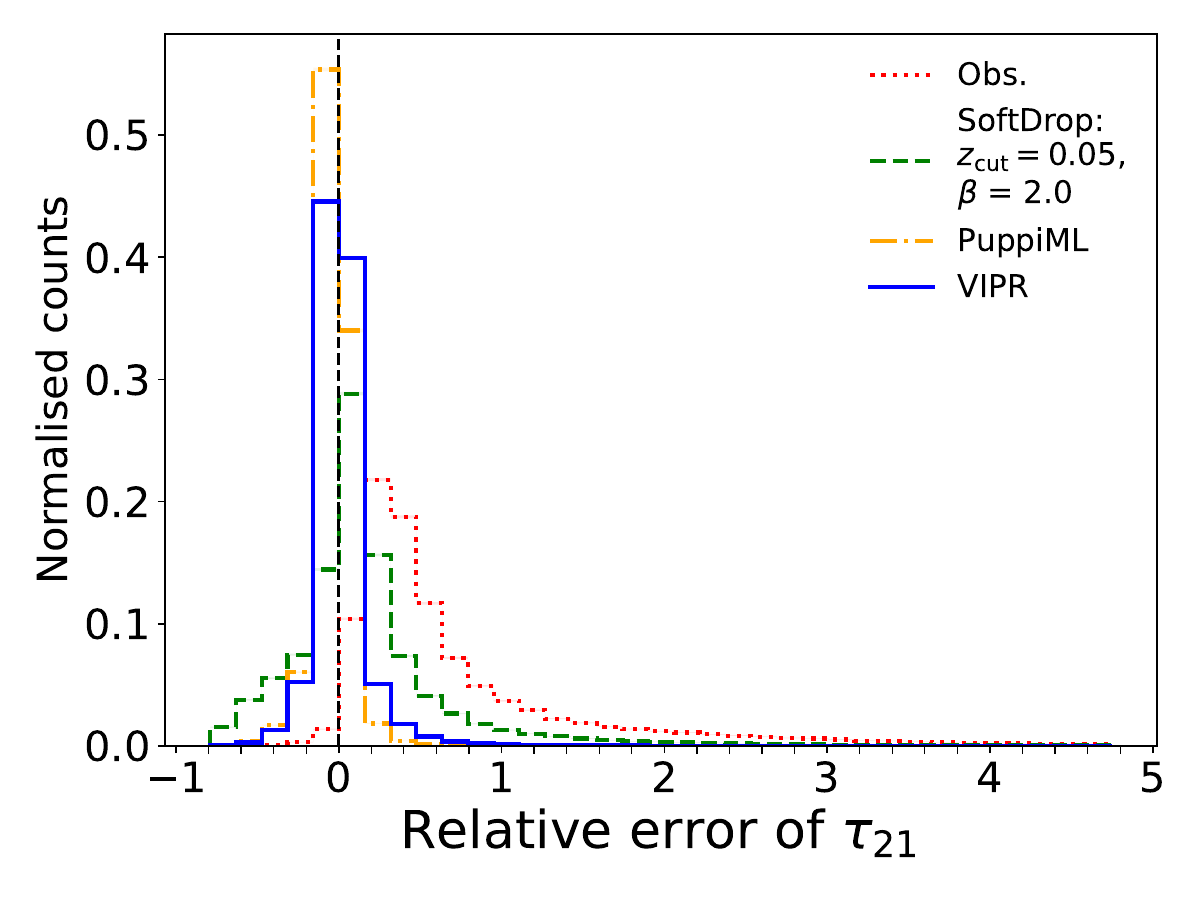}}}
    \caption{
    Comparisons of the RE between observed (red), \softdrop (green), \vipr (blue)
    and \puppiml (orange) jets in $d_{12}$, $d_{23}$, and $\tau_{21}$.
    Ideally, the RE distribution should be symmetric
    and as close to a delta peak at zero as possible.
    The observed jet is generated using a pile-up distribution of 
    \normal{200}{50}.
    }
    \label{fig:marginal_resolution_of_single_app}
\end{figure}

\begin{figure*}[htpb]
    \centering
    \subfloat[]{{\includegraphics[width=0.33\textwidth]{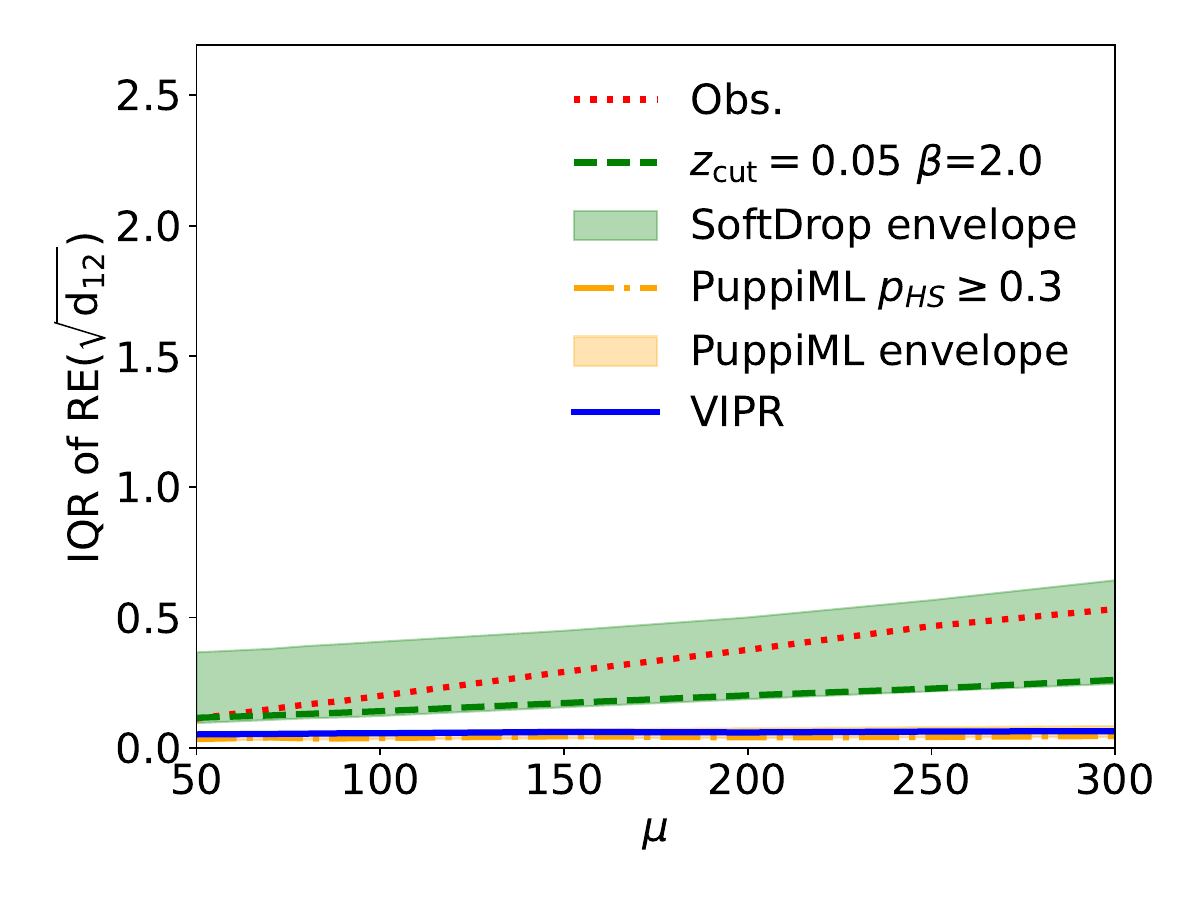}}}
   \subfloat[]{{\includegraphics[width=0.33\textwidth]{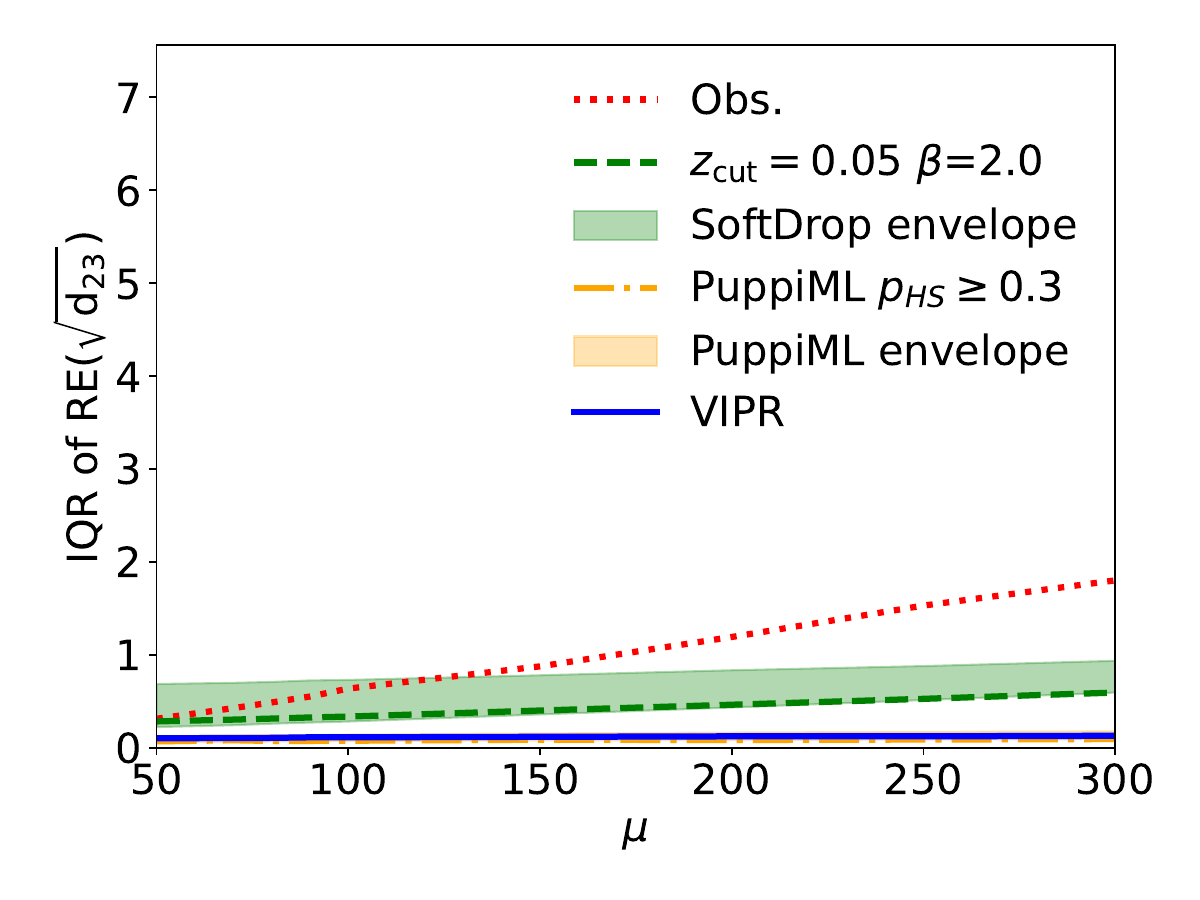}}}
    \subfloat[]{{\includegraphics[width=0.33\textwidth]{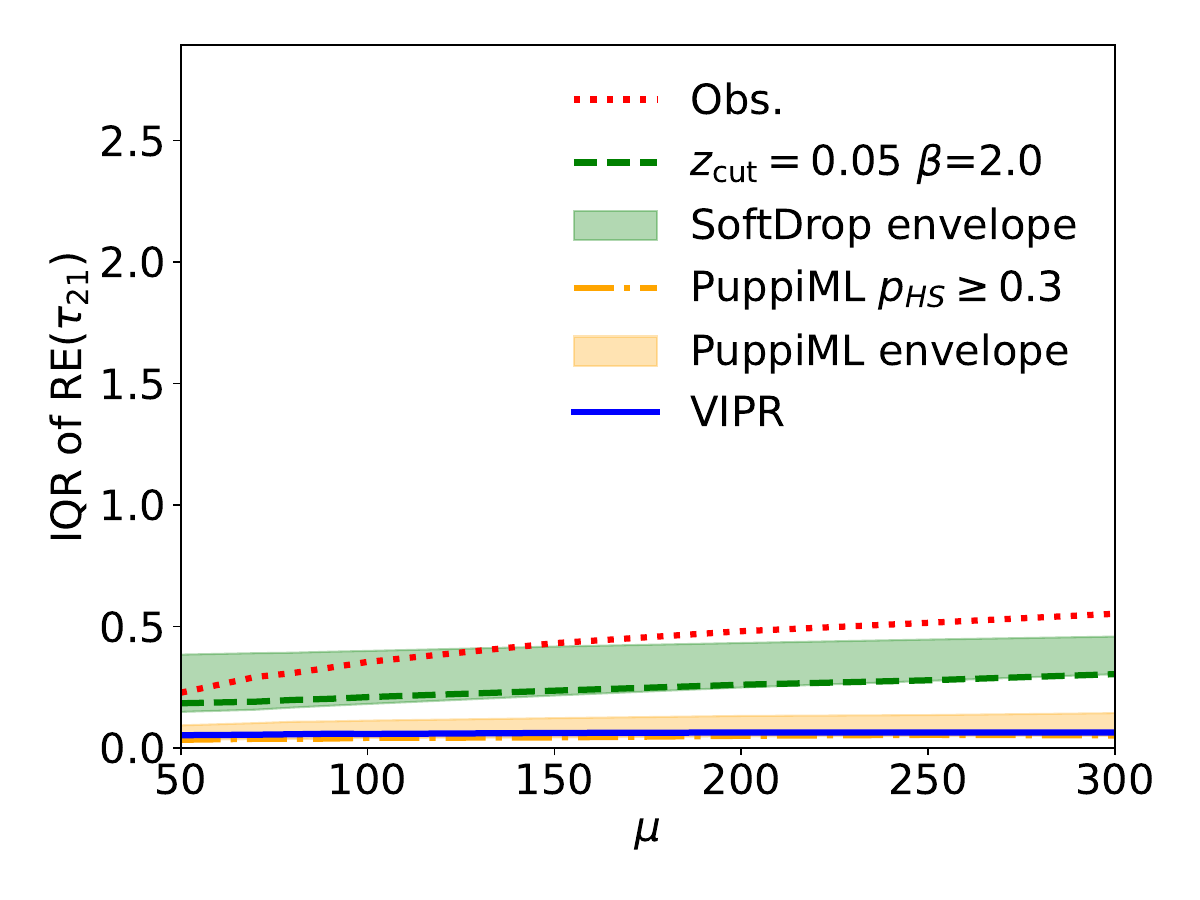}}}
    \caption{
    Comparisons of the IQR of the RE as a function of $\mu$
    for $d_{12}$, $d_{23}$, and $\tau_{21}$.
    The IQR measures the width of the distribution.
    Ideally, the IQR should be as close to zero as possible
    and remain constant as a function of $\mu$.
    If the IQR remains constant as a function of $\mu$, 
    it indicates that the pile-up removal method is robust towards increasing pile-up.
    The envelope of a scan of \softdrop settings is shown
    with the best resulting parameters in dashed green.
    The envelope of a scan of \puppiml cuts is also shown.
    }
    \label{fig:missing_response_variables_at_different_iqr_app}
\end{figure*}

\cref{fig:marginal_resolution_of_single_app} shows the RE of some additional substructure variables. 
The RE is ideally symmetric and close to a delta peak at zero.
The additional substructure variables are the $d_{12}$, $d_{23}$, and $\tau_{21}$.
\cref{fig:missing_response_variables_at_different_iqr_app} shows the IQR of the RE as a function of $\mu$ for various \pt, mass, and substructure variables.

\begin{figure*}[htpb]
    \centering
    \subfloat[]{{\includegraphics[width=0.33\textwidth]{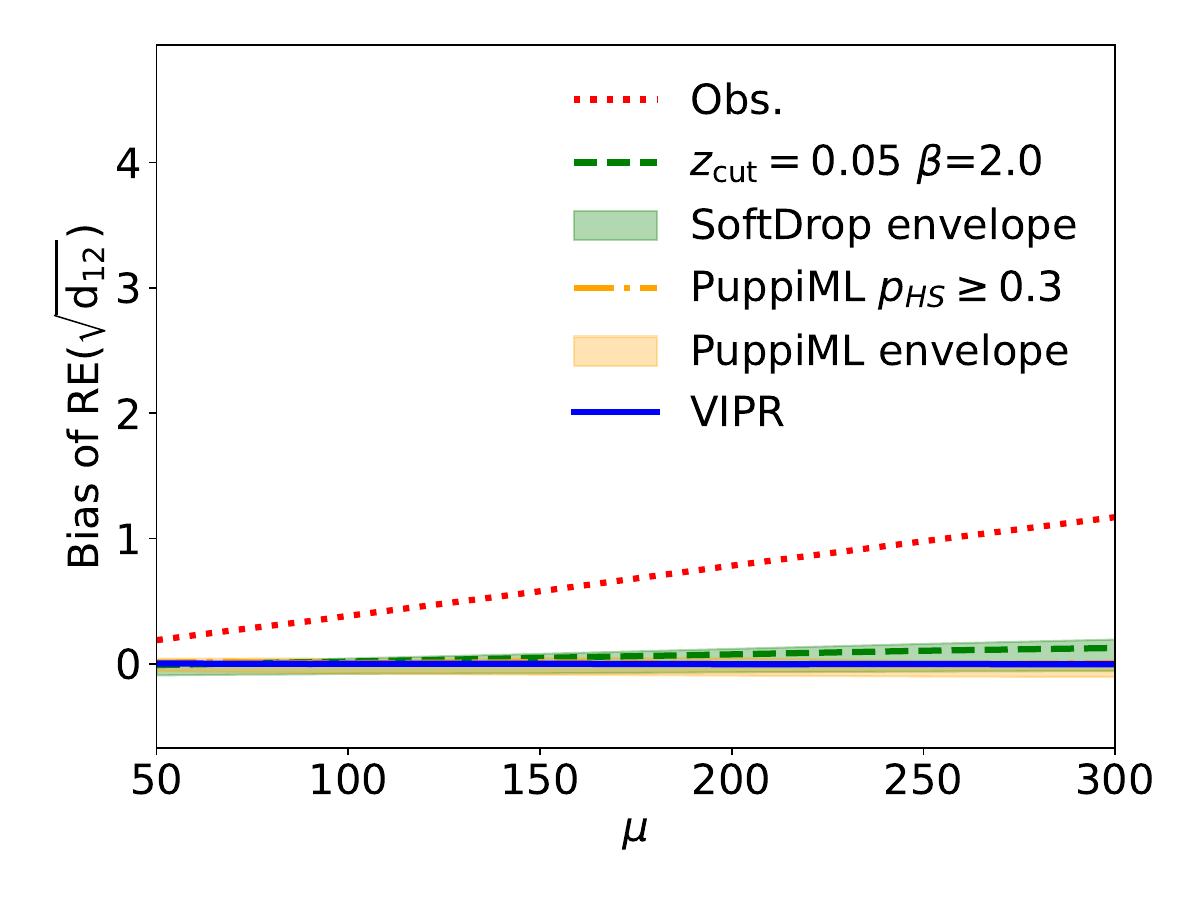}}}
    \subfloat[]{{\includegraphics[width=0.33\textwidth]{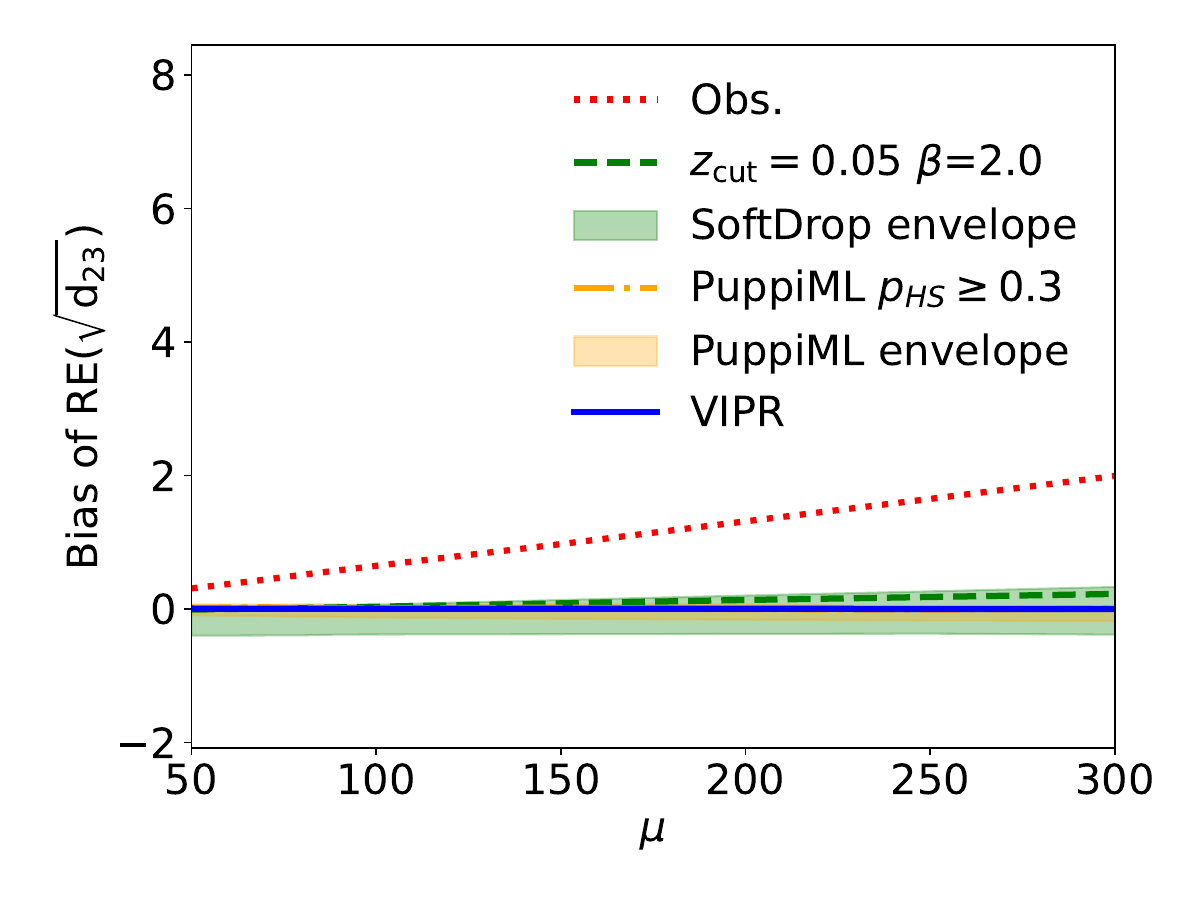}}}
    \subfloat[]{{\includegraphics[width=0.33\textwidth]{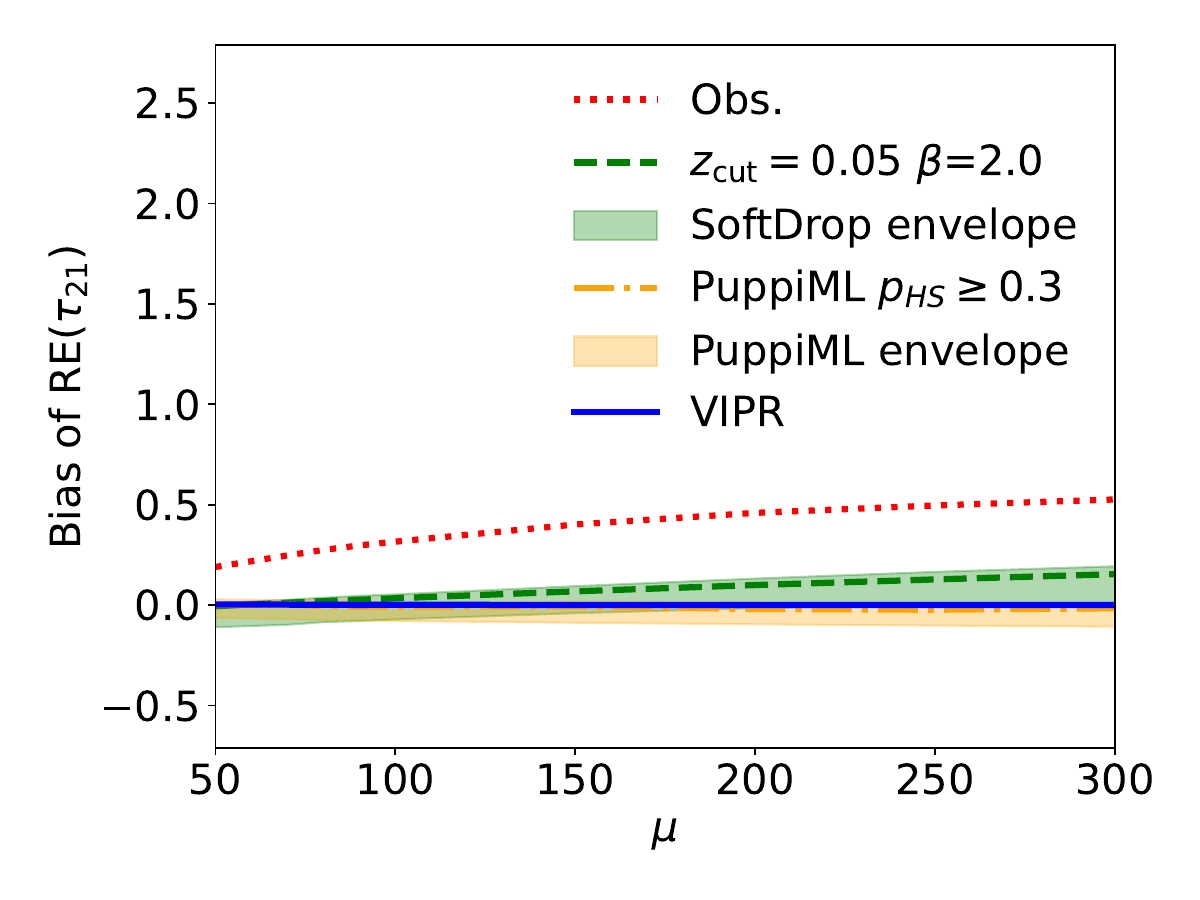}}}
    \caption{
        Comparisons of the bias of the RE as a function of $\mu$
        for $d_{12}$, $d_{23}$, and $\tau_{21}$.
        The bias measures the median of the distribution.
        Ideally, the bias should be as close to zero as possible
        and remain constant as a function of $\mu$.
        If the bias remains constant as a function of $\mu$, 
        it indicates that the pile-up removal method is robust towards increasing pile-up.
        The envelope of a scan of \softdrop settings is shown
        with the best resulting parameters in dashed green.
        The envelope of a scan of \puppiml cuts is also shown.
    }
    \label{fig:missing_response_variables_at_different_mu_app}
\end{figure*}

\FloatBarrier

\subsection{Performance as a function of $\mu$}
\begin{figure*}[htpb]
    \centering
    \subfloat[]{{\includegraphics[width=0.33\textwidth]{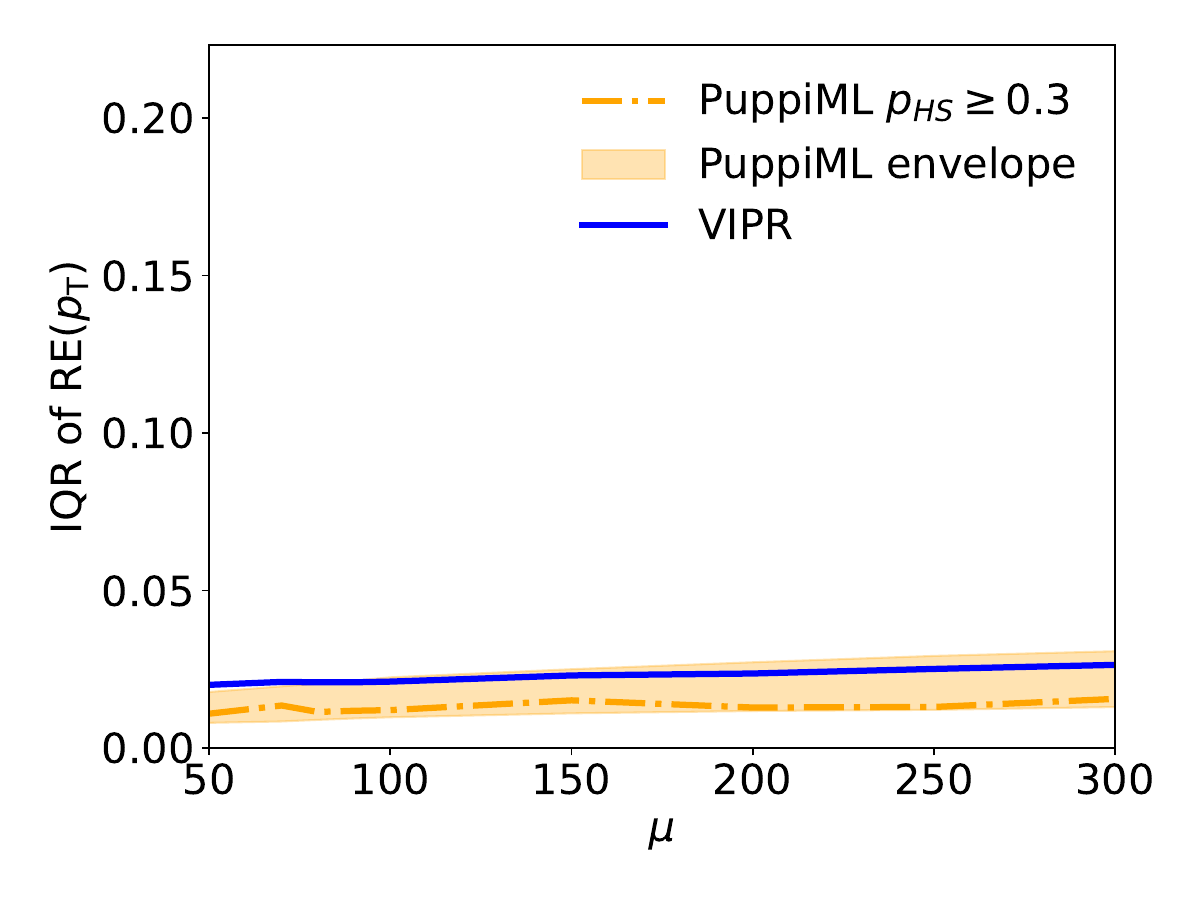}}}
    \subfloat[]{{\includegraphics[width=0.33\textwidth]{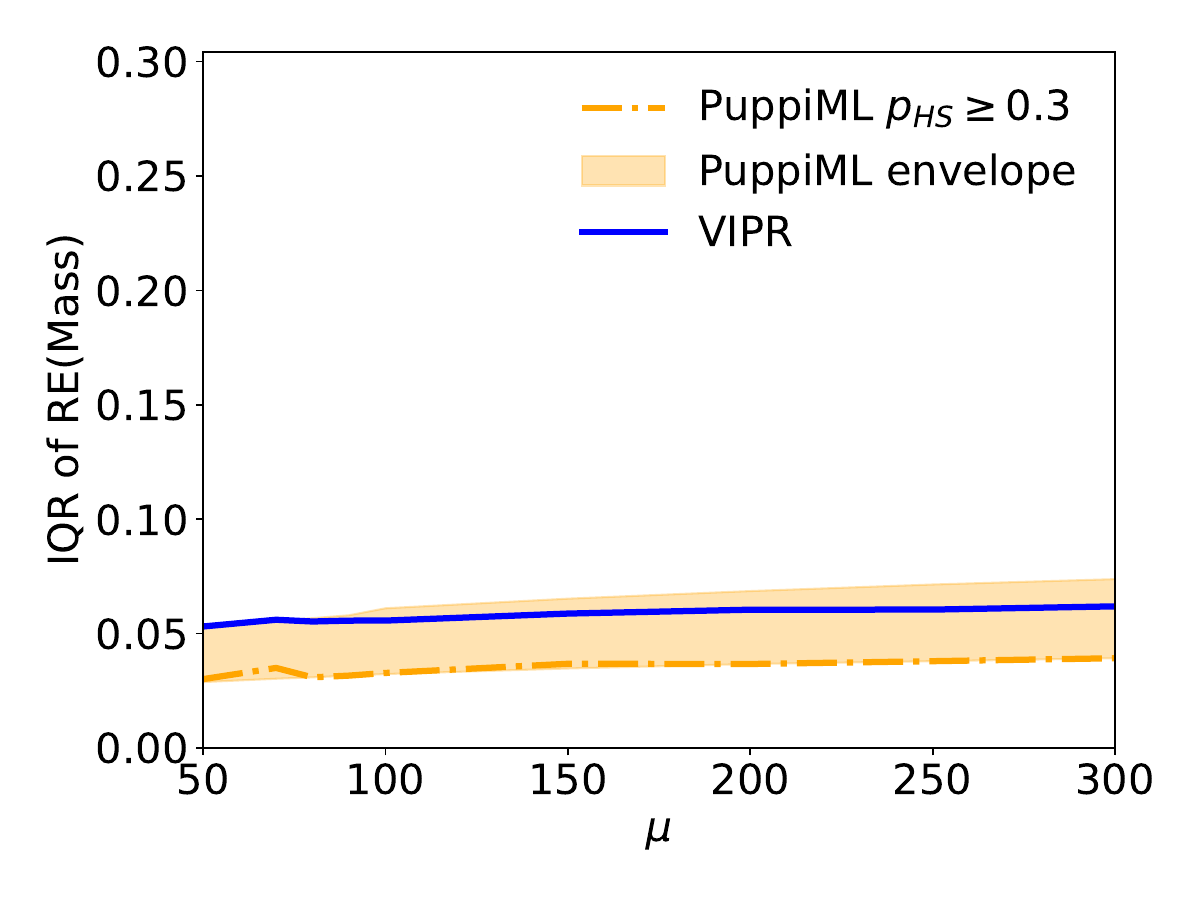}}}
    \subfloat[]{{\includegraphics[width=0.33\textwidth]{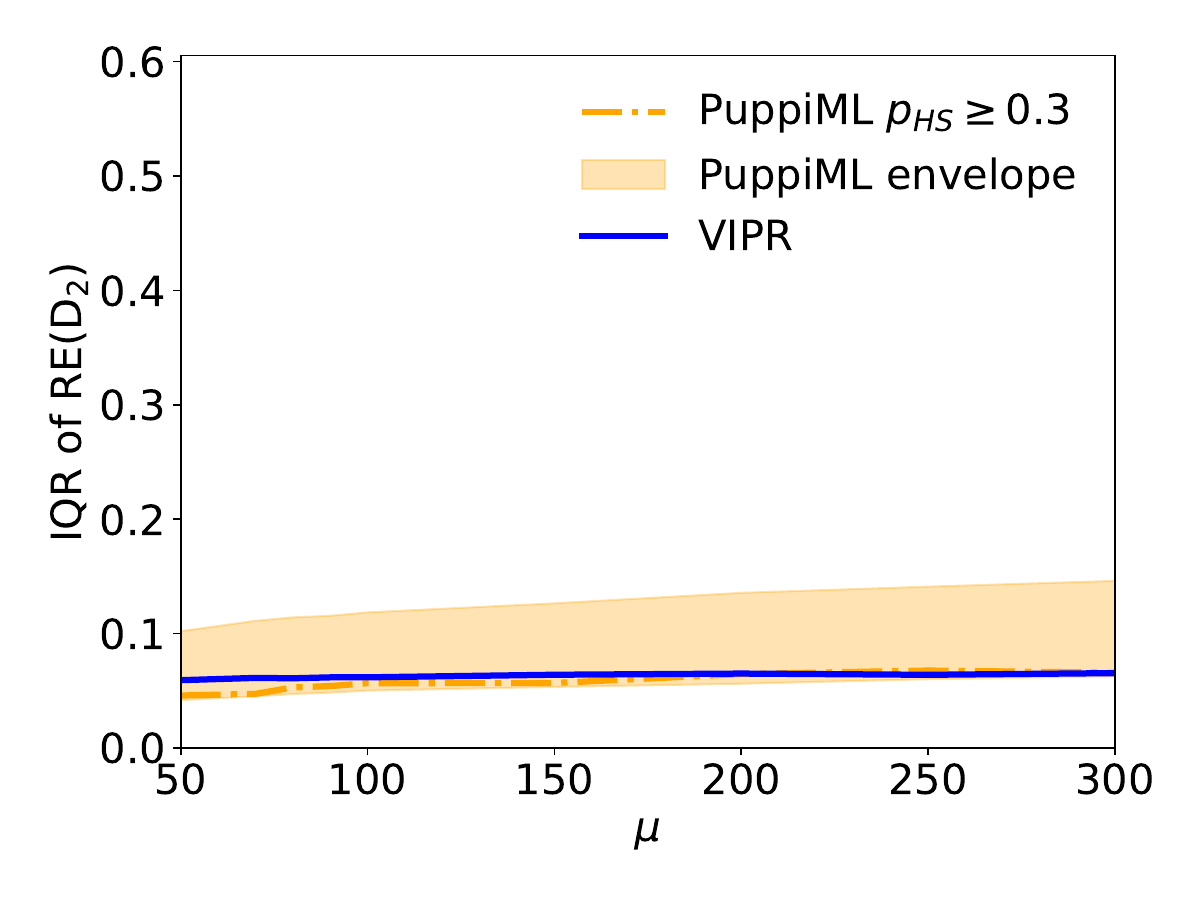}}}
    \\
    \subfloat[]{{\includegraphics[width=0.33\textwidth]{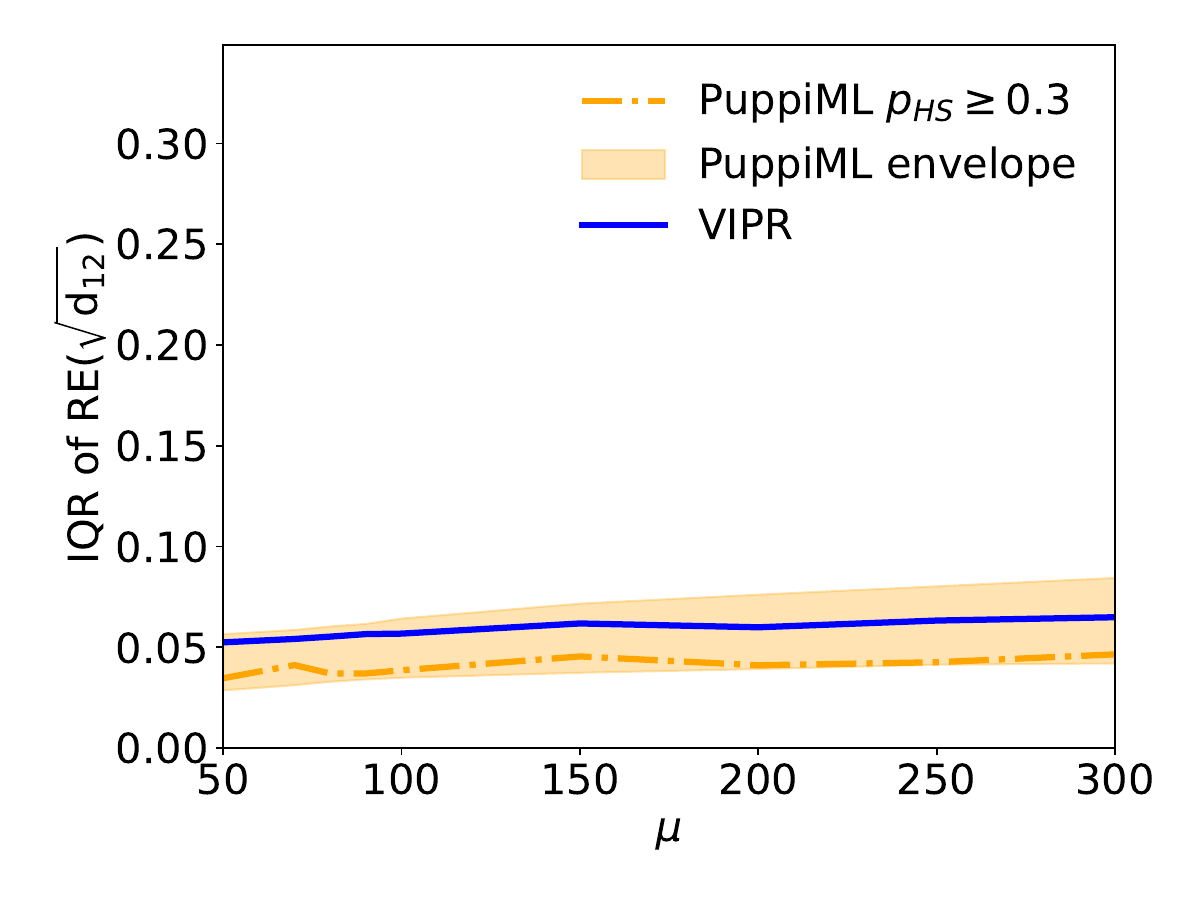}}}
   \subfloat[]{{\includegraphics[width=0.33\textwidth]{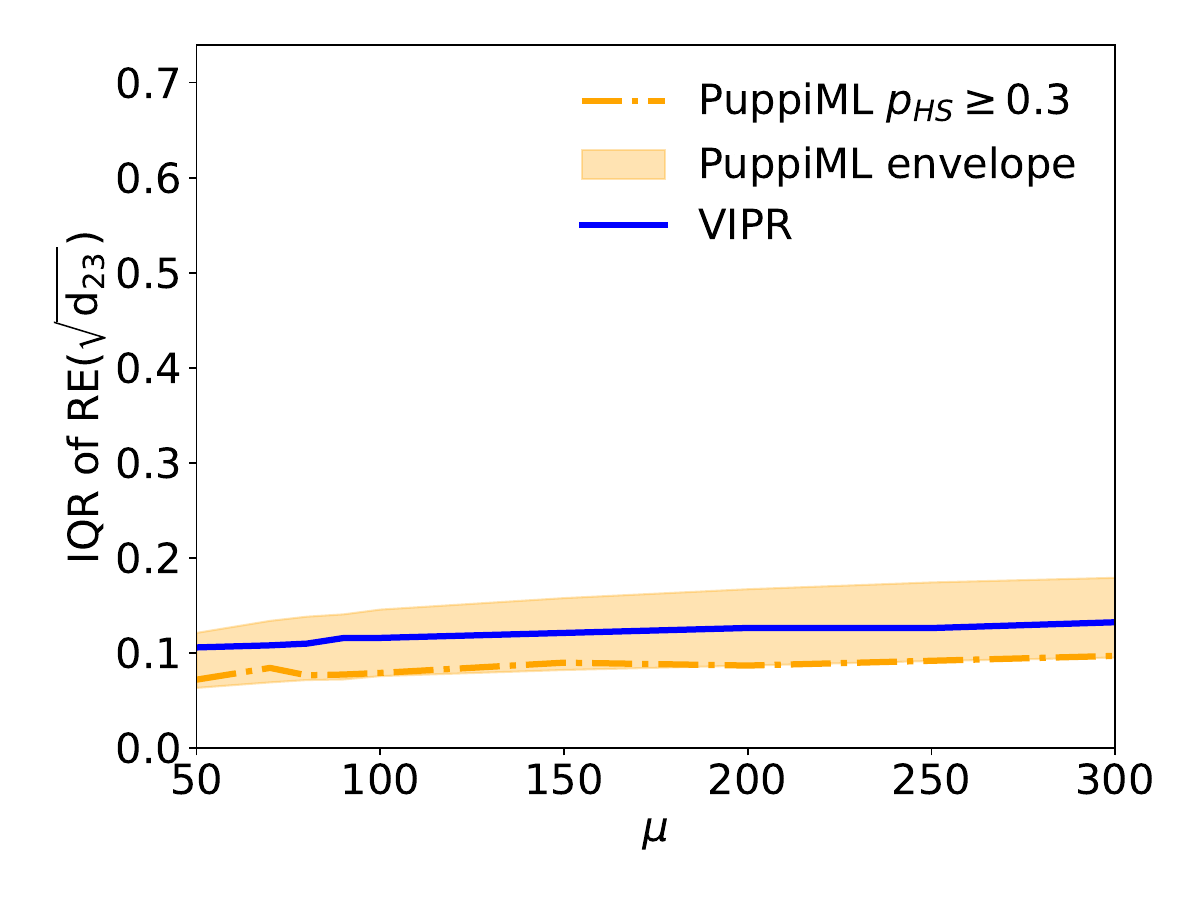}}}
    \subfloat[]{{\includegraphics[width=0.33\textwidth]{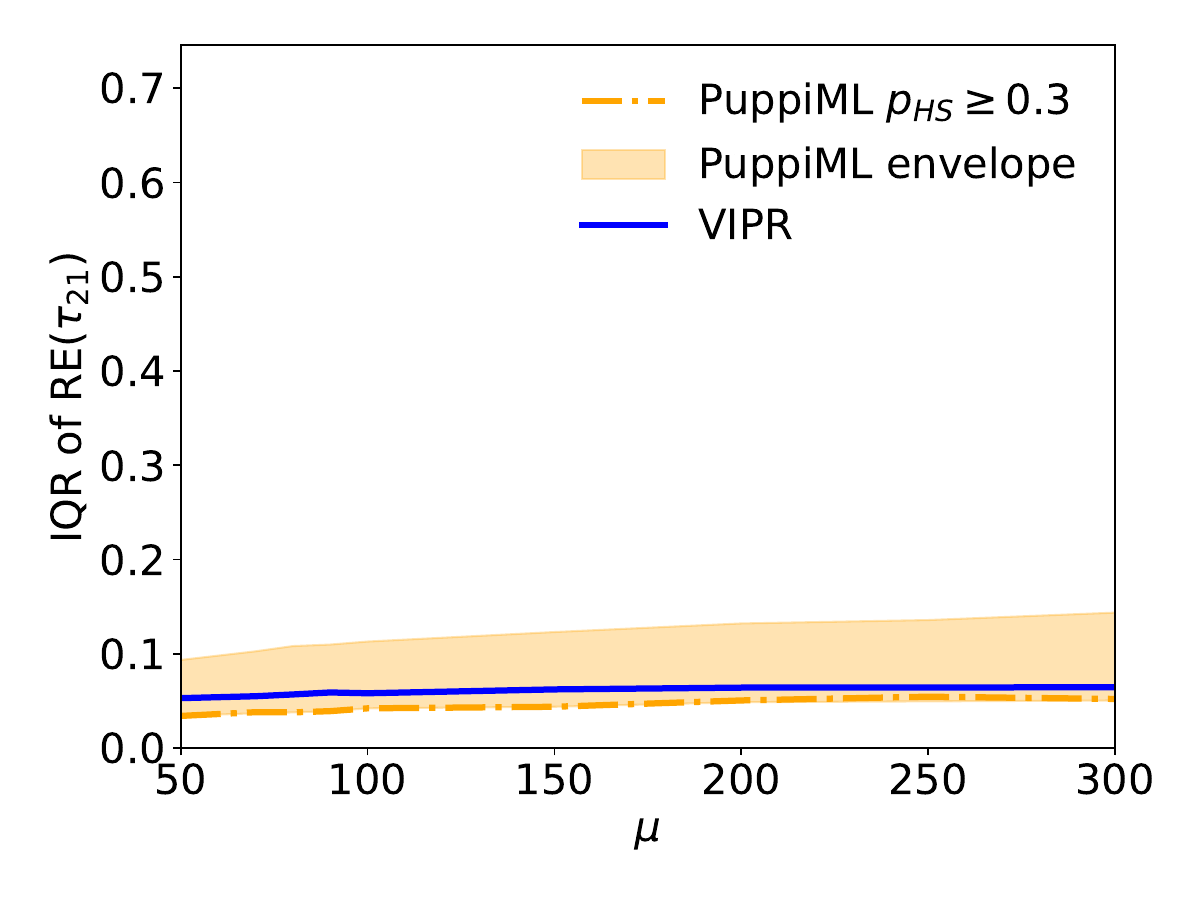}}}
    \\
    \subfloat[]{{\includegraphics[width=0.33\textwidth]{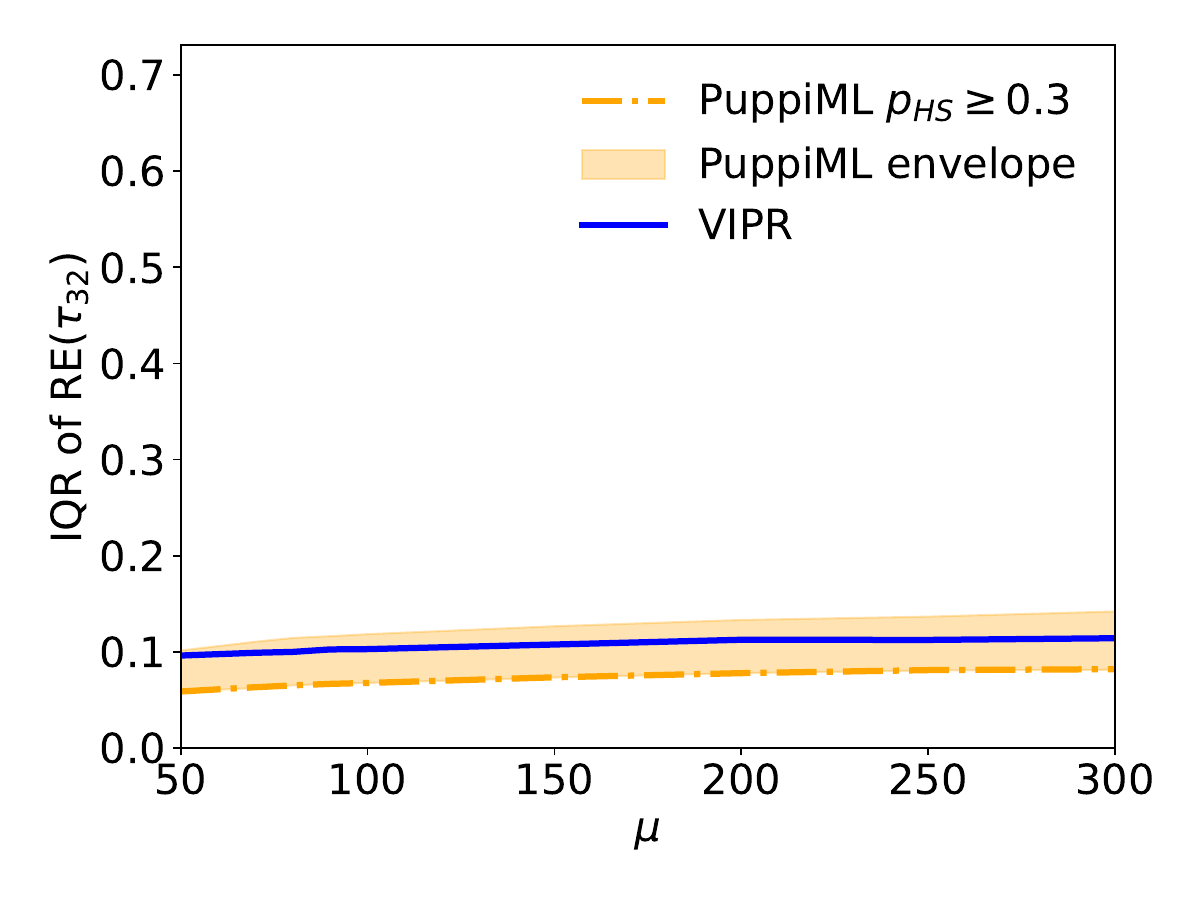}}}
    \caption{
    Comparisons of the IQR of the RE as a function of $\mu$
    for \pt, mass and substructure variables.
    The IQR measures the width of the distribution.
    Ideally, the IQR should be as close to zero as possible
    and remain constant as a function of $\mu$.
    If the IQR remains constant as a function of $\mu$, 
    it indicates that the pile-up removal method is robust towards increasing pile-up.
    The envelope of a scan of \softdrop settings is shown
    with the best resulting parameters in dashed green.
    }
    \label{fig:response_at_different_iqr_app}
\end{figure*}

\begin{figure*}[htpb]
    \centering
    \subfloat[]{{\includegraphics[width=0.33\textwidth]{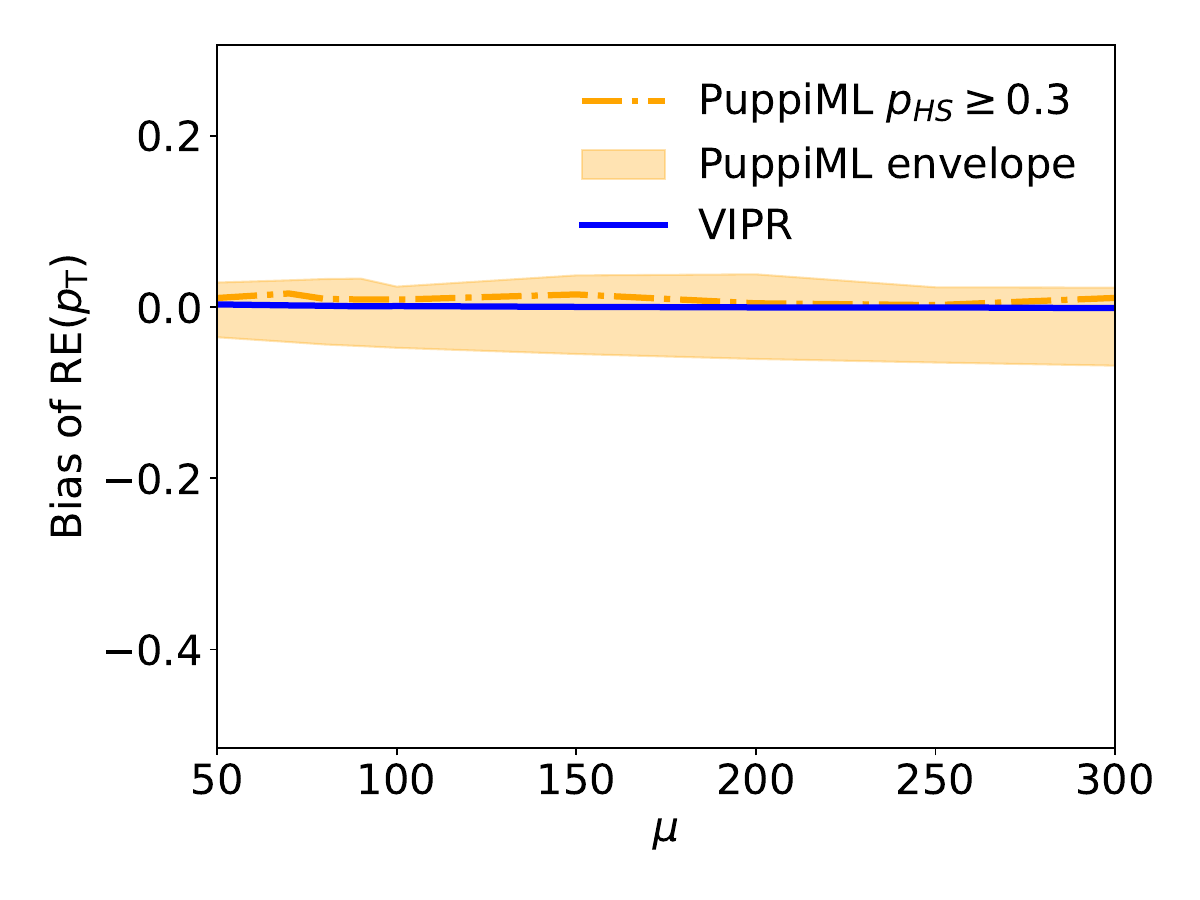}}}
    \subfloat[]{{\includegraphics[width=0.33\textwidth]{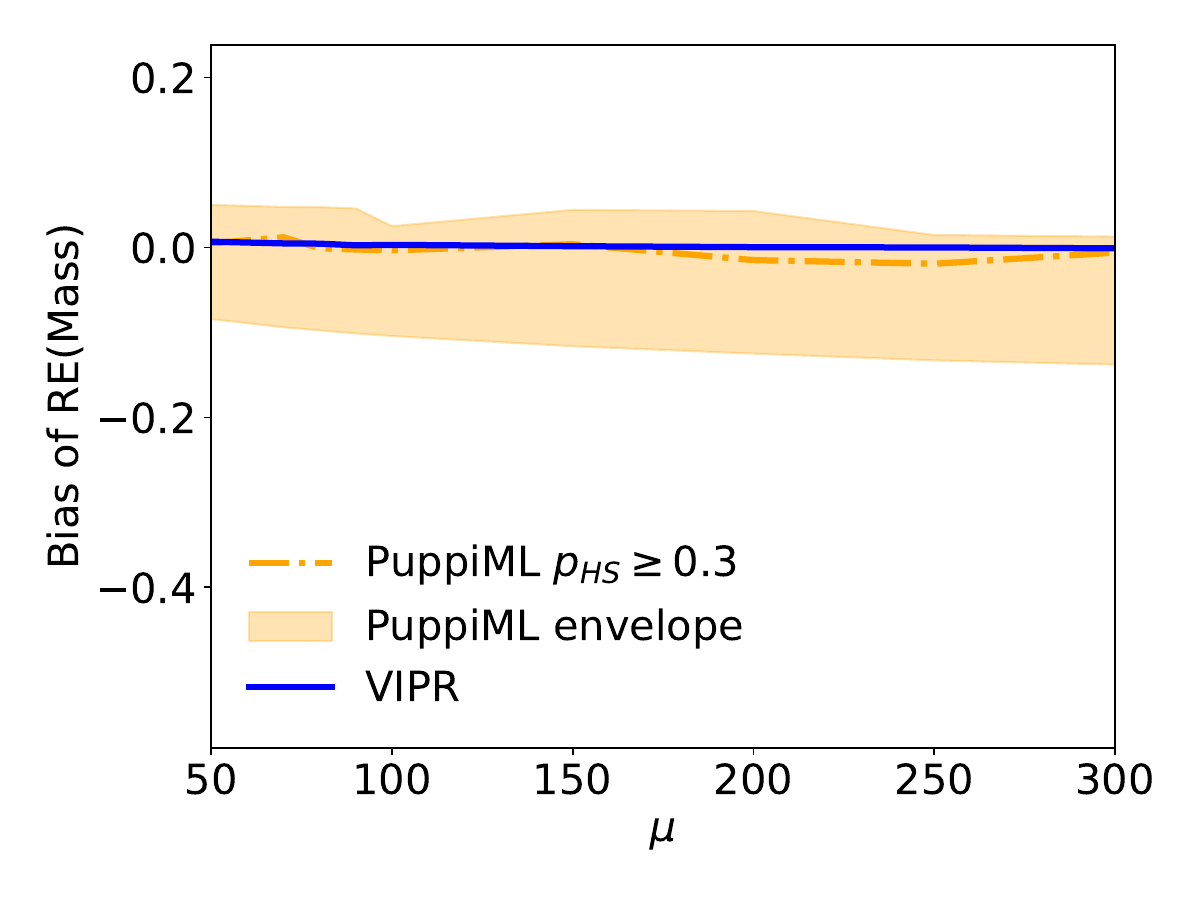}}}
    \subfloat[]{{\includegraphics[width=0.33\textwidth]{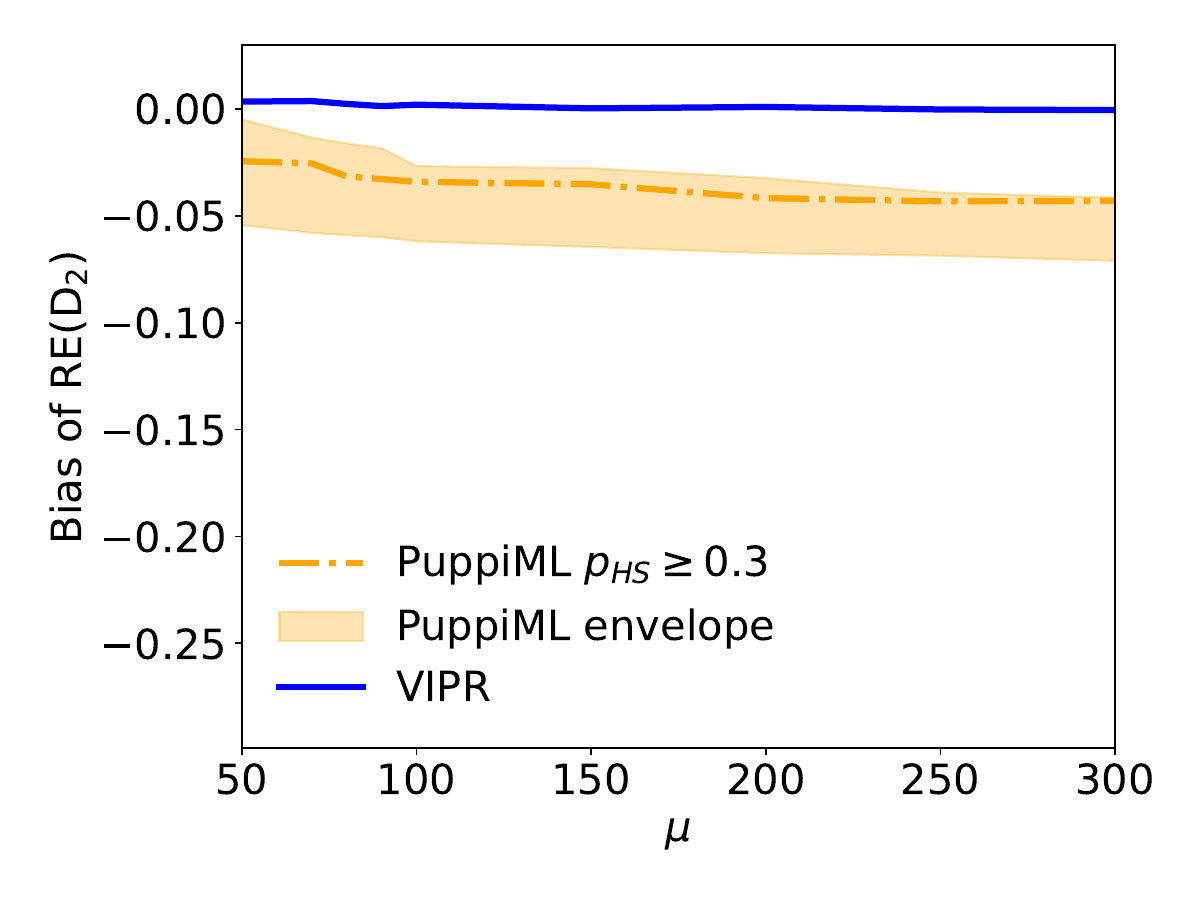}}}
    \\
    \subfloat[]{{\includegraphics[width=0.33\textwidth]{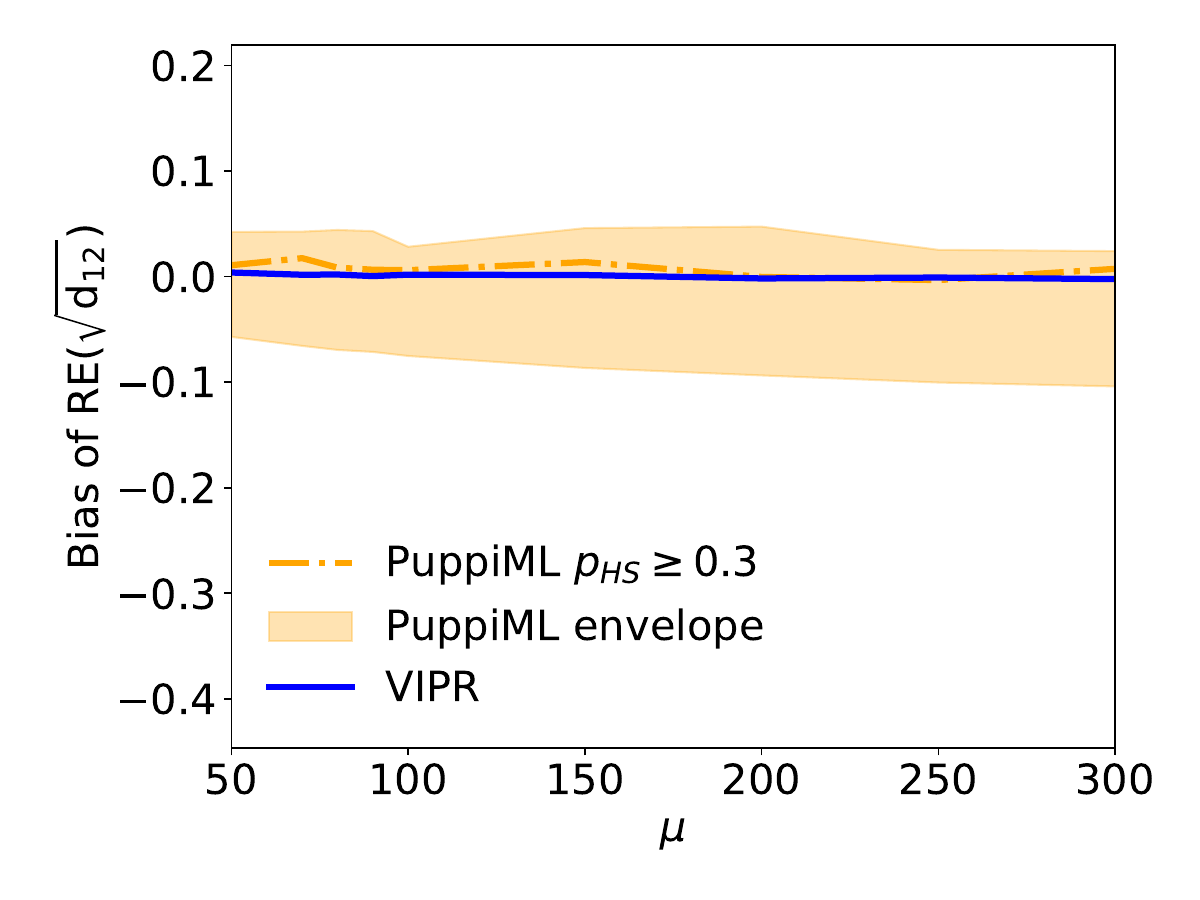}}}
   \subfloat[]{{\includegraphics[width=0.33\textwidth]{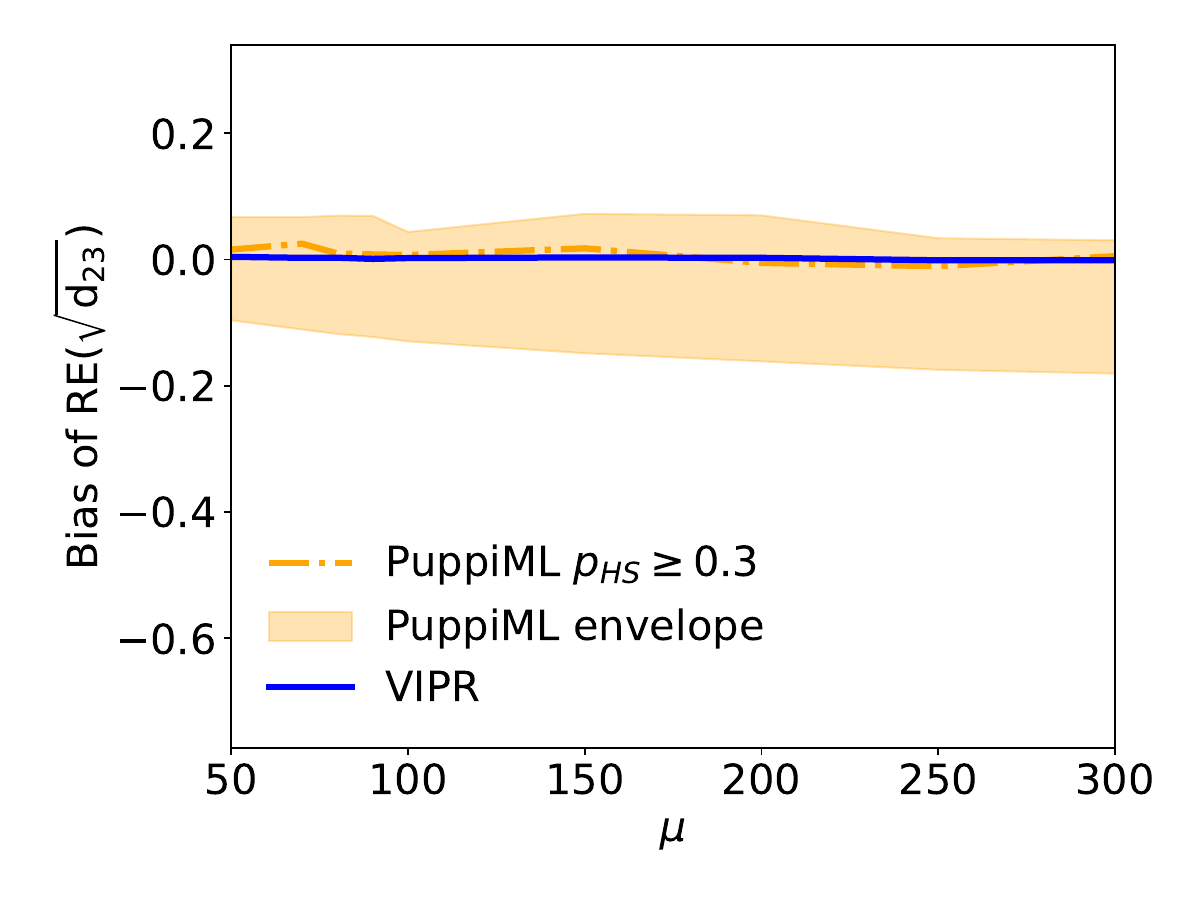}}}
    \subfloat[]{{\includegraphics[width=0.33\textwidth]{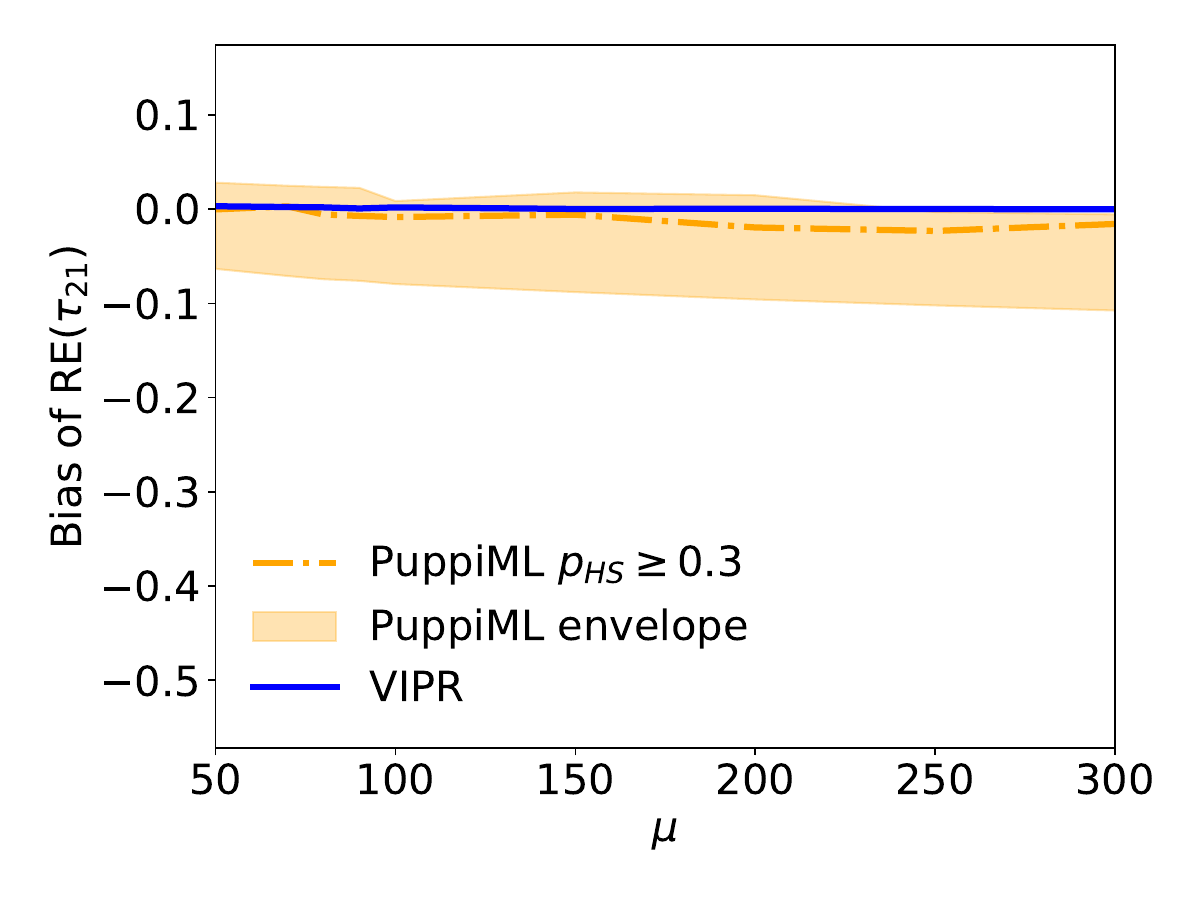}}}
    \\
    \subfloat[]{{\includegraphics[width=0.33\textwidth]{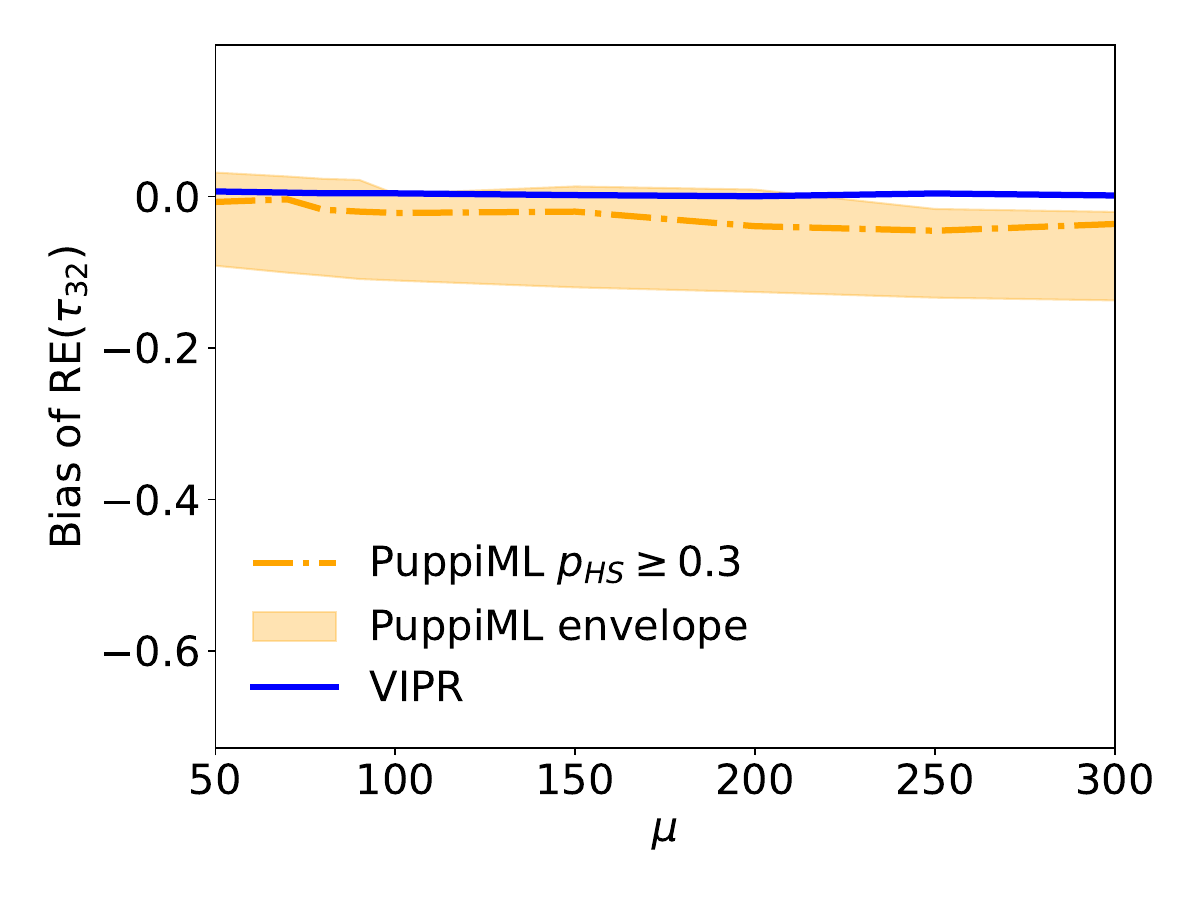}}}
    \caption{
        Comparisons of the bias of the RE as a function of $\mu$
        for various substructure variables.
        The bias measures the median of the distribution.
        Ideally, the bias should be as close to zero as possible
        and remain constant as a function of $\mu$.
        If the bias remains constant as a function of $\mu$, 
        it indicates that the pile-up removal method is robust towards increasing pile-up.
        The envelope of a scan of \softdrop settings is shown
        with the best resulting parameters in dashed green.
    }
    \label{fig:response_at_different_mu_app}
\end{figure*}

\cref{fig:response_at_different_iqr_app,fig:response_at_different_mu_app} shows the IQR and bias of the 
RE as a function of $\mu$ for \pt, mass and substructure variables for only \vipr and \puppiml.
Only \vipr and \puppiml are being shown here, so that the plots 
are not too crowded and easier to interpret.

\FloatBarrier

\subsubsection{Performance vs $\mu$ with a simulated detector efficiency of $\epsilon_{\mathrm{det}} = 90\%$}
\begin{figure*}[htpb]
    \centering
    \subfloat[]{{\includegraphics[width=0.33\textwidth]{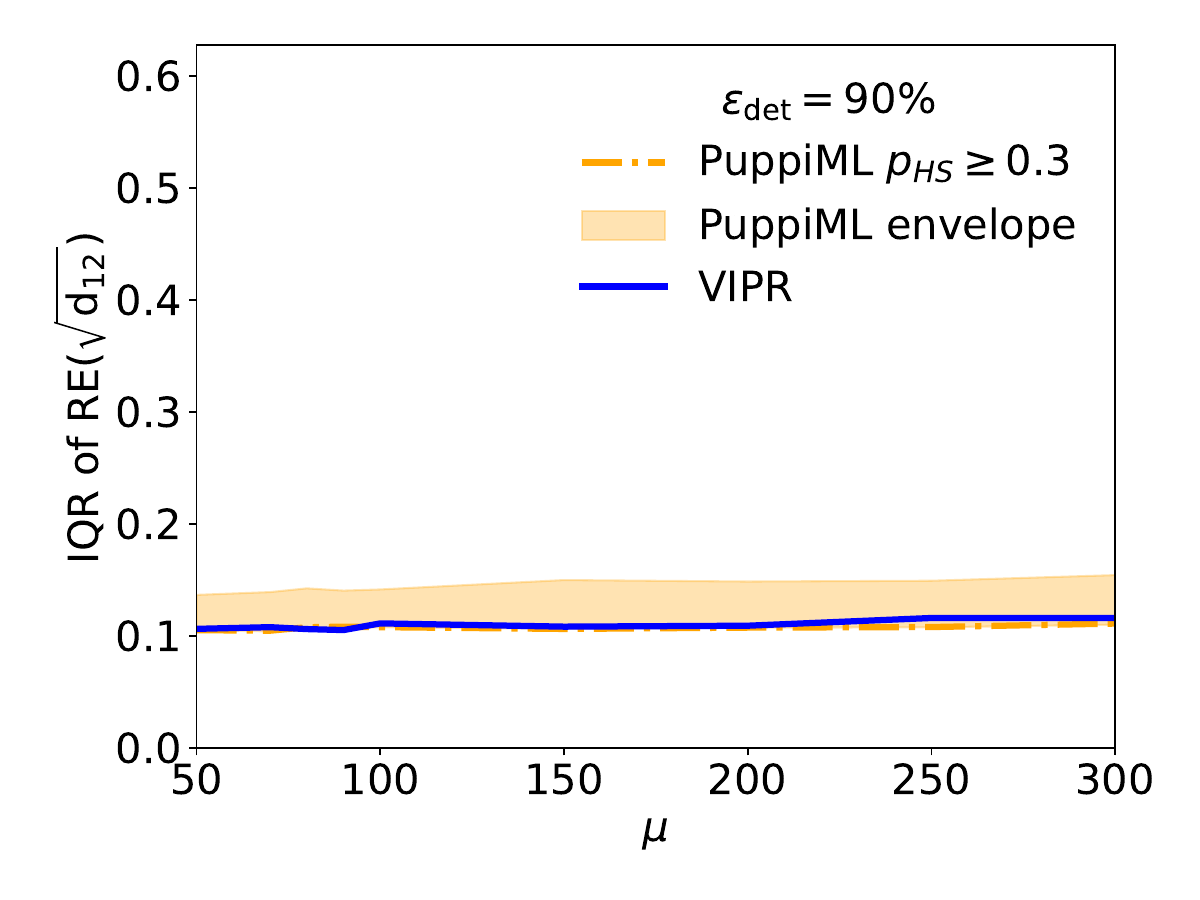}}}
   \subfloat[]{{\includegraphics[width=0.33\textwidth]{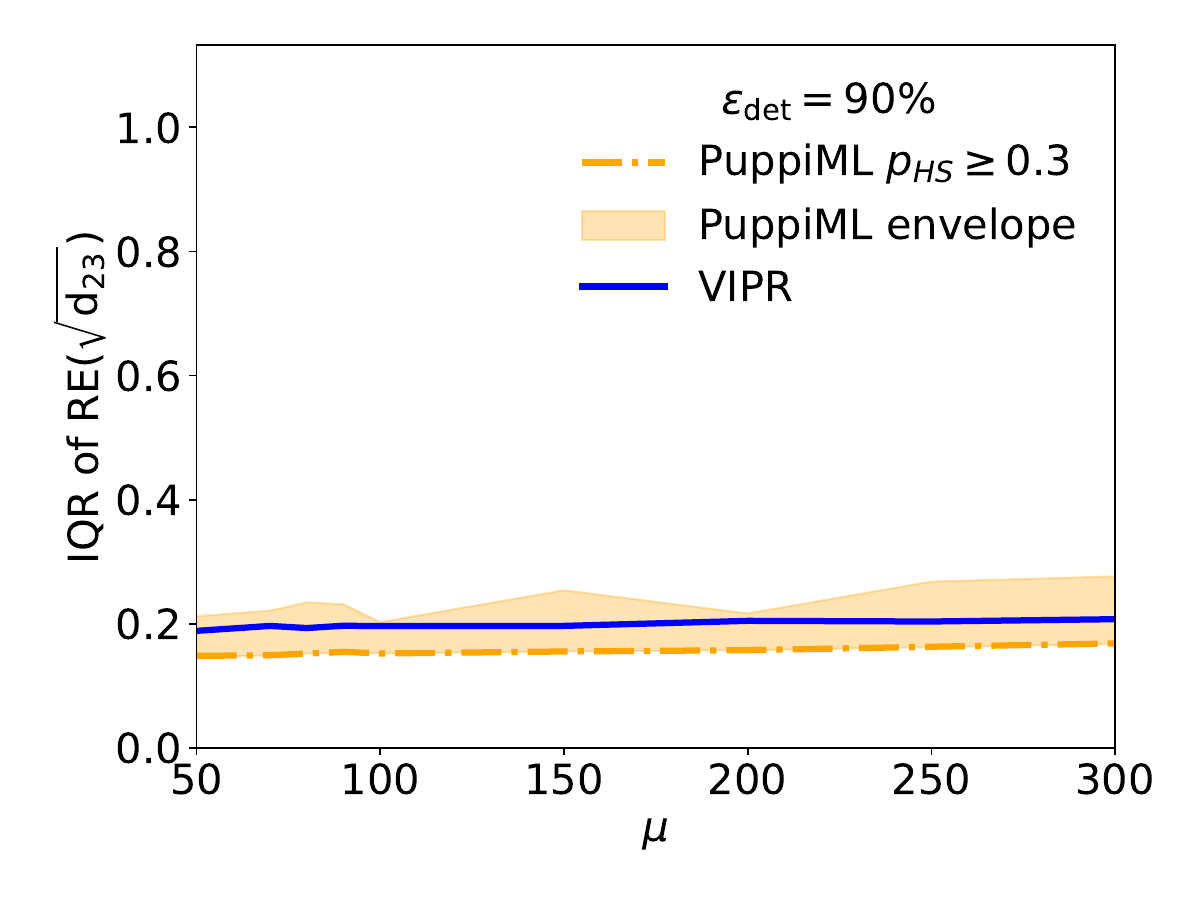}}}
    \subfloat[]{{\includegraphics[width=0.33\textwidth]{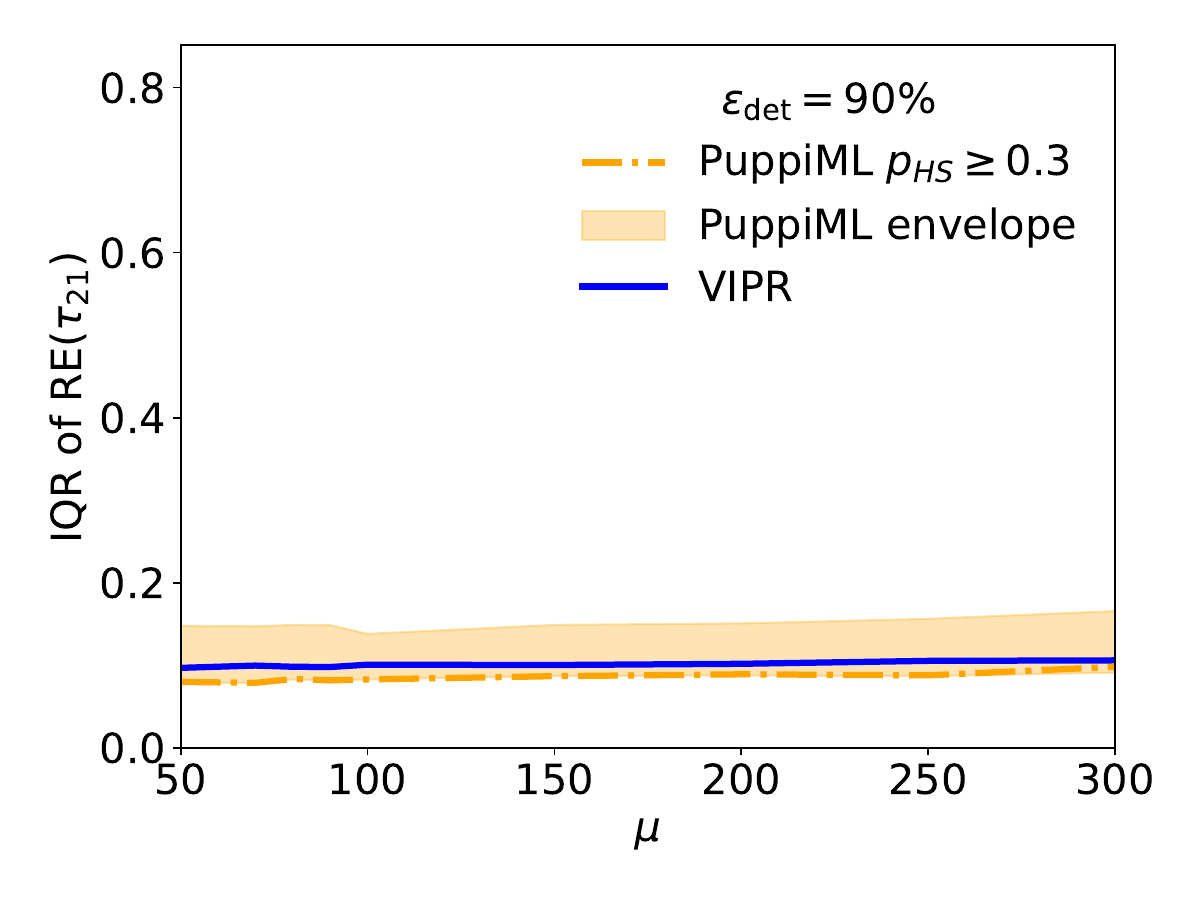}}}
    \caption{
    Comparisons of the IQR of the RE as a function of $\mu$
    for various \pt, mass, and substructure variables.
    The IQR measures the width of the distribution.
    Ideally, the IQR should be as close to zero as possible
    and remain constant as a function of $\mu$.
    If the IQR remains constant as a function of $\mu$, 
    it indicates that the pile-up removal method is robust towards increasing pile-up.
    The envelope of a scan of \puppiml cuts is also shown
    with the best resulting parameters in orange.
    Both \vipr and \puppiml have been trained on sample with $\epsilon_{\mathrm{det}} = 90\%$.  
    }
    \label{fig:response_at_different_iqr_app_eff_90}
\end{figure*}

\begin{figure*}[htpb]
    \centering
    \subfloat[]{{\includegraphics[width=0.33\textwidth]{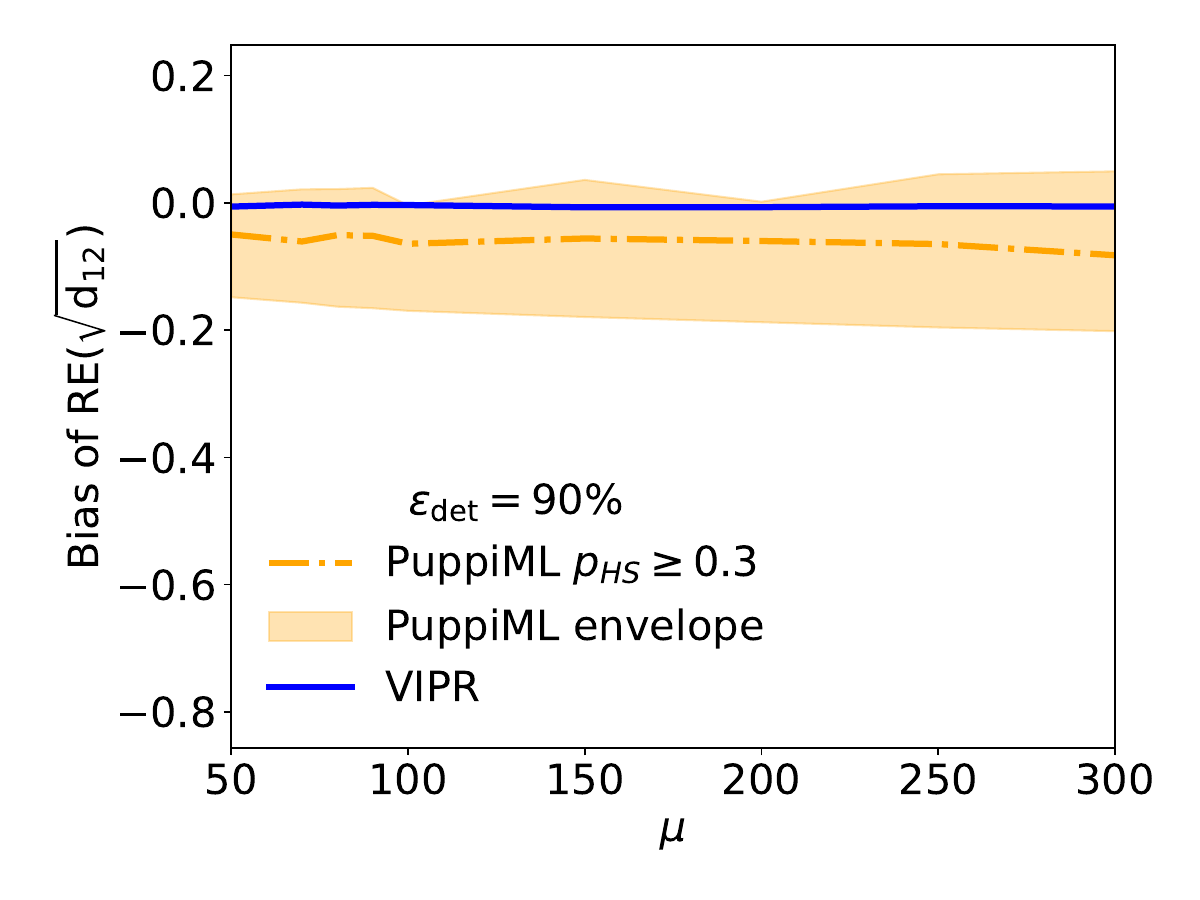}}}
    \subfloat[]{{\includegraphics[width=0.33\textwidth]{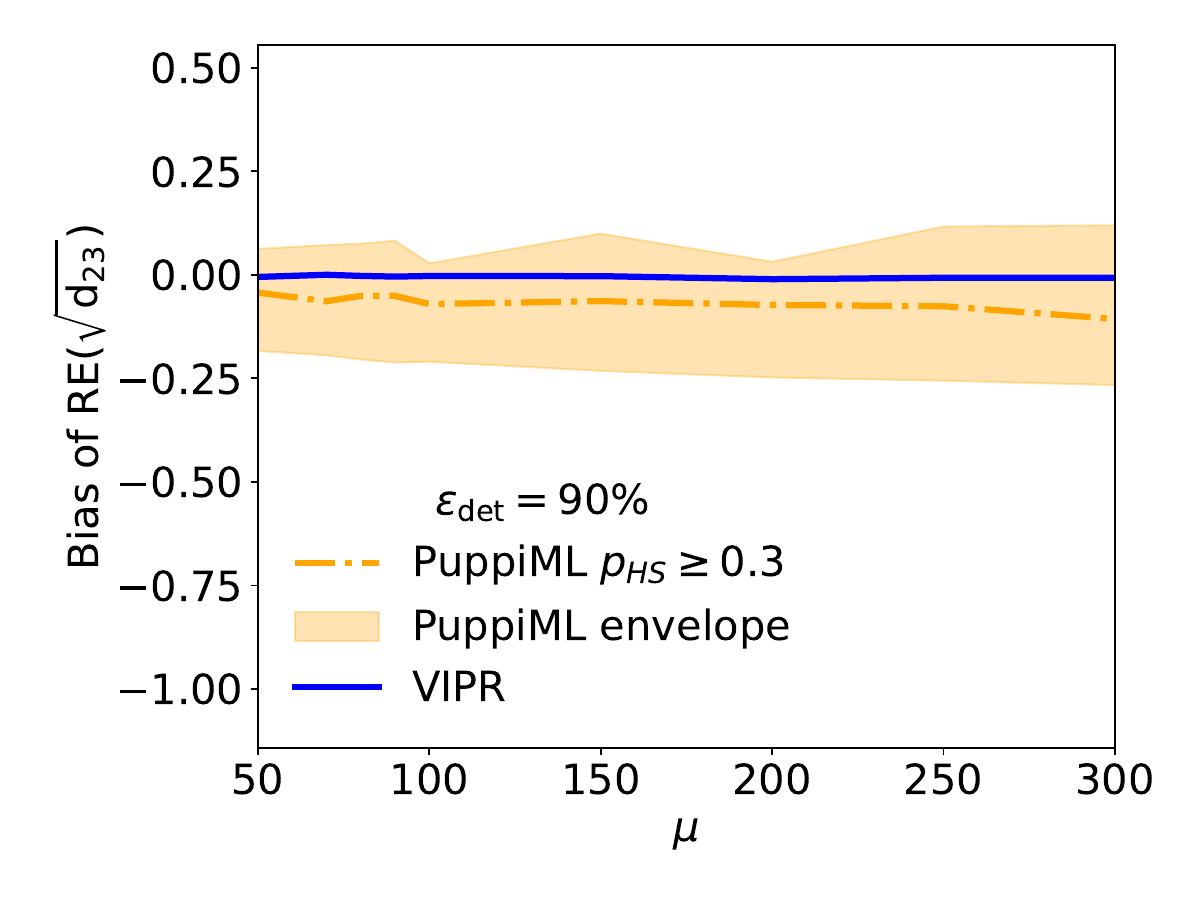}}}
    \subfloat[]{{\includegraphics[width=0.33\textwidth]{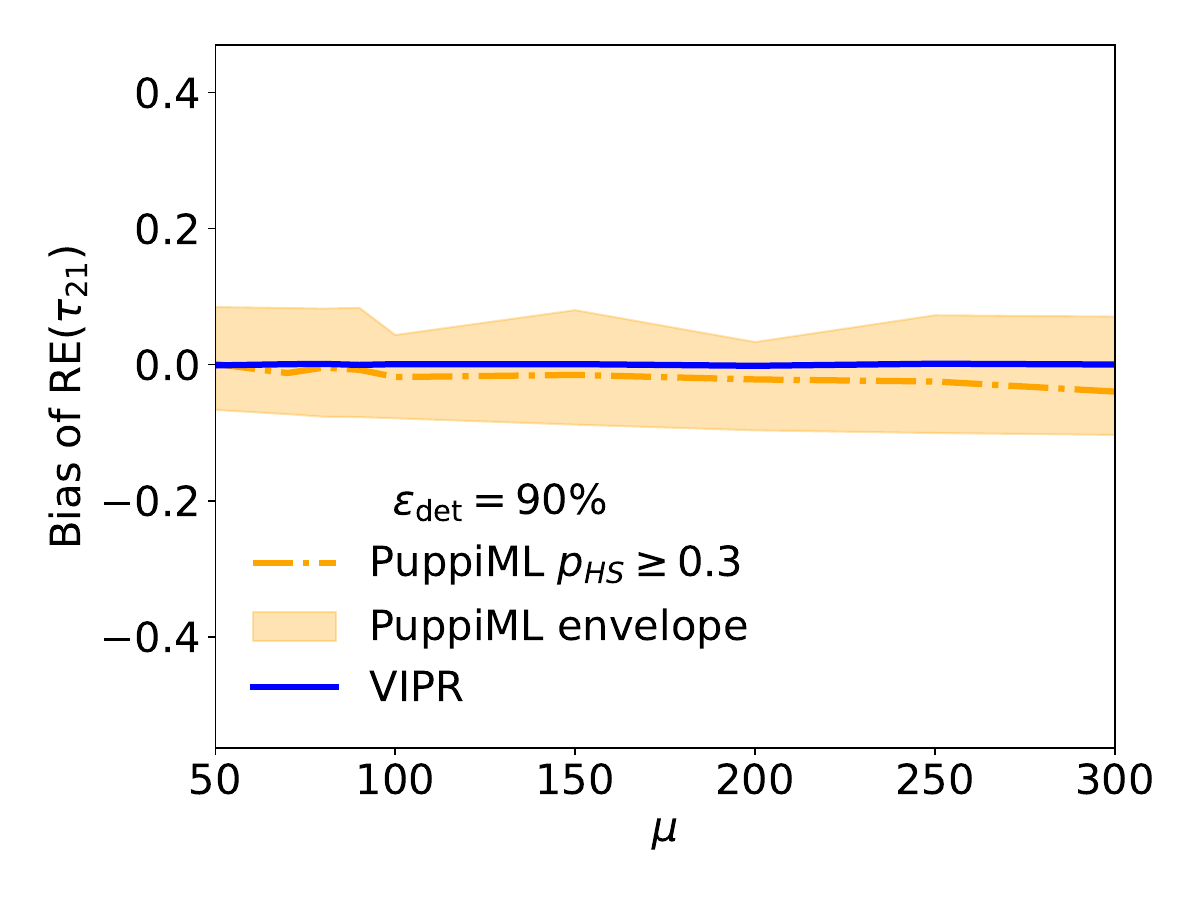}}}
    \caption{
        Comparisons of the bias of the RE as a function of $\mu$
        for various \pt, mass, and substructure variables.
        The bias measures the median of the distribution.
        Ideally, the bias should be as close to zero as possible
        and remain constant as a function of $\mu$.
        If the bias remains constant as a function of $\mu$, 
        it indicates that the pile-up removal method is robust towards increasing pile-up.
        The envelope of a scan of \puppiml cuts is also shown
        with the best resulting parameters in dashed orange.
        Both \vipr and \puppiml have been trained on sample with $\epsilon_{\mathrm{det}} = 90\%$.  
    }
    \label{fig:response_at_different_mu_app_eff_90}
\end{figure*}

By reducing the detector efficiency to $\epsilon_{\mathrm{det}} = 90\%$, we can assess the performance of \vipr in a more realistic scenario.
\cref{fig:response_at_different_iqr_app_eff_90,fig:response_at_different_mu_app_eff_90} 
shows the bias and IQR for $d_{12}$, $d_{23}$, and $\tau_{21}$, 
where the detector efficiency has been set to $\epsilon_{\mathrm{det}} = 90\%$.

\FloatBarrier

\subsection{Hyperparameter}
Hyperparameter for the diffusion model is shown in \cref{table:vipr},
and the hyperparameters for the normalising flow are shown in \cref{table:flow}.
\begin{table}[htpb]
    \centering
    \resizebox{0.4\textwidth}{!}{\begin{tabular}{lll}
        \hline
        \hline
        \multirow{3}{*}{Diffusion} & Time embedding               & Sinusoidal          \\
        & Dimension           & 64                  \\
        & Frequency range         & {[}0.001, 80{]}     \\ \hline
        \multirow{4}{*}{Train}     & EMA                 & 0.999               \\
        & LR                  & 0.005               \\
        & LR scheduler        & warmup(step=100000) \\
        & Batch size          & 256                 \\ \hline
        \multirow{5}{*}{Model}     & Hidden dimension    & 256                 \\
        & Attention heads     & 16                  \\
        & Activation function & GELU                \\
        & MLP upscale         & 2                   \\
        & Number of CAE       & 2                   \\ 
        \hline
        \hline
    \end{tabular}}
    \caption{Table of hyperparameters used for \vipr. 
    The MLP is the standard MLP from Ref~\cite{attn, preLN}}
    \label{table:vipr}
\end{table}
\begin{table}[htpb]
    \centering
    \resizebox{0.4\textwidth}{!}{\begin{tabular}{lll}
        \hline
        \hline
        \multirow{4}{*}{CAE} & Hidden dimension    & 256                     \\
        & Number of encoders  & 2                       \\
        & Number of decoders  & 2                       \\
        & Attention heads     & 8                       \\ \hline
        \multirow{4}{*}{Train}           & Gradient clip       & 10                      \\
        & LR                  & [0.0001, 1e-7]                  \\
        & LR scheduler        & CosineAnnealingLR \\
        & Batch size          & 256                     \\ \hline
        \multirow{7}{*}{Flow}            & Context dimension   & 512                     \\
        & Number of stacks    & 4                       \\
        & Activation function & GELU                    \\
        & Base distribution   & Normal distribution     \\
        & Transformation      & RQS                     \\
        & Number of bins      & 12                      \\
        & Tail                & Linear[-4,4]            \\ 
        \hline
        \hline
        \end{tabular}}
    \caption{Table of hyperparameters used for the flow used to estimate $p(N;j)$. 
    The MLP hyperparameters are the same as in \cref{table:vipr}.}
    \label{table:flow}
\end{table}
 
\end{document}